\documentclass[11pt]{article}
\pdfoutput=1
\usepackage{graphicx}	
\usepackage{amsmath}	
\usepackage{dcolumn}
\usepackage{bm}
\usepackage{graphics}
\usepackage{afterpage}
\usepackage{float}
\usepackage{rotating}
\usepackage{multirow}
\usepackage{tabularx}
\usepackage{booktabs}
\usepackage{multirow}
\usepackage{fancyhdr}
\usepackage[utf8]{inputenc}
\usepackage{theorem}
\usepackage{moreverb}
\usepackage{euscript}
\usepackage{psfrag}
\usepackage{slashed}
\usepackage{mathtools}
\usepackage{makecell}
\usepackage[flushleft]{threeparttable}
\usepackage{adjustbox}

\usepackage{dcolumn}
\usepackage{bm}
\usepackage[mathlines]{lineno}
\usepackage{lettrine}
\usepackage[dvipsnames]{xcolor}
\usepackage{jhep}
\usepackage{listings}
\usepackage{color,xcolor}
\usepackage{natbib}
\usepackage{bm}
\usepackage{color}
\usepackage{wasysym}
\usepackage{slashed}
\usepackage{lipsum}
\usepackage{array}
\usepackage{varwidth}
\usepackage{lettrine}
\usepackage{subcaption}
\usepackage[object=vectorian]{pgfornament}

\usepackage{booktabs}
\newcommand{\ra}[1]{\renewcommand{\arraystretch}{#1}}

\newcommand{\refjnl}[1]{{\rm#1}}

\def\apj{\refjnl{ApJ}}                 % Astrophysical Journal
\input Zallman.fd

\LettrineTextFont{\itshape}
\definecolor{nice}{rgb}{0.8,0, 0.8}
\definecolor{calirvoyant}{rgb}{0.34, 0.015, 0.42}
\definecolor{paradiso}{rgb}{0.21568627450980393, 0.5058823529411764, 0.5254901960784314}

\hypersetup{
     colorlinks   = true,
     citecolor    = paradiso,
     urlcolor     = paradiso,
     linkcolor    = paradiso
}

\preprint{SLAC-PUB-17773}

    \title{Optimal Celestial Bodies for Dark Matter Detection}

\author[a,b]{Rebecca K. Leane,}
\emailAdd{rleane@slac.stanford.edu}
\author[a,b]{Joshua Tong}
\emailAdd{joshtong@stanford.edu}

\affiliation[a]{Particle Theory Group, SLAC National Accelerator Laboratory, Stanford, CA 94035, USA}
\affiliation[b]{Kavli Institute for Particle Astrophysics and Cosmology, Stanford University, Stanford, CA 94035, USA}

\date{\today}
\abstract{A wide variety of celestial bodies have been considered as dark matter detectors. Which stands the best chance of delivering the discovery of dark matter? Which is the most powerful dark matter detector? We investigate a range of objects, including the Sun, Earth, Jupiter, Brown Dwarfs, White Dwarfs, Neutron Stars, Stellar populations, and Exoplanets. We quantify how different objects are optimal dark matter detectors in different regimes by deconstructing some of the in-built assumptions in these search sensitivities, including observation potential and particle model assumptions. We find new constraints and future sensitivities across a range of dark matter annihilation final states. We quantify mediator properties leading to detectable celestial-body energy injection or Standard Model fluxes, and show how different objects can be expected to deliver corroborating signals. We discuss different search strategies, their opportunities and limitations, and the interplay of regimes where different celestial objects are optimal dark matter detectors. Deconstructing the assumptions of these searches leads us to point out a new search using the Galactic center stellar population that can provide greater sensitivity to the dark matter-nucleon scattering cross section than the Sun, despite being significantly further away in our Galaxy.}

\begin{document}
\maketitle

\newpage
\section{Introduction}

\lettrine{P}{eppered throughout the galaxy}, our sky shines with a wide spectrum of unique dark matter detectors. Some have been known since time immemorial, such as the Sun and Jupiter, while others were only detected for the first time in the last few decades, such as exoplanets. Stars and planets have become of increasing interest recently to detect dark matter, due to their powerful sensitivities to untested dark matter parameter space. By now, just about all stars and planets have been considered in varying contexts to test dark matter's properties, with an extensive list including the Earth~\cite{Freese:1985qw,Gould:1991va,Mack:2007xj,Chauhan:2016joa,Bramante:2019fhi,Feng:2015hja,Das:2022srn,Das:2024jdz,Neufeld:2018slx,Pospelov:2020ktu, Pospelov:2019vuf, Rajendran:2020tmw,Xu:2021lmg,Budker:2021quh, McKeen:2022poo,Billard:2022cqd,Li:2022idr,Bramante:2022pmn,Pospelov:2023mlz,McKeen:2023ztq,Ema:2024oce,Moore:2024mot}, the Sun~\cite{Batell:2009zp,Schuster:2009au,Schuster:2009fc,Bell_2011,Feng:2015hja,Kouvaris:2010,Feng:2016ijc,Allahverdi:2016fvl,Leane:2017vag,Arina:2017sng,Garani:2017jcj,Albert:2018jwh, Albert:2018vcq,Nisa:2019mpb,Niblaeus:2019gjk,Cuoco:2019mlb,Serini:2020yhb,Acevedo:2020gro,Mazziotta:2020foa,Bell:2021pyy,Bose:2021cou,Maity:2023rez}, Jupiter~\cite{Batell:2009zp,Leane:2021tjj,Li:2022wix,French:2022ccb,Blanco:2023qgi,Ray:2023auh,Yan:2023kdg,Croon:2023bmu,Ansarifard:2024fan,Linden:2024uph}, Brown Dwarfs~\cite{Leane:2020wob,Leane:2021ihh,Bhattacharjee:2023qfi,Linden:2024uph,Acevedo:2024zkg,Ilie:2023lbi,Benito:2024yki}, White Dwarfs~\cite{Mochkovitch:1985vi, Moskalenko:2006mk,Bertone:2007ae,McCullough:2010ai,Hooper:2010es,
Kouvaris:2010jy,
Bramante:2015cua,
Graham:2015apa,
Amaro-Seoane:2015uny,
Graham:2018efk,
Cermeno:2018qgu,
Acevedo:2019gre,
Acevedo:2020avd,
Acevedo:2021kly,
Janish:2019nkk,
Krall:2017xij,
Panotopoulos:2020kuo,
Curtin:2020tkm,
Bell:2021fye,
DeRocco:2022rze,
Ramirez-Quezada:2022uou,
Smirnov:2022zip,
Garani:2023esk,Acevedo:2023xnu}, Neutron Stars~\cite{Goldman:1989nd,
Gould:1989gw,
Kouvaris:2007ay,
Bertone:2007ae,
deLavallaz:2010wp,
Kouvaris:2010,
McDermott:2011jp,
Kouvaris:2011fi,
Guver:2012ba,
Bramante:2013hn,
Bell:2013xk,
Bramante:2013nma,
Bertoni:2013bsa,
Kouvaris:2010jy,
McCullough:2010ai,
Perez-Garcia:2014dra,
Bramante:2015cua,
Graham:2015apa,
Cermeno:2016olb,
Graham:2018efk,
Acevedo:2019gre,
Janish:2019nkk,
Krall:2017xij,
McKeen:2018xwc,
Baryakhtar:2017dbj,
Raj:2017wrv,
Bell:2018pkk,
Chen:2018ohx,
Hamaguchi:2019oev,
Camargo:2019wou,
Garani:2019fpa,
Bell:2019pyc,
Acevedo:2019agu,
Joglekar:2019vzy,
Joglekar:2020liw,
Bell:2020jou,
Garani:2020wge,
Leane:2021ihh,Bose:2021yhz,Collier:2022cpr,Nguyen:2022zwb,Bose:2023yll,Alvarez:2023fjj,Acevedo:2024ttq,Dasgupta:2020dik,Dasgupta:2020mqg,Bhattacharya:2023stq}, other Stars~\cite{1989ApJ...338...24S,Fairbairn:2007bn, Scott:2008ns, Iocco:2008xb,Freese:2008hb, Taoso:2008kw, Sivertsson:2010zm, Freese:2015mta,Ilie:2020iup, Ilie:2020nzp, Lopes:2021jcy,Ellis:2021ztw,John:2023knt,Croon:2023trk,Ilie:2023zfv,Freese:2010re,Bhattacharya:2024pmp,John:2024thz}, and Exoplanets~\cite{Leane:2020wob,Blanco:2023qgi,Acevedo:2024zkg,Benito:2024yki}. 

The first works in this area in the 1980s focused on the scenario where dark matter scatters with Standard Model (SM) particles inside the Sun or the Earth, loses sufficient energy to become gravitationally captured and thermalizes, before promptly annihilating into SM particles which produce neutrinos. Given the ghostly interactions of neutrinos, if they do not have too high energy, neutrinos can stream directly out of the Earth or the Sun, and be detectable with neutrino telescopes on Earth. Soon after, it was realized that dark matter annihilation into prompt SM particles could lead to the dark matter rest-energy being absorbed by the Earth, increasing the internal temperature of the Earth, which is detectable by deep-underground measurements. At this point in time, the particle physics community was largely interested in classes of models where dark matter-SM interactions were mediated by existing SM particles, such as $W$ or $Z$ bosons, motivated by the so-called ``WIMP miracle". This naturally meant that short-lived or prompt dark matter annihilation signals, such as those expected from dark matter-Earth heating, or neutrinos produced in the core, were exclusively considered.

As pure $W$ or $Z$ boson mediation of dark matter-SM interactions became increasingly constrained, new ideas for how dark matter could interact with the SM were required. The consideration of a new dark mediator became increasingly standard in dark matter particle physics. As the potential properties of this particle such as mass and interaction strengths are not yet known, the discovery space and range of signatures was expanded. For example, if the captured dark matter annihilates to mediators that are either sufficiently kinematically boosted or long-lived, the annihilation products can be produced outside the celestial body and be detectable.  In terms of solar searches, this sparked interest in a new search strategy: detecting the annihilation products from the Sun directly, such as with gamma-ray telescopes which can have exquisite sensitivity to this scenario~\cite{Batell:2009zp,Schuster:2009au}. Since then, a wide variety of searches and signatures have been considered, ranging from effects of dark matter energy injection, igniting or destroying objects with ultra-heavy dark matter, and even dark matter inducing $H_3^+$ spectral features in planetary atmospheres~\cite{Blanco:2023qgi}.

Just as the range of signatures considered nowadays is vast, the range of celestial bodies considered is vast. Traditionally, going back about thirty to forty years ago, searches using the Earth or the Sun almost exclusively reserved the community's interest, due to the fact that they are local, and therefore potential dark matter signals are relatively easy to detect. In some sense, they are the most obvious targets. It was also somewhat where telescope or detector technology placed us at the time. Within the last decade or so, neutron stars and white dwarfs appeared more prominently on the dark matter scene, making a splash due to their extended expected reach for weaker dark matter interactions. In the last few years, new detectable searches have been proposed with planets such as Jupiter~\cite{Leane:2021tjj}, and more exotic objects, such as brown dwarfs and exoplanets~\cite{Leane:2020wob}, especially due to new telescope technologies.

Given the wide and growing list of stars and planets used to search for dark matter, the goal of this work is to answer the timely question: which celestial body is the \textit{best} dark matter detector, and under what circumstances? And importantly, why are we considering such a range of celestial bodies? Are we gaining anything new by considering more than one or two types of stars or planets? The answers to these questions are non-trivial, and rely on assumptions made across a multi-dimensional parameter space. Some assumptions are more obvious than others, and some assumptions can dramatically impact how realistic a given search actually is. For example, neutron stars may naively appear to be the best detectors due to their superior cross section reach, but they are very small objects and so can be very difficult to detect. The Earth or Sun might naively appear to be the best in the sense they are the most local, but they do not generally retain lighter dark matter. In some cases, surprising targets can be the most optimal dark matter detectors, even over traditionally considered objects. The goal of this work is to break down and discuss subtle assumptions across the wide range of existing searches, and demonstrate why and how particular celestial targets are optimal in varying scenarios. We will discuss and demonstrate what makes a celestial body optimal, across a multi-dimensional space, given signal detectability and probability of search completion, and given a particular class of particle physics models.

Our paper is organized as follows. We begin in Sec.~\ref{sec:review} by reviewing dark matter capture in celestial objects, and detailing the required inputs. We discuss what makes a celestial body optimal in Sec.~\ref{sec:optimal}, including its physical properties and location, as well as the properties of a given search strategy. In Sec.~\ref{sec:energy} and Sec.~\ref{sec:prod} we calculate new bounds across a range of final states applicable to many dark matter models, focusing largely on dark matter heating and gamma-ray searches and their interplay, detailing telescopes and detection prospects for these signals, and discussing a range of considered dark matter annihilation signals. In Sec.~\ref{sec:map} we map out the dark matter discovery and exclusion space, demonstrating the first model independent analysis of the range of dark sector parameters that can be probed by a given search or object, and showing that some objects probe more parameter space than naively expected. As a consequence, we point out a previously overlooked Galactic center stellar search which can be optimal over even the Sun in some parameter space, and detail this search in Sec.~\ref{sec:stellar}. We present our concluding remarks in Sec.~\ref{sec:conclusion}.

\section{Review of Dark Matter Capture}
\label{sec:review}

Stars and planets in our Galaxy are embedded in an extended dark matter halo. Assuming interactions between the dark matter and SM particles inside a given celestial body, dark matter can scatter, lose energy, and become gravitationally captured. The rate at which this process occurs is called the capture rate, and is the fundamental quantity required to determine any signal strength. Once the dark matter is captured, there is a range of possible signatures.
We will discuss potential signals shortly in the following section, but begin first here with the necessary inputs to calculate the dark matter capture rate.

\subsection{Dark Matter Density Distribution}

The density distribution of dark matter in our Galaxy determines how much dark matter is available to be captured, and thus is an essential input for capture rates. Currently, the dark matter density at the local position is fairly well measured, and is thought to be about $0.4$~GeV/cm$^3$~\cite{Read:2014qva}. Further away from the local position is known with less certainty, and especially towards the Galactic center there are very large systematic error bars on the dark matter density value. There are a few commonly considered benchmarks, with results varying between more cored profiles ($i.e.$ flattening towards the Galactic center), or more cuspy profiles (rapidly growing towards the Galactic center). Cored-like profiles, such as Einasto profiles, are predicted for Milky Way-like halos in the dark matter-only Aquarius simulations~\cite{Pieri:2009je}. Cuspy profiles, including the standard Navarro-Frenk-White (NFW) profile, are well predicted once baryons are included in simulations. Adiabatic contraction is expected in the inner Galaxy, and provides good motivation for even cuspier profiles than the standard NFW~\cite{2011arXiv1108.5736G,DiCintio:2014xia}, which can be described by generalized NFW (gNFW) profiles.

The general form of a Galactic dark matter density distribution $\rho(r)$ at a distance $r$ from the Galactic center can be parametrically described by 
\begin{equation}
\rho_{\mathrm{NFW}}(r)=\frac{\rho_{0}}{\left(r / r_{s}\right)^{\gamma}\left(1+\left(r / r_{s}\right)\right)^{3-\gamma}}\,,    
\end{equation}
where $r_s$ is the scale radius and $\gamma$ determines the inner slope of the dark matter profile. For a standard NFW profile, $\gamma = 1$~\cite{Navarro:1996gj}, and for a generalized NFW (gNFW) profile benchmark we will consider, $\gamma = 1.5$. An increasing value of $\gamma$ corresponds to an increasingly cuspy dark matter profile. We take $r_s=12$ kpc in all cases. The normalization factor $\rho_0$ is adjusted so that the dark matter profile provides the expected $\rho_\chi^\odot \simeq 0.4 \ \rm GeV / cm^{3}$ in the Solar neighborhood~\cite{Read:2014qva,Salucci:2010qr}. We pick gNFW as a benchmark, but our findings of comparing which targets are optimal will not change significantly with this choice. As an example of the effect of DM profile on these results see Ref.~\cite{Acevedo:2023xnu}.

\subsection{Dark Matter Capture by a Single Celestial Body}

Equipped with a dark matter density input throughout the Galaxy, we can now determine the rate at which a single celestial body will capture dark matter.

\subsubsection*{Maximum Capture Rates}

As celestial objects travel through the Galactic dark matter halo, they traverse substantial dark matter densities. If all of the dark matter that passes through the body is captured, this geometric capture rate is given by
\begin{equation}
C_{\mathrm{geo} }=\pi R^{2} n_{\chi}(r) v_{0}\left(1+\frac{3}{2} \frac{v_{\mathrm{esc}}^{2}}{\bar{v}(r)^{2}}\right) \xi\left(v_{p}, \bar{v}(r)\right),
\label{eq:geo}
\end{equation}
 where $\bar{v}$ is the dark matter velocity dispersion, $R$ is the radius of the object, $n_\chi(r) = \rho_\chi(r)/m_\chi$ is the dark matter number density at a distance $r$ from the Galactic center (and $m_\chi$ is the dark matter mass), $v_0 = \sqrt{8/3\pi}\bar{v}$ where the dark matter velocity dispersion $\bar{v}$ is related to its circular velocity $v_c$ by $\bar{v} = \sqrt{3/2}v_c$. For the local position, we take a local velocity dispersion of 237 km/s, towards the Galactic center we use Ref.~\cite{Sofue:2013kja}. In Eq.~(\ref{eq:geo}), $\xi\left(v_{p}, \bar{v},(r)\right)$ accounts for the motion of the compact object relative to the dark matter halo ($\sim$ 1 for our purposes). The escape velocity for an object at its surface is given by~\cite{Bramante:2017xlb}
\begin{equation}
    v_{\rm esc}=
\begin{cases}
  \sqrt{\dfrac{2\,G_NM}{R}}, \quad \text{non-relativistic}\\    
  \sqrt{2\chi}, \quad \text{relativistic}    
\end{cases}
\end{equation}
 where $M$ is the celestial-body mass, $G_N$ is the gravitational constant, and 
\begin{equation}
\chi = 1-\sqrt{1-\frac{2\,G_NM}{R}}\,.
\end{equation}
In some cases, even the maximum capture rate detailed in Eq.~(\ref{eq:geo}), cannot be reached. This can occur for example if dark matter is sufficiently light that it is reflected out of the object, even when it has a very large interaction cross section. This was first pointed out in Ref.~\cite{Zaharijas:2004jv} and formulated for the Earth for very light dark matter in Ref.~\cite{Neufeld:2018slx}, and simulations to determine the reflection factor across a wider dark matter mass range and in multiple celestial bodies have been presented in Ref.~\cite{Leane:2023woh}. We do not include effects of reflection in this work, as it can be model dependent. On the other mass end, if dark matter is ultra-heavy, it can blast through objects even with large cross sections, due to limited stopping power, see the discussion in $e.g.$ Ref.~\cite{Leane:2023woh}.

The maximum capture rate in Eq.~(\ref{eq:geo}) also cannot be reached if the object is so dense that any fermionic dark matter scattering suffers from Pauli blocking. For a neutron star and dark matter mass less than about a GeV, the maximum capture cross section increases with dark matter mass due to Pauli blocking given by
\begin{equation}
    \sigma_{\text{Pauli}} =  \frac{\pi R^2 p_{F,n}}{\gamma m_\chi v_{\text{esc}} N_n},
\end{equation}
where $p_{F,n} \sim 0.45$ GeV is the Fermi momentum of neutrons in a neutron star, $N_n$ is the number of neutrons in the neutron star, and $\gamma \sim 1.4$ is the Lorentz factor of the dark matter particle incident at a neutron star's surface~\cite{Baryakhtar:2017dbj}. For neutron stars, the maximal capture rate can also be decreased due to nuclear effects which we will neglect for simplicity, see Ref.~\cite{Bell:2020obw} for details.

\subsubsection*{Linking Capture Rates to Scattering Cross Sections}

The simple expression in Eq.~(\ref{eq:geo}) above only applies for a geometric maximum possible capture rate, and is only technically reached with an infinite dark matter-SM scattering cross section. In order to make contact with more physical scenarios and arbitrary cross sections which do not lead to all of the dark matter being captured, we need a treatment of dark matter scattering, which we now review.

The probability for a particle to undergo $N$ scatters is~\cite{Bramante:2017xlb}
\begin{equation}
\mathbb{P}_{N} = 2 \int_{0}^{1} d y \frac{y e^{-y \tau}(y \tau)^{N}}{N !}\,,
\label{eq:pN}
\end{equation}
where the optical depth is
\begin{equation}
    \tau=\frac{3}{2}\frac{\sigma_{\chi j}}{\sigma_{\text{tr}}}\,,
\end{equation}
with $\sigma_{\chi j}$ the dark matter-SM scattering cross section, and the transition cross section of dark matter scattering with nucleons is
\begin{equation}
    \sigma_{\mathrm{tr}} = \frac{\pi R^2}{N_j}\,,
\end{equation}
where $N_j = f_j M / m_j$ is the number of targets composing the object, $m_j$ is the mass of the target, and $f_j$ is the associated mass fraction. If the dark matter-SM cross section is larger than the transition cross section, then dark matter is expected on average to scatter multiple times (known as the ``multiscatter" or ``optically thick" regime); cross sections less than this correspond to dark matter scattering on average once or less while transiting the object (known as the ``single scatter" or ``optically thin" regime). 

Eq.~\eqref{eq:pN} can also be written as~\cite{Ilie:2020vec,Ilie:2021iyh}  
\begin{equation}
\mathbb{P}_N=2(N+1)\,\frac{1 - \text{G}(N+2, \tau )}{\tau^2}\,,
\label{eq:pN-gamma}
\end{equation}
where $\text{G}(N+2,\tau)$ is the regularized upper incomplete gamma function defined as 
\begin{equation}
G(a, x)=\frac{1}{\Gamma(a)} \int_x^{\infty} t^{a-1} e^{-t} d t\,,   
\label{eq:upper}
\end{equation}
where $\Gamma(a)$ is the complete gamma function. Eq.~(\ref{eq:pN-gamma}) is useful to speed up the calculation when $N$ is large.
The capture rate for $N$ scatters is given by~\cite{Bramante:2017xlb}
\begin{equation}
\begin{aligned}
C_{N}= \frac{\pi R^{2} \mathbb{P}_{N}}{\left(1-2 G_{N} M / R\right)} \frac{\sqrt{6} n_{\chi}}{3 \sqrt{\pi} \bar{v}} \left\{\left(2 \bar{v}^{2}+3 v_{\mathrm{esc}}^{2}\right)- \left(2 \bar{v}^{2}+3 v_{N}^{2}\right) \exp \left[\frac{-3(v_{N}^{2}-v_{\mathrm{esc}}^{2})}{2 \bar{v}^{2}}\right] \right\},
\label{eqn:multiscatter-capture}
\end{aligned}
\end{equation}
where 
\begin{equation}
v_N = v_{\rm esc} \,\left(1-\frac{\beta_+}{2}\right)^{-N/2},   
\end{equation}
and 
\begin{equation}
\beta_+ = \frac{4m_\chi m_j}{(m_\chi + m_j)^2}  \,.
\end{equation}
If the factor $\xi\left(v_{p}, \bar{v}(r)\right) \sim 1$, which is true for our case, then $C_N \sim \mathbb{P}_N C_{\text{geo}}$ when $N$ is large~\cite{Leane:2021ihh}. Summing over all scatters, the total capture rate is given by
\begin{equation}
C = \sum^\infty_{N=1}C_N,
\label{eqn:multiscatter-capture-sum}
\end{equation}
where in practice, we sum until a finite maximum $N \gg \tau$ as a good approximation. For large $\tau$, dark matter undergoes approximately $N \sim \tau$ scatters. A recent description of capture across all mass and kinematic regimes (extending to more regimes than reflected in the expressions here), along with a public code package, is given in Ref.~\cite{Leane:2023woh}.

\subsubsection*{Cross Section Types and Elemental Compositions}

Dark matter-SM scattering cross sections can be grouped as spin-dependent or spin-independent. Spin-independent scattering leads to particles with any spin contributing to capturing dark matter. The spin-independent scattering cross section is given by~\cite{Bramante:2019fhi}
\begin{equation}
\sigma_{\chi j}^{(S I)}=A_j^2\left(\frac{\mu\left(m_j\right)}{\mu\left(m_N\right)}\right)^2 \sigma_{\chi N}^{(S I)}\,,
\label{eq:SI-cross-section-scaling}
\end{equation}
where $\sigma_{\chi N}^{(S I)}$ is the dark matter-nucleon cross section, $m_N$ is the nucleon mass, $\mu(m_j)$ and $\mu(m_N)$ is the dark matter-nuclei and dark matter-nucleon reduced mass respectively, and $A_j$ is the number of nucleons in atom $j$. Note however the Born approximation here breaks down when the cross section is larger than about $10^{-32}$ cm$^2$~\cite{Digman:2019wdm, Xu:2020qjk}. 

For our spin-independent calculations, we for simplicity assume gas giants and brown dwarfs are pure hydrogen, and for neutron stars we assume pure neutrons. For the Earth, we use results from Ref.~\cite{Bramante:2019fhi} where the Earth's core, mantle, and crust elemental composition is used. For white dwarfs, we calculate the capture rate assuming the composition is pure carbon-12. For average main sequence stars, and our Sun, we use solar elemental abundances~\cite{Asplund:2009fu} of the greatest contributing elements to the capture rate: O, Fe, Si, Ne, Mg, He, S, and N, in descending order of their contribution~\cite{Feng:2016ijc}. We apply these abundances to stellar searches as the Sun has average metallicity~\cite{Bellinger_2019}.

Spin-dependent scattering classifies the interactions where the cross section depends on the particle's spin. Spin-dependent scattering is described by~\cite{Bramante:2019fhi}
\begin{equation}
\sigma_{\chi j}^{(S D)}=\left(\frac{\mu\left(m_j\right)}{\mu\left(m_N\right)}\right)^2 \frac{4\left(J_j+1\right)}{3 J_j}\left[a_p\left\langle S_p\right\rangle_j+a_n\left\langle S_n\right\rangle_j\right]^2 \sigma_{\chi N}^{(S D)}\,,
\end{equation}
where $J_j$ is the total nuclear spin, $\left\langle S_p\right\rangle_j$ and $\left\langle S_n\right\rangle_j$ are the average proton and neutron spin respectively, $a_p$ and $a_n$ are dark matter proton and neutron couplings respectively. For spin-dependent searches using nuclear-burning stars, exoplanets, brown dwarfs, and Jupiter we assume for simplicity that the object's composition is pure hydrogen. Neutron stars are assumed to be purely neutrons. White dwarfs are considered to have no spin-dependent dark matter scattering, due to our assumption they are purely carbon-12, which has zero nuclear spin. For the Earth's spin-dependent constraints, we use results from Ref.~\cite{Bramante:2019fhi} with Earth's composition.

\subsection{Population Searches}

With the capture rate for a single celestial body described above, we can now generalize the capture rate for whole celestial-body populations, which in some particle model cases we will want to study in later sections. This will be because some searches with some objects require the boost of a full population to obtain a detectable signal.

\subsubsection*{Abundances}

To determine a population-level dark matter signal, we need to first provide modeling for the number density distribution of celestial bodies. After presenting their abundances, we will then discuss the reason for our choice of these objects---brown dwarfs, neutron stars, white dwarfs, and Galactic center stars---later in Sec.~\ref{sec:optimal}. 

\textit{For brown dwarfs}, the distribution can be given by a simple power law~\cite{Leane:2021ihh}
\begin{equation}
    n_{\mathrm{BD}}=\ \ 7.5 \times 10^4 \left(\frac{r}{1 \mathrm{pc}}\right)^{-1.5} \mathrm{pc}^{-3},
\end{equation}
where $r$ is the distance to the central black hole Sgr A*. 

\textit{For white dwarfs}, the distribution can similarly be given by~\cite{Acevedo:2023xnu}
\begin{equation}
    n_{\mathrm{WD}}=\ \ 3.28 \times 10^5 \left(\frac{r}{1.5 \mathrm{pc}}\right)^{-1.4} \mathrm{pc}^{-3}\,.
\end{equation}

\textit{For neutron stars}, the distribution can be given by~\cite{Leane:2021ihh}
\begin{equation}
    n_{\mathrm{NS}} = 
        \begin{cases}
      \ 5.98 \times 10^3\left(\frac{r}{1 \mathrm{pc}}\right)^{-1.7} \mathrm{pc}^{-3}  & \, \text{$0.1 \mathrm{pc}<r<2 \mathrm{pc}$}\,, \\
      \\
      \ 2.08 \times 10^4\left(\frac{r}{1 \mathrm{pc}}\right)^{-3.5} \mathrm{pc}^{-3} & \, \text{$r>2 \mathrm{pc}$}\,. \\
    \end{cases}
    \label{eq:ns_numden}
\end{equation}

\textit{For nuclear-burning stars in the nuclear star cluster}, the distribution can be given by~\cite{Schodel:2017vjf}
\begin{equation}
n_{\text{stars}}=2^{(\beta-\gamma) / \alpha} n_0\left(\frac{r}{r_0}\right)^{-\gamma}\left(1+\left(\frac{r}{r_0}\right)^\alpha\right)^{(\gamma-\beta) / \alpha},
\end{equation}
where $r_0$ is the scale radius, and $n_0$ is the normalization given by $n_{\text{stars}} = 1.25\times 10^5$ pc$^{-3}$ at 1 parsec from Sgr A*~\cite{Generozov:2018niv}. The shape of the broken power law is determined by $\gamma$, $\beta$, and $\alpha$.
We use the parameters $\gamma = 1.16 \pm 0.02$, $\beta = 3.2 \pm 0.3$, $r_0 = 3.2\pm0.2$ pc, and $\alpha = 10$, which were found in Ref.~\cite{Generozov:2018niv} to provide the best fit to the observed stellar distribution.
While one can also add the contribution from the stars in the bulge, we have checked that this only enhances the total stellar capture rate by about a factor of two. We follow Ref.~\cite{Generozov:2018niv} in modeling all Galactic center stars as having their average mass, 0.3 $M_\odot$. The contribution of more massive stars boosts the Galactic capture rate by about a factor of two.

\subsubsection*{Galactic Capture}

The capture rate of a population of celestial bodies is given by integrating over the capture rate of all relevant bodies in the Milky Way~\cite{Leane:2021ihh}
\begin{equation}
C_{\mathrm{gal}}=4 \pi \int_{r_{1}}^{r_{2}} dr C(r) n(r) r^{2}\,,
\label{eqn:Ctot}
\end{equation}
where $n(r)$ is the number density of objects in the integration region. We integrate a Galactic population from 0.1 pc, the minimum radius of the velocity dispersion data~\cite{Sofue:2013kja}, to the Earth's local position at $\sim8$~kpc from the Galactic center, though the dominant signal region is the inner few parsecs.

\section{What Makes a Celestial Body Optimal?}
\label{sec:optimal}

We have outlined how dark matter is captured inside a celestial object. Now, to move to detectable signals, we need to consider what object is optimal to use. We detail what makes a celestial body optimal in this section.

\subsection{Properties About the Celestial Body Itself}

The main properties to optimize over for a celestial body search are:

\begin{itemize}
    \item \textit{Radius:} Bigger objects are generally better in terms of the maximum amount of dark matter they can capture, due to the geometric capture rate increasing with surface area. In addition, larger objects are generally easier to detect with telescopes, due to luminosity scaling with surface area.
    \item \textit{Density:} Denser objects have higher escape velocities, which makes a high-density object efficient at capturing dark matter with low interaction cross sections.
    \item \textit{Core temperature:} If the core temperature is too high, the kinetic energy of lighter dark matter can reach escape velocity speeds such that it escapes the body's gravitational potential, removing most detectable signals.
\end{itemize}

Generally speaking, the most optimal object possible would possess the largest radius, with the highest density, and the lowest core temperature. However, this is of course not often how objects are made in nature. An object with a large radius is often not very dense, and conversely, a dense object usually has a smaller radius. The coolest objects are also not usually the densest objects, as high densities naturally lead to higher core temperatures. Therefore in order to determine which celestial object is optimal for a given search, these properties need to be maximised to best address the search strategy at hand. Obviously, a candidate that fails multiple of these criteria relative to another object, is not the best choice to maximize dark matter sensitivity for a given search. For example, objects such as the Moon or asteroids are almost always useless for any signal relying on dark matter capture, because they are both smaller and less dense than many other objects that could be considered.

\begin{table*}\centering
\ra{1.3}
\begin{tabular}{@{}lp{0.5in}p{0.5in}p{0.5in}p{0.5in}p{0.5in}@{}}\toprule
& \multicolumn{5}{c}{Celestial Body Properties}
\\\cmidrule(lr){2-6}
           & \multicolumn{1}{c}{Escape velocity [$c$]} & \multicolumn{1}{c}{Mass [$\mathcal{M}_\odot$]} &  \multicolumn{1}{c}{Radius [$R_\odot$]} & \multicolumn{1}{c}{T$_\mathrm{core}$ [K]} & \multicolumn{1}{c}{$\sigma_{\text{tr}}$ [cm$^2$]} \\\midrule
\textbf{Neutron Star} & \multicolumn{1}{r}{0.7} & \multicolumn{1}{r}{1.4} & \multicolumn{1}{r}{10$^{-5}$} &  \multicolumn{1}{r}{10$^5$} & \multicolumn{1}{c}{10$^{-45}$}  \\
\textbf{White Dwarf} & \multicolumn{1}{r}{10$^{-2}$} & \multicolumn{1}{r}{0.6} & \multicolumn{1}{r}{10$^{-2}$} & \multicolumn{1}{r}{10$^5$} & \multicolumn{1}{c}{10$^{-41}$} \\
\textbf{Average MS Star} & \multicolumn{1}{r}{10$^{-3}$} & \multicolumn{1}{r}{0.3} & \multicolumn{1}{r}{0.3} & \multicolumn{1}{r}{10$^7$} & \multicolumn{1}{c}{10$^{-36}$}  \\
\textbf{Sun} & \multicolumn{1}{r}{10$^{-3}$} & \multicolumn{1}{r}{1} & \multicolumn{1}{r}{1}  & \multicolumn{1}{r}{10$^7$} & \multicolumn{1}{c}{10$^{-35}$} \\
\textbf{Brown Dwarf}    & \multicolumn{1}{r}{10$^{-3}$} & \multicolumn{1}{r}{10$^{-2}$} & \multicolumn{1}{r}{0.1} & \multicolumn{1}{r}{10$^4$--10$^6$} & \multicolumn{1}{c}{10$^{-35}$} \\
\textbf{Jupiter} & \multicolumn{1}{r}{10$^{-4}$} & \multicolumn{1}{r}{10$^{-3}$} & \multicolumn{1}{r}{0.1} & \multicolumn{1}{r}{10$^4$} & \multicolumn{1}{c}{10$^{-34}$} \\
\textbf{Earth}   & \multicolumn{1}{r}{10$^{-5}$} & \multicolumn{1}{r}{10$^{-6}$} & \multicolumn{1}{r}{10$^{-2}$} & \multicolumn{1}{r}{10$^4$} & \multicolumn{1}{c}{10$^{-33}$} \\\bottomrule
\end{tabular}
\caption{Approximate properties of the range of celestial bodies considered for dark matter searches, including their escape velocity, mass, radius, core temperature, and transition cross section.}
\label{tab:celestial-body}
\end{table*}

Table~\ref{tab:celestial-body} summarizes celestial bodies used for dark matter searches, and their approximate properties such as escape velocity, mass, radius, core temperature, and transition cross section. The transition cross section represents the switch between the single and multiscatter regimes, and while it does not generically correspond to the cross section at which all incoming dark matter is captured, it is a proxy for the capture efficiency of the object. From this table, the scenarios where some of these objects will be optimal is already clear, given the bullet points above. To reach the lowest cross sections, the densest objects will often be the best, which is a neutron star followed by a white dwarf. Similarly, to reach the lightest dark matter masses, high escape velocities relative to core temperatures will be necessary, rendering also neutron stars followed by white dwarfs superior. However, calling these ``superior" has the in-built assumption that these objects are detectable to begin with; we therefore shortly discuss properties about location which will render these objects detectable or not. If fluxes need to be maximized for a signal to be detectable, the largest object will often instead be superior, which depending on the signal may be the Sun or an average main sequence (MS) star. If fluxes need to be only somewhat larger for a signal to be detectable, brown dwarfs or Jupiters may be optimal, as they are larger than neutron stars and white dwarfs and so can have larger capture rates and consequently larger fluxes, and also extend their sensitivity to lighter dark matter masses than the Sun, due to their cooler core temperatures despite lower escape velocities.

\subsection{Properties About the Celestial-Body Location}

As discussed above, what makes a celestial body optimal depends on its intrinsic properties such as radius, density, and core temperature. However, if we want to detect a signal from a celestial body, it also matters where it is and what its environment is like. The main properties to optimize over for a celestial body search location are:

\begin{itemize}
    \item \textit{Dark Matter Mass Density:} Regions with high dark matter mass density have more dark matter mass available to be captured, leading to higher capture rates and larger consequent signals.
    \item \textit{Dark Matter Velocity:} The slower the dark matter is moving, the easier it is to capture. Therefore, low dark matter velocity environments are optimal.
    \item \textit{Distance Away:} The closer the object is to the telescope or detector, the larger the flux will be. Flux is inversely proportional to the square of the distance to the object, rapidly decreasing the further away the object is from the observer.
\end{itemize}

As with the intrinsic properties of the celestial body, we unfortunately don't get all these location properties at once. At the closest distance is the Earth or the Sun, which means fluxes can be particularly large, but the dark matter density is very low relative to other positions such as the Galactic center. Conversely, the Galactic center has very high dark matter content, but signals will suffer a large suppression from the inverse distance squared. We instead must optimize over these properties depending on what particular celestial body advantages or limitations already exist. For example, dark matter capture signatures from small objects such as neutron stars can be very difficult to detect unless they are extremely close by, or if they exist in a region with $e.g.$ massively enhanced dark matter densities, in order to overcome limitations from their small radii (low surface area leading to low luminosities, and lower capture rates than other objects in the same region).

Table~\ref{tab:locations} details four main locations that have been considered in the past for celestial body searches: the local position, globular clusters, the Galactic bulge, and the nuclear stellar cluster. We have defined the local position to be within 100 pc of Earth. Globular clusters are old spheroidal stellar systems. The Galactic bulge is the central region of our Galaxy made up of stars and dust, and is thought to be mostly older stars. The nuclear stellar cluster is even further into the Galactic center, 
and is a compact stellar nucleus with high stellar density and high luminosity surrounding the innermost part of our Galaxy. Alongside these locations, we detail the distance away these locations are from Earth, and the dark matter density and velocity in these locations. While there are substantial systematic uncertainties in both the dark matter density and velocity in systems other than the local position, ranges can be taken to bracket the systematics. Dark matter density values for the Galactic bulge and nuclear cluster are quoted by taking a range of gNFW and Einasto profiles ($i.e.$ cored or cuspy profiles). Globular clusters may have dark matter content but the numbers are generally not robust~\cite{Moore:1995pb,Saitoh:2005tt}; we show 1000 GeV/cm$^3$ as the nominal value taken for the Messier 4 globular cluster in previous dark matter capture studies, but write it with an asterisk due to the substantial systematics in this number.

\begin{table*}\centering
\ra{1.3}
\begin{tabular}{@{}lp{0.5in}p{0.5in}p{0.5in}@{}}\toprule
& \multicolumn{3}{c}{Location Properties}
\\\cmidrule(lr){2-4}
           & \multicolumn{1}{c}{DM Density [GeV/cm$^3$]} & \multicolumn{1}{c}{DM Velocity [km/s]} & \multicolumn{1}{c}{Distance [kpc]}  \\\midrule
\textbf{Local Position}    & \multicolumn{1}{r}{0.4} & \multicolumn{1}{r}{270} & \multicolumn{1}{r}{$<0.1$}   \\
\textbf{Globular Clusters} & \multicolumn{1}{r}{1000*} & \multicolumn{1}{r}{10}  & \multicolumn{1}{r}{$\sim2$}   \\
\textbf{Galactic Bulge} & \multicolumn{1}{r}{$\sim100-1000$} & \multicolumn{1}{r}{$\sim100-200$}   & \multicolumn{1}{r}{$\sim7$}  \\
\textbf{Nuclear Cluster} & \multicolumn{1}{r}{$\sim10^3-10^6$} & \multicolumn{1}{r}{$\sim100-800$} & \multicolumn{1}{r}{$\sim8$}   \\\bottomrule
\end{tabular}
\caption{Search locations used for celestial-body dark matter signals, with the location properties including dark matter density, dark matter velocity, and distance from Earth.}
\label{tab:locations}
\end{table*}

Of these locations, the local position is the best for objects with low luminosities, as its main advantage is that as fluxes drop off inversely with the square of the distance, the local position can enable lower fluxes to be detected. If a largely enhanced signal is needed for detection, globular clusters may be optimal, as their low dark matter velocities can lead to higher capture rates, as do their large dark matter densities, though these systems suffer the caveat of robustness of dark matter content. If large signals are needed for detection, moving further and further into the Galactic center may be optimal due to large dark matter densities, provided that the celestial-body's intrinsic properties (such as its radius, which is relevant for its luminosity) do not limit signal detection.

There are some more specific properties about the location that can be important, depending on the wavelength of the particular signature. One potentially important property is the amount of dust present in the environment, as dust extinction can significantly reduce the detectability for particular wavelengths. Another can be stellar crowding, which can occur when too many bright stars overshine the potential signal in a given telescope pixel, and can be particularly important for signals towards the Galactic center. We also have so far not mentioned properties about the specific detection signature, which will depend on the particle physics model. We now therefore discuss a range of potential dark matter detection signals arising after dark matter is captured, and put these in the context of classes of particle physics models. We will refer back to the optimal properties discussed in this section about the celestial body and its location when discussing a range of search strategies below.

\section{Dark Matter Energy Injection Searches}
\label{sec:energy}

Once dark matter is captured, there are many potential possible signals. We focus on scenarios where dark matter annihilates, see Refs.~\cite{Ray:2023auh, Goldman:1989nd, Gould:1989gw, Bertone:2007ae, deLavallaz:2010wp, McDermott:2011jp, Kouvaris:2010jy, Kouvaris:2011fi, Bell:2013xk, Guver:2012ba, Bramante:2013hn, Bramante:2013nma, Kouvaris:2013kra, Bramante:2014zca, Garani:2018kkd, Kouvaris:2018wnh, Lin:2020zmm, Takhistov:2020vxs, Garani:2021gvc, Steigerwald:2022pjo, Singh:2022wvw, Starkman:1990nj, Kurita:2015vga, Acevedo:2020gro, Banks:2021sba, Steigman:1978wqb, Spergel:1984re, Faulkner:1985rm, Gould:1989hm, Lopes:2001ra, Lopes:2002gp, Bottino:2002pd, Lopes:2012af, Lopes:2013xua, Lopes:2014aoa, Geytenbeek:2016nfg, Frandsen:2010yj, Cumberbatch:2010hh, Taoso:2010tg, Vincent:2014jia, Vincent:2015gqa, Vincent:2016dcp} for examples of signatures of non-annihilating dark matter, such as the impact on energy transport and black hole formation and destruction.

\begin{figure}[t!]
    \centering
    \includegraphics[width=0.47\columnwidth]{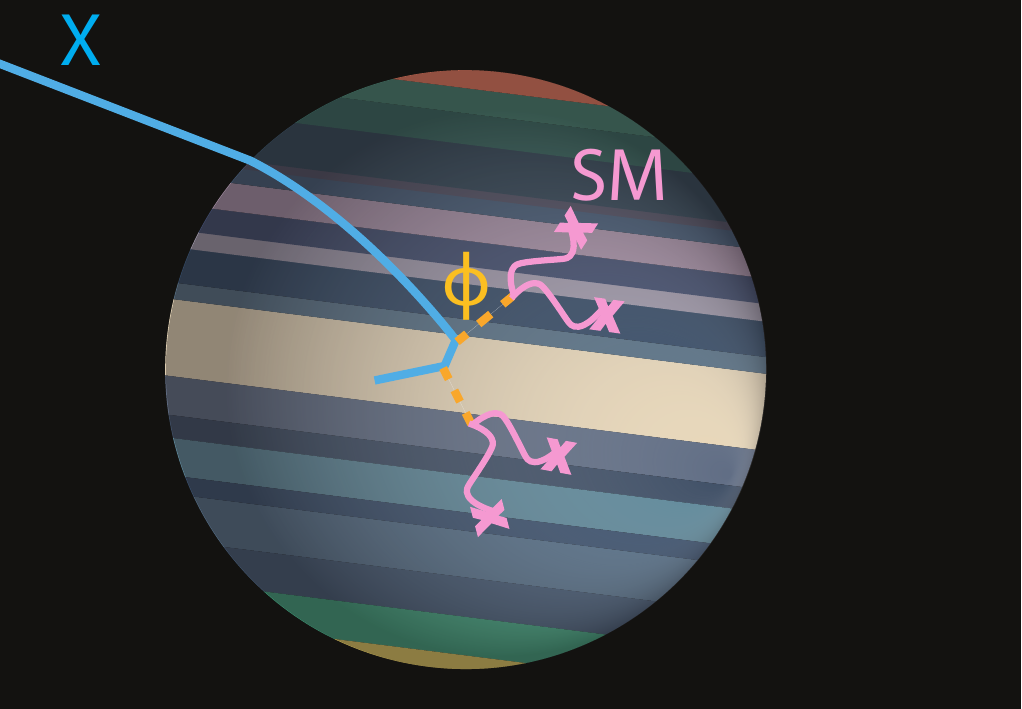}
    \includegraphics[width=0.47\columnwidth]{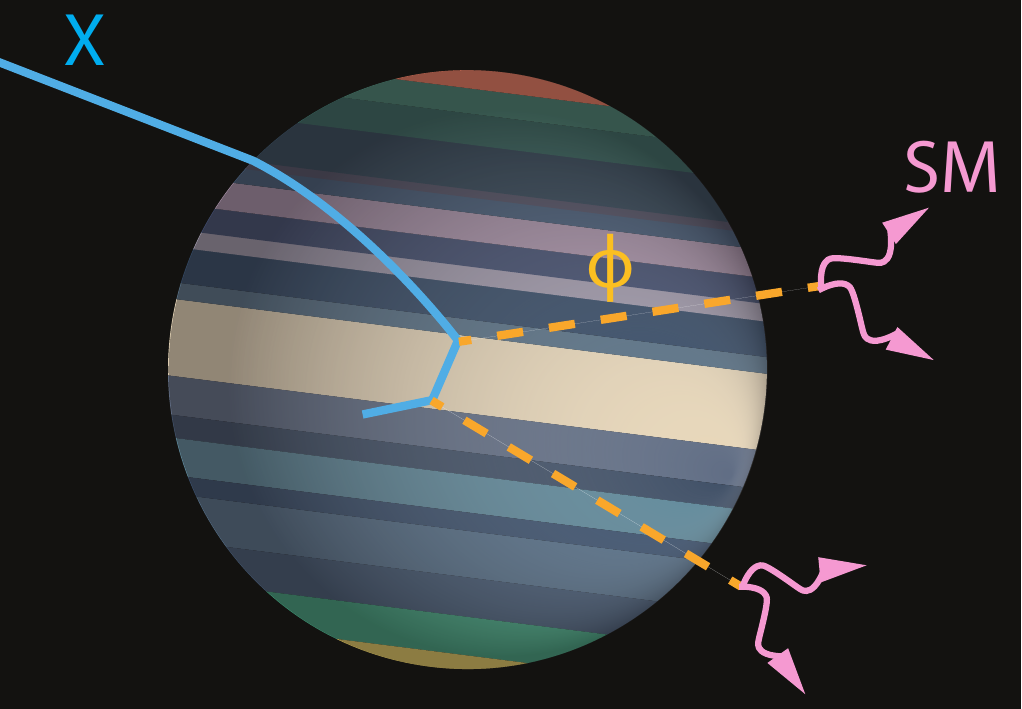}
    \caption{ Schematic of dark matter capture and annihilation. Dark matter scatters off constituents of celestial bodies. Dark matter may then annihilate and produce SM products, mediated by a dark mediator. If the mediator decays in the body's interior, as in the left panel, the annihilation event can inject energy into the object, or produce detectable neutrinos. If the decay length is long enough, as in the right panel, annihilation products other than neutrinos can be detected directly.}
    \label{fig:schematic}
\end{figure}

Figure~\ref{fig:schematic} schematically shows the two classes of signatures that we will consider. Firstly in this section, we will discuss where dark matter energy injection heats objects or alters the evolution of objects (left). In the next section, we will discuss those where SM annihilation products are detected directly (right).

\subsection{Calculation of Dark Matter Induced Luminosity}
 
Given the capture rate of dark matter, we can calculate the effects of dark matter energy injection on a celestial body. If dark matter annihilation decay products are deposited inside the body, they transfer their energy into the star or planet, which can change the luminosity of the object. The most simple case can be calculated when there are no significant non-dark matter internal heat flow processes, as is the case for objects with no nuclear fusion such as planets or brown dwarfs. In this scenario, assuming capture-annihilation equilibrium, we find the heat power produced by dark matter is given by
\begin{equation}
    \Gamma_{\text {heat }}^{\mathrm{DM}} = \mathbb{P}_{\text{heat}} m_\chi C  + (\gamma-1) m_\chi C\,,
\end{equation}
where $C$ is the capture rate, the $(\gamma - 1)$ factor is due to dark matter depositing kinetic energy as it becomes captured by the object~\cite{Baryakhtar:2017dbj}, and $\mathbb{P}_{\text{heat}}$ is the probability for the dark matter annihilation to result in heating, which depends on the mediator decay length. For a neutron star $\gamma \sim 1.4$ due to its large escape velocity, for other bodies $\gamma \sim 1$ nullifies the kinetic heating term, unless additional long-range forces are present~\cite{Acevedo:2024zkg} which we do not consider in this work. We see that the heat power is proportional to the capture rate, which is simply because the captured dark matter releases its mass-energy to heat the body. We can compare the dark matter heat power to the total heat power given by~\cite{Leane:2020wob}
\begin{equation}
\Gamma_{\text {heat }}^{\mathrm{tot}}=\Gamma_{\text {heat }}^{\mathrm{ext}}+
\Gamma_{\text {heat }}^{\mathrm{int}}+
\Gamma_{\text {heat }}^{\mathrm{DM}}=
4 \pi R^2 \sigma_{\mathrm{SB}} T^4 \epsilon \text {,}
\label{eq:heatlum}
\end{equation}
where $\Gamma_{\text {heat }}^{\mathrm{ext}}$ is the total external heat power (assumed to be zero for free-floating objects with no host star), $\Gamma_{\text {heat }}^{\mathrm{int}}$ is the internal heat power, and $\Gamma_{\text {heat }}^{\mathrm{DM}}$ is the dark matter heat power, $R$ is the radius of the object, $T$ is its temperature, $\sigma_{\mathrm{SB}}$ is the Stefan-Boltzmann constant, and $\epsilon \in [0,1]$ is the emissivity which is a measure of the radiation efficiency of the object.

In the case of dark matter energy injection in stars undergoing nuclear fusion such as main sequence stars, the energy transport and effect on the celestial body is more complicated. In that scenario, determining the luminosity of a star due to dark matter energy injection generally requires the use of simulations with $e.g.$ stellar evolution codes such as \texttt{MESA}~\cite{2011ApJS..192....3P, 2013ApJS..208....4P, 2015ApJS..220...15P, 2018ApJS..234...34P, 2019ApJS..243...10P, 2023ApJS..265...15J}. Note that even for the scenario discussed above where nuclear fusion is not occurring in planets, even then there can be a mild effect on the evolution of the planet ($e.g.$ a change in planetary radius), though this is not expected to be large given the amount of dark matter heating expected, and at first order can be estimated with the Eq.~(\ref{eq:heatlum}) above.

\subsection{Telescopes and Detection Prospects}
\label{section:telescope}

A range of objects and locations have been considered for dark matter energy injection detection. We now detail what telescopes or measurements are used to search for dark matter energy injection, and in the case of upcoming observations discuss the prospects for future detection. We will consider the subtleties in these searches and show existing limits and projections at the end of this section.

\subsubsection{Borehole Temperature Gradient Measurements}

To study Earth's geological processes, about 20,000 boreholes have already been drilled throughout Earth's crust. The temperature gradient in these boreholes is recorded, which when multiplied by the thermal
conductivity of the relevant material yields a heat flux~\cite{beardsmore2001crustal,pollack1993heat,se-1-5-2010,global_heat_flow_data_assessment_group_global_2024}. Use of these boreholes to test any dark matter contribution to Earth's internal heat flow was first considered in Ref.~\cite{Mack:2007xj} (see also Refs.~\cite{Starkman:1990nj,Bramante:2019fhi}). As the dark matter heat flow in some cases exceeds the measured values, these internal temperature gradients have been used to constrain dark matter already.

\subsubsection{Infrared Radiation}

The coldest existing objects are planets, due to their lack of nuclear fusion, and due to sometimes being quite far from any external radiation source. Planets have surface temperatures as low as about $100$~K. Therefore, any dark matter heating signal must overpower at least this heating threshold. A temperature as low as $100$~K corresponds to the far infrared, and therefore the best telescopes to measure the smallest dark matter heating rates (and therefore smallest dark matter scattering cross sections), are infrared or near-infrared telescopes.

Current telescopes and near-future telescopes that are excellent for detecting dark matter heating are JWST, the Rubin Telescope/LSST, and the Roman Telescope (previously WFIRST).
JWST is an absolute powerhouse, and was launched recently on Christmas Day 2021. 
JWST can observe infrared and optical wavelengths from 0.6 -- 28 microns~\cite{Windhorst:2005as}.
Rubin is slated to achieve first light in 2024, and can observe near-infrared, as well as optical wavelengths, with sensitivity to about $0.32-1.05$ microns~\cite{LSSTScience:2009jmu}.
Roman is awaiting launch, planned around 2027, and can observe near-infrared, as well as partially into optical wavelengths, with sensitivity to about $0.48 - 2.3$ microns~\cite{Wang:2022qov}. JWST is superior at detecting deepest into the infrared of any existing or past telescope, which is an important advantage for testing very cold objects. 
However, JWST will have limited survey time, and in this regard Roman may be superior, with Galactic center surveys with large exposure times expected. In addition, while the current design of Roman only extends to 2.0 microns, an additional $K$-filter would provide infrared sensitivities~\cite{stauffer2018science}. On other fronts, it is also possible that Gaia Near Infra-Red (GaiaNIR), a proposed successor of Gaia in the near-infrared, may improve sensitivities to the infrared, along with other potential future telescopes.

\subsubsection{Optical Radiation}

Celestial bodies that capture sufficient dark matter can become hot enough to radiate in the optical wavelengths. A superior telescope for optical wavelengths remains (even after more than 30 years) the Hubble Space Telescope. Hubble can observe the near-infrared, optical, and ultraviolet, with sensitivities between about $0.12-2$ microns. Already Hubble has observed white dwarfs in the globular cluster Messier 4~\cite{Bedin:2009it}. 

The Very Large Telescope (VLT) of the European Southern
Observatory in Chile can detect near-infrared and optical wavelengths, and alongside the W. M. Keck Observatory
10-meter telescopes in Hawaii has been used to study a population of stars called S-stars orbiting the supermassive blackhole Sgr A* in the center of our Galaxy~\cite{Ghez:2003qj, Martins:2007rv,  
 Ghez:2008ms, Gillessen:2008qv, 2010RvMP...82.3121G, 
2017ApJ...847..120H, Pei_ker_2020}. These spectral measurements of S-stars are so remarkable that they led to the award of the Nobel Prize, and are a testament to the precision in which some celestial bodies can be measured.

\subsubsection{Existing Limits, Interplay, and Feasibility}

After discussing optimal celestial body and location properties, alongside some searches, we can finally detail some existing and projected search sensitivities. Thanks to the previous discussion, we can now understand why these limits sit where they do in the dark matter parameter space, and why some objects are used in particular circumstances.

\begin{figure}[t!]
    \centering
    \includegraphics[width=0.49\linewidth]{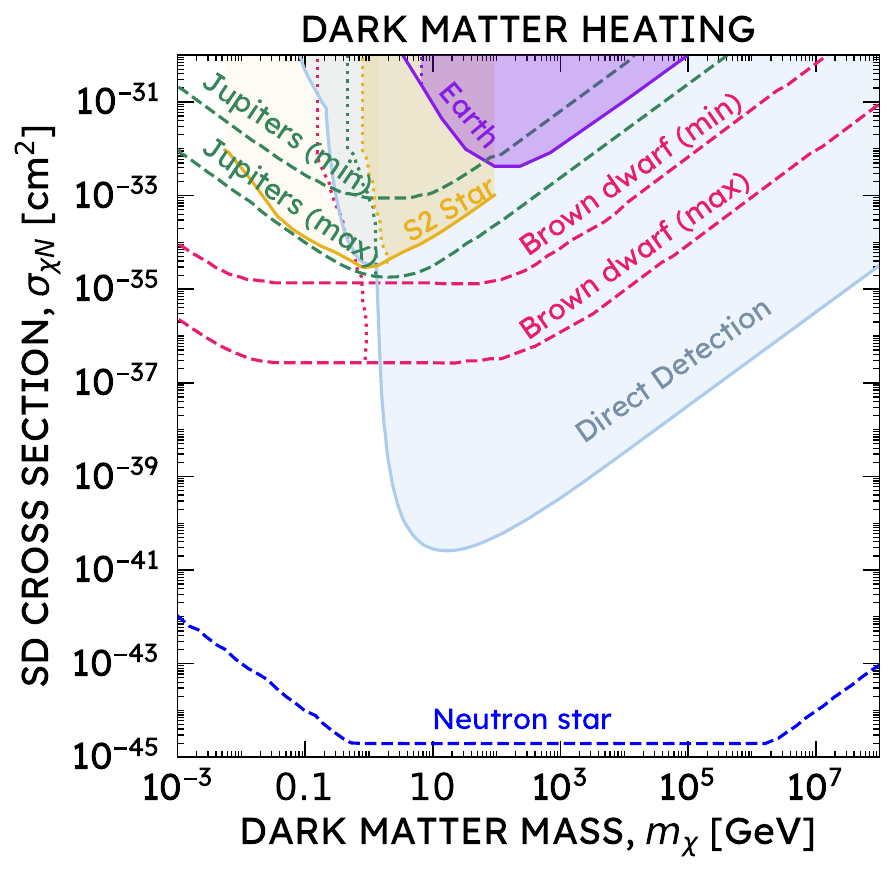}
%    \hspace{5mm}
    \includegraphics[width=0.49\linewidth]{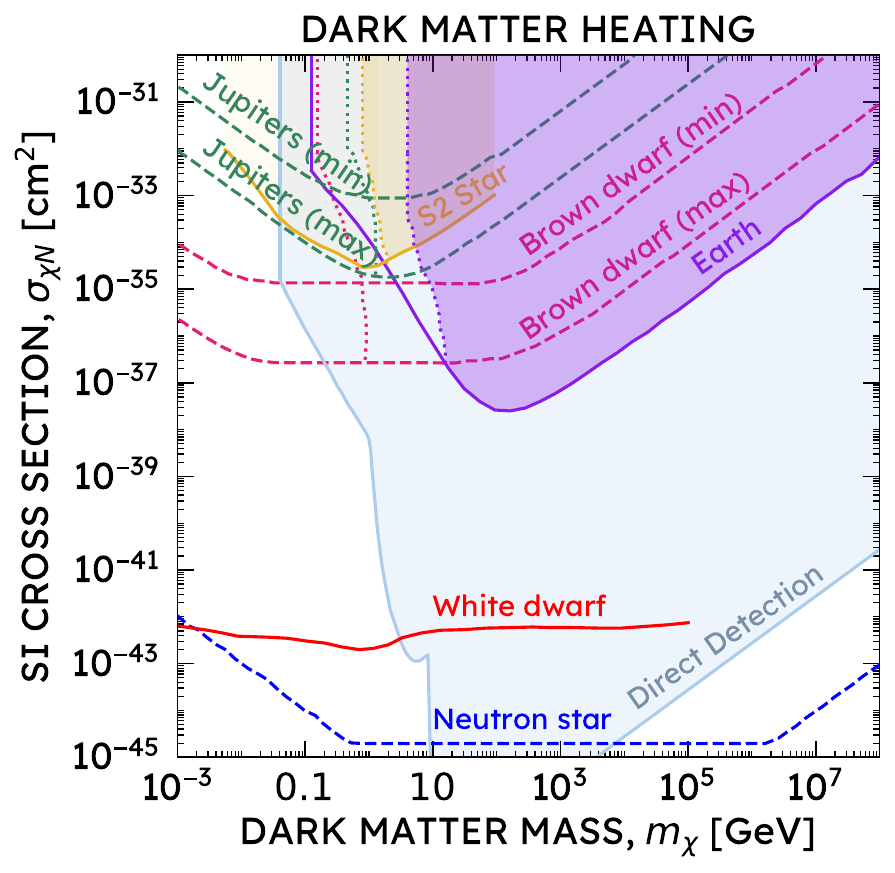}
    \caption{Sensitivities to the dark matter-nucleon scattering spin-dependent (left) or spin-independent (right) cross section as a function of dark matter mass, arising from dark matter energy injection, for a range of celestial objects as labeled. Solid lines indicate existing limits, dashed lines indicate future projections. The left of the dotted lines indicate where dark matter is evaporated at light masses if only contact interactions are present; long-range interactions may probe lower dark matter masses. Complementary limits from direct detection experiments are also shown. See text for additional details.}
    \label{fig:heating}
\end{figure}

Figure~\ref{fig:heating} shows existing limits and future sensitivities to dark matter energy injection, including limits from the Earth~\cite{Bramante:2019fhi}, S-star S2~\cite{John:2023knt}, exoplanets and brown dwarfs~\cite{Leane:2020wob}, neutron stars~\cite{Baryakhtar:2017dbj}, and white dwarfs~\cite{Bell:2021fye}. We also show for comparison existing direct detection experiments which also probe the dark matter-nucleon scattering rate, with the strongest limits arising from Collar~\cite{Collar:2018ydf}, CRESST~\cite{CRESST:2022dtl}, PICO~\cite{PICO:2019vsc}, LZ~\cite{LZ:2022lsv}, and DarkSide-50~\cite{Franco:2023sjx}. Dark matter boosted by cosmic rays can constrain some of this space~\cite{Super-Kamiokande:2022ncz,Bringmann:2018cvk,Maity:2022exk}, however, these limits are model dependent because the energy transfer is large and a mediator must be chosen. Therefore, we do not present boosted dark matter limits in these plots or hereafter. For the sensitivities shown, the lower dark matter mass threshold is set by the evaporation process. This occurs when the object's core temperature is too high relative to its escape velocity. To the left of the dotted lines, particle physics models with contact interactions will evaporate (to obtain these, we use Refs.~\cite{Gould:1989hm,Leane:2022hkk}, see also Ref.~\cite{Garani:2021feo}). However, if there are long-range forces present, the left of the dotted lines can also be probed, due to the long-range forces forming an evaporation barrier~\cite{Acevedo:2023owd}. We now discuss the results of this figure and the interpretation of each of these searches one by one.

\subsubsection*{\centering Neutron Stars}

We start with neutron stars, as Fig.~\ref{fig:heating} implies they provide the superior reach. The sensitivity in Fig.~\ref{fig:heating} corresponds to all the dark matter passing through the neutron star being captured (though note this neglects nuclear effects~\cite{Bell:2020obw}, which can weaken this line by a few orders of magnitude). This significant sensitivity is precisely why neutron stars have dominated the recent literature, and arises because neutron stars are so dense, that they are extremely efficient at capture and so can probe very small scattering cross sections. However, achieving this sensitivity requires actually detecting a dark matter signal inside a neutron star to begin with. Unfortunately, the flux at this cross section is relatively small, because the radius of a neutron star is tiny compared to other objects. This means that this sensitivity is only achievable if a neutron star is found very close by (within about 10 pc~\cite{Baryakhtar:2017dbj}), because the distance squared suppression on top of the small flux is a large suppression in the signal size. In addition, if a neutron star very close by is found, it also needs to be very old~\cite{Baryakhtar:2017dbj}, so that it has cooled sufficiently such that a dark matter heating signal can overpower any internal heat processes, which are quite large when the neutron star is young. Both of these factors together imply that finding such an object will be a significant challenge.

An additional difficulty in achieving this sensitivity line, is that even if we do find a sufficiently old and close-by neutron star, we have to get the observational data. Searches for neutron stars in our neighborhood require dedicated observing time. JWST has been estimated to detect dark matter heating of a neutron star at 10~pc in $\sim 10^5$s with SNR 2~\cite{Baryakhtar:2017dbj}, but for further (and potentially more realistic) distances, this quickly blows out of reasonable observing time limits.

A strong advantage of neutron stars if they are detectable, apart from their scattering cross section reach, is that they can also probe a broader range of particle physics models~\cite{Baryakhtar:2017dbj}. Because they have such high escape velocities, the incoming dark matter particles can have very high energies, and potentially overcome any mass splittings to upscatter into heavier states, as is the case for inelastic dark matter models. Some dark matter models also have velocity-suppressed scattering, and in that case as the incoming dark matter moves so quickly, this suppression can be lifted. Neutron stars also afford an additional heat source almost regardless of the dark matter model, which is dark kinetic heating (though see also Ref.~\cite{Acevedo:2024zkg}). This occurs as neutron stars absorb dark matter kinetic energy after it is accelerated to relativistic speeds by the gravitational potential of the neutron star~\cite{Baryakhtar:2017dbj}.

\subsubsection*{\centering Exoplanets and Brown Dwarfs}

Dark matter energy injection can affect brown dwarfs and Jupiters, and future sensitivities to this are shown in Fig.~\ref{fig:heating} assuming near maximal capture rates (labeled ``min") or about 10 percent of maximal capture rates (labeled ``max"). Brown dwarfs and Jupiters are similar objects, with brown dwarfs displaying superior sensitivity due to being about the same size as Jupiters in radius, but being much more dense. However, brown dwarfs being more dense leads to higher internal temperatures, meaning that they need larger dark matter capture rates relative to their size than Jupiters, for dark matter heating to overcome their internal heat background. Apart from neutron stars and white dwarfs, exoplanets and brown dwarfs provide the highest possible sensitivity to light dark matter, due to the interplay of their low core temperatures and relatively high escape velocities. This makes them ideal for probing light dark matter, and due to their densities provide a good reach to the scattering cross section, and due to their radii can be much easier to detect than neutron stars or white dwarfs.

To obtain the labeled brown dwarf and Jupiters sensitivities, there are two potential search strategies. One is to consider exoplanets in the local position~\cite{Leane:2020wob}, where they are easy to detect. In this case, luminosities are high and require very little observing time; detection of many exoplanets locally with some non-negligible dark matter heating is very likely. The drawback to searches in the local position is that the dark matter density is low, so temperature increases from dark matter may be non-negligible but are not stark; therefore large statistics would need to be obtained to make any robust statement. However, in addition to the many known exoplanet candidates, an extensive suite of current and future telescopes will examine our local neighborhood to identify and measure more candidates. For example, Gaia is expected to find $21,000\pm6,000$ long-period Jupiters and brown dwarfs within 500 pc from Earth, within a few years of operating~\cite{Perryman_2014}. Within 10 years, it is estimated to detect $70,000\pm20,000$ new exoplanets of interest~\cite{Perryman_2014}. This will substantially increase candidates and statistics for this search. For the local position search, it is important to note that for brown dwarfs, given their higher internal temperatures, a local search will not be feasible, and Jupiter-like planets are required.

Another search strategy to realize the Jupiters and brown dwarfs energy injection sensitivities shown in Fig.~\ref{fig:heating}, is to use the Galactic bulge~\cite{Leane:2020wob}. As detailed in the optimal locations section, the bulge provides a much larger dark matter density, at the cost of a reduction in luminosity from the squared distance suppression. However, as the surface area of Jupiters and brown dwarfs is larger than most celestial objects, their luminosity is not reduced so much that they cannot still be detectable with JWST or another infrared telescope. Therefore using exoplanets and brown dwarfs the Galactic bulge provides potentially strong sensitivity to the dark matter scattering cross section. The detectability criteria of the lines shown for a bulge search is at least SNR 2 after $10^5$s of integration time; shorter integration times can still provide the same sensitivity for exoplanets that are maximally heated. The bulge search also offers a benefit of probing the dark matter density distribution, as exoplanets should be heated proportionally to their local dark matter density, and so should rise in temperature (scaling with the dark matter density distribution) towards the Galactic center. This allows for an additional handle on a potential dark matter origin of any anomalous heating. Note that for both of the search strategies, even larger sensitivities can be obtained if there is dark kinetic heating present~\cite{Acevedo:2024zkg}, which is not included in the plots.

For both the exoplanet/brown dwarf search strategies discussed above, JWST was originally identified as the main potential telescope of interest~\cite{Leane:2020wob}. However, other instruments such as the Roman Telescope are likely more fruitful, due to their planned Galactic bulge surveys. This gives a much higher likelihood of dark matter heat detection, as large exposure times will be much easier to obtain than with JWST, which does not plan to survey the bulge for long periods.

\subsubsection*{\centering White Dwarfs}

Dark matter energy injection has been studied for white dwarfs, and is shown for the spin-independent bounds in Fig.~\ref{fig:heating}. In contrast to the other celestial objects shown, there are no spin-dependent bounds for white dwarfs. This is because white dwarfs are dominantly made of carbon or oxygen, which have zero nuclear spin, leading to a negligible spin-dependent scattering rate; spin-dependent scattering only occurs with elements that have non-zero nuclear spin. The sensitivity line shown corresponds to the lowest cross section with the maximum capture rate for a white dwarf, which we see is the next closest potential probe to neutron stars in sensitivity strength. This is due to the white dwarfs high density and therefore reaching a lower cross section with a maximal capture rate. However, while the maximal capture cross section can be low, white dwarfs have very high internal heat. This means that in order to perform a heating search, the dark matter capture rate needs to be very large to overcome the hot backgrounds. For this reason, white dwarf heating has been previously studied in the globular cluster Messier 4 (M4), due to the high potential dark matter density and low dark matter velocity, which can therefore lead to white dwarfs radiating in the optical due to extreme dark matter heating. The Hubble Space Telescope has measured the optical radiation of a population of cold white dwarfs in M4~\cite{Bedin:2009it}, allowing their luminosities to constrain the dark matter scattering rate as shown~\cite{Bell:2021fye}. 

As discussed earlier, the dark matter density in Messier 4 must be verified to confirm this constraint, and it is not currently robust. While one would naturally consider Galactic center heating searches for white dwarfs to overcome the large uncertainty in globular cluster dark matter densities, this direction is not currently plausible, as no white dwarfs have been identified in the inner Galaxy. This is because white dwarfs are roughly Earth-sized and so are not as large as $e.g.$ main sequence stars, taking a hit in their luminosities due to small surface areas, as well as that white dwarfs emitting in the optical greatly suffer from dust extinction. 

\subsubsection*{\centering S-Stars}

Nuclear burning stars naively are not ideal for energy injection searches, as their heat backgrounds are very high. For this reason, the Sun is not generally considered for dark matter energy injection searches (though for asymmetric dark matter where the dark matter build-up can be higher, some interesting signatures are possible~\cite{Banks:2021sba, Steigman:1978wqb, Spergel:1984re, Faulkner:1985rm, Gould:1989hm, Lopes:2001ra, Lopes:2002gp, Bottino:2002pd, Lopes:2012af, Lopes:2013xua, Lopes:2014aoa, Geytenbeek:2016nfg, Frandsen:2010yj, Cumberbatch:2010hh, Taoso:2010tg, Vincent:2014jia, Vincent:2015gqa, Vincent:2016dcp}). Main sequence stars can however be used to search for dark matter energy injection, provided they reside in environments with extreme dark matter enhancements. One such group of stars, called S-Stars, have had their spectral properties measured by VLT and Keck while orbiting Sgr A*, and encounter densities at least seven orders of magnitude larger than the local position. In this case, the dark matter energy injection can be so extreme that its evolution is disrupted~\cite{John:2023knt}. The ``S2 Star" limit in Fig.~\ref{fig:heating} is set by observing that the S-star S2 in the Galactic center has not had its evolution disrupted by dark matter energy injection, which would change its measured luminosity by VLT and Keck~\cite{John:2023knt}. We show S2 rather than other S-Stars, as it is the most robust limit of those previously considered~\cite{John:2023knt}.

We see that the Fig.~\ref{fig:heating} limit from S2 reaches further into some of the parameter space than the Earth. It reaches lower dark matter masses despite its higher core temperature, due to its much higher density and consequent escape velocity more efficiently trapping lighter dark matter particles. However, compared to the brown dwarf or Jupiters sensitivities, S2 does not reach as far into the light dark matter space for pure contact interactions, and instead requires the evaporation barrier to retain light particles as light as exoplanets can. Neglecting any coherent enhancement, as per the spin-dependent bounds, main sequence stars reach further into the scattering cross section space than the Earth also due to their higher densities. Note that for the spin-independent bounds, any potential coherent enhancement has been neglected for the S2 star.

\subsubsection*{\centering Earth}

The Earth heat flow bounds in Fig.~\ref{fig:heating} are set using borehole measurements as discussed earlier. The main strength of the Earth search is that it is a local measurement of a very well-studied object, making the results fairly robust. However, we see that in the cross section reach, compared to other objects probing dark matter energy injection, the Earth generally provides the weaker constraints. Referencing the optimal bodies discussion above, this is due to the Earth being relatively small in radius, such that its capture rate is smaller than other objects, though its heavy elements such as iron can give its rates a boost for the spin-independent scattering case. The Earth also is not as dense as other objects, and so has a lower escape velocity, such that lighter dark matter particles are not as easily retained. The low core temperature of the Earth does not overcome the fact that the escape velocity is so low. Therefore, we see that the Earth's sensitivity does not reach to masses lower than other objects, even in the contact interaction case.

\section{Dark Matter Standard Model Particle Production Searches}
\label{sec:prod}

We now consider classes of models and scenarios where the SM annihilation products are detectable directly, in contrast to the energy injection scenarios discussed above.

\subsection{Standard Model Particle Flux}

We can evaluate the flux of SM particles incident at Earth given by~\cite{Leane:2017vag}
\begin{equation}
E^{2} \frac{d \Phi}{d E}=\frac{\Gamma}{4 \pi D^{2}} \times E^{2} \frac{d N}{d E} \times  \mathbb{P}_{\mathrm{surv}} \times \text{BR}(\phi \rightarrow \text{SM}),    
\label{eq:DM-gamma-ray-flux}
\end{equation}
where $\Gamma=C/2$ is the annihilation rate when annihilation is in equilibrium with capture $C$, $D$ is the distance from the body to the Earth, $E^2 dN/dE$ is the energy spectrum, $\text{BR}(\phi \rightarrow \text{SM})$ is the branching ratio of the mediator to SM particles, and $\mathbb{P}_{\mathrm{surv}}$ is the probability for the particles to survive to the telescope, which we will assume for now to be one as per the common assumption of the literature; we will relax this assumption in the upcoming Sec.~\ref{sec:map}.

In energy-injection searches, dark matter annihilation deposits the energy inside the body, with the main detectable usually being a change in the luminosity. On the other hand, if the annihilation products survive outside the body, a spectrum $E^2 dN/dE$ of the SM final states can be detected. This spectrum is a distribution of energies with a shape characteristic of the particular SM final states, depending on the particle physics model of interest. 

To investigate a wide range of particle physics models, we simulate using \texttt{Pythia}~\cite{Sjostrand:2014zea} the spectrum of dark matter annihilating to two mediators which then decay to a pair of SM particles: $\gamma \gamma$, $e^+e^-$, $\mu^+\mu^-$, $\tau^+\tau^-$, $b\overline{b}$, or $q\overline{q}$. These are then showered and hadronized, such that we produce the fully decayed spectra in vacuum.
As we describe shortly, we will mostly focus on the most promising detection channel, gamma rays. 
This means that the SM final states above decay and hadronize to produce a detectable gamma-ray spectrum. 
To broadly probe a wide range of mediator masses, we note that sensitivities are independent of the mediator when the limit is set with the spectrum's peak when the mediator is boosted~\cite{Leane:2017vag}. 
We assume the mediator mass is either $m_\phi = 0.9 m_\chi$ or $0.1 m_\chi$ and present the best case to highlight where potential detectability exits; when $m_\phi = 0.1 m_\chi$ would not kinematically allow the SM final state of interest to be produced, the $m_\phi = 0.9 m_\chi$ case is considered.

\subsection{Telescopes and Detection Prospects}

\subsubsection{Gamma Rays}

Gamma rays propagate directly from the source, and as they couple to electromagnetism can be straightforward to detect with telescopes. Generally, gamma rays provide the best sensitivity for celestial-body searches for dark matter annihilation to almost any SM final state.

A key benefit to studying celestial bodies in gamma rays is the wealth of existing telescopes and datasets that can test outcomes \textit{now}. Existing telescopes we consider for dark matter searches include Fermi Gamma-Ray Space Telescope (Fermi)~\cite{Tang:2018wqp, Malyshev:2015hqa, Fermi-LAT:2011nwz}, the High Energy Stereoscopic System (H.E.S.S.)~\cite{Aharonian:2009zk}, and the High-Altitude Water Cherenkov Observatory (HAWC)~\cite{Albert:2018jwh}. Fermi detects gamma rays through pair conversion and sits in low-Earth orbit observing the whole gamma-ray sky. The Large Area Telescope (LAT) instrument on Fermi is sensitive to gamma-ray energies from around 10 MeV to 300 GeV. H.E.S.S. is an Imaging Atmospheric Cherenkov telescope that detects Cherenkov radiation due to gamma rays in the atmosphere with energies 100 GeV to 100 TeV. Its location in the southern hemisphere gives it excellent sensitivity to the Galactic center. However, it cannot observe solar gamma rays because the Sun is too bright in the ultraviolet for this detection approach~\cite{Maggio:2021pca}. HAWC is a water Cherenkov telescope that detects Cherenkov radiation from charged particles in water tanks, sourced from gamma rays in the atmosphere with energies of 500 GeV to 100 TeV. This technique of detecting Cherenkov radiation from charged particles instead of directly imaging atmospheric Cherenkov radiation allows it to observe Solar gamma rays. It is located in the northern hemisphere, making it sub-optimal for observing the Galactic center. 

For almost all objects, we set limits on dark matter sourced fluxes by requiring that the dark matter annihilation flux is less than the observed flux for the given search in a given energy bin. This flux corresponds to a limit on the dark matter-SM scattering cross section, through the capture rate $C$ in Eq.~(\ref{eq:DM-gamma-ray-flux}), which is detailed in the earlier Sec.~\ref{sec:review}. For Jupiter, our limits are set differently, and we instead use the recent measurement of Jupiter in gamma rays~\cite{Leane:2021tjj} to set limits at a 95\% confidence level, as the likelihoods are available. We calculate the gamma-ray flux from dark matter annihilation in each of sixty log-spaced bins in the energy range of 0.01 GeV to 10 GeV. We exclude the signal if the test statistic~\cite{Read:2002hq, Cowan:2010js}
\begin{equation}
    t_{\sigma_{\chi N}} = -2\ln\left[\frac{L(\sigma_{\chi N})}{L(\sigma_{\chi N} = 0)}\right]
\end{equation}
is larger than 3.84, corresponding to a 95\% confidence level limit on the dark matter-SM scattering cross section. $L(\sigma_{\chi N} = 0)$ is the background likelihood and $L(\sigma_{\chi N})$ is the signal plus background likelihood. We follow Ref.~\cite{Leane:2021tjj} with the bin's background flux equal to zero if the maximum likelihood flux is negative.

\subsubsection{Neutrinos}

Neutrinos produced in the core of celestial bodies, and those produced outside celestial bodies from boosted or long-lived mediators, have been searched for in the past. Telescopes which can detect such signals include Super-Kamiokande (Super-K)~\cite{Super-Kamiokande:2015xms}, Hyper-Kamiokande (Hyper-K)~\cite{Abe:2011ts}, ANTARES~\cite{ANTARES:2016xuh,ANTARES:2016bxz}, and IceCube~\cite{IceCube:2016yoy,IceCube:2020wxa}. While neutrinos are an interesting final state to study, they in almost all cases provide weaker sensitivities than those found with gamma-ray telescopes. The only exception to this is if dark matter produces neutrino lines, such that a sharp box spectrum is produced. However, generally due to gauge invariance particles coupled to neutrinos also couple to charged leptons, which would also lead to either a heating or gamma-ray signal. Therefore, given the inferior sensitivity to SM final states, we do not show the weaker neutrino bounds on our plots but refer the interested reader to $e.g.$ Refs.~\cite{Super-Kamiokande:2015xms, Abe:2011ts, ANTARES:2016xuh,ANTARES:2016bxz,IceCube:2016yoy,IceCube:2020wxa}.

\subsubsection{Electrons and Positrons}

Electrons and positrons can be detected with cosmic-ray telescopes, such as the Alpha Magnetic Spectrometer (AMS)~\cite{AMS:2021nhj} or Dark Matter Particle Explorer (DAMPE)~\cite{DAMPE:2017fbg}. A significant difficulty in using electrons or positrons directly is that they are deflected in magnetic fields, generally weakening any potential sensitivity compared to direct gamma-ray searches. As searches using these telescopes yield weaker sensitivities than gamma-ray telescopes to all SM final states, we do not show the weaker direct electron/positron detection constraints on our plots, but refer the interested reader to $e.g.$ Refs.~\cite{Schuster:2009au, Schuster:2009fc, Feng:2016ijc, FermiLAT:2011ozd, Li:2022wix, DAMPE:2017fbg, AMS:2021nhj}.

One interesting exception to this scenario is for dark matter induced $H_3^+$ production~\cite{Blanco:2023qgi}, which involves electrons/positrons but does not involve detecting them directly. It has been shown that charged particles can ionize molecular hydrogen in planetary atmospheres, creating an excess of $H_3^+$ which can be detected with fly-by data or $e.g.$ JWST~\cite{Blanco:2023qgi}. This may be the most promising detection channel in some cases, and so we will include this in our parameter space plots.

\subsubsection{New Limits, Interplay, and Assumptions}

\begin{figure*}[t!]
\centering
\includegraphics[width=0.9\linewidth]{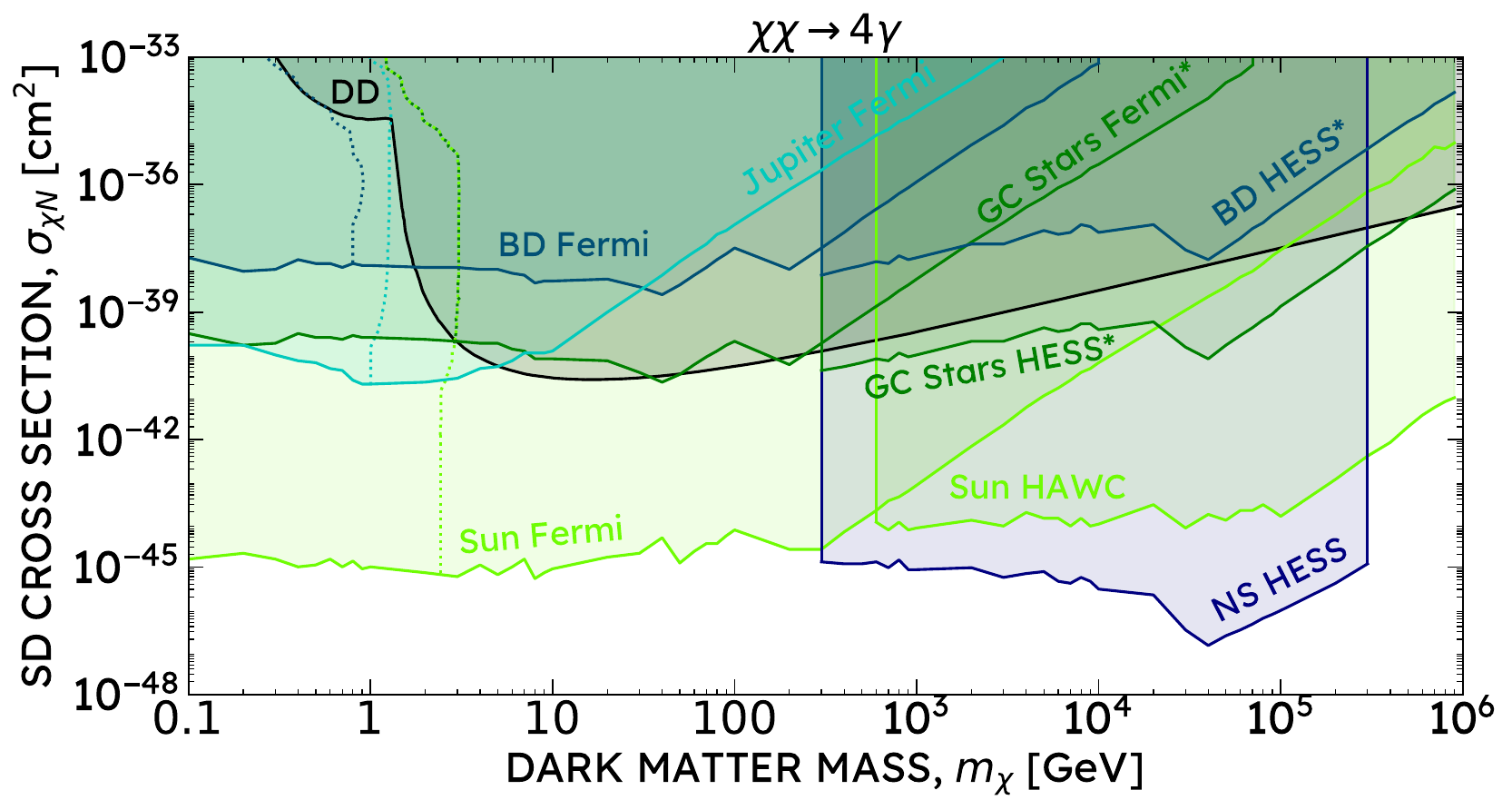} \\
\includegraphics[width=0.9\linewidth]{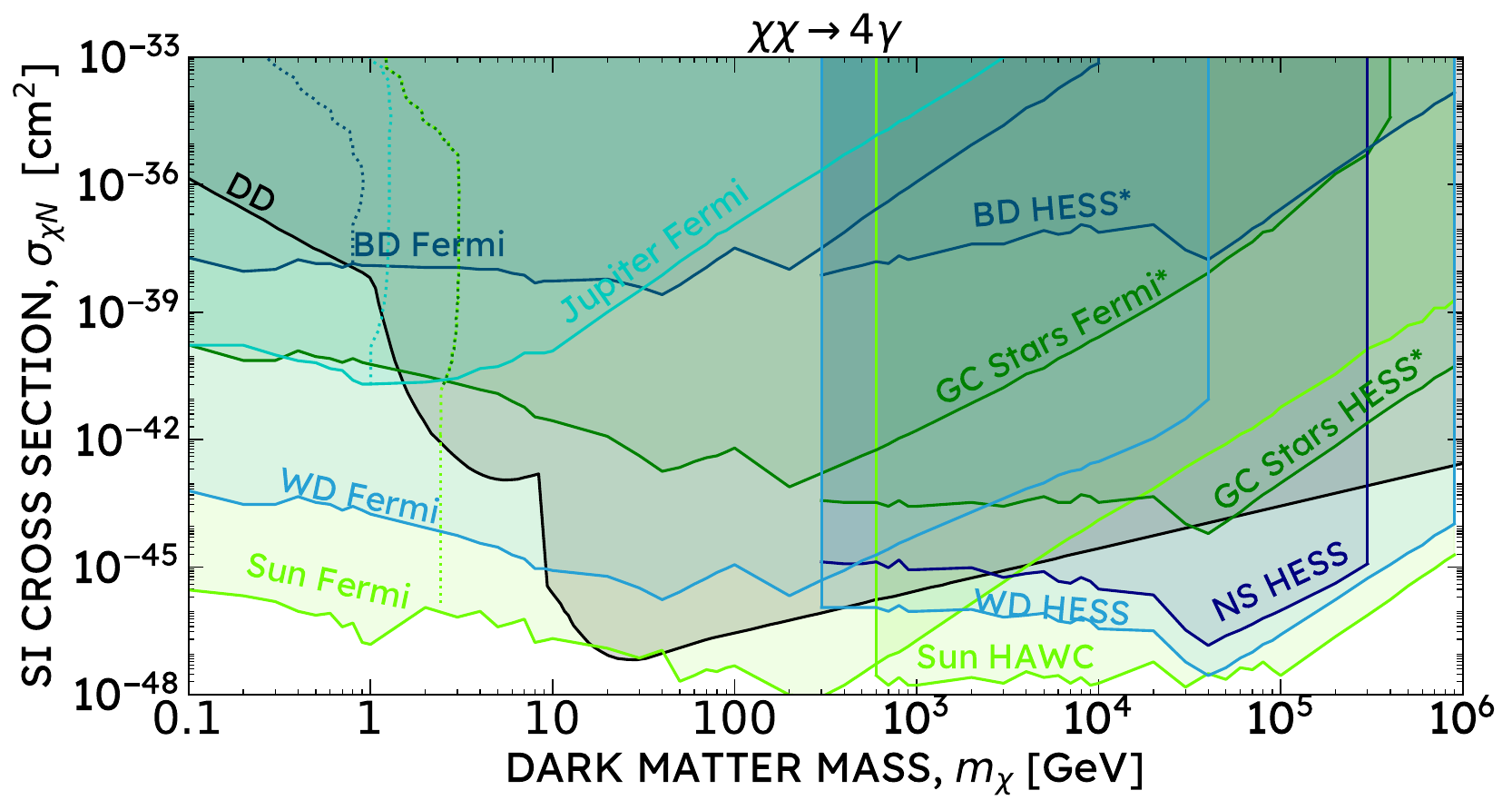}
\caption{Constraints on dark matter-SM scattering cross section, assuming dark matter annihilates to gamma rays. Limits are shown for celestial bodies alongside the utilized dataset as labeled, dotted lines indicate evaporation for contact interactions. Asterisk indicates new results in this work.}
\label{fig:cross-section-gammas}
\end{figure*}

\begin{figure*}[t!]
\centering
\includegraphics[width=0.9\linewidth]{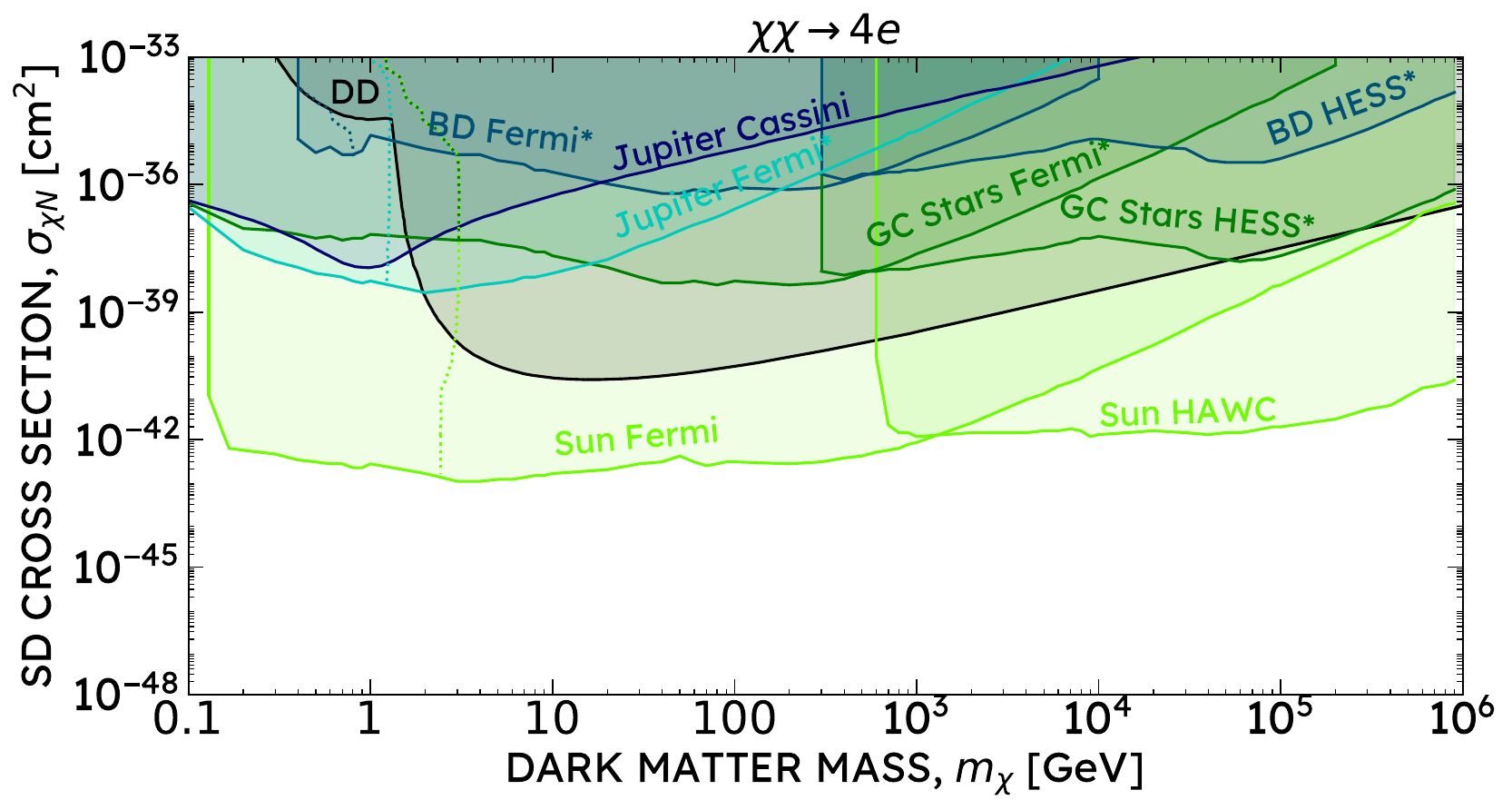} \\
\includegraphics[width=0.9\linewidth]{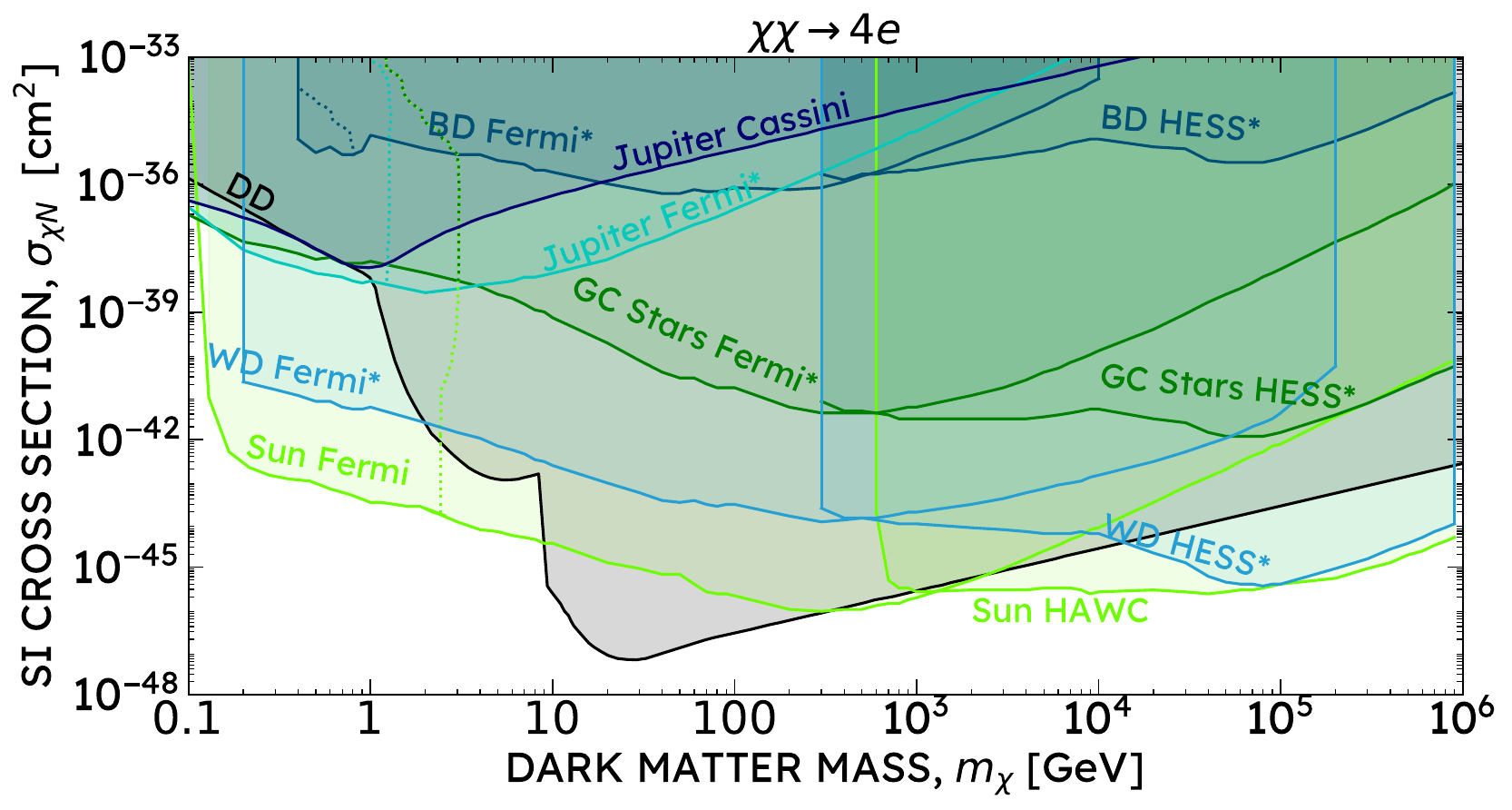}
\caption{Constraints on dark matter-SM scattering cross section, assuming dark matter annihilates to electrons. Limits are shown for celestial bodies alongside the utilized dataset as labeled, dotted lines indicate evaporation for contact interactions. Asterisk indicates new results in this work.}
\label{fig:cross-section-e}
\end{figure*}

\begin{figure*}[t!]
\centering
\includegraphics[width=0.9\linewidth]{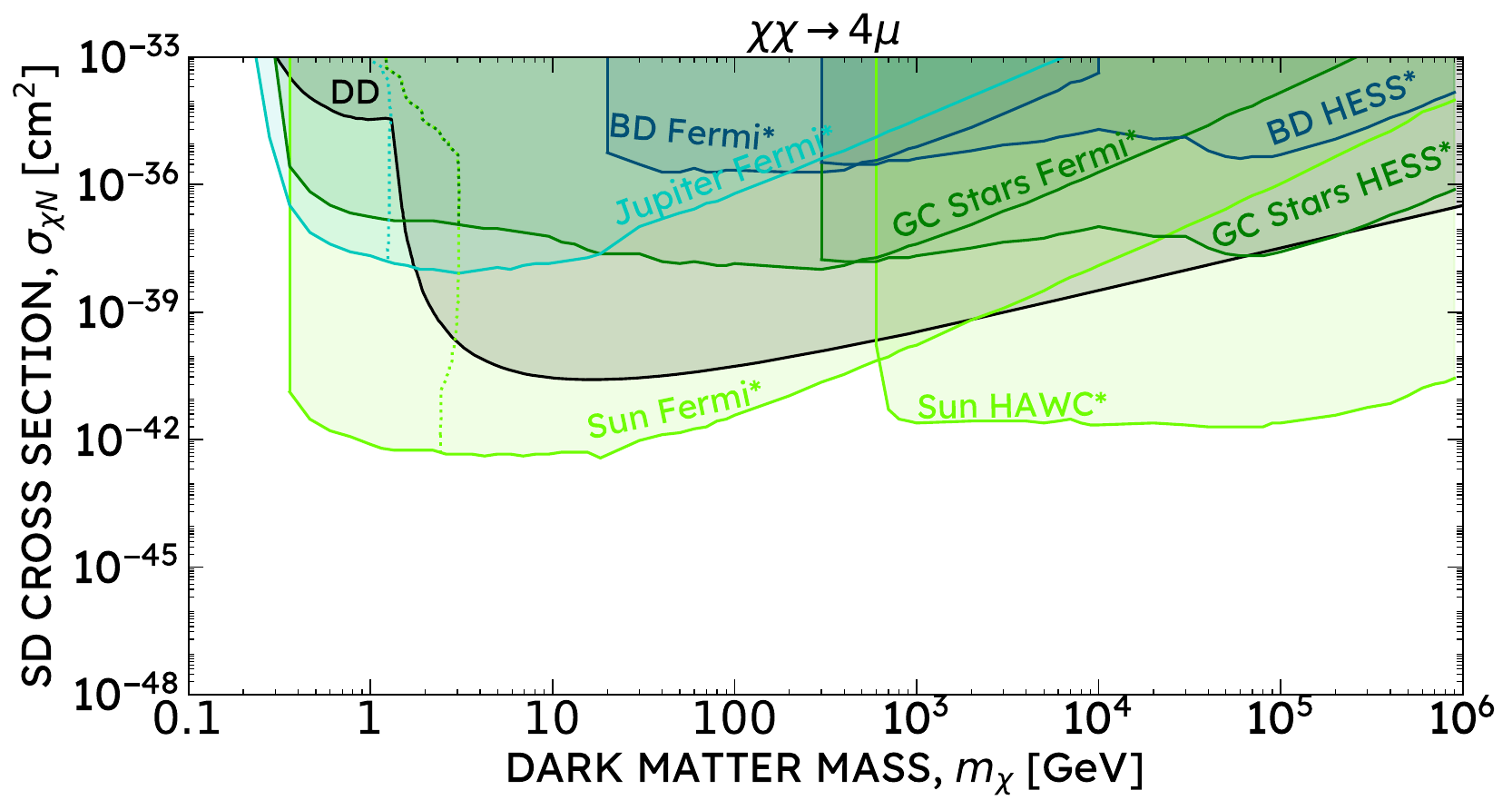} \\
\includegraphics[width=0.9\linewidth]{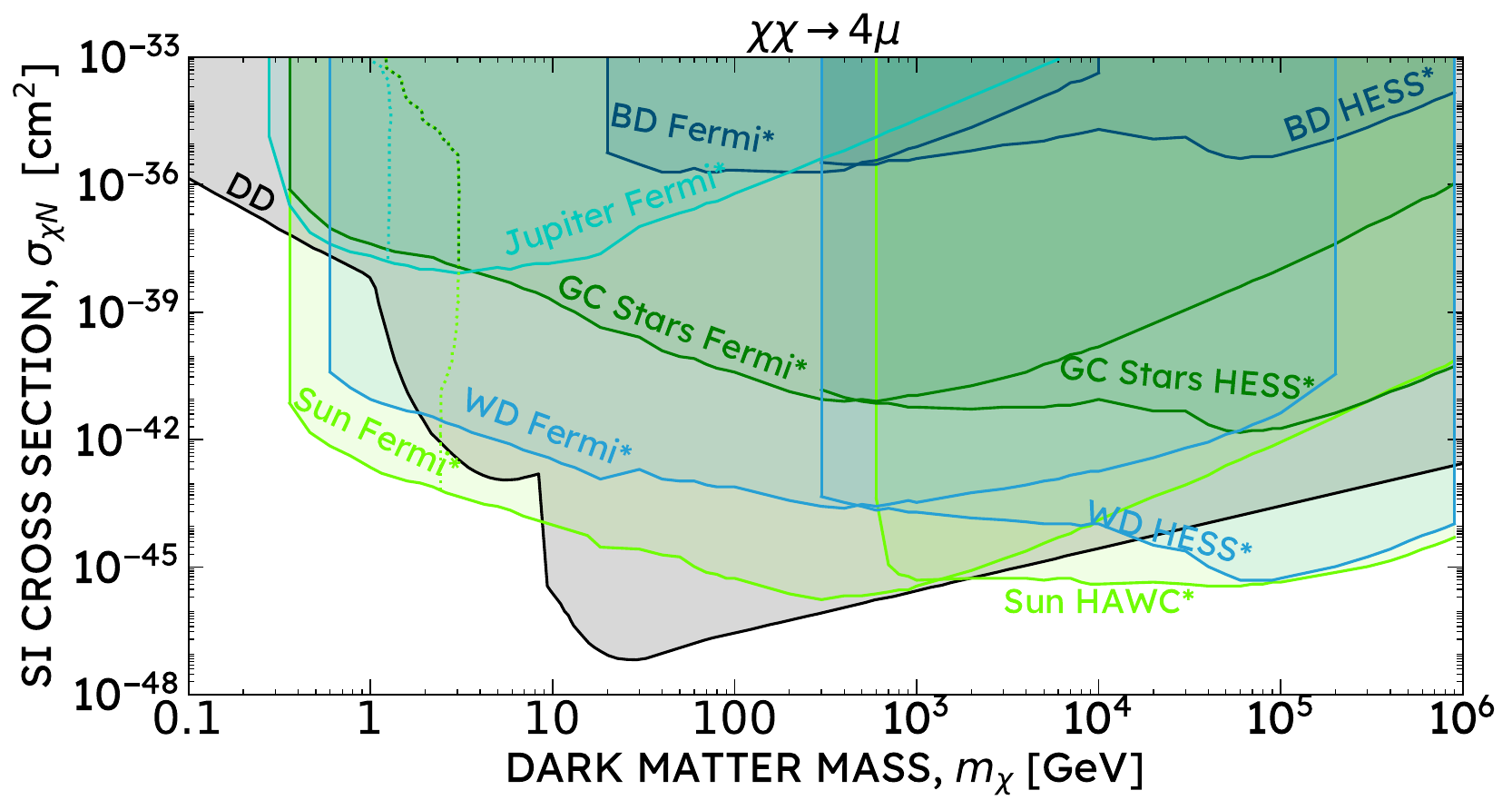}
\caption{Constraints on dark matter-SM scattering cross section, assuming dark matter annihilates to muons. Limits are shown for celestial bodies alongside the utilized dataset as labeled, dotted lines indicate evaporation for contact interactions. Asterisk indicates new results in this work.}
\label{fig:cross-section-mu}
\end{figure*}

\begin{figure*}[t!]
\centering
\includegraphics[width=0.9\linewidth]{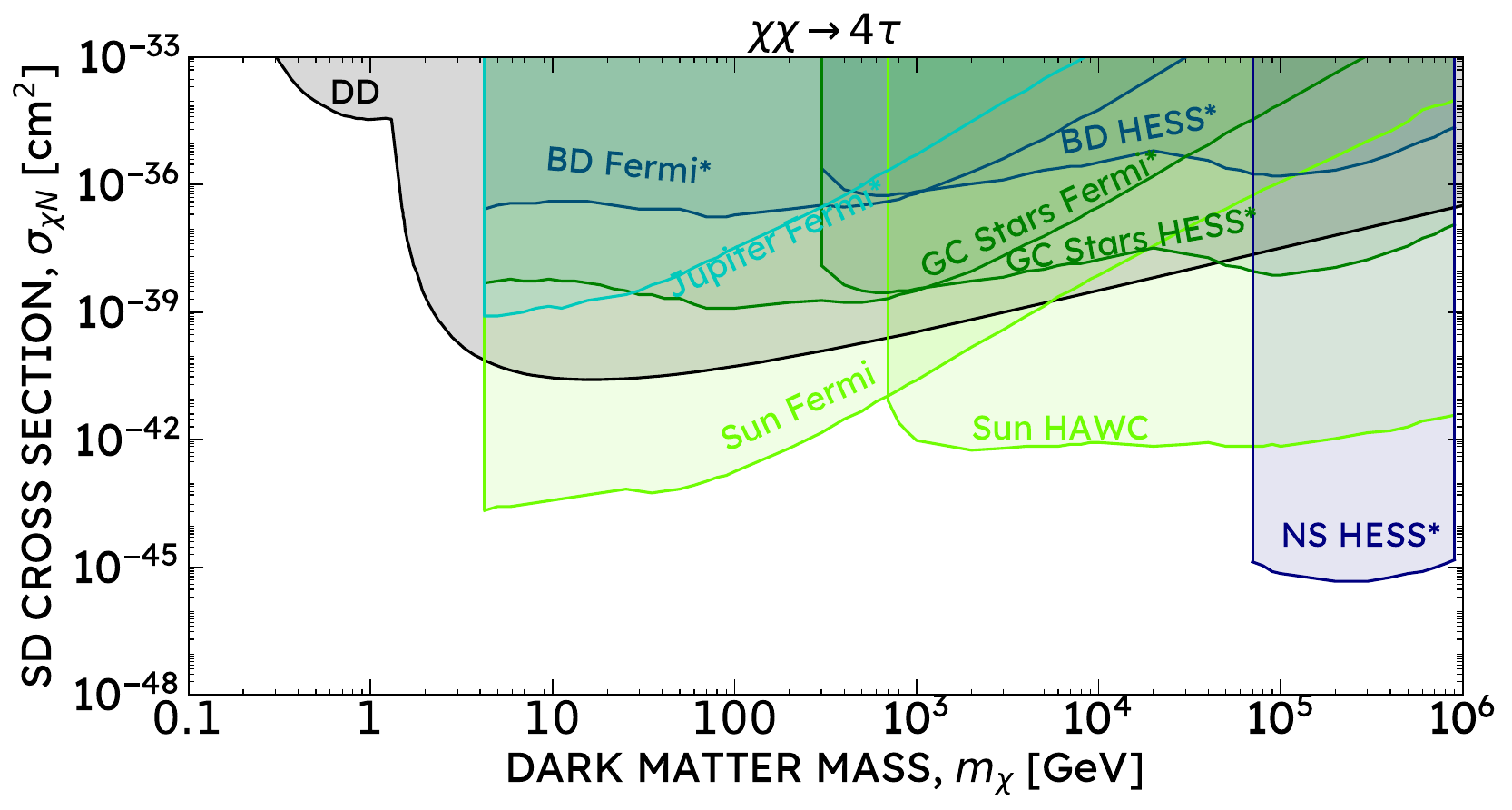} \\
\includegraphics[width=0.9\linewidth]{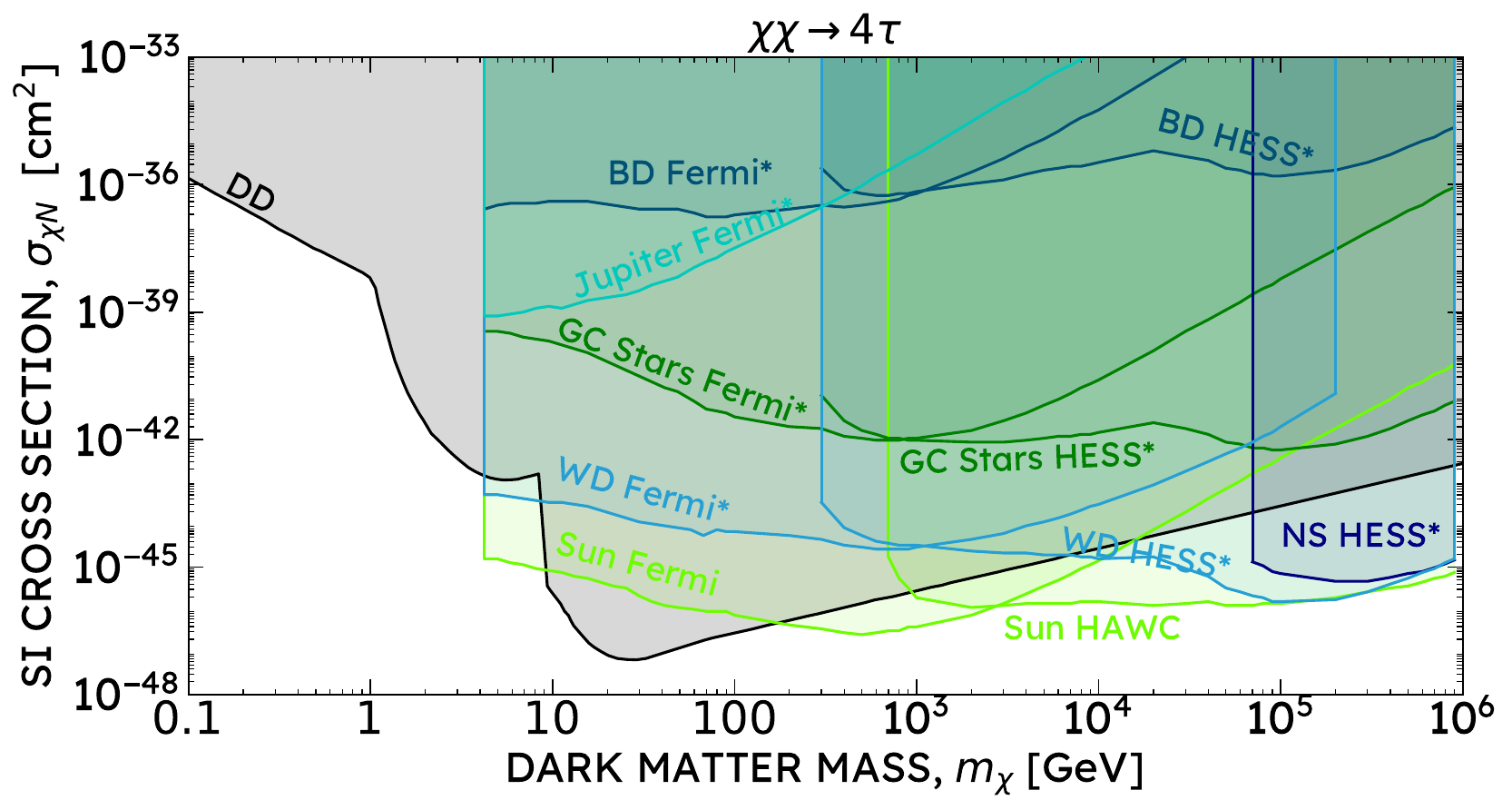}
\caption{Constraints on dark matter-SM scattering cross section, assuming dark matter annihilates to $\tau$ leptons. Limits are shown for celestial bodies alongside the utilized dataset as labeled, dotted lines indicate evaporation for contact interactions. Asterisk indicates new results in this work.}
\label{fig:cross-section-tau}
\end{figure*}

\begin{figure*}[t!]
\centering
\includegraphics[width=0.9\linewidth]{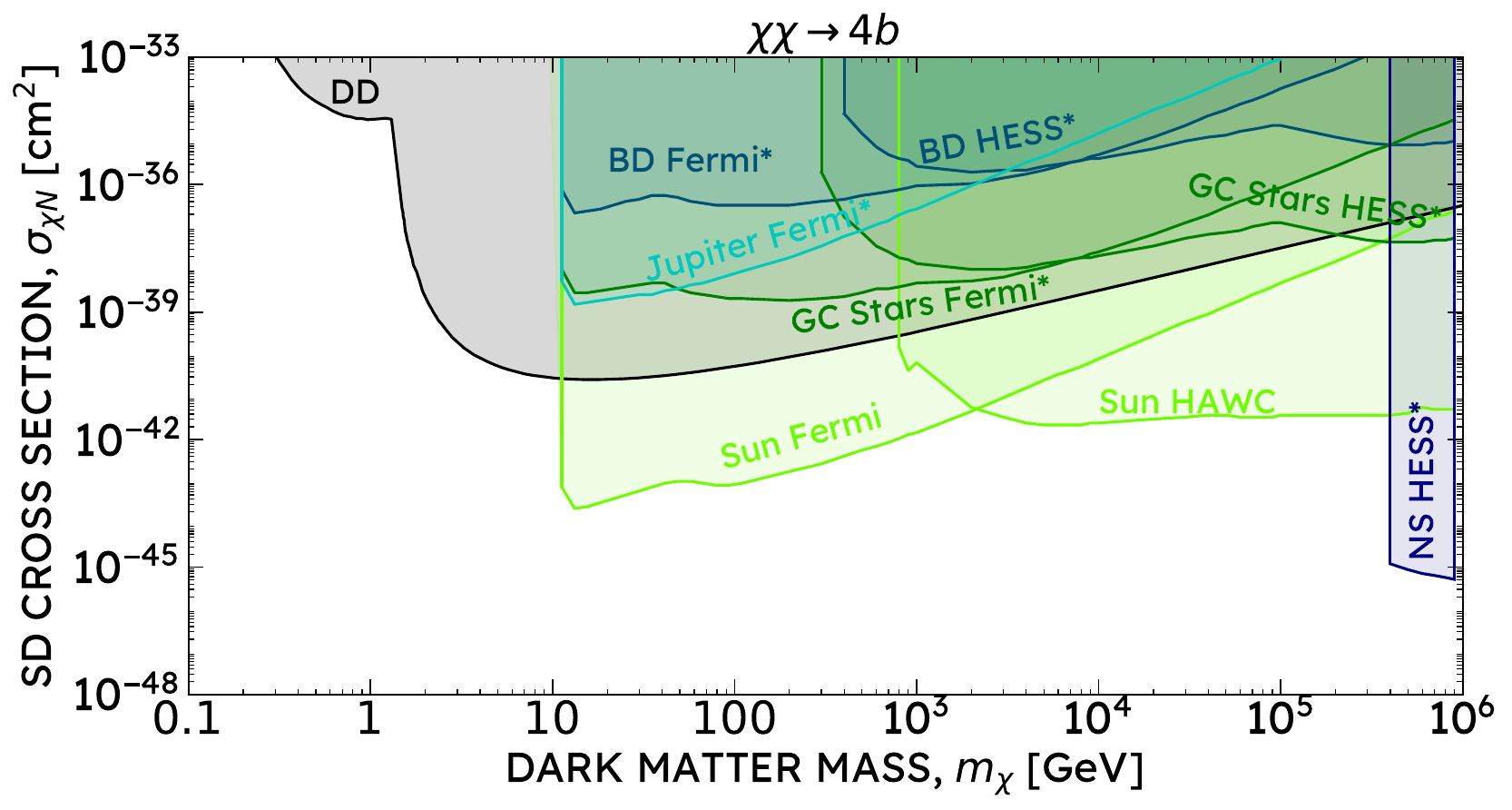} \\
\includegraphics[width=0.9\linewidth]{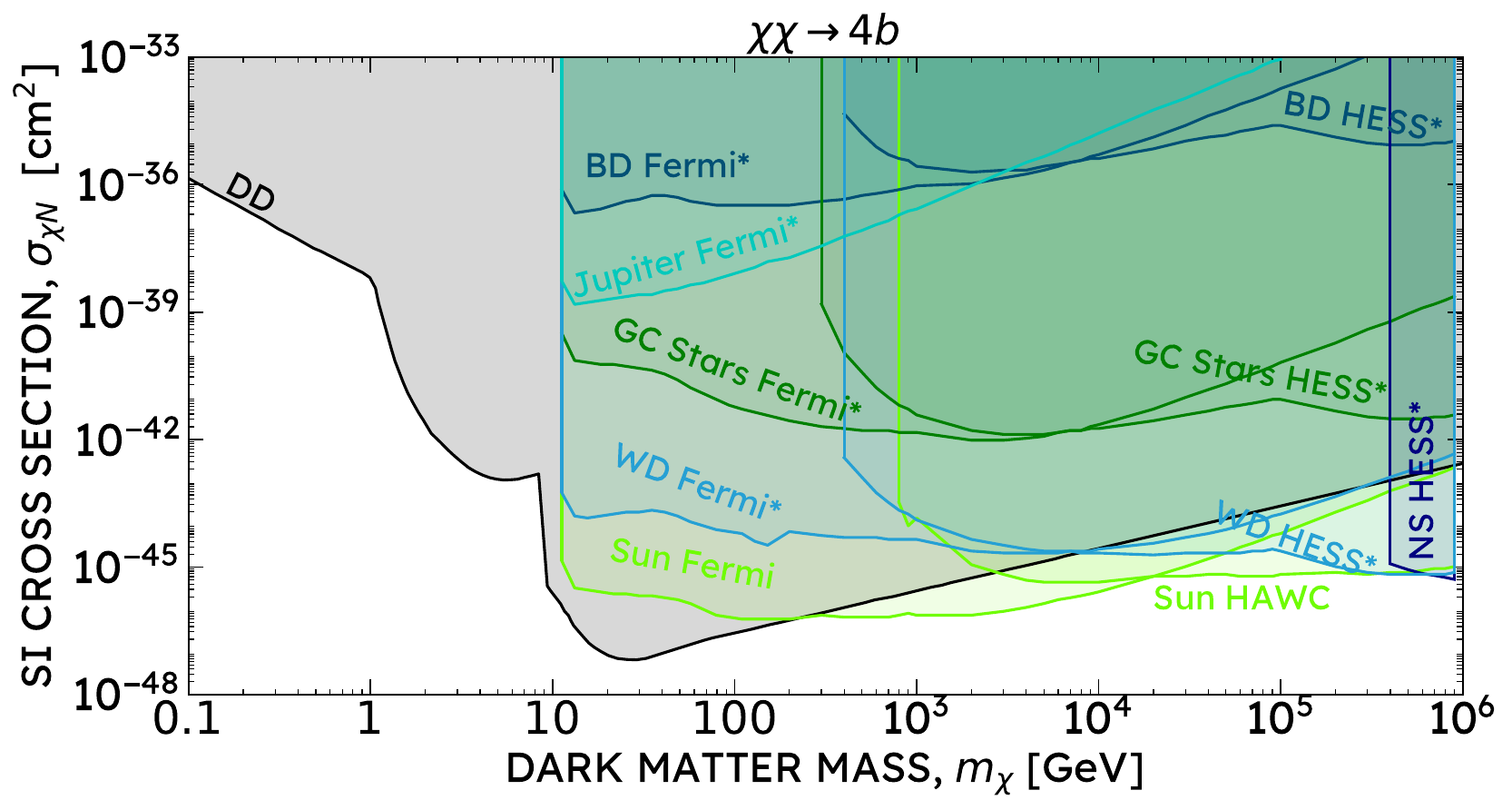}
\caption{Constraints on dark matter-SM scattering cross section, assuming dark matter annihilates to $b$ quarks. Limits are shown for celestial bodies alongside the utilized dataset as labeled, dotted lines indicate evaporation for contact interactions. Asterisk indicates new results in this work.}
\label{fig:cross-section-b}
\end{figure*}

\begin{figure*}[t!]
\centering
\includegraphics[width=0.9\linewidth]{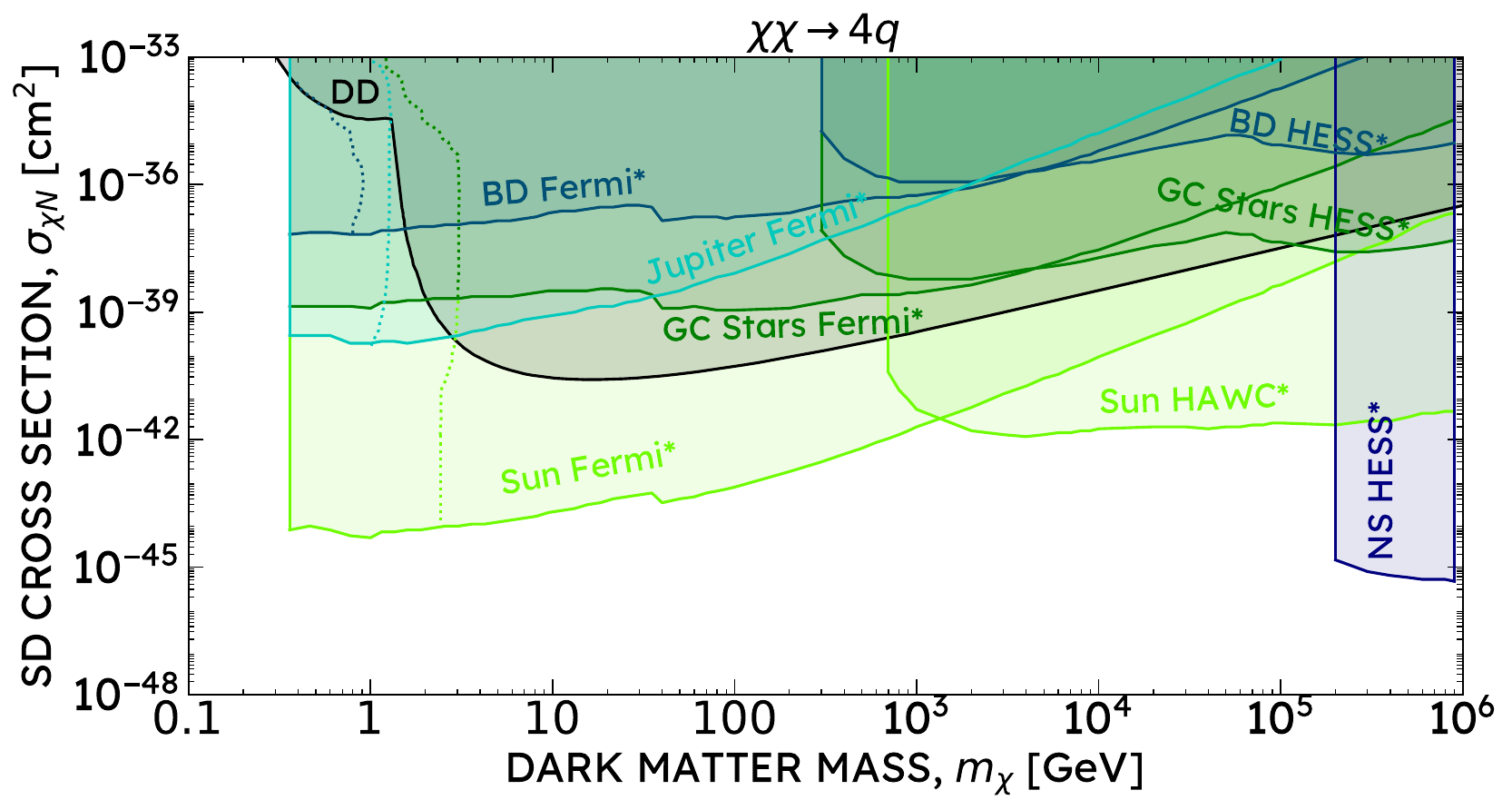} \\
\includegraphics[width=0.9\linewidth]{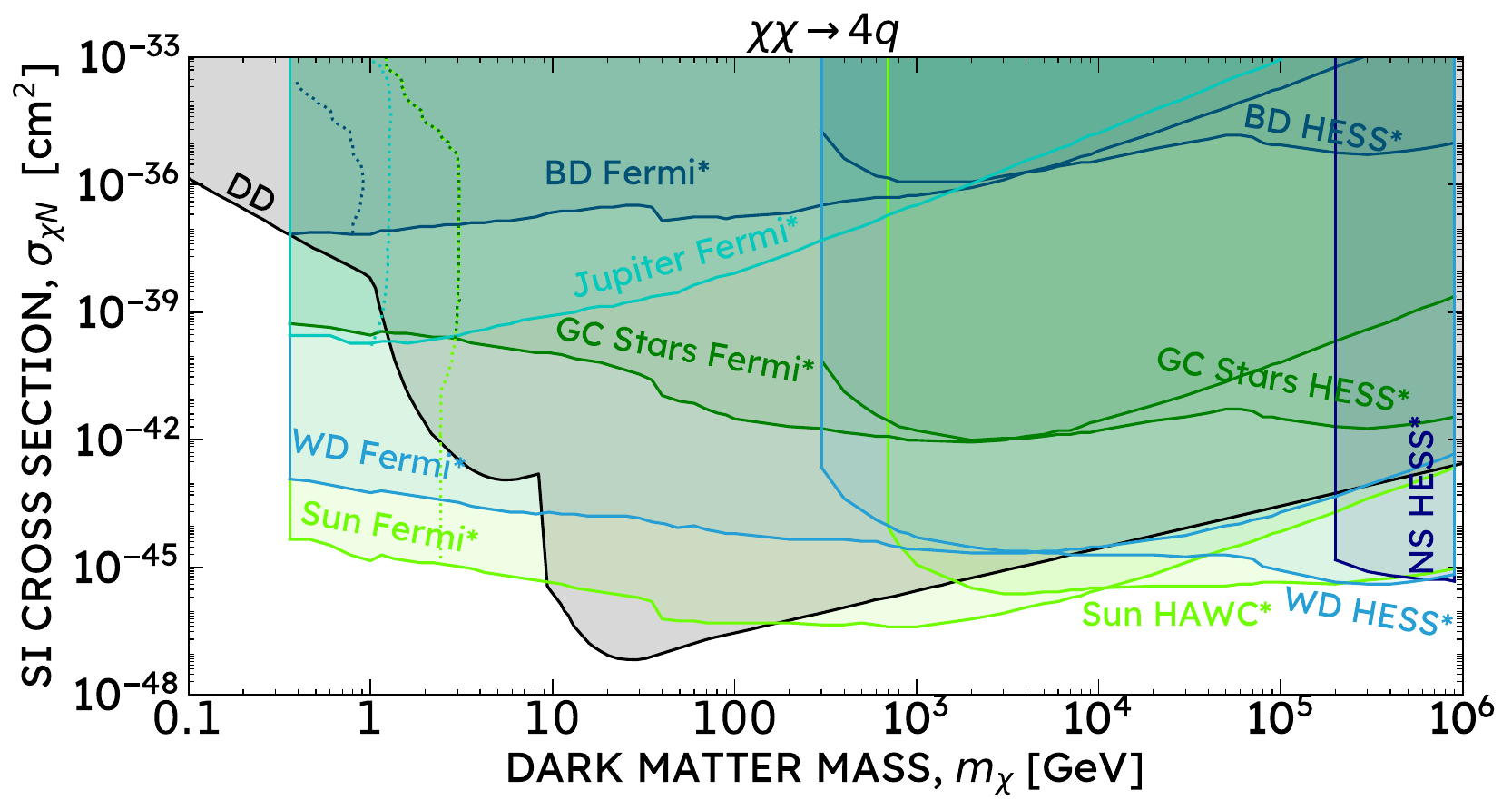}
\caption{Constraints on dark matter-SM scattering cross section, assuming dark matter annihilates to light quarks ($q=u,d,s,c$). Limits are shown for celestial bodies alongside the utilized dataset as labeled, dotted lines indicate evaporation for contact interactions. Asterisk indicates new results in this work.}
\label{fig:cross-section-q}
\end{figure*}

Figures~\ref{fig:cross-section-gammas}, \ref{fig:cross-section-e}, \ref{fig:cross-section-mu}, \ref{fig:cross-section-tau}, \ref{fig:cross-section-b}, \ref{fig:cross-section-q}, show the spin-independent and spin-dependent constraints on annihilation to final states $\gamma$, electrons, muons, taus, $b$-quarks and light quarks (up, down, charm, strange) respectively, for a range of celestial bodies including the Sun~\cite{Batell:2009zp, Leane:2017vag}, Jupiter~\cite{Leane:2021tjj,Blanco:2023qgi}, brown dwarfs~\cite{Leane:2021ihh}, white dwarfs~\cite{Acevedo:2023xnu}, neutron stars~\cite{Leane:2021ihh}, and nuclear burning Galactic center stars (``GC Stars"). We have calculated multiple new limits in these plots; our analyses with new limits are indicated with an asterisk. In all these plots, the limits are set under the assumption that all of the annihilation events escape the celestial body, and that the probability of detecting these final states with the telescope as labeled is approximately one. These are the standard assumptions usually made in the literature for SM final state searches; we will show in the next section how these results change with differing assumptions. The dotted lines show where dark matter would be evaporated for only contact interactions, the left of the dotted lines can be probed if long-range interactions are present. 
We also show for comparison existing direct detection experiments which also probe the dark matter-nucleon scattering rate, with the strongest limits arising from Collar~\cite{Collar:2018ydf}, CRESST~\cite{CRESST:2022dtl}, PICO~\cite{PICO:2019vsc}, LZ~\cite{LZ:2022lsv}, and DarkSide-50~\cite{Franco:2023sjx}. 

Across Figs.~\ref{fig:cross-section-gammas}, \ref{fig:cross-section-e}, \ref{fig:cross-section-mu}, \ref{fig:cross-section-tau}, \ref{fig:cross-section-b}, \ref{fig:cross-section-q}, we see that direct decays into gamma rays provide the strongest constraints. The reason for this is, in addition to the supreme sensitivity of gamma-ray observatories, the energy spectrum for gamma rays using a gamma-ray telescope is a sharp box spectrum, such that the annihilation events are sharply peaked in a leading energy bin close to the dark matter mass. For other final states, the spectra are softened, with significant energy spread across a wider range of energy bins. At larger dark matter masses than the telescope's maximum energy sensitivity, the limits weaken for each final state as the tail end of the spectrum sets the limit. Note that for $\tau$ leptons and $b$ quarks, their telescope sensitivities are limited by kinematically being able to produce them, rather than the telescope sensitivity itself.

Across Figs.~\ref{fig:cross-section-gammas}, \ref{fig:cross-section-e}, \ref{fig:cross-section-mu}, \ref{fig:cross-section-tau}, \ref{fig:cross-section-b}, \ref{fig:cross-section-q}, we also see as expected that spin-independent interactions provide stronger constraints than spin-dependent interactions. 
This is because coherent enhancement of dark matter-SM scattering leads to larger scattering rates, which can be larger than orders of magnitude in some cases (see Eq.~(\ref{eq:SI-cross-section-scaling})). 
The shapes of the constraints, however, are similar regardless of spin-dependent or independent interactions.
Note that for white dwarfs, as they are dominantly made of either oxygen or carbon, only spin-independent constraints are expected due to the even spin of their target nuclei.

We now discuss the results of these figures and the interpretation of each of these searches one by one.

\subsubsection*{\centering The Sun}

We begin with discussing the solar SM final state results, as according to Figs.~\ref{fig:cross-section-gammas}, \ref{fig:cross-section-e}, \ref{fig:cross-section-mu}, \ref{fig:cross-section-tau}, \ref{fig:cross-section-b}, \ref{fig:cross-section-q} they appear to generally give the strongest constraints. 

The reason for the power of the solar constraints is that the sensitivity of the Fermi telescope and HAWC Observatory to solar gamma rays is so powerful, combined with the fact that the expected dark matter flux from the Sun can be many orders of magnitude above the solar gamma-ray flux already observed. The underlying reason for this is that the Sun is so close by. This means that, while the cross section for the Sun corresponding to effectively all the dark matter being captured is around $10^{-35}~$cm$^2$ or so, cross sections substantially smaller than this can be probed, as only very small dark matter fluxes are required for potential detection. Further, because the Sun has such a large radius, the maximum capture rate and therefore maximum solar dark matter flux, is much higher than other objects, such that the ``starting point" of a dark matter flux at the maximum capture cross section of around $10^{-35}~$cm$^2$ is already extremely large. (In contrast, a neutron star which is already maxed out in its capture rate at this cross section, is multiple orders of magnitude lower in its corresponding capture rate due to being so small.)
Overall, this means that cross sections substantially smaller than its maximum capture cross section can still be probed, leading to excellent sensitivity down to very small cross sections. The Sun is a perfect example of how both the intrinsic properties of the celestial body, as well as the telescope detectability, are vital for considering an optimal search strategy. When it comes to detecting SM final states, other objects that may be naively better such as neutron stars are clearly not very optimal.

\subsubsection*{\centering Neutron Stars}

Compared to the Sun, neutron stars are much further away, suffering from flux suppression. This means that SM final states from a single neutron star will not generally be detectable. The signal is substantially increased if we instead consider the full population of neutron stars in the Galactic center, as first pointed out in Ref.~\cite{Leane:2021ihh}. Thanks to their high density, their maximum capture rate occurs at exceptionally low cross sections, down to as low as about $10^{-45}$~cm$^2$. However, as neutron stars are so small, the maximum flux possible from a neutron star is very small compared to other objects. This means that for the Galactic center gamma-ray search, their flux is only just above the detection threshold when the capture rate is at a maximum, and even then only with the most optimal energy spectrum (direct decay to gamma rays). Therefore, in Fig.~\ref{fig:cross-section-gammas}, we see that strong sensitivity to dark matter final states from neutron stars is possible if dark matter decays directly into gamma rays. However, as the flux is only just above the detection threshold, any other final state drops below the detection threshold, and cannot be improved by simply increasing the cross section, because the maximum capture rate is already required to get any detection ($i.e.$ increasing the cross section above the maximum does not lead to any additional dark matter being significantly captured). This is why the ``NS H.E.S.S." limit does not appear on the other final state plots. This is also why only the H.E.S.S. data can constrain dark matter sourced gamma rays in neutron stars; in the TeV energy range the gamma-ray background in the Galactic center is lower than in the GeV energy range; once focusing on GeV-scale gamma rays with Fermi, the Galactic center background is too large for neutron stars to overcome it.

Additional assumptions about the neutron star search include the dark matter density distribution, as well as the neutron star density distribution. The dark matter density has the largest uncertainty and can vary drastically by several orders of magnitude depending on the distribution (see earlier discussion about dark matter density profiles). In fact, because the neutron star search requires maximal dark matter capture rates, using a profile other than the cuspy gNFW does not lead to any constraint on the parameter space~\cite{Leane:2021ihh}.

\subsubsection*{\centering White Dwarfs}

In Figs.~\ref{fig:cross-section-gammas}, \ref{fig:cross-section-e}, \ref{fig:cross-section-mu}, \ref{fig:cross-section-tau}, \ref{fig:cross-section-b}, \ref{fig:cross-section-q}, we see that white dwarfs appear to be generally the next strongest for probing SM final states, after the solar searches. 
White dwarfs are ideal because of the interplay of their low core temperatures and high densities leading to lower dark matter masses being retained compared to the Sun. They are furthermore more efficient at capture compared to the Sun due to their higher densities. However compared to the Sun, they are much further away, and require exploitation of the large Galactic center white dwarf population to enhance their rate. The white dwarfs generally perform better than the neutron star version of the similar Galactic center search, as white dwarfs are Earth-sized, their maximum capture rate is larger than the smaller neutron stars. This means that the white dwarf gamma-ray search does not drop below detection thresholds for a wide range of dark matter masses, and is therefore detectable at both GeV-scales with Fermi, and TeV-scales with H.E.S.S.. Similarly, the white dwarfs do not drop below detection thresholds for softer spectra than direct decay into gamma rays, unlike the neutron star search. This search is a good example of not only considering the cross section corresponding to the maximum capture rate of an object when identifying what the ideal object or search strategy is -- white dwarfs have worse cross sections than neutron stars when considering saturation cross sections alone (about 3 or so orders of magnitude weaker), but they still outperform neutron stars at the same search for the reasons outlined above. Note however that white dwarf constraints only appear for the spin-independent limits due to the even spin of their targets. The heavier elements present in a white dwarf, however, lead to large enhancements to the spin-independent rate, which are not obtained for other objects. 

In addition, again for the Galactic center search, there are large uncertainties in the dark matter density profile, but due to the white dwarfs having more parameter space available to not drop below detection thresholds, more dark matter density profiles than just a cuspy gNFW can be probed (we only show a gNFW in the plots here for brevity, for other profile choices in the case of direct decays to gamma rays, see Ref.~\cite{Acevedo:2023xnu}).

\subsubsection*{\centering Brown Dwarfs}

In Figs.~\ref{fig:cross-section-gammas}, \ref{fig:cross-section-e}, \ref{fig:cross-section-mu}, \ref{fig:cross-section-tau}, \ref{fig:cross-section-b}, \ref{fig:cross-section-q}, results for final state searches from gamma-rays sourced by dark matter in the Galactic center population of brown dwarfs are shown. Compared to the Sun, the brown dwarf limits extend to lower dark matter masses, due to a more optimal competition between the cooler brown dwarf core and its escape velocity leading to brown dwarfs retaining lighter dark matter. While brown dwarfs have larger maximum capture rates than both white dwarfs and neutron stars, they are less dense, leading to higher cross sections corresponding to their maximum capture rate. As such, the bounds do not extend to as low cross sections as the white dwarfs or neutron stars (where the neutron star limits appear). As the brown dwarf number density population is lower than that expected for white dwarfs, this also contributes to the lower rates.

In addition, again for the Galactic center search, there are large uncertainties in the dark matter density profile, but like white dwarfs, as brown dwarfs have more parameter space available than neutron stars to not drop below detection thresholds, more dark matter density profiles than just a cuspy gNFW can be probed (we only show a gNFW in the plots here for brevity, for other profile choices in the case of direct decays to gamma rays, see Ref.~\cite{Leane:2021ihh}).

\subsubsection*{\centering Main Sequence Stars}

Figs.~\ref{fig:cross-section-gammas}, \ref{fig:cross-section-e}, \ref{fig:cross-section-mu}, \ref{fig:cross-section-tau}, \ref{fig:cross-section-b}, \ref{fig:cross-section-q}, also show results for final state searches from gamma-rays sourced by dark matter in the Galactic center population of nuclear-burning stars. This search has not been previously considered, and naively appears to be weaker than the solar gamma-ray search, with perhaps no added benefit. As mentioned above, all of the constraints in these figures assume the final states produced all make it to the telescope for detection. However, in reality, this statement is highly dependent on the mediator decay length and the particle physics model properties. We will therefore shortly investigate how these bounds change removing the standard assumptions of a survival probability of $\sim1$ in the literature, and show how this search can outperform the Sun.

\subsubsection*{\centering Jupiter}

Finally, we show in Figs.~\ref{fig:cross-section-gammas}, \ref{fig:cross-section-e}, \ref{fig:cross-section-mu}, \ref{fig:cross-section-tau}, \ref{fig:cross-section-b}, \ref{fig:cross-section-q} our results for SM final states produced in Jupiter, using Fermi telescope data. As the Fermi telescope's sensitivity to Jovian gamma rays is so high, even very small dark matter fluxes from Jupiter can be detected. In fact, the threshold sensitivity for Jovian dark matter fluxes is about six or so orders of magnitude below the maximum capture rate possible for Jupiter, leaving a wide range of cross sections testable with this search. Note that the Jovian gamma-ray limits in Ref.~\cite{Leane:2021tjj} differ by a constant factor due to the assumption there that the mediator decays at the surface, giving a survival probability factor $\sim2$ lower.

Compared to the Galactic center searches, the local dark matter conditions near Jupiter are fairly well known, and so the uncertainties in these constraints are much lower. Considering local objects, under the assumption that all of the SM products escape all objects of interest, Jupiter limits are always weaker than those expected from the Sun, due to Jupiter being smaller than the Sun and further away. However, Jupiter has a cooler core than the Sun, such that it retains lighter dark matter particles, and outperforms the Sun at lower dark matter masses~\cite{Leane:2021tjj}. In addition, as noted above, the assumption that the mediator escapes is not guaranteed across a more general parameter space; it could be that shorter decay lengths are more favorable to escape Jupiter but not the Sun (or a brown dwarf for that matter).  

For the electron final state plot in Fig.~\ref{fig:cross-section-e}, we also show limits from dark matter induced $H_3^+$ production using Cassini data on the night side of Jupiter~\cite{Blanco:2023qgi}. While any ionizing species will be efficient at producing this signal, we only show the original bounds as quoted from Ref.~\cite{Blanco:2023qgi}. Compared to the gamma-ray search with Fermi, there are different assumptions in the Cassini ionization search, as the ionization signal is not sensitive to the shape of the annihilation spectrum; all that is required is that the bulk of the ionizing energy is above hydrogen's ionization threshold of 13.6 eV. This means that different model classes may be more optimal for either search.

We now discuss a broader view of the dark matter discovery and exclusion space, considering variations in mediator decay lengths and properties, as well as the interplay of the SM final state searches discussed in this section with the energy injection searches discussed in the previous section.

\section{Mapping Out the Dark Matter Discovery and Exclusion Space}
\label{sec:map}

\subsection{Interplay of Survival Probability and Cross Section Reach}

One assumption commonly made in works studying celestial body searches for SM particles from dark matter annihilation is that the survival probability of these products reaching the telescope is of order unity. However, this is a highly model-dependent statement. This can be seen by writing the survival probability for the particles to survive to the telescope (which is related to the flux via Eq.~(\ref{eq:DM-gamma-ray-flux})),
\begin{equation}
\mathbb{P}_{\text{surv}} = e^{-R/L}-e^{-D/L},
\label{eq:Psurv}
\end{equation}
where $R$ is the radius of the body of interest, $D$ is the distance between the body and the telescope, and $L$ is the decay length of the mediator which is given by
\begin{equation}
    L = \gamma \tau,
\end{equation}
where $\gamma = m_\chi/m_\phi$ is the boost of the mediator and $\tau$ is the mediator's lifetime. Clearly, the decay length depends on both the boost of the mediator and its lifetime, such that fixing the chance the mediator decays before the telescope with probability $\sim 1$ implicitly fixes these parameters as well. Fixing the survival probability to be one also makes directly comparing multiple celestial-body searches difficult, as bodies have different radii $R$ (leading to different decay lengths escaping), as well as different positions in the Galaxy $D$ (leading to different decay lengths actually being detectable by the telescope).

\begin{figure}[t]
    \centering
    \includegraphics[width=0.65\columnwidth]{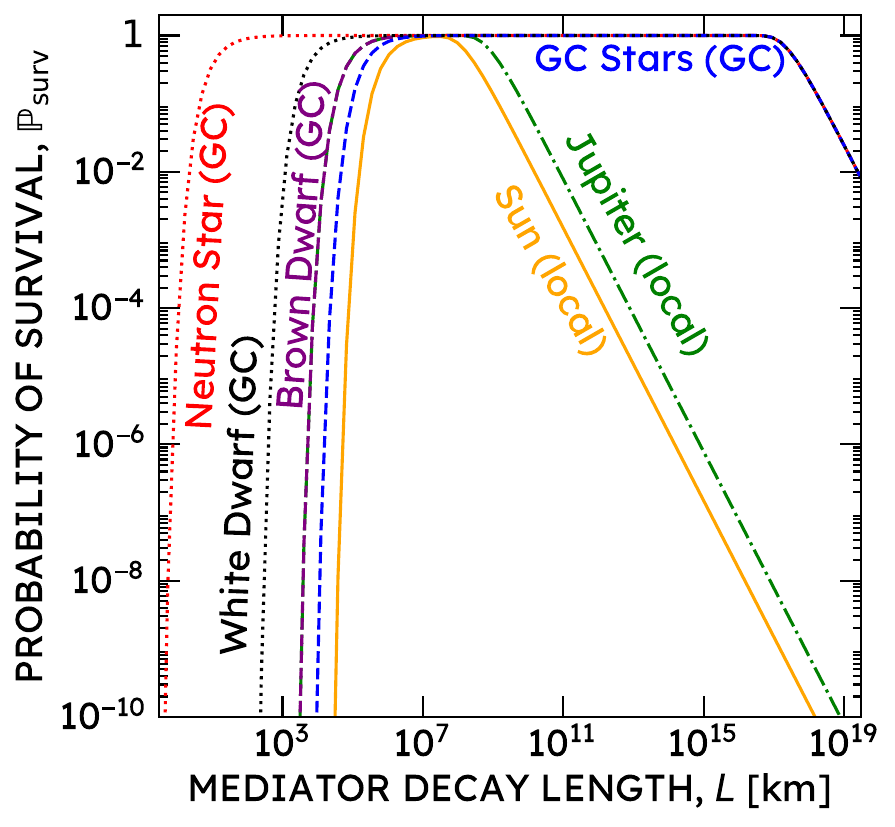}
    \caption{Survival probability $\mathbb{P}_\mathrm{surv}$ as a function of mediator decay length for a range of celestial body searches that require detection of SM products for a dark matter annihilation signal. Objects which are considered for Galactic center searches are labeled ``GC", while those studied in the local position are labeled ``local".}
    \label{fig:psurv}
\end{figure}

Figure~\ref{fig:psurv} makes this point explicit, and shows how the survival probability varies as a function of mediator decay length $L$, for multiple celestial body searches. The objects used for Galactic center population searches are labeled with ``GC", and clearly allow for a survival probability of $\sim1$ across the widest range of mediator decay lengths. This is simply because they are so far away, that very large mediator decay lengths have a high probability of decaying before the telescope and being detected. On the other hand, the local objects have a much smaller region where the decay probability is one, as they are close to our telescopes, and longer decay lengths are less likely to decay before the telescope. 

For all objects, for decay lengths smaller than the radius of the object, $e^{-R/L}$ exponentially suppresses $\mathbb{P}_\mathrm{surv}$. When the decay length is much larger than the object to telescope distance, $i.e.$ $L\gg D$, $\mathbb{P}_\mathrm{surv} \sim (D-R)/L$. Neutron stars are tiny, with a radius of about 10~km; this means that $\mathbb{P}_\mathrm{surv}$ is not suppressed at the smallest decay lengths, while the mediator cannot escape other objects. The shortest decay lengths are clearly probed for the smallest objects, ordered from neutron stars, white dwarfs, brown dwarfs and Jupiter, an average main sequence star (labeled ``GC Stars"), and lastly the Sun. The Sun has the largest radius of the objects considered and is also the closest to the Earth, giving it the smallest region where $\mathbb{P}_\mathrm{surv} \sim$ 1.

In Fig.~\ref{fig:psurv}, a clear consequence of fixing $\mathbb{P}_\mathrm{surv} \sim$ 1 is that the mediator properties are also implicitly fixed as mentioned above. In addition, fixing $\mathbb{P}_\mathrm{surv} \sim$ 1 when showing a plot of sensitivities in the cross section vs dark matter mass plane, as is done often in the literature as well as our previous section, then does not include information about the trade-off between very powerful cross section sensitivities for a given search, with potentially weaker survival probability values, or vice versa. For example, looking at Fig.~\ref{fig:psurv}, we see that for larger decay lengths past the Earth's distance, both the Sun and Jupiter have a diminishing $\mathbb{P}_\mathrm{surv}$, but if their cross section sensitivity is so extreme that a signal can be still detected even with a penalty of a small $\mathbb{P}_\mathrm{surv}$, they might still be better probes than other searches which have poor cross section sensitivity and actually require $\mathbb{P}_\mathrm{surv}\sim1$ for a signal to be detected. In addition, while for example the Sun would be naively thought to be a better target than main sequence stars in the Galactic center (our ``GC Stars" line in Fig.~\ref{fig:psurv}) due to its proximity, from Fig.~\ref{fig:psurv} at very long decay lengths the Sun is suppressed in survival probability compared to the Galactic center stellar search. Therefore, an important quantity to consider is the interplay of the cross section sensitivity with the survival probability, which reveals additional information about which body and search is optimal for dark mediator properties. This can be parametrized in the mediator decay length vs dark matter mass plane for fixed cross sections, which we will show at the end of this section.

\subsection{Intersecting Heating and Gamma-Ray Searches}
\label{section:intersecting-heating-gamma}

Another important point for considering the mediator decay length dependence of our search sensitivities, is that for some decay lengths, both heating and gamma-ray searches are viable. Assuming that the dark matter is annihilating at the core of the body, with a mediator decay length of $L$, we find the probability that the decay products will be absorbed by the body with radius $R$ (leading to heating), is
\begin{equation}
\mathbb{P}_{\text{heat}} = 1-e^{-R/L}\,.
\label{eq:pheat}
\end{equation}
There is an interplay between the probability of survival of SM products to the telescope, $\mathbb{P}_{\mathrm{surv}}$, and the probability of absorption within the celestial body $\mathbb{P}_{\text{heat}}$. There are decay lengths where some dark matter energy will be injected, as well as produce a significant flux of SM particles, from the same object. In addition, given the range of celestial-body properties, there will be different objects that have heating or SM product detection for the same decay length, offering a complementary test of the dark sector parameter space across a spectrum of objects.

To demonstrate the interplay and overlap of both heating and SM product searches, we set limits on the mediator decay length in two ways. For the exoplanet/brown dwarf heating search, we require the heating rate to be at least 10\% of the maximum capture rate, corresponding to the minimum detectable heating with JWST's sensitivity~\cite{Leane:2020wob}. This can be seen equivalently as requiring about $10\%$ of the total available dark matter's rest-mass energy being required for the search to be viable.
For heating searches with other objects, their sensitivity to energy injection is lower and they need to capture almost all of the incident dark matter to have a detectable heating signal. For these objects, we consider mediators where 95\% of the decay products are deposited in the object. Assuming $X\%$ of the time the mediator decays inside the object, then Eq. (\ref{eq:pheat}) implies
\begin{equation}
    L_{\mathrm{heat}} = \frac{-R}{\ln(10^{-2}X)},
\end{equation}
which gives us the decay length threshold $L_{\mathrm{heat}}$ where beyond that value, a heating search no longer renders limits because too many mediators escape the body. For example, for neutron stars $L_{\mathrm{heat}}$ = 3.3 km,  white dwarfs $L_{\mathrm{heat}}$ = 2322 km, and Earth $L_{\mathrm{heat}}$ = 2126 km. Above these values, more than 5\% of the mediators escape the object and because the objects need almost all of the dark matter energy injected to see an effect, this is enough to make dark matter heating too low. These values assume that the objects need to capture and annihilate almost all of the dark matter to detect a heating signal.

For gamma-ray searches, $\mathbb{P}_{\mathrm{surv}}$ in Eq.~(\ref{eq:Psurv}) suppresses the dark matter annihilation signal of short decay length mediators ($i.e.$ decays inside the object). Still, we find gamma-ray sensitivity to mediator decay lengths much smaller than the radius of the object, since there is a non-zero probability for the mediator to sometimes decay outside the body. For example, Solar gamma rays can set limits on mediator decay lengths down to 0.04 $R_{\odot}$ because the SM flux from dark matter annihilation can be suppressed by many orders of magnitude before dropping below Fermi's sensitivity. We will see a similar intersection of energy injection and SM flux parameter space for brown dwarfs, white dwarfs, and Galactic center stars.

\subsection{Dark Matter Mediator Parameter Space}

We now quantify the celestial body sensitivity to a model independent parameterization of dark sector model space for fixed cross sections. The survival probability or heating probability given mediator properties will be given by Eq.~(\ref{eq:Psurv}) or Eq.~(\ref{eq:pheat}), rather than set to one, to capture more dark sector information. In this subsection, we will consider sensitivities to the parameter space from energy injection searches, and for SM product searches will focus on direct annihilation into gamma rays for simplicitly; qualitatively similar features will appear for other SM final states. The searches plotted in this subsection are the same as those shown in Sec.~\ref{sec:energy} and Sec.~\ref{sec:prod}, albeit investigated in a different parameter space. The considerations outlined in earlier sections about each object's detectability still apply to the projections here and should be kept in mind. Note that our assumption of a boosted mediator ensures the gamma rays point directly from the center of the object. In this case, the objects will be point sources even with mediators decaying at long distances from the object. However, for non-boosted mediators with decay lengths very far from the radius of the object, the telescope will observe an extended source. This decreases the sensitivity and would require a re-analysis of the data which is outside the focus of our paper, see Ref.~\cite{Leane:2017vag} for discussion. We only consider parameter space where a point source is a good description, though the complementarity we aim to highlight will be similar in the extended source scenario.

\begin{figure}[t!]
    \centering
    \includegraphics[width=\columnwidth]{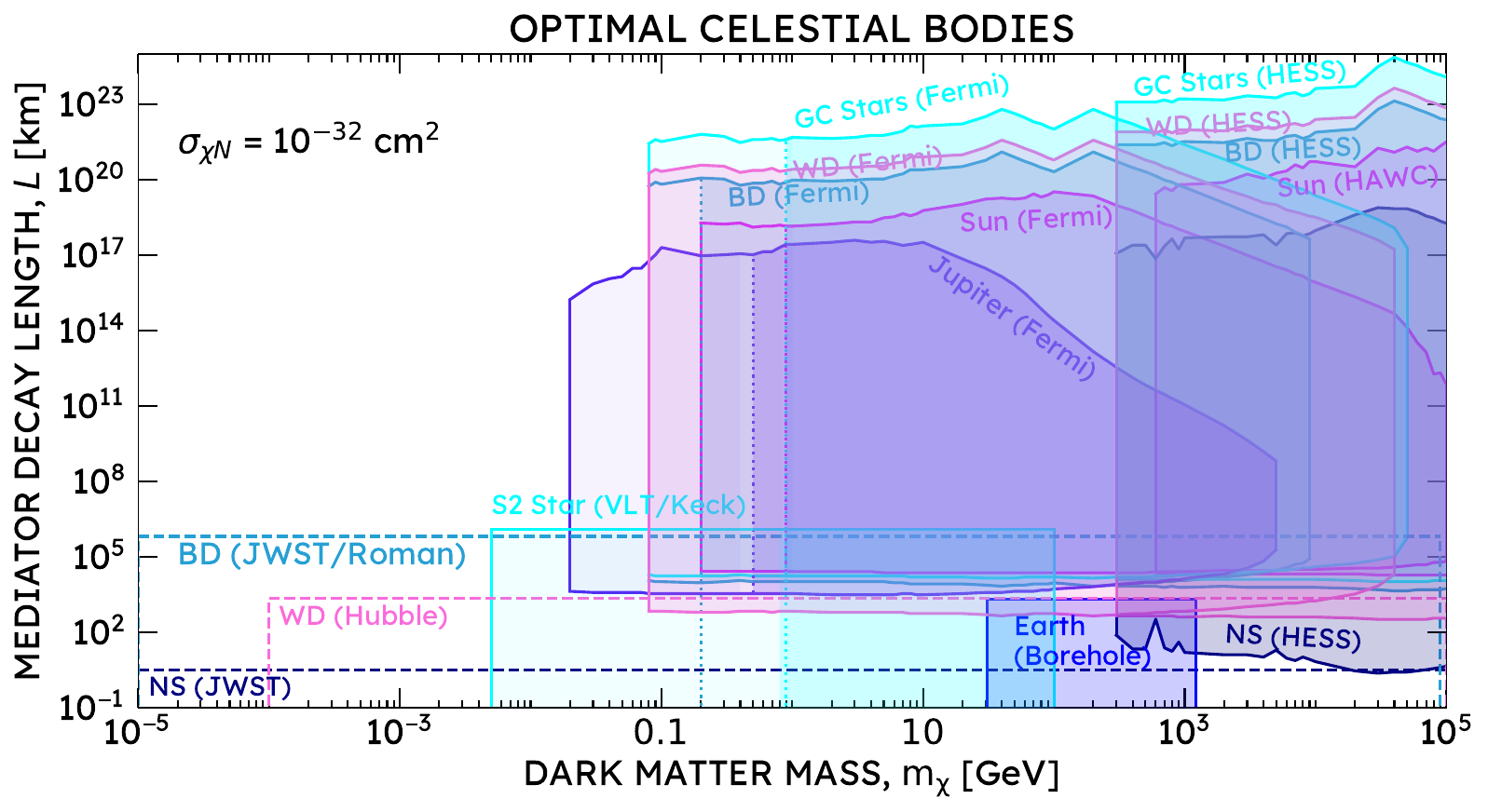}\vspace{6mm}
    \includegraphics[width=\columnwidth]{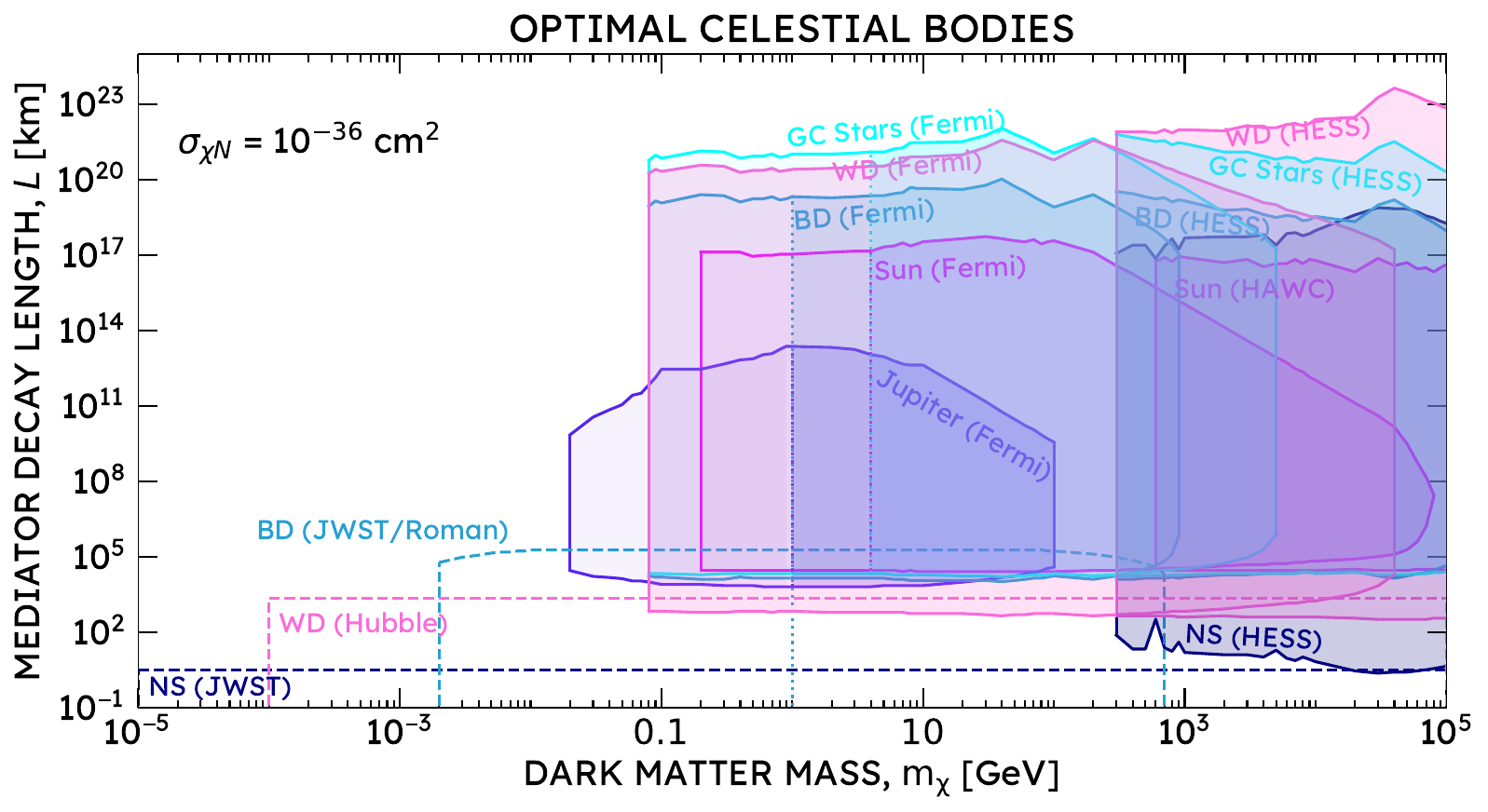}
    \caption{Dark mediator decay length and dark matter mass parameter space probed by a range of celestial body searches, for fixed dark matter-SM scattering cross sections of 10$^{-32}$ cm$^2$ (top) and 10$^{-36}$ cm$^2$ (bottom). Shaded regions are constraints with existing data, non-shaded dashed regions are projections for future searches, see text for details.}
    \label{fig:sigma-32-36}
\end{figure}

\begin{figure}
    \centering
    \includegraphics[width=\columnwidth]{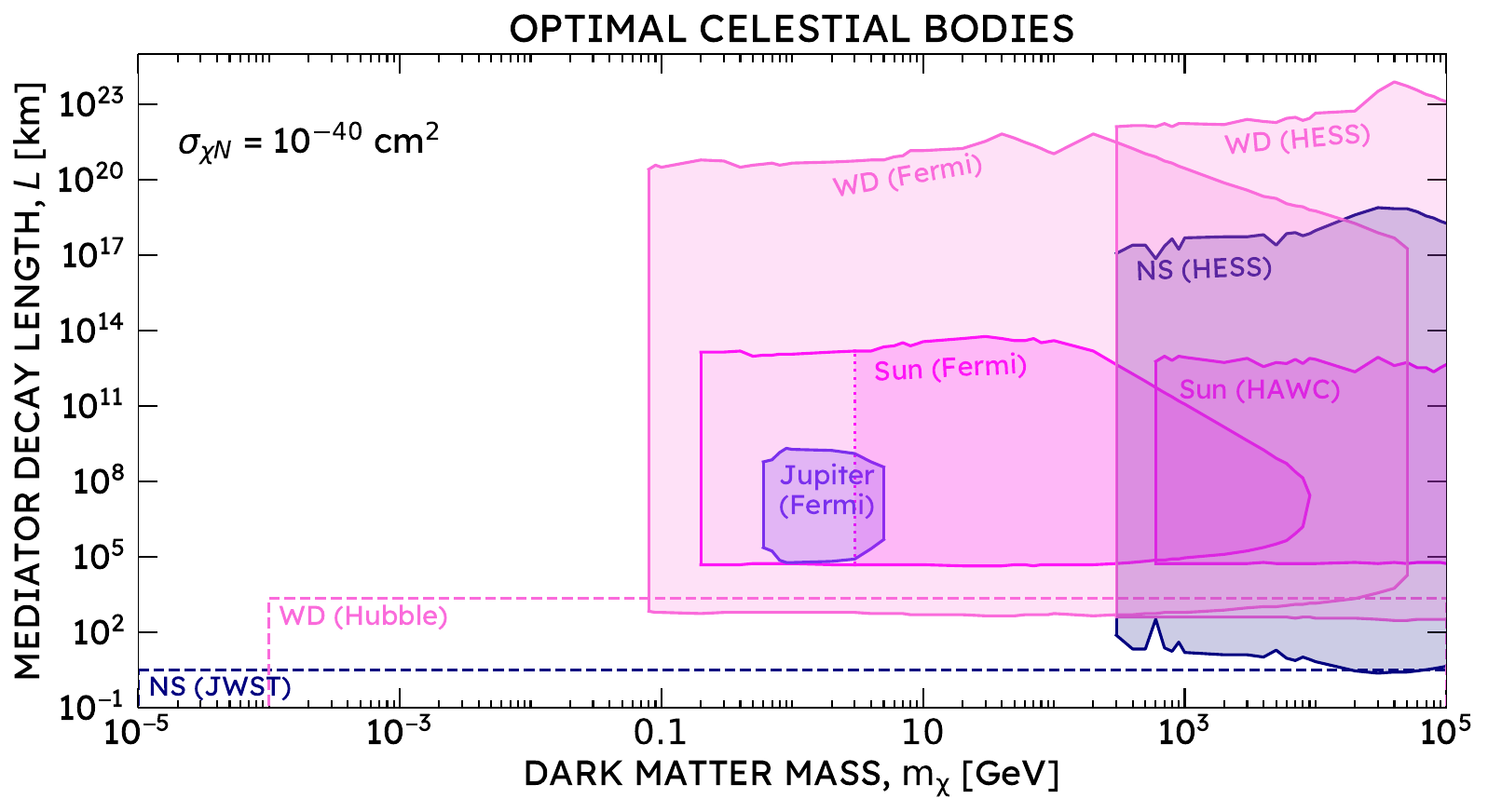}\vspace{6mm}
    \includegraphics[width=\columnwidth]{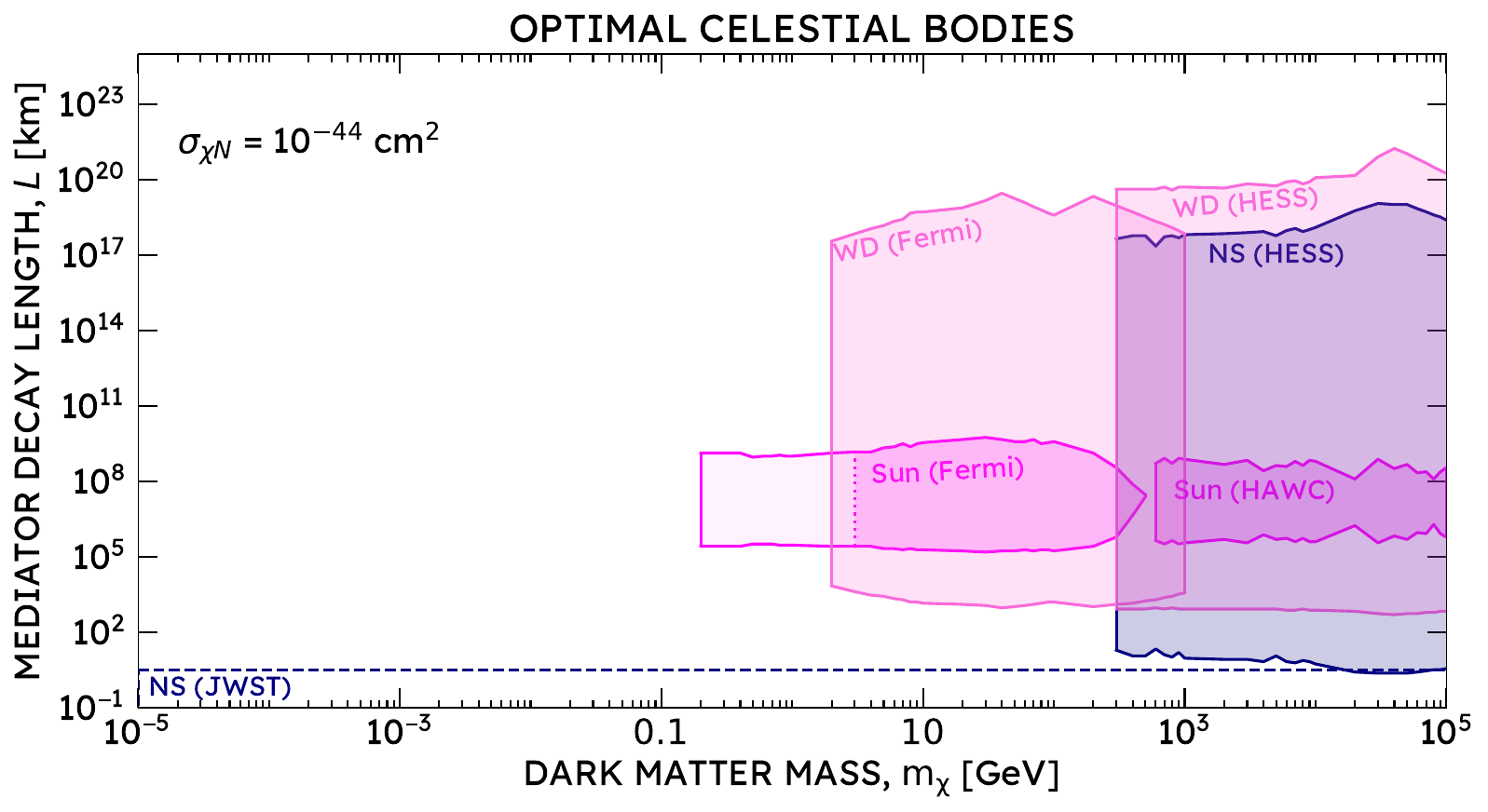}
    \caption{Dark mediator decay length and dark matter mass parameter space probed by a range of celestial body searches, for fixed dark matter-SM scattering cross sections of 10$^{-40}$ cm$^2$ (top) and 10$^{-44}$ cm$^2$ (bottom). Shaded regions are constraints with existing data, non-shaded dashed regions are projections for future searches, see text for details.}
    \label{fig:sigma-40-44}
\end{figure}

Figures~\ref{fig:sigma-32-36} and \ref{fig:sigma-40-44} show our calculated limits and projected sensitivities to mediator decay length as a function of dark matter mass for fixed cross sections of $10^{-32}$~cm$^2$, $10^{-36}$~cm$^2$, $10^{-40}$~cm$^2$ and $10^{-44}$~cm$^2$, for multiple celestial-body searches, with the object and relevant telescope for detection labeled. Solid lines with shaded areas correspond to limits we set with current data, dashed lines without shaded areas correspond to future sensitivities. The dotted lines within each search's area correspond to when dark matter evaporates if its scattering with the SM is via a contact interaction, to the left of the same search region is testable if a long-range force prevents evaporation, and the most-left part of the regions are truncated by the fundamental search sensitivity. In these figures, no coherent enhancement in the scattering is shown, and most objects are assumed to be purely nucleons for simplicity. The two exceptions are white dwarfs where there is only a spin-independent signal possible, and so they are taken to be purely carbon-12 (including the coherent enhancement), and the Earth, which is the spin-dependent result, and includes the spin-dependent scattering with Earth's nuclei as detailed in Ref.~\cite{Bramante:2019fhi}.

In Figs~\ref{fig:sigma-32-36} and \ref{fig:sigma-40-44}, there are multiple features that were not apparent when considering the sensitivity to the cross section as a function of dark matter mass as per the previous sections and the previous literature. Firstly, objects which have the most reach in the cross section vs dark matter mass plane do not necessarily have the most reach in the mediator decay length vs dark matter mass plane. For example, we see that the span of the neutron star's reach when considering a lone local heated neutron star with JWST (blue dashed at the bottom of the plots) is substantially limited compared to other searches. At high cross sections of $10^{-32}$~cm$^2$, it is not the ``optimal" target for most dark sector parameters. For the lower cross section shown of $10^{-44}$~cm$^2$, we see that its reach does not change; this is because all these cross sections are above its threshold cross section corresponding to maximum capture rates\footnote{Note however that we have only considered the case where the neutron star signal also includes dark matter annihilation heat, in order to make the signal more detectable. If a neutron star search is executed despite the difficulties discussed in the earlier heating section, and even the low rates expected from dark kinetic heating alone are sufficient, then the neutron star would cover the whole parameter space.}. On the other hand, larger objects have larger capture rates in the regime the cross section is sufficiently large that all the dark matter is captured that passes through them. 

Heating searches in general probe shorter mediator decay lengths, as they require SM products to be absorbed by the object, and are shown in regions at the bottom of the plots in Figs~\ref{fig:sigma-32-36} and \ref{fig:sigma-40-44}. When the cross section is large enough for all objects to be detectable, we clearly see a hierarchy in the decay length the object can probe based on its radius. This is simply due to the size of the objects; a larger object will more readily absorb decay products that are produced from a longer decay length. The neutron star, white dwarf, and S2 bounds all assume that effectively all the dark matter passing through is captured, such that decay lengths that are longer than the radius of the object are not easily accommodated in the sensitivity, leading to the flat features for each line, corresponding to a mediator decay length at or less than the object's radius. For the brown dwarf sensitivity however, because the JWST sensitivity for this search is estimated to be down to about 10\% of the maximum capture rate, that means that about 90\% of the time the mediator can decay outside the surface of the brown dwarf, and a heating signal will still be detectable. This means that decay lengths longer than the radius of the brown dwarf are testable even via heating, and is why the brown dwarf heating line is curved at some cross sections; it is taking into account a diminished mediator production rate inside the brown dwarf by increasing the capture rate proportionally. This also means that simultaneously from a given brown dwarf, there can be detectable heating as well as gamma rays. This is seen from the fact that the Galactic center gamma-ray search using brown dwarfs, labeled ``BD (Fermi)", overlaps in the parameter space it can probe from JWST measured brown dwarf heating. Similarly, the white dwarf heating has some overlap with the white dwarf Galactic center gamma-ray search (where the latter is sufficiently sensitive to lose some signal from mediators not escaping), even though these are searches for different signals from different white dwarf populations.

In Figs~\ref{fig:sigma-32-36} and \ref{fig:sigma-40-44}, the Galactic center gamma-ray searches generally probe the widest part of the parameter space. This occurs because the Galactic center is so far away, leading to a wide range of decay lengths directly before the telescope, see the survival probability figure, Fig.~\ref{fig:psurv}. However, it is also interesting to see that both the local objects considered with gamma rays, the Sun and Jupiter, are not substantially far off the sensitivities for the Galactic center searches at higher cross sections, despite having suppressed survival probabilities for most decay lengths as per Fig.~\ref{fig:psurv}. The reason for this is twofold: (1) the gamma-ray backgrounds for both the Sun and Jupiter are extremely low (or non-existent), such that (2) the maximum dark matter signal arising from these objects is many orders of magnitude above this background due to them being so close to the telescope. This means that they can take a large penalty in their mediator survival probabilities, if the cross section remains sufficiently high, and still have excellent sensitivity -- there is always a trade off between the cross section value and the survival probability of the mediator. This is not captured in the standard cross section vs dark matter mass plane plots, and means that both the Sun and Jupiter can readily probe decay lengths much larger than their distance to the Earth, as their strong signal vs background sensitivity means the low chance that a mediator with a long decay length happens to decay on a shorter distance on occasion still provides sensitivity.

Overall, in Figs~\ref{fig:sigma-32-36} and \ref{fig:sigma-40-44}, we see that the dark matter constraint and discovery space is highly complementary. If one type of signal is discovered in one object, there can be other objects with different or similar signals that can corroborate discovery. Furthermore, for dark matter heating searches with celestial bodies, when considering one search alone the exact dark matter mass or cross section giving rise to the signal cannot be distinguished, all that can be extracted is what size capture rate produced the signal, as many dark matter masses and cross sections can produce the same size capture rate. However, in our Figs~\ref{fig:sigma-32-36} and \ref{fig:sigma-40-44}, an extra handle on the potential dark matter mass and cross section can actually be obtained by considering complementary objects, as the overlapping signal sensitivity regions for given objects in the parameter space correspond to fixed dark sector parameters.

The main complementary constraint on mediator lifetimes shown in Figs~\ref{fig:sigma-32-36} and \ref{fig:sigma-40-44} is given by Big Bang nucleosynthesis (BBN)~\cite{Kawasaki:2004yh,Jedamzik:2006xz}. Generally it is required that a new mediator lifetime is less than about a second, though in some models this constraint can be relaxed~\cite{Pospelov:2010hj}. Considering a mediator with decay length $L = m_\chi \tau / m_\phi$ implies a dark sector particle with a prescribed lifetime and mass. To generate the gamma-ray spectra for Figs~\ref{fig:sigma-32-36} and \ref{fig:sigma-40-44}, we assumed a boost of $m_\chi/m_\phi = $ 10. This choice is arbitrarily made just to ensure that the mediator is boosted sufficiently; as far as the limits are concerned, the gamma-ray spectrum produced is independent of the boost for sufficiently large boosts when the limit is set with the spectrum’s peak (see $e.g.$ discussion in Ref.~\cite{Leane:2017vag}), as per the bulk of our parameter space. This means that, while a boost of 10 implies that to get long decay lengths we need long lifetimes that could conflict with BBN, the boost can in fact instead be made much larger with no significant change in the spectrum, and therefore no change in the limits, such that there is no conflict with BBN. For example, fixing our mediator lifetime at one second and varying the mediator mass can lead to extremely long decay lengths while not affecting the elemental synthesis in the early Universe.  Therefore, our results can apply to regimes where the boost is large and the lifetime small such that there is no impact on nucleosynthesis. In other words, for the limits in Figs~\ref{fig:sigma-32-36} and \ref{fig:sigma-40-44}, it is only the product of the boost and the lifetime that is important, such that there is flexibility in the dark sector parameters to not be constrained by BBN, depending on the specific model.

In considering the dark sector parameter space as per Figs~\ref{fig:sigma-32-36} and \ref{fig:sigma-40-44}, we notice a previously overlooked search that appears to dominate the parameter space: a search for gamma rays from main sequence stars in the Galactic center (labeled ``GC Stars"). Comparing to the earlier figures in Sec.~\ref{sec:prod}, this search naively appeared to be orders of magnitude weaker than using gamma-ray emission from our Sun. This was only a consequence of fixing the survival probability everywhere to one, which then gave the Galactic center stellar search a penalty in flux due to its distance away, without giving the Sun the penalty in the gamma-ray survival probability for long decay lengths. Given this new search with new sensitivity, we now briefly investigate the benefits of considering Galactic center gamma-ray main sequence stellar emission.

\section{Application to a New Stellar Search}
\label{sec:stellar}

\subsection{Faster Dark Matter Equilibrium Timescales}

Our searches detailed in the previous sections all assume that celestial bodies' dark matter capture and annihilation processes are in equilibrium. Another complementary strength for any search in the Galactic center is that because the dark matter capture rates are so high (due to higher dark matter densities), Galactic center bodies fall into capture-annihilation equilibrium in a shorter timescale. This means that they can probe smaller annihilation rates more easily. We now demonstrate the annihilation rates leading to equilibrium for the population of stars in the Galactic center. The equilibrium timescale is given by~\cite{Acevedo:2023xnu}
\begin{align}
    t_{\rm eq} = \left(\frac{V}{\langle \sigma v \rangle_{\mathrm{ann}} \, C}\right)^{1/2}~,
    \label{eq:t_eq1}
\end{align}
where $C$ is the capture rate, $V$ is the volume in which annihilation proceeds, $v$ is the relative velocity between dark matter particles, and $\langle \sigma v \rangle_{\mathrm{ann}}$ is the thermally averaged annihilation cross section. The partial wave expansion is given by
\begin{equation}
    \langle \sigma v \rangle_{\mathrm{ann}} = \langle \sigma v \rangle_{\mathrm{ann}}^0 \, \sum_{\ell=0}^{\infty} a_\ell \, v^{2\ell}\, ,
\end{equation}
where the term with $\ell=0$ gives the leading $s$-wave term, and $\ell=1$ gives the leading $p$-wave term. The contribution of each $\ell$-wave is found by setting $a_\ell = 1$ while keeping the other modes equal to zero. In a conservative approach, we use the volume of the celestial body as the annihilation volume. The relative velocity of the annihilating dark matter particles is estimated as the velocity in the core, given by
\begin{equation}
    v = \sqrt{3 T_\mathrm{core} / m_\chi} \,,
\end{equation}
where the $T_{\rm core}$ is the core temperature. After being captured by a star similar to the Sun, the dark matter will have relative velocities $\mathcal{O}(10^{-3} c)$. Assuming an example dark matter mass of 1 GeV and scattering cross section of 10$^{-38}$ cm$^2$, we find the timescales for stellar equilibrium for stars at ten pc from the Galactic center are
\begin{equation}
    t_{\rm eq} \simeq 
        \begin{cases}
      \ 2 \times 10^{-4} \ {\rm Gyr}  & \ \text{$s$-wave} \\
      \\
      \ 0.2 \ {\rm Gyr} \left(\dfrac{10^{-3} \, c}{v}\right) & \ \text{$p$-wave}, \\
    \end{cases}
    \label{eq:t_eq2}
\end{equation}
where we have normalized $\langle \sigma v \rangle_{\mathrm{ann}}^0 = 3 \times 10^{-26} \ \rm cm^3 / s$. The capture-annihilation equilibrium timescales we find in Eq.~\eqref{eq:t_eq2} are less than the age of an average star, even for $p$-wave dark matter annihilation. Stars even closer to the Galactic center will equilibrate faster because of increased dark matter density and, therefore, increased capture rates. In comparison, for the Sun which is at the local position, the equilibrium timescales are 0.2 Gyr for $s$-wave and 200 Gyr for $p$-wave annihilation for the same dark matter mass and cross section, highlighting an additional benefit in considering the Galactic center stellar population over the local Sun, which is not apparent in any of our plots due to an implicit assumption made. For simplicity, when comparing the stellar signal to the solar signal, we keep the assumption that the Sun is in equilibrium, but in reality at the lower cross sections it will not be in equilibrium and would be even weaker in comparison to the Galactic center stellar signal. 

\begin{figure}
    \centering
    \includegraphics[width=0.6\columnwidth]{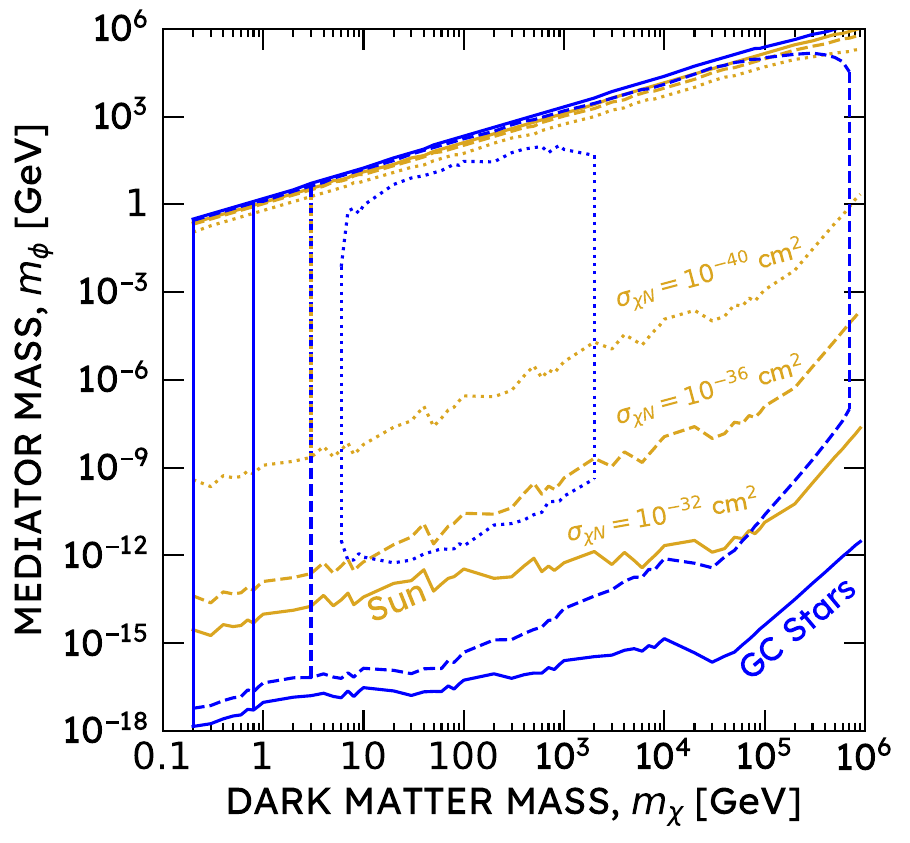}
   \caption{GC Stars (blue) and Solar (gold) constraints on the dark matter and mediator mass with $\tau = $ 0.1 s. The solid, dashed, and dotted lines correspond to fixed dark matter-nucleon scattering cross sections of $10^{-32}$, $10^{-36}$, $10^{-40}$ cm$^2$ respectively. The vertical lines inside the contour indicate the evaporation mass.} 
    \label{fig:solar-stellar-my-vs-mx-combined}
\end{figure}

\subsection{Dark Sector Constraints}

Figure~\ref{fig:solar-stellar-my-vs-mx-combined} shows our calculated limits on Galactic center stars compared to the Sun, as a function of mediator mass and dark matter mass for three different fixed cross sections as labeled. This shows that when the lifetime and cross section are fixed, due to suppression of long decay lengths, the Sun's flux is reduced compared to the Galactic center's stellar flux. The contours of constant cross section reveal that the stars in the Galactic center are more sensitive to dark matter capture and annihilation for lighter mediator masses. The lines inside the contours indicate the evaporation mass under the assumption of contact interactions, for the cross section as labeled. These lines are approximately overlapping for the Sun and Galactic center stars as they are similar objects.

\begin{figure}
    \centering
    \includegraphics[width=0.6\columnwidth]{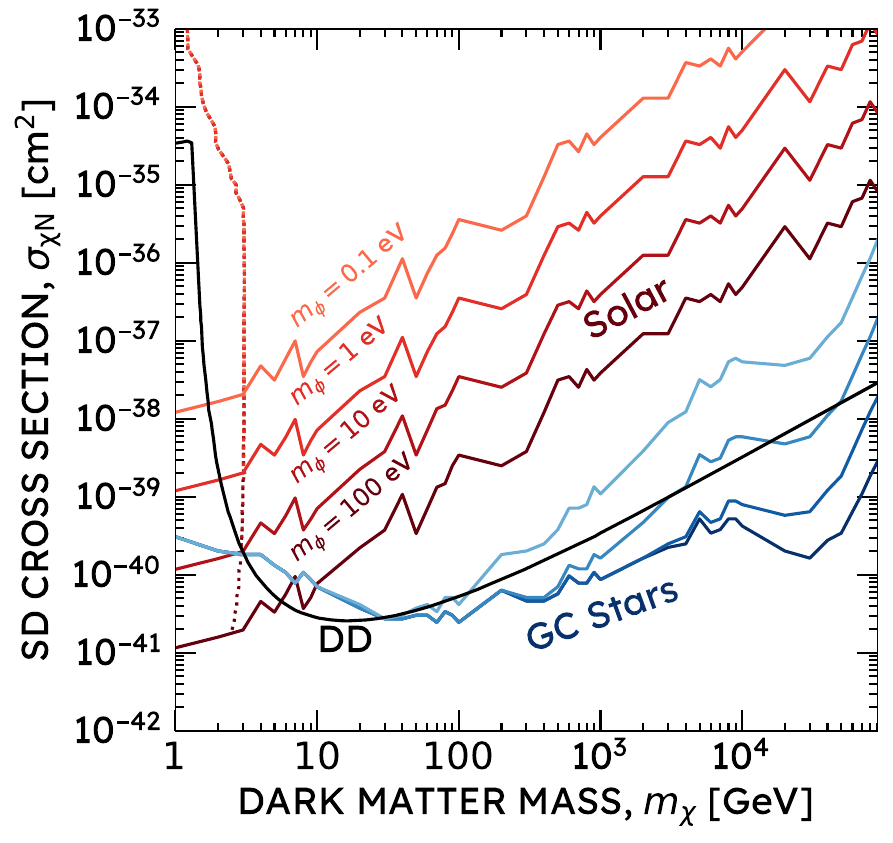}
    \caption{Constraints on the dark matter-SM scattering cross section as a function of dark matter mass for the Sun and the Galactic center stars (``GC Stars"), using Fermi and HAWC data. Here we assume mediator lifetimes of $\tau=1\,$s. The stellar and Solar curves, in order of increasing brightness, correspond to mediator masses of $10^{-7}$, $10^{-8}$, $10^{-9}$, $10^{-10}$ GeV, with the evaporation limit drawn as a dotted line. Direct detection constraints are also shown.
    }
    \label{fig:solar-stellar-constraints-fixed-my}
\end{figure}

Figure \ref{fig:solar-stellar-constraints-fixed-my} shows our calculated constraints on the dark matter-SM scattering cross section as a function of dark matter mass for the Sun and the Galactic center stars, using Fermi and HAWC data. Here we have lifted the assumption of order unity survival probability of the mediator's decay to gamma rays, for a fixed example lifetime of one second. 
As the mediator mass increases beyond 100 eV (and is still light enough to escape), the Sun constrains smaller cross sections. This is because the decay length is longer than the Solar-Earth distance, suppressing the flux as only a fraction of the mediators decay before passing the Earth. As the mediator mass increases, its decay length reduces and a larger fraction of mediators decay between the Sun and the Earth. $\mathbb{P}_{\text{surv}}$, and therefore the flux, continues to increase with increasing mediator mass, giving the Sun increasing sensitivity to small cross sections. 
However, the Galactic stellar population constraints remain constant as the mediator mass increases beyond 100 eV (and is still light enough to escape). The mediator decay length increases but stays below the Galactic center-Earth distance, therefore the probability of survival remains in the $\mathbb{P}_{\text{surv}} \approx 1$ regime, and the flux remains constant. For mediator masses above 100 eV, the Sun constrains smaller cross sections while the stellar population constraints are unchanged. Overall, in this figure the Galactic center stellar population constrains new parameter space relative to direct detection and the Sun, for mediator masses between 1 eV and 100 eV, due to larger kinematic boosts increasing the mediator decay length. 

This new Galactic center stellar search shows the significance of the assumptions behind each search strategy. Our result shows that, even though Galactic center stars are further away than a similar object in the local position (the Sun), they can be more optimal for dark matter sensitivity than the Sun depending on the particle physics assumptions.

\section{Summary and Conclusions}
\label{sec:conclusion}

A wide range of celestial bodies have been considered as dark matter detectors in the literature. We have surveyed a substantial range of the searches, including using the Sun, Earth, Jupiter, Brown Dwarfs, White Dwarfs, Neutron Stars, Stellar populations, and Exoplanets. We detailed what properties make these objects intrinsically optimal, such as their radius, density, and core temperature. We also detailed what properties about the location in which any of these objects are searched for are important to optimize over, which is tied directly to the properties of the object as well as the advantages of a given location. We therefore provided a qualitative understanding of why some objects are used and not others for a given search, and that the wide range of celestial objects available along with the wide range of potential signatures all afford different opportunities to understand the nature of dark matter, and test the dark sector parameter space in complementary ways.

All of the celestial body searches have different in built assumptions. One of these is the feasibility of the search. While some projected sensitivities to the dark matter parameter space appear superior with particular objects or search strategies, they in reality might be or already are hampered by difficulties in the search. We discussed multiple cases, including $e.g.$ neutron stars (whose heating detection prospects can be difficult given their size and required candidates), as well as white dwarfs (whose heating limits have substantial systematics in the dark matter densities currently assumed). Searches such as those with exoplanets or brown dwarfs, which are not as extensive in the cross section vs dark matter mass plane, may in the end be more likely to be realized. On the other hand, searches with $e.g.$ neutron stars if carried out truly provide strong particle model and cross section sensitivity.

Another assumption vital for the searches is the underlying class of particle physics models to which the search applies. We investigated the interplay of Standard Model production searches, with dark matter energy injection searches. Usually these two classes of searches are considered separately, where generally the former corresponds to a long-lived or boosted dark mediator, and the latter corresponds to prompt dark matter annihilation via a short-lived or not very boosted mediator. We compared and calculated new limits on the dark matter-nucleon scattering cross section as a function of dark matter mass for a variety of dark matter annihilation channels, which can be mapped to a specific particle physics model. In addition, we calculated and showed how these two search classes are not completely distinct, and in fact in some cases, both the signatures can be produced simultaneously even in the same object, where the mediator can with some frequency decay both in and outside the celestial body. We also showed the interplay with a wide range of celestial objects, where for a given parameter combination, corroborating signals can be expected across multiple objects. In the mediator mass and dark matter mass plane, the consequences of relaxing some assumptions previously made in the literature (such as mediator decay probability dependence of the signal) made clear that some searches actually cover more dark sector parameters than naively expected. We consequently identified a new search using gamma rays from Galactic center nuclear-burning stars, which can outperform gamma-ray searches using our own local Sun.

Going forward, as telescope technology continues to improve, and as the latest telescopes launch and obtain increasing data on a diversity of celestial objects, our targets and prospects for dark matter detection will only improve. Exploiting the complementarity of celestial-body signals across the broader dark sector parameter space will afford us new opportunities to gain insights to the unknown nature of the dark sector.
 
\section*{Acknowledgments}

We thank J. Acevedo, C. Bravo, A. Das, N. Desai, S. Gaiser, T. Linden, C. O'Hare, M. Peskin, and F. Tanedo for helpful discussions and comments. RKL and JT are supported in part by the U.S. Department of Energy under Contract DE-AC02-76SF00515. JT was also supported by a Stanford Physics Department fellowship and an EDGE fellowship.

\bibliography{main.bib}

\begin{thebibliography}{240}%
\makeatletter
\providecommand \@ifxundefined [1]{%
 \@ifx{#1\undefined}
}%
\providecommand \@ifnum [1]{%
 \ifnum #1\expandafter \@firstoftwo
 \else \expandafter \@secondoftwo
 \fi
}%
\providecommand \@ifx [1]{%
 \ifx #1\expandafter \@firstoftwo
 \else \expandafter \@secondoftwo
 \fi
}%
\providecommand \natexlab [1]{#1}%
\providecommand \enquote  [1]{``#1''}%
\providecommand \bibnamefont  [1]{#1}%
\providecommand \bibfnamefont [1]{#1}%
\providecommand \citenamefont [1]{#1}%
\providecommand \href@noop [0]{\@secondoftwo}%
\providecommand \href [0]{\begingroup \@sanitize@url \@href}%
\providecommand \@href[1]{\@@startlink{#1}\@@href}%
\providecommand \@@href[1]{\endgroup#1\@@endlink}%
\providecommand \@sanitize@url [0]{\catcode `\\12\catcode `\$12\catcode
  `\&12\catcode `\#12\catcode `\^12\catcode `\_12\catcode `\%12\relax}%
\providecommand \@@startlink[1]{}%
\providecommand \@@endlink[0]{}%
\providecommand \url  [0]{\begingroup\@sanitize@url \@url }%
\providecommand \@url [1]{\endgroup\@href {#1}{\urlprefix }}%
\providecommand \urlprefix  [0]{URL }%
\providecommand \Eprint [0]{\href }%
\providecommand \doibase [0]{http://dx.doi.org/}%
\providecommand \selectlanguage [0]{\@gobble}%
\providecommand \bibinfo  [0]{\@secondoftwo}%
\providecommand \bibfield  [0]{\@secondoftwo}%
\providecommand \translation [1]{[#1]}%
\providecommand \BibitemOpen [0]{}%
\providecommand \bibitemStop [0]{}%
\providecommand \bibitemNoStop [0]{.\EOS\space}%
\providecommand \EOS [0]{\spacefactor3000\relax}%
\providecommand \BibitemShut  [1]{\csname bibitem#1\endcsname}%
\let\auto@bib@innerbib\@empty
%</preamble>
\bibitem [{\citenamefont {Freese}(1986)}]{Freese:1985qw}%
  \BibitemOpen
  \bibfield  {author} {\bibinfo {author} {\bibfnamefont {K.}~\bibnamefont
  {Freese}},\ }\href {\doibase 10.1016/0370-2693(86)90349-7} {\bibfield
  {journal} {\bibinfo  {journal} {Phys. Lett. B}\ }\textbf {\bibinfo {volume}
  {167}},\ \bibinfo {pages} {295} (\bibinfo {year} {1986})}\BibitemShut
  {NoStop}%
\bibitem [{\citenamefont {Gould}(1992)}]{Gould:1991va}%
  \BibitemOpen
  \bibfield  {author} {\bibinfo {author} {\bibfnamefont {A.}~\bibnamefont
  {Gould}},\ }\href {\doibase 10.1086/171057} {\bibfield  {journal} {\bibinfo
  {journal} {Astrophys. J.}\ }\textbf {\bibinfo {volume} {387}},\ \bibinfo
  {pages} {21} (\bibinfo {year} {1992})}\BibitemShut {NoStop}%
\bibitem [{\citenamefont {Mack}\ \emph {et~al.}(2007)\citenamefont {Mack},
  \citenamefont {Beacom},\ and\ \citenamefont {Bertone}}]{Mack:2007xj}%
  \BibitemOpen
  \bibfield  {author} {\bibinfo {author} {\bibfnamefont {G.~D.}\ \bibnamefont
  {Mack}}, \bibinfo {author} {\bibfnamefont {J.~F.}\ \bibnamefont {Beacom}}, \
  and\ \bibinfo {author} {\bibfnamefont {G.}~\bibnamefont {Bertone}},\ }\href
  {\doibase 10.1103/PhysRevD.76.043523} {\bibfield  {journal} {\bibinfo
  {journal} {Phys. Rev. D}\ }\textbf {\bibinfo {volume} {76}},\ \bibinfo
  {pages} {043523} (\bibinfo {year} {2007})},\ \Eprint
  {http://arxiv.org/abs/0705.4298} {arXiv:0705.4298 [astro-ph]} \BibitemShut
  {NoStop}%
\bibitem [{\citenamefont {Chauhan}\ and\ \citenamefont
  {Mohanty}(2016)}]{Chauhan:2016joa}%
  \BibitemOpen
  \bibfield  {author} {\bibinfo {author} {\bibfnamefont {B.}~\bibnamefont
  {Chauhan}}\ and\ \bibinfo {author} {\bibfnamefont {S.}~\bibnamefont
  {Mohanty}},\ }\href {\doibase 10.1103/PhysRevD.94.035024} {\bibfield
  {journal} {\bibinfo  {journal} {Phys. Rev. D}\ }\textbf {\bibinfo {volume}
  {94}},\ \bibinfo {pages} {035024} (\bibinfo {year} {2016})},\ \Eprint
  {http://arxiv.org/abs/1603.06350} {arXiv:1603.06350 [hep-ph]} \BibitemShut
  {NoStop}%
\bibitem [{\citenamefont {Bramante}\ \emph {et~al.}(2020)\citenamefont
  {Bramante}, \citenamefont {Buchanan}, \citenamefont {Goodman},\ and\
  \citenamefont {Lodhi}}]{Bramante:2019fhi}%
  \BibitemOpen
  \bibfield  {author} {\bibinfo {author} {\bibfnamefont {J.}~\bibnamefont
  {Bramante}}, \bibinfo {author} {\bibfnamefont {A.}~\bibnamefont {Buchanan}},
  \bibinfo {author} {\bibfnamefont {A.}~\bibnamefont {Goodman}}, \ and\
  \bibinfo {author} {\bibfnamefont {E.}~\bibnamefont {Lodhi}},\ }\href
  {\doibase 10.1103/PhysRevD.101.043001} {\bibfield  {journal} {\bibinfo
  {journal} {Phys. Rev. D}\ }\textbf {\bibinfo {volume} {101}},\ \bibinfo
  {pages} {043001} (\bibinfo {year} {2020})},\ \Eprint
  {http://arxiv.org/abs/1909.11683} {arXiv:1909.11683 [hep-ph]} \BibitemShut
  {NoStop}%
\bibitem [{\citenamefont {Feng}\ \emph
  {et~al.}(2016{\natexlab{a}})\citenamefont {Feng}, \citenamefont {Smolinsky},\
  and\ \citenamefont {Tanedo}}]{Feng:2015hja}%
  \BibitemOpen
  \bibfield  {author} {\bibinfo {author} {\bibfnamefont {J.~L.}\ \bibnamefont
  {Feng}}, \bibinfo {author} {\bibfnamefont {J.}~\bibnamefont {Smolinsky}}, \
  and\ \bibinfo {author} {\bibfnamefont {P.}~\bibnamefont {Tanedo}},\ }\href
  {\doibase 10.1103/PhysRevD.93.015014} {\bibfield  {journal} {\bibinfo
  {journal} {Phys. Rev. D}\ }\textbf {\bibinfo {volume} {93}},\ \bibinfo
  {pages} {015014} (\bibinfo {year} {2016}{\natexlab{a}})},\ \bibinfo {note}
  {[Erratum: Phys.Rev.D 96, 099901 (2017)]},\ \Eprint
  {http://arxiv.org/abs/1509.07525} {arXiv:1509.07525 [hep-ph]} \BibitemShut
  {NoStop}%
\bibitem [{\citenamefont {Das}\ \emph {et~al.}(2024{\natexlab{a}})\citenamefont
  {Das}, \citenamefont {Kurinsky},\ and\ \citenamefont {Leane}}]{Das:2022srn}%
  \BibitemOpen
  \bibfield  {author} {\bibinfo {author} {\bibfnamefont {A.}~\bibnamefont
  {Das}}, \bibinfo {author} {\bibfnamefont {N.}~\bibnamefont {Kurinsky}}, \
  and\ \bibinfo {author} {\bibfnamefont {R.~K.}\ \bibnamefont {Leane}},\ }\href
  {\doibase 10.1103/PhysRevLett.132.121801} {\bibfield  {journal} {\bibinfo
  {journal} {Phys. Rev. Lett.}\ }\textbf {\bibinfo {volume} {132}},\ \bibinfo
  {pages} {121801} (\bibinfo {year} {2024}{\natexlab{a}})},\ \Eprint
  {http://arxiv.org/abs/2210.09313} {arXiv:2210.09313 [hep-ph]} \BibitemShut
  {NoStop}%
\bibitem [{\citenamefont {Das}\ \emph {et~al.}(2024{\natexlab{b}})\citenamefont
  {Das}, \citenamefont {Kurinsky},\ and\ \citenamefont {Leane}}]{Das:2024jdz}%
  \BibitemOpen
  \bibfield  {author} {\bibinfo {author} {\bibfnamefont {A.}~\bibnamefont
  {Das}}, \bibinfo {author} {\bibfnamefont {N.}~\bibnamefont {Kurinsky}}, \
  and\ \bibinfo {author} {\bibfnamefont {R.~K.}\ \bibnamefont {Leane}},\ }\href
  {\doibase 10.1007/JHEP07(2024)233} {\bibfield  {journal} {\bibinfo  {journal}
  {JHEP}\ }\textbf {\bibinfo {volume} {07}},\ \bibinfo {pages} {233} (\bibinfo
  {year} {2024}{\natexlab{b}})},\ \Eprint {http://arxiv.org/abs/2405.00112}
  {arXiv:2405.00112 [hep-ph]} \BibitemShut {NoStop}%
\bibitem [{\citenamefont {Neufeld}\ \emph {et~al.}(2018)\citenamefont
  {Neufeld}, \citenamefont {Farrar},\ and\ \citenamefont
  {McKee}}]{Neufeld:2018slx}%
  \BibitemOpen
  \bibfield  {author} {\bibinfo {author} {\bibfnamefont {D.~A.}\ \bibnamefont
  {Neufeld}}, \bibinfo {author} {\bibfnamefont {G.~R.}\ \bibnamefont {Farrar}},
  \ and\ \bibinfo {author} {\bibfnamefont {C.~F.}\ \bibnamefont {McKee}},\
  }\href {\doibase 10.3847/1538-4357/aad6a4} {\bibfield  {journal} {\bibinfo
  {journal} {Astrophys. J.}\ }\textbf {\bibinfo {volume} {866}},\ \bibinfo
  {pages} {111} (\bibinfo {year} {2018})},\ \Eprint
  {http://arxiv.org/abs/1805.08794} {arXiv:1805.08794 [astro-ph.CO]}
  \BibitemShut {NoStop}%
\bibitem [{\citenamefont {Pospelov}\ and\ \citenamefont
  {Ramani}(2021)}]{Pospelov:2020ktu}%
  \BibitemOpen
  \bibfield  {author} {\bibinfo {author} {\bibfnamefont {M.}~\bibnamefont
  {Pospelov}}\ and\ \bibinfo {author} {\bibfnamefont {H.}~\bibnamefont
  {Ramani}},\ }\href {\doibase 10.1103/PhysRevD.103.115031} {\bibfield
  {journal} {\bibinfo  {journal} {Phys. Rev. D}\ }\textbf {\bibinfo {volume}
  {103}},\ \bibinfo {pages} {115031} (\bibinfo {year} {2021})},\ \Eprint
  {http://arxiv.org/abs/2012.03957} {arXiv:2012.03957 [hep-ph]} \BibitemShut
  {NoStop}%
\bibitem [{\citenamefont {Pospelov}\ \emph {et~al.}(2020)\citenamefont
  {Pospelov}, \citenamefont {Rajendran},\ and\ \citenamefont
  {Ramani}}]{Pospelov:2019vuf}%
  \BibitemOpen
  \bibfield  {author} {\bibinfo {author} {\bibfnamefont {M.}~\bibnamefont
  {Pospelov}}, \bibinfo {author} {\bibfnamefont {S.}~\bibnamefont {Rajendran}},
  \ and\ \bibinfo {author} {\bibfnamefont {H.}~\bibnamefont {Ramani}},\ }\href
  {\doibase 10.1103/PhysRevD.101.055001} {\bibfield  {journal} {\bibinfo
  {journal} {Phys. Rev. D}\ }\textbf {\bibinfo {volume} {101}},\ \bibinfo
  {pages} {055001} (\bibinfo {year} {2020})},\ \Eprint
  {http://arxiv.org/abs/1907.00011} {arXiv:1907.00011 [hep-ph]} \BibitemShut
  {NoStop}%
\bibitem [{\citenamefont {Rajendran}\ and\ \citenamefont
  {Ramani}(2021)}]{Rajendran:2020tmw}%
  \BibitemOpen
  \bibfield  {author} {\bibinfo {author} {\bibfnamefont {S.}~\bibnamefont
  {Rajendran}}\ and\ \bibinfo {author} {\bibfnamefont {H.}~\bibnamefont
  {Ramani}},\ }\href {\doibase 10.1103/PhysRevD.103.035014} {\bibfield
  {journal} {\bibinfo  {journal} {Phys. Rev. D}\ }\textbf {\bibinfo {volume}
  {103}},\ \bibinfo {pages} {035014} (\bibinfo {year} {2021})},\ \Eprint
  {http://arxiv.org/abs/2008.06061} {arXiv:2008.06061 [hep-ph]} \BibitemShut
  {NoStop}%
\bibitem [{\citenamefont {Xu}\ and\ \citenamefont {Farrar}(2021)}]{Xu:2021lmg}%
  \BibitemOpen
  \bibfield  {author} {\bibinfo {author} {\bibfnamefont {X.}~\bibnamefont
  {Xu}}\ and\ \bibinfo {author} {\bibfnamefont {G.~R.}\ \bibnamefont
  {Farrar}},\ }\href@noop {} {\  (\bibinfo {year} {2021})},\ \Eprint
  {http://arxiv.org/abs/2112.00707} {arXiv:2112.00707 [hep-ph]} \BibitemShut
  {NoStop}%
\bibitem [{\citenamefont {Budker}\ \emph {et~al.}(2022)\citenamefont {Budker},
  \citenamefont {Graham}, \citenamefont {Ramani}, \citenamefont
  {Schmidt-Kaler}, \citenamefont {Smorra},\ and\ \citenamefont
  {Ulmer}}]{Budker:2021quh}%
  \BibitemOpen
  \bibfield  {author} {\bibinfo {author} {\bibfnamefont {D.}~\bibnamefont
  {Budker}}, \bibinfo {author} {\bibfnamefont {P.~W.}\ \bibnamefont {Graham}},
  \bibinfo {author} {\bibfnamefont {H.}~\bibnamefont {Ramani}}, \bibinfo
  {author} {\bibfnamefont {F.}~\bibnamefont {Schmidt-Kaler}}, \bibinfo {author}
  {\bibfnamefont {C.}~\bibnamefont {Smorra}}, \ and\ \bibinfo {author}
  {\bibfnamefont {S.}~\bibnamefont {Ulmer}},\ }\href {\doibase
  10.1103/PRXQuantum.3.010330} {\bibfield  {journal} {\bibinfo  {journal} {PRX
  Quantum}\ }\textbf {\bibinfo {volume} {3}},\ \bibinfo {pages} {010330}
  (\bibinfo {year} {2022})},\ \Eprint {http://arxiv.org/abs/2108.05283}
  {arXiv:2108.05283 [hep-ph]} \BibitemShut {NoStop}%
\bibitem [{\citenamefont {McKeen}\ \emph {et~al.}(2022)\citenamefont {McKeen},
  \citenamefont {Moore}, \citenamefont {Morrissey}, \citenamefont {Pospelov},\
  and\ \citenamefont {Ramani}}]{McKeen:2022poo}%
  \BibitemOpen
  \bibfield  {author} {\bibinfo {author} {\bibfnamefont {D.}~\bibnamefont
  {McKeen}}, \bibinfo {author} {\bibfnamefont {M.}~\bibnamefont {Moore}},
  \bibinfo {author} {\bibfnamefont {D.~E.}\ \bibnamefont {Morrissey}}, \bibinfo
  {author} {\bibfnamefont {M.}~\bibnamefont {Pospelov}}, \ and\ \bibinfo
  {author} {\bibfnamefont {H.}~\bibnamefont {Ramani}},\ }\href@noop {} {\
  (\bibinfo {year} {2022})},\ \Eprint {http://arxiv.org/abs/2202.08840}
  {arXiv:2202.08840 [hep-ph]} \BibitemShut {NoStop}%
\bibitem [{\citenamefont {Billard}\ \emph {et~al.}(2022)\citenamefont
  {Billard}, \citenamefont {Pyle}, \citenamefont {Rajendran},\ and\
  \citenamefont {Ramani}}]{Billard:2022cqd}%
  \BibitemOpen
  \bibfield  {author} {\bibinfo {author} {\bibfnamefont {J.}~\bibnamefont
  {Billard}}, \bibinfo {author} {\bibfnamefont {M.}~\bibnamefont {Pyle}},
  \bibinfo {author} {\bibfnamefont {S.}~\bibnamefont {Rajendran}}, \ and\
  \bibinfo {author} {\bibfnamefont {H.}~\bibnamefont {Ramani}},\ }\href@noop {}
  {\  (\bibinfo {year} {2022})},\ \Eprint {http://arxiv.org/abs/2208.05485}
  {arXiv:2208.05485 [hep-ph]} \BibitemShut {NoStop}%
\bibitem [{\citenamefont {Li}\ \emph {et~al.}(2023)\citenamefont {Li},
  \citenamefont {Liu},\ and\ \citenamefont {Xue}}]{Li:2022idr}%
  \BibitemOpen
  \bibfield  {author} {\bibinfo {author} {\bibfnamefont {Y.}~\bibnamefont
  {Li}}, \bibinfo {author} {\bibfnamefont {Z.}~\bibnamefont {Liu}}, \ and\
  \bibinfo {author} {\bibfnamefont {Y.}~\bibnamefont {Xue}},\ }\href {\doibase
  10.1088/1475-7516/2023/05/060} {\bibfield  {journal} {\bibinfo  {journal}
  {JCAP}\ }\textbf {\bibinfo {volume} {05}},\ \bibinfo {pages} {060} (\bibinfo
  {year} {2023})},\ \Eprint {http://arxiv.org/abs/2209.04387} {arXiv:2209.04387
  [hep-ph]} \BibitemShut {NoStop}%
\bibitem [{\citenamefont {Bramante}\ \emph {et~al.}(2022)\citenamefont
  {Bramante}, \citenamefont {Kumar}, \citenamefont {Mohlabeng}, \citenamefont
  {Raj},\ and\ \citenamefont {Song}}]{Bramante:2022pmn}%
  \BibitemOpen
  \bibfield  {author} {\bibinfo {author} {\bibfnamefont {J.}~\bibnamefont
  {Bramante}}, \bibinfo {author} {\bibfnamefont {J.}~\bibnamefont {Kumar}},
  \bibinfo {author} {\bibfnamefont {G.}~\bibnamefont {Mohlabeng}}, \bibinfo
  {author} {\bibfnamefont {N.}~\bibnamefont {Raj}}, \ and\ \bibinfo {author}
  {\bibfnamefont {N.}~\bibnamefont {Song}},\ }\href@noop {} {\  (\bibinfo
  {year} {2022})},\ \Eprint {http://arxiv.org/abs/2210.01812} {arXiv:2210.01812
  [hep-ph]} \BibitemShut {NoStop}%
\bibitem [{\citenamefont {Pospelov}\ and\ \citenamefont
  {Ray}(2024)}]{Pospelov:2023mlz}%
  \BibitemOpen
  \bibfield  {author} {\bibinfo {author} {\bibfnamefont {M.}~\bibnamefont
  {Pospelov}}\ and\ \bibinfo {author} {\bibfnamefont {A.}~\bibnamefont {Ray}},\
  }\href {\doibase 10.1088/1475-7516/2024/01/029} {\bibfield  {journal}
  {\bibinfo  {journal} {JCAP}\ }\textbf {\bibinfo {volume} {2024}},\ \bibinfo
  {pages} {029} (\bibinfo {year} {2024})},\ \Eprint
  {http://arxiv.org/abs/2309.10032} {arXiv:2309.10032 [hep-ph]} \BibitemShut
  {NoStop}%
\bibitem [{\citenamefont {McKeen}\ \emph {et~al.}(2023)\citenamefont {McKeen},
  \citenamefont {Morrissey}, \citenamefont {Pospelov}, \citenamefont {Ramani},\
  and\ \citenamefont {Ray}}]{McKeen:2023ztq}%
  \BibitemOpen
  \bibfield  {author} {\bibinfo {author} {\bibfnamefont {D.}~\bibnamefont
  {McKeen}}, \bibinfo {author} {\bibfnamefont {D.~E.}\ \bibnamefont
  {Morrissey}}, \bibinfo {author} {\bibfnamefont {M.}~\bibnamefont {Pospelov}},
  \bibinfo {author} {\bibfnamefont {H.}~\bibnamefont {Ramani}}, \ and\ \bibinfo
  {author} {\bibfnamefont {A.}~\bibnamefont {Ray}},\ }\href {\doibase
  10.1103/PhysRevLett.131.011005} {\bibfield  {journal} {\bibinfo  {journal}
  {Phys. Rev. Lett.}\ }\textbf {\bibinfo {volume} {131}},\ \bibinfo {pages}
  {011005} (\bibinfo {year} {2023})},\ \Eprint
  {http://arxiv.org/abs/2303.03416} {arXiv:2303.03416 [hep-ph]} \BibitemShut
  {NoStop}%
\bibitem [{\citenamefont {Ema}\ \emph {et~al.}(2024)\citenamefont {Ema},
  \citenamefont {Pospelov},\ and\ \citenamefont {Ray}}]{Ema:2024oce}%
  \BibitemOpen
  \bibfield  {author} {\bibinfo {author} {\bibfnamefont {Y.}~\bibnamefont
  {Ema}}, \bibinfo {author} {\bibfnamefont {M.}~\bibnamefont {Pospelov}}, \
  and\ \bibinfo {author} {\bibfnamefont {A.}~\bibnamefont {Ray}},\ }\href@noop
  {} {\  (\bibinfo {year} {2024})},\ \Eprint {http://arxiv.org/abs/2402.03431}
  {arXiv:2402.03431 [hep-ph]} \BibitemShut {NoStop}%
\bibitem [{\citenamefont {Moore}\ and\ \citenamefont
  {Slatyer}(2024)}]{Moore:2024mot}%
  \BibitemOpen
  \bibfield  {author} {\bibinfo {author} {\bibfnamefont {M.}~\bibnamefont
  {Moore}}\ and\ \bibinfo {author} {\bibfnamefont {T.~R.}\ \bibnamefont
  {Slatyer}},\ }\href@noop {} {\  (\bibinfo {year} {2024})},\ \Eprint
  {http://arxiv.org/abs/2403.03972} {arXiv:2403.03972 [hep-ph]} \BibitemShut
  {NoStop}%
\bibitem [{\citenamefont {Batell}\ \emph {et~al.}(2010)\citenamefont {Batell},
  \citenamefont {Pospelov}, \citenamefont {Ritz},\ and\ \citenamefont
  {Shang}}]{Batell:2009zp}%
  \BibitemOpen
  \bibfield  {author} {\bibinfo {author} {\bibfnamefont {B.}~\bibnamefont
  {Batell}}, \bibinfo {author} {\bibfnamefont {M.}~\bibnamefont {Pospelov}},
  \bibinfo {author} {\bibfnamefont {A.}~\bibnamefont {Ritz}}, \ and\ \bibinfo
  {author} {\bibfnamefont {Y.}~\bibnamefont {Shang}},\ }\href {\doibase
  10.1103/PhysRevD.81.075004} {\bibfield  {journal} {\bibinfo  {journal} {Phys.
  Rev. D}\ }\textbf {\bibinfo {volume} {81}},\ \bibinfo {pages} {075004}
  (\bibinfo {year} {2010})},\ \Eprint {http://arxiv.org/abs/0910.1567}
  {arXiv:0910.1567 [hep-ph]} \BibitemShut {NoStop}%
\bibitem [{\citenamefont {Schuster}\ \emph
  {et~al.}(2010{\natexlab{a}})\citenamefont {Schuster}, \citenamefont {Toro},\
  and\ \citenamefont {Yavin}}]{Schuster:2009au}%
  \BibitemOpen
  \bibfield  {author} {\bibinfo {author} {\bibfnamefont {P.}~\bibnamefont
  {Schuster}}, \bibinfo {author} {\bibfnamefont {N.}~\bibnamefont {Toro}}, \
  and\ \bibinfo {author} {\bibfnamefont {I.}~\bibnamefont {Yavin}},\ }\href
  {\doibase 10.1103/PhysRevD.81.016002} {\bibfield  {journal} {\bibinfo
  {journal} {Phys. Rev. D}\ }\textbf {\bibinfo {volume} {81}},\ \bibinfo
  {pages} {016002} (\bibinfo {year} {2010}{\natexlab{a}})},\ \Eprint
  {http://arxiv.org/abs/0910.1602} {arXiv:0910.1602 [hep-ph]} \BibitemShut
  {NoStop}%
\bibitem [{\citenamefont {Schuster}\ \emph
  {et~al.}(2010{\natexlab{b}})\citenamefont {Schuster}, \citenamefont {Toro},
  \citenamefont {Weiner},\ and\ \citenamefont {Yavin}}]{Schuster:2009fc}%
  \BibitemOpen
  \bibfield  {author} {\bibinfo {author} {\bibfnamefont {P.}~\bibnamefont
  {Schuster}}, \bibinfo {author} {\bibfnamefont {N.}~\bibnamefont {Toro}},
  \bibinfo {author} {\bibfnamefont {N.}~\bibnamefont {Weiner}}, \ and\ \bibinfo
  {author} {\bibfnamefont {I.}~\bibnamefont {Yavin}},\ }\href {\doibase
  10.1103/PhysRevD.82.115012} {\bibfield  {journal} {\bibinfo  {journal} {Phys.
  Rev. D}\ }\textbf {\bibinfo {volume} {82}},\ \bibinfo {pages} {115012}
  (\bibinfo {year} {2010}{\natexlab{b}})},\ \Eprint
  {http://arxiv.org/abs/0910.1839} {arXiv:0910.1839 [hep-ph]} \BibitemShut
  {NoStop}%
\bibitem [{\citenamefont {Bell}\ and\ \citenamefont
  {Petraki}(2011)}]{Bell_2011}%
  \BibitemOpen
  \bibfield  {author} {\bibinfo {author} {\bibfnamefont {N.~F.}\ \bibnamefont
  {Bell}}\ and\ \bibinfo {author} {\bibfnamefont {K.}~\bibnamefont {Petraki}},\
  }\href {\doibase 10.1088/1475-7516/2011/04/003} {\bibfield  {journal}
  {\bibinfo  {journal} {Journal of Cosmology and Astroparticle Physics}\
  }\textbf {\bibinfo {volume} {2011}},\ \bibinfo {pages} {003–003} (\bibinfo
  {year} {2011})}\BibitemShut {NoStop}%
\bibitem [{\citenamefont {Kouvaris}\ and\ \citenamefont
  {Tinyakov}(2010)}]{Kouvaris:2010}%
  \BibitemOpen
  \bibfield  {author} {\bibinfo {author} {\bibfnamefont {C.}~\bibnamefont
  {Kouvaris}}\ and\ \bibinfo {author} {\bibfnamefont {P.}~\bibnamefont
  {Tinyakov}},\ }\href {\doibase 10.1103/PhysRevD.82.063531} {\bibfield
  {journal} {\bibinfo  {journal} {Phys. Rev. D}\ }\textbf {\bibinfo {volume}
  {82}},\ \bibinfo {pages} {063531} (\bibinfo {year} {2010})},\ \Eprint
  {http://arxiv.org/abs/1004.0586} {arXiv:1004.0586 [astro-ph.GA]} \BibitemShut
  {NoStop}%
\bibitem [{\citenamefont {Feng}\ \emph
  {et~al.}(2016{\natexlab{b}})\citenamefont {Feng}, \citenamefont {Smolinsky},\
  and\ \citenamefont {Tanedo}}]{Feng:2016ijc}%
  \BibitemOpen
  \bibfield  {author} {\bibinfo {author} {\bibfnamefont {J.~L.}\ \bibnamefont
  {Feng}}, \bibinfo {author} {\bibfnamefont {J.}~\bibnamefont {Smolinsky}}, \
  and\ \bibinfo {author} {\bibfnamefont {P.}~\bibnamefont {Tanedo}},\ }\href
  {\doibase 10.1103/PhysRevD.93.115036} {\bibfield  {journal} {\bibinfo
  {journal} {Phys. Rev. D}\ }\textbf {\bibinfo {volume} {93}},\ \bibinfo
  {pages} {115036} (\bibinfo {year} {2016}{\natexlab{b}})},\ \bibinfo {note}
  {[Erratum: Phys.Rev.D 96, 099903 (2017)]},\ \Eprint
  {http://arxiv.org/abs/1602.01465} {arXiv:1602.01465 [hep-ph]} \BibitemShut
  {NoStop}%
\bibitem [{\citenamefont {Allahverdi}\ \emph {et~al.}(2017)\citenamefont
  {Allahverdi}, \citenamefont {Gao}, \citenamefont {Knockel},\ and\
  \citenamefont {Shalgar}}]{Allahverdi:2016fvl}%
  \BibitemOpen
  \bibfield  {author} {\bibinfo {author} {\bibfnamefont {R.}~\bibnamefont
  {Allahverdi}}, \bibinfo {author} {\bibfnamefont {Y.}~\bibnamefont {Gao}},
  \bibinfo {author} {\bibfnamefont {B.}~\bibnamefont {Knockel}}, \ and\
  \bibinfo {author} {\bibfnamefont {S.}~\bibnamefont {Shalgar}},\ }\href
  {\doibase 10.1103/PhysRevD.95.075001} {\bibfield  {journal} {\bibinfo
  {journal} {Phys. Rev. D}\ }\textbf {\bibinfo {volume} {95}},\ \bibinfo
  {pages} {075001} (\bibinfo {year} {2017})},\ \Eprint
  {http://arxiv.org/abs/1612.03110} {arXiv:1612.03110 [hep-ph]} \BibitemShut
  {NoStop}%
\bibitem [{\citenamefont {Leane}\ \emph {et~al.}(2017)\citenamefont {Leane},
  \citenamefont {Ng},\ and\ \citenamefont {Beacom}}]{Leane:2017vag}%
  \BibitemOpen
  \bibfield  {author} {\bibinfo {author} {\bibfnamefont {R.~K.}\ \bibnamefont
  {Leane}}, \bibinfo {author} {\bibfnamefont {K.~C.~Y.}\ \bibnamefont {Ng}}, \
  and\ \bibinfo {author} {\bibfnamefont {J.~F.}\ \bibnamefont {Beacom}},\
  }\href {\doibase 10.1103/PhysRevD.95.123016} {\bibfield  {journal} {\bibinfo
  {journal} {Phys. Rev. D}\ }\textbf {\bibinfo {volume} {95}},\ \bibinfo
  {pages} {123016} (\bibinfo {year} {2017})},\ \Eprint
  {http://arxiv.org/abs/1703.04629} {arXiv:1703.04629 [astro-ph.HE]}
  \BibitemShut {NoStop}%
\bibitem [{\citenamefont {Arina}\ \emph {et~al.}(2017)\citenamefont {Arina},
  \citenamefont {Backovi\'c}, \citenamefont {Heisig},\ and\ \citenamefont
  {Lucente}}]{Arina:2017sng}%
  \BibitemOpen
  \bibfield  {author} {\bibinfo {author} {\bibfnamefont {C.}~\bibnamefont
  {Arina}}, \bibinfo {author} {\bibfnamefont {M.}~\bibnamefont {Backovi\'c}},
  \bibinfo {author} {\bibfnamefont {J.}~\bibnamefont {Heisig}}, \ and\ \bibinfo
  {author} {\bibfnamefont {M.}~\bibnamefont {Lucente}},\ }\href {\doibase
  10.1103/PhysRevD.96.063010} {\bibfield  {journal} {\bibinfo  {journal} {Phys.
  Rev. D}\ }\textbf {\bibinfo {volume} {96}},\ \bibinfo {pages} {063010}
  (\bibinfo {year} {2017})},\ \Eprint {http://arxiv.org/abs/1703.08087}
  {arXiv:1703.08087 [astro-ph.HE]} \BibitemShut {NoStop}%
\bibitem [{\citenamefont {Garani}\ and\ \citenamefont
  {Palomares-Ruiz}(2017)}]{Garani:2017jcj}%
  \BibitemOpen
  \bibfield  {author} {\bibinfo {author} {\bibfnamefont {R.}~\bibnamefont
  {Garani}}\ and\ \bibinfo {author} {\bibfnamefont {S.}~\bibnamefont
  {Palomares-Ruiz}},\ }\href {\doibase 10.1088/1475-7516/2017/05/007}
  {\bibfield  {journal} {\bibinfo  {journal} {JCAP}\ }\textbf {\bibinfo
  {volume} {05}},\ \bibinfo {pages} {007} (\bibinfo {year} {2017})},\ \Eprint
  {http://arxiv.org/abs/1702.02768} {arXiv:1702.02768 [hep-ph]} \BibitemShut
  {NoStop}%
\bibitem [{\citenamefont {Albert}\ \emph
  {et~al.}(2018{\natexlab{a}})\citenamefont {Albert} \emph
  {et~al.}}]{Albert:2018jwh}%
  \BibitemOpen
  \bibfield  {author} {\bibinfo {author} {\bibfnamefont {A.}~\bibnamefont
  {Albert}} \emph {et~al.} (\bibinfo {collaboration} {HAWC}),\ }\href {\doibase
  10.1103/PhysRevD.98.123012} {\bibfield  {journal} {\bibinfo  {journal} {Phys.
  Rev.}\ }\textbf {\bibinfo {volume} {D98}},\ \bibinfo {pages} {123012}
  (\bibinfo {year} {2018}{\natexlab{a}})},\ \Eprint
  {http://arxiv.org/abs/1808.05624} {arXiv:1808.05624 [hep-ph]} \BibitemShut
  {NoStop}%
%%CITATION = ARXIV:1808.05624;%%
\bibitem [{\citenamefont {Albert}\ \emph
  {et~al.}(2018{\natexlab{b}})\citenamefont {Albert} \emph
  {et~al.}}]{Albert:2018vcq}%
  \BibitemOpen
  \bibfield  {author} {\bibinfo {author} {\bibfnamefont {A.}~\bibnamefont
  {Albert}} \emph {et~al.} (\bibinfo {collaboration} {HAWC}),\ }\href {\doibase
  10.1103/PhysRevD.98.123011} {\bibfield  {journal} {\bibinfo  {journal} {Phys.
  Rev. D}\ }\textbf {\bibinfo {volume} {98}},\ \bibinfo {pages} {123011}
  (\bibinfo {year} {2018}{\natexlab{b}})},\ \Eprint
  {http://arxiv.org/abs/1808.05620} {arXiv:1808.05620 [astro-ph.HE]}
  \BibitemShut {NoStop}%
\bibitem [{\citenamefont {Nisa}\ \emph {et~al.}(2019)\citenamefont {Nisa},
  \citenamefont {Beacom}, \citenamefont {BenZvi}, \citenamefont {Leane},
  \citenamefont {Linden}, \citenamefont {Ng}, \citenamefont {Peter},\ and\
  \citenamefont {Zhou}}]{Nisa:2019mpb}%
  \BibitemOpen
  \bibfield  {author} {\bibinfo {author} {\bibfnamefont {M.~U.}\ \bibnamefont
  {Nisa}}, \bibinfo {author} {\bibfnamefont {J.~F.}\ \bibnamefont {Beacom}},
  \bibinfo {author} {\bibfnamefont {S.~Y.}\ \bibnamefont {BenZvi}}, \bibinfo
  {author} {\bibfnamefont {R.~K.}\ \bibnamefont {Leane}}, \bibinfo {author}
  {\bibfnamefont {T.}~\bibnamefont {Linden}}, \bibinfo {author} {\bibfnamefont
  {K.~C.~Y.}\ \bibnamefont {Ng}}, \bibinfo {author} {\bibfnamefont {A.~H.~G.}\
  \bibnamefont {Peter}}, \ and\ \bibinfo {author} {\bibfnamefont
  {B.}~\bibnamefont {Zhou}},\ }\href@noop {} {\  (\bibinfo {year} {2019})},\
  \Eprint {http://arxiv.org/abs/1903.06349} {arXiv:1903.06349 [astro-ph.HE]}
  \BibitemShut {NoStop}%
%%CITATION = ARXIV:1903.06349;%%
\bibitem [{\citenamefont {Niblaeus}\ \emph {et~al.}(2019)\citenamefont
  {Niblaeus}, \citenamefont {Beniwal},\ and\ \citenamefont
  {Edsjo}}]{Niblaeus:2019gjk}%
  \BibitemOpen
  \bibfield  {author} {\bibinfo {author} {\bibfnamefont {C.}~\bibnamefont
  {Niblaeus}}, \bibinfo {author} {\bibfnamefont {A.}~\bibnamefont {Beniwal}}, \
  and\ \bibinfo {author} {\bibfnamefont {J.}~\bibnamefont {Edsjo}},\ }\href
  {\doibase 10.1088/1475-7516/2019/11/011} {\bibfield  {journal} {\bibinfo
  {journal} {JCAP}\ }\textbf {\bibinfo {volume} {11}},\ \bibinfo {pages} {011}
  (\bibinfo {year} {2019})},\ \Eprint {http://arxiv.org/abs/1903.11363}
  {arXiv:1903.11363 [astro-ph.HE]} \BibitemShut {NoStop}%
\bibitem [{\citenamefont {Cuoco}\ \emph {et~al.}(2020)\citenamefont {Cuoco},
  \citenamefont {De~La Torre~Luque}, \citenamefont {Gargano}, \citenamefont
  {Gustafsson}, \citenamefont {Loparco}, \citenamefont {Mazziotta},\ and\
  \citenamefont {Serini}}]{Cuoco:2019mlb}%
  \BibitemOpen
  \bibfield  {author} {\bibinfo {author} {\bibfnamefont {A.}~\bibnamefont
  {Cuoco}}, \bibinfo {author} {\bibfnamefont {P.}~\bibnamefont {De~La
  Torre~Luque}}, \bibinfo {author} {\bibfnamefont {F.}~\bibnamefont {Gargano}},
  \bibinfo {author} {\bibfnamefont {M.}~\bibnamefont {Gustafsson}}, \bibinfo
  {author} {\bibfnamefont {F.}~\bibnamefont {Loparco}}, \bibinfo {author}
  {\bibfnamefont {M.}~\bibnamefont {Mazziotta}}, \ and\ \bibinfo {author}
  {\bibfnamefont {D.}~\bibnamefont {Serini}},\ }\href {\doibase
  10.1103/PhysRevD.101.022002} {\bibfield  {journal} {\bibinfo  {journal}
  {Phys. Rev. D}\ }\textbf {\bibinfo {volume} {101}},\ \bibinfo {pages}
  {022002} (\bibinfo {year} {2020})},\ \Eprint
  {http://arxiv.org/abs/1912.09373} {arXiv:1912.09373 [astro-ph.HE]}
  \BibitemShut {NoStop}%
\bibitem [{\citenamefont {Serini}\ \emph {et~al.}(2020)\citenamefont {Serini},
  \citenamefont {Loparco},\ and\ \citenamefont {Mazziotta}}]{Serini:2020yhb}%
  \BibitemOpen
  \bibfield  {author} {\bibinfo {author} {\bibfnamefont {D.}~\bibnamefont
  {Serini}}, \bibinfo {author} {\bibfnamefont {F.}~\bibnamefont {Loparco}}, \
  and\ \bibinfo {author} {\bibfnamefont {M.~N.}\ \bibnamefont {Mazziotta}}
  (\bibinfo {collaboration} {Fermi-LAT}),\ }\href {\doibase
  10.22323/1.358.0544} {\bibfield  {journal} {\bibinfo  {journal} {PoS}\
  }\textbf {\bibinfo {volume} {ICRC2019}},\ \bibinfo {pages} {544} (\bibinfo
  {year} {2020})}\BibitemShut {NoStop}%
\bibitem [{\citenamefont {Acevedo}\ \emph
  {et~al.}(2021{\natexlab{a}})\citenamefont {Acevedo}, \citenamefont
  {Bramante}, \citenamefont {Goodman}, \citenamefont {Kopp},\ and\
  \citenamefont {Opferkuch}}]{Acevedo:2020gro}%
  \BibitemOpen
  \bibfield  {author} {\bibinfo {author} {\bibfnamefont {J.~F.}\ \bibnamefont
  {Acevedo}}, \bibinfo {author} {\bibfnamefont {J.}~\bibnamefont {Bramante}},
  \bibinfo {author} {\bibfnamefont {A.}~\bibnamefont {Goodman}}, \bibinfo
  {author} {\bibfnamefont {J.}~\bibnamefont {Kopp}}, \ and\ \bibinfo {author}
  {\bibfnamefont {T.}~\bibnamefont {Opferkuch}},\ }\href {\doibase
  10.1088/1475-7516/2021/04/026} {\bibfield  {journal} {\bibinfo  {journal}
  {JCAP}\ }\textbf {\bibinfo {volume} {04}},\ \bibinfo {pages} {026} (\bibinfo
  {year} {2021}{\natexlab{a}})},\ \Eprint {http://arxiv.org/abs/2012.09176}
  {arXiv:2012.09176 [hep-ph]} \BibitemShut {NoStop}%
\bibitem [{\citenamefont {Mazziotta}\ \emph {et~al.}(2020)\citenamefont
  {Mazziotta}, \citenamefont {Loparco}, \citenamefont {Serini}, \citenamefont
  {Cuoco}, \citenamefont {De~La Torre~Luque}, \citenamefont {Gargano},\ and\
  \citenamefont {Gustafsson}}]{Mazziotta:2020foa}%
  \BibitemOpen
  \bibfield  {author} {\bibinfo {author} {\bibfnamefont {M.}~\bibnamefont
  {Mazziotta}}, \bibinfo {author} {\bibfnamefont {F.}~\bibnamefont {Loparco}},
  \bibinfo {author} {\bibfnamefont {D.}~\bibnamefont {Serini}}, \bibinfo
  {author} {\bibfnamefont {A.}~\bibnamefont {Cuoco}}, \bibinfo {author}
  {\bibfnamefont {P.}~\bibnamefont {De~La Torre~Luque}}, \bibinfo {author}
  {\bibfnamefont {F.}~\bibnamefont {Gargano}}, \ and\ \bibinfo {author}
  {\bibfnamefont {M.}~\bibnamefont {Gustafsson}},\ }\href {\doibase
  10.1103/PhysRevD.102.022003} {\bibfield  {journal} {\bibinfo  {journal}
  {Phys. Rev. D}\ }\textbf {\bibinfo {volume} {102}},\ \bibinfo {pages}
  {022003} (\bibinfo {year} {2020})},\ \Eprint
  {http://arxiv.org/abs/2006.04114} {arXiv:2006.04114 [astro-ph.HE]}
  \BibitemShut {NoStop}%
\bibitem [{\citenamefont {Bell}\ \emph
  {et~al.}(2021{\natexlab{a}})\citenamefont {Bell}, \citenamefont {Dent},\ and\
  \citenamefont {Sanderson}}]{Bell:2021pyy}%
  \BibitemOpen
  \bibfield  {author} {\bibinfo {author} {\bibfnamefont {N.~F.}\ \bibnamefont
  {Bell}}, \bibinfo {author} {\bibfnamefont {J.~B.}\ \bibnamefont {Dent}}, \
  and\ \bibinfo {author} {\bibfnamefont {I.~W.}\ \bibnamefont {Sanderson}},\
  }\href {\doibase 10.1103/PhysRevD.104.023024} {\bibfield  {journal} {\bibinfo
   {journal} {Phys. Rev. D}\ }\textbf {\bibinfo {volume} {104}},\ \bibinfo
  {pages} {023024} (\bibinfo {year} {2021}{\natexlab{a}})},\ \Eprint
  {http://arxiv.org/abs/2103.16794} {arXiv:2103.16794 [hep-ph]} \BibitemShut
  {NoStop}%
\bibitem [{\citenamefont {Bose}\ \emph
  {et~al.}(2022{\natexlab{a}})\citenamefont {Bose}, \citenamefont {Maity},\
  and\ \citenamefont {Ray}}]{Bose:2021cou}%
  \BibitemOpen
  \bibfield  {author} {\bibinfo {author} {\bibfnamefont {D.}~\bibnamefont
  {Bose}}, \bibinfo {author} {\bibfnamefont {T.~N.}\ \bibnamefont {Maity}}, \
  and\ \bibinfo {author} {\bibfnamefont {T.~S.}\ \bibnamefont {Ray}},\ }\href
  {\doibase 10.1103/PhysRevD.105.123013} {\bibfield  {journal} {\bibinfo
  {journal} {Phys. Rev. D}\ }\textbf {\bibinfo {volume} {105}},\ \bibinfo
  {pages} {123013} (\bibinfo {year} {2022}{\natexlab{a}})},\ \Eprint
  {http://arxiv.org/abs/2112.08286} {arXiv:2112.08286 [hep-ph]} \BibitemShut
  {NoStop}%
\bibitem [{\citenamefont {Maity}\ \emph {et~al.}(2023)\citenamefont {Maity},
  \citenamefont {Saha}, \citenamefont {Mondal},\ and\ \citenamefont
  {Laha}}]{Maity:2023rez}%
  \BibitemOpen
  \bibfield  {author} {\bibinfo {author} {\bibfnamefont {T.~N.}\ \bibnamefont
  {Maity}}, \bibinfo {author} {\bibfnamefont {A.~K.}\ \bibnamefont {Saha}},
  \bibinfo {author} {\bibfnamefont {S.}~\bibnamefont {Mondal}}, \ and\ \bibinfo
  {author} {\bibfnamefont {R.}~\bibnamefont {Laha}},\ }\href@noop {} {\
  (\bibinfo {year} {2023})},\ \Eprint {http://arxiv.org/abs/2308.12336}
  {arXiv:2308.12336 [hep-ph]} \BibitemShut {NoStop}%
\bibitem [{\citenamefont {Leane}\ and\ \citenamefont
  {Linden}(2023)}]{Leane:2021tjj}%
  \BibitemOpen
  \bibfield  {author} {\bibinfo {author} {\bibfnamefont {R.~K.}\ \bibnamefont
  {Leane}}\ and\ \bibinfo {author} {\bibfnamefont {T.}~\bibnamefont {Linden}},\
  }\href {\doibase 10.1103/PhysRevLett.131.071001} {\bibfield  {journal}
  {\bibinfo  {journal} {Phys. Rev. Lett.}\ }\textbf {\bibinfo {volume} {131}},\
  \bibinfo {pages} {071001} (\bibinfo {year} {2023})},\ \Eprint
  {http://arxiv.org/abs/2104.02068} {arXiv:2104.02068 [astro-ph.HE]}
  \BibitemShut {NoStop}%
\bibitem [{\citenamefont {Li}\ and\ \citenamefont {Fan}(2022)}]{Li:2022wix}%
  \BibitemOpen
  \bibfield  {author} {\bibinfo {author} {\bibfnamefont {L.}~\bibnamefont
  {Li}}\ and\ \bibinfo {author} {\bibfnamefont {J.}~\bibnamefont {Fan}},\
  }\href {\doibase 10.1007/JHEP10(2022)186} {\bibfield  {journal} {\bibinfo
  {journal} {JHEP}\ }\textbf {\bibinfo {volume} {10}},\ \bibinfo {pages} {186}
  (\bibinfo {year} {2022})},\ \Eprint {http://arxiv.org/abs/2207.13709}
  {arXiv:2207.13709 [hep-ph]} \BibitemShut {NoStop}%
\bibitem [{\citenamefont {French}\ and\ \citenamefont
  {Sher}(2022)}]{French:2022ccb}%
  \BibitemOpen
  \bibfield  {author} {\bibinfo {author} {\bibfnamefont {G.~M.}\ \bibnamefont
  {French}}\ and\ \bibinfo {author} {\bibfnamefont {M.}~\bibnamefont {Sher}},\
  }\href {\doibase 10.1103/PhysRevD.106.115037} {\bibfield  {journal} {\bibinfo
   {journal} {Phys. Rev. D}\ }\textbf {\bibinfo {volume} {106}},\ \bibinfo
  {pages} {115037} (\bibinfo {year} {2022})},\ \Eprint
  {http://arxiv.org/abs/2210.04761} {arXiv:2210.04761 [hep-ph]} \BibitemShut
  {NoStop}%
\bibitem [{\citenamefont {Blanco}\ and\ \citenamefont
  {Leane}(2024)}]{Blanco:2023qgi}%
  \BibitemOpen
  \bibfield  {author} {\bibinfo {author} {\bibfnamefont {C.}~\bibnamefont
  {Blanco}}\ and\ \bibinfo {author} {\bibfnamefont {R.~K.}\ \bibnamefont
  {Leane}},\ }\href {\doibase 10.1103/PhysRevLett.132.261002} {\bibfield
  {journal} {\bibinfo  {journal} {Phys. Rev. Lett.}\ }\textbf {\bibinfo
  {volume} {132}},\ \bibinfo {pages} {261002} (\bibinfo {year} {2024})},\
  \Eprint {http://arxiv.org/abs/2312.06758} {arXiv:2312.06758 [hep-ph]}
  \BibitemShut {NoStop}%
\bibitem [{\citenamefont {Ray}(2023)}]{Ray:2023auh}%
  \BibitemOpen
  \bibfield  {author} {\bibinfo {author} {\bibfnamefont {A.}~\bibnamefont
  {Ray}},\ }\href {\doibase 10.1103/PhysRevD.107.083012} {\bibfield  {journal}
  {\bibinfo  {journal} {Phys. Rev. D}\ }\textbf {\bibinfo {volume} {107}},\
  \bibinfo {pages} {083012} (\bibinfo {year} {2023})},\ \Eprint
  {http://arxiv.org/abs/2301.03625} {arXiv:2301.03625 [hep-ph]} \BibitemShut
  {NoStop}%
\bibitem [{\citenamefont {Yan}\ \emph {et~al.}(2023)\citenamefont {Yan},
  \citenamefont {Li},\ and\ \citenamefont {Fan}}]{Yan:2023kdg}%
  \BibitemOpen
  \bibfield  {author} {\bibinfo {author} {\bibfnamefont {S.}~\bibnamefont
  {Yan}}, \bibinfo {author} {\bibfnamefont {L.}~\bibnamefont {Li}}, \ and\
  \bibinfo {author} {\bibfnamefont {J.}~\bibnamefont {Fan}},\ }\href@noop {} {\
   (\bibinfo {year} {2023})},\ \Eprint {http://arxiv.org/abs/2312.06746}
  {arXiv:2312.06746 [hep-ph]} \BibitemShut {NoStop}%
\bibitem [{\citenamefont {Croon}\ and\ \citenamefont
  {Smirnov}(2023)}]{Croon:2023bmu}%
  \BibitemOpen
  \bibfield  {author} {\bibinfo {author} {\bibfnamefont {D.}~\bibnamefont
  {Croon}}\ and\ \bibinfo {author} {\bibfnamefont {J.}~\bibnamefont
  {Smirnov}},\ }\href@noop {} {\  (\bibinfo {year} {2023})},\ \Eprint
  {http://arxiv.org/abs/2309.02495} {arXiv:2309.02495 [hep-ph]} \BibitemShut
  {NoStop}%
\bibitem [{\citenamefont {Ansarifard}\ and\ \citenamefont
  {Farzan}(2024)}]{Ansarifard:2024fan}%
  \BibitemOpen
  \bibfield  {author} {\bibinfo {author} {\bibfnamefont {S.}~\bibnamefont
  {Ansarifard}}\ and\ \bibinfo {author} {\bibfnamefont {Y.}~\bibnamefont
  {Farzan}},\ }\href@noop {} {\  (\bibinfo {year} {2024})},\ \Eprint
  {http://arxiv.org/abs/2401.13043} {arXiv:2401.13043 [hep-ph]} \BibitemShut
  {NoStop}%
\bibitem [{\citenamefont {Linden}\ \emph {et~al.}(2024)\citenamefont {Linden},
  \citenamefont {Nguyen},\ and\ \citenamefont {Tait}}]{Linden:2024uph}%
  \BibitemOpen
  \bibfield  {author} {\bibinfo {author} {\bibfnamefont {T.}~\bibnamefont
  {Linden}}, \bibinfo {author} {\bibfnamefont {T.~T.~Q.}\ \bibnamefont
  {Nguyen}}, \ and\ \bibinfo {author} {\bibfnamefont {T.~M.~P.}\ \bibnamefont
  {Tait}},\ }\href@noop {} {\  (\bibinfo {year} {2024})},\ \Eprint
  {http://arxiv.org/abs/2402.01839} {arXiv:2402.01839 [hep-ph]} \BibitemShut
  {NoStop}%
\bibitem [{\citenamefont {Leane}\ and\ \citenamefont
  {Smirnov}(2021)}]{Leane:2020wob}%
  \BibitemOpen
  \bibfield  {author} {\bibinfo {author} {\bibfnamefont {R.~K.}\ \bibnamefont
  {Leane}}\ and\ \bibinfo {author} {\bibfnamefont {J.}~\bibnamefont
  {Smirnov}},\ }\href {\doibase 10.1103/PhysRevLett.126.161101} {\bibfield
  {journal} {\bibinfo  {journal} {Phys. Rev. Lett.}\ }\textbf {\bibinfo
  {volume} {126}},\ \bibinfo {pages} {161101} (\bibinfo {year} {2021})},\
  \Eprint {http://arxiv.org/abs/2010.00015} {arXiv:2010.00015 [hep-ph]}
  \BibitemShut {NoStop}%
\bibitem [{\citenamefont {Leane}\ \emph {et~al.}(2021)\citenamefont {Leane},
  \citenamefont {Linden}, \citenamefont {Mukhopadhyay},\ and\ \citenamefont
  {Toro}}]{Leane:2021ihh}%
  \BibitemOpen
  \bibfield  {author} {\bibinfo {author} {\bibfnamefont {R.~K.}\ \bibnamefont
  {Leane}}, \bibinfo {author} {\bibfnamefont {T.}~\bibnamefont {Linden}},
  \bibinfo {author} {\bibfnamefont {P.}~\bibnamefont {Mukhopadhyay}}, \ and\
  \bibinfo {author} {\bibfnamefont {N.}~\bibnamefont {Toro}},\ }\href {\doibase
  10.1103/PhysRevD.103.075030} {\bibfield  {journal} {\bibinfo  {journal}
  {Phys. Rev. D}\ }\textbf {\bibinfo {volume} {103}},\ \bibinfo {pages}
  {075030} (\bibinfo {year} {2021})},\ \Eprint
  {http://arxiv.org/abs/2101.12213} {arXiv:2101.12213 [astro-ph.HE]}
  \BibitemShut {NoStop}%
\bibitem [{\citenamefont {Bhattacharjee}\ and\ \citenamefont
  {Calore}(2023)}]{Bhattacharjee:2023qfi}%
  \BibitemOpen
  \bibfield  {author} {\bibinfo {author} {\bibfnamefont {P.}~\bibnamefont
  {Bhattacharjee}}\ and\ \bibinfo {author} {\bibfnamefont {F.}~\bibnamefont
  {Calore}},\ }\href@noop {} {\  (\bibinfo {year} {2023})},\ \Eprint
  {http://arxiv.org/abs/2311.18455} {arXiv:2311.18455 [astro-ph.HE]}
  \BibitemShut {NoStop}%
\bibitem [{\citenamefont {Acevedo}\ \emph
  {et~al.}(2024{\natexlab{a}})\citenamefont {Acevedo}, \citenamefont {Leane},\
  and\ \citenamefont {Reilly}}]{Acevedo:2024zkg}%
  \BibitemOpen
  \bibfield  {author} {\bibinfo {author} {\bibfnamefont {J.~F.}\ \bibnamefont
  {Acevedo}}, \bibinfo {author} {\bibfnamefont {R.~K.}\ \bibnamefont {Leane}},
  \ and\ \bibinfo {author} {\bibfnamefont {A.~J.}\ \bibnamefont {Reilly}},\
  }\href@noop {} {\  (\bibinfo {year} {2024}{\natexlab{a}})},\ \Eprint
  {http://arxiv.org/abs/2405.02393} {arXiv:2405.02393 [astro-ph.EP]}
  \BibitemShut {NoStop}%
\bibitem [{\citenamefont {Ilie}\ \emph {et~al.}(2024)\citenamefont {Ilie},
  \citenamefont {Levy},\ and\ \citenamefont {Diks}}]{Ilie:2023lbi}%
  \BibitemOpen
  \bibfield  {author} {\bibinfo {author} {\bibfnamefont {C.}~\bibnamefont
  {Ilie}}, \bibinfo {author} {\bibfnamefont {C.}~\bibnamefont {Levy}}, \ and\
  \bibinfo {author} {\bibfnamefont {J.}~\bibnamefont {Diks}},\ }\href {\doibase
  10.1088/1475-7516/2024/04/082} {\bibfield  {journal} {\bibinfo  {journal}
  {JCAP}\ }\textbf {\bibinfo {volume} {04}},\ \bibinfo {pages} {082} (\bibinfo
  {year} {2024})},\ \Eprint {http://arxiv.org/abs/2312.13979} {arXiv:2312.13979
  [astro-ph.CO]} \BibitemShut {NoStop}%
\bibitem [{\citenamefont {Benito}\ \emph {et~al.}(2024)\citenamefont {Benito},
  \citenamefont {Karchev}, \citenamefont {Leane}, \citenamefont {P\~oder},
  \citenamefont {Smirnov},\ and\ \citenamefont {Trotta}}]{Benito:2024yki}%
  \BibitemOpen
  \bibfield  {author} {\bibinfo {author} {\bibfnamefont {M.}~\bibnamefont
  {Benito}}, \bibinfo {author} {\bibfnamefont {K.}~\bibnamefont {Karchev}},
  \bibinfo {author} {\bibfnamefont {R.~K.}\ \bibnamefont {Leane}}, \bibinfo
  {author} {\bibfnamefont {S.}~\bibnamefont {P\~oder}}, \bibinfo {author}
  {\bibfnamefont {J.}~\bibnamefont {Smirnov}}, \ and\ \bibinfo {author}
  {\bibfnamefont {R.}~\bibnamefont {Trotta}},\ }\href {\doibase
  10.1088/1475-7516/2024/07/038} {\bibfield  {journal} {\bibinfo  {journal}
  {JCAP}\ }\textbf {\bibinfo {volume} {07}},\ \bibinfo {pages} {038} (\bibinfo
  {year} {2024})},\ \Eprint {http://arxiv.org/abs/2405.09578} {arXiv:2405.09578
  [astro-ph.IM]} \BibitemShut {NoStop}%
\bibitem [{\citenamefont {Mochkovitch}\ \emph {et~al.}(1985)\citenamefont
  {Mochkovitch}, \citenamefont {Olive},\ and\ \citenamefont
  {Silk}}]{Mochkovitch:1985vi}%
  \BibitemOpen
  \bibfield  {author} {\bibinfo {author} {\bibfnamefont {R.}~\bibnamefont
  {Mochkovitch}}, \bibinfo {author} {\bibfnamefont {K.~A.}\ \bibnamefont
  {Olive}}, \ and\ \bibinfo {author} {\bibfnamefont {J.}~\bibnamefont {Silk}},\
  }\href@noop {} {\enquote {\bibinfo {title} {{White Dwarf Cooling and Cold
  Dark Matter}},}\ } (\bibinfo {year} {1985})\BibitemShut {NoStop}%
\bibitem [{\citenamefont {Moskalenko}\ and\ \citenamefont
  {Wai}(2006)}]{Moskalenko:2006mk}%
  \BibitemOpen
  \bibfield  {author} {\bibinfo {author} {\bibfnamefont {I.~V.}\ \bibnamefont
  {Moskalenko}}\ and\ \bibinfo {author} {\bibfnamefont {L.}~\bibnamefont
  {Wai}},\ }\href@noop {} {\  (\bibinfo {year} {2006})},\ \Eprint
  {http://arxiv.org/abs/astro-ph/0608535} {arXiv:astro-ph/0608535} \BibitemShut
  {NoStop}%
\bibitem [{\citenamefont {Bertone}\ and\ \citenamefont
  {Fairbairn}(2008)}]{Bertone:2007ae}%
  \BibitemOpen
  \bibfield  {author} {\bibinfo {author} {\bibfnamefont {G.}~\bibnamefont
  {Bertone}}\ and\ \bibinfo {author} {\bibfnamefont {M.}~\bibnamefont
  {Fairbairn}},\ }\href {\doibase 10.1103/PhysRevD.77.043515} {\bibfield
  {journal} {\bibinfo  {journal} {Phys. Rev. D}\ }\textbf {\bibinfo {volume}
  {77}},\ \bibinfo {pages} {043515} (\bibinfo {year} {2008})},\ \Eprint
  {http://arxiv.org/abs/0709.1485} {arXiv:0709.1485 [astro-ph]} \BibitemShut
  {NoStop}%
\bibitem [{\citenamefont {McCullough}\ and\ \citenamefont
  {Fairbairn}(2010)}]{McCullough:2010ai}%
  \BibitemOpen
  \bibfield  {author} {\bibinfo {author} {\bibfnamefont {M.}~\bibnamefont
  {McCullough}}\ and\ \bibinfo {author} {\bibfnamefont {M.}~\bibnamefont
  {Fairbairn}},\ }\href {\doibase 10.1103/PhysRevD.81.083520} {\bibfield
  {journal} {\bibinfo  {journal} {Phys. Rev. D}\ }\textbf {\bibinfo {volume}
  {81}},\ \bibinfo {pages} {083520} (\bibinfo {year} {2010})},\ \Eprint
  {http://arxiv.org/abs/1001.2737} {arXiv:1001.2737 [hep-ph]} \BibitemShut
  {NoStop}%
\bibitem [{\citenamefont {Hooper}\ \emph {et~al.}(2010)\citenamefont {Hooper},
  \citenamefont {Spolyar}, \citenamefont {Vallinotto},\ and\ \citenamefont
  {Gnedin}}]{Hooper:2010es}%
  \BibitemOpen
  \bibfield  {author} {\bibinfo {author} {\bibfnamefont {D.}~\bibnamefont
  {Hooper}}, \bibinfo {author} {\bibfnamefont {D.}~\bibnamefont {Spolyar}},
  \bibinfo {author} {\bibfnamefont {A.}~\bibnamefont {Vallinotto}}, \ and\
  \bibinfo {author} {\bibfnamefont {N.~Y.}\ \bibnamefont {Gnedin}},\ }\href
  {\doibase 10.1103/PhysRevD.81.103531} {\bibfield  {journal} {\bibinfo
  {journal} {Phys. Rev. D}\ }\textbf {\bibinfo {volume} {81}},\ \bibinfo
  {pages} {103531} (\bibinfo {year} {2010})},\ \Eprint
  {http://arxiv.org/abs/1002.0005} {arXiv:1002.0005 [hep-ph]} \BibitemShut
  {NoStop}%
\bibitem [{\citenamefont {Kouvaris}\ and\ \citenamefont
  {Tinyakov}(2011{\natexlab{a}})}]{Kouvaris:2010jy}%
  \BibitemOpen
  \bibfield  {author} {\bibinfo {author} {\bibfnamefont {C.}~\bibnamefont
  {Kouvaris}}\ and\ \bibinfo {author} {\bibfnamefont {P.}~\bibnamefont
  {Tinyakov}},\ }\href {\doibase 10.1103/PhysRevD.83.083512} {\bibfield
  {journal} {\bibinfo  {journal} {Phys. Rev. D}\ }\textbf {\bibinfo {volume}
  {83}},\ \bibinfo {pages} {083512} (\bibinfo {year} {2011}{\natexlab{a}})},\
  \Eprint {http://arxiv.org/abs/1012.2039} {arXiv:1012.2039 [astro-ph.HE]}
  \BibitemShut {NoStop}%
\bibitem [{\citenamefont {Bramante}(2015)}]{Bramante:2015cua}%
  \BibitemOpen
  \bibfield  {author} {\bibinfo {author} {\bibfnamefont {J.}~\bibnamefont
  {Bramante}},\ }\href {\doibase 10.1103/PhysRevLett.115.141301} {\bibfield
  {journal} {\bibinfo  {journal} {Phys. Rev. Lett.}\ }\textbf {\bibinfo
  {volume} {115}},\ \bibinfo {pages} {141301} (\bibinfo {year} {2015})},\
  \Eprint {http://arxiv.org/abs/1505.07464} {arXiv:1505.07464 [hep-ph]}
  \BibitemShut {NoStop}%
%%CITATION = ARXIV:1505.07464;%%
\bibitem [{\citenamefont {Graham}\ \emph {et~al.}(2015)\citenamefont {Graham},
  \citenamefont {Rajendran},\ and\ \citenamefont {Varela}}]{Graham:2015apa}%
  \BibitemOpen
  \bibfield  {author} {\bibinfo {author} {\bibfnamefont {P.~W.}\ \bibnamefont
  {Graham}}, \bibinfo {author} {\bibfnamefont {S.}~\bibnamefont {Rajendran}}, \
  and\ \bibinfo {author} {\bibfnamefont {J.}~\bibnamefont {Varela}},\ }\href
  {\doibase 10.1103/PhysRevD.92.063007} {\bibfield  {journal} {\bibinfo
  {journal} {Phys. Rev.}\ }\textbf {\bibinfo {volume} {D92}},\ \bibinfo {pages}
  {063007} (\bibinfo {year} {2015})},\ \Eprint
  {http://arxiv.org/abs/1505.04444} {arXiv:1505.04444 [hep-ph]} \BibitemShut
  {NoStop}%
%%CITATION = ARXIV:1505.04444;%%
\bibitem [{\citenamefont {Amaro-Seoane}\ \emph {et~al.}(2016)\citenamefont
  {Amaro-Seoane}, \citenamefont {Casanellas}, \citenamefont {Sch\"odel},
  \citenamefont {Davidson},\ and\ \citenamefont
  {Cuadra}}]{Amaro-Seoane:2015uny}%
  \BibitemOpen
  \bibfield  {author} {\bibinfo {author} {\bibfnamefont {P.}~\bibnamefont
  {Amaro-Seoane}}, \bibinfo {author} {\bibfnamefont {J.}~\bibnamefont
  {Casanellas}}, \bibinfo {author} {\bibfnamefont {R.}~\bibnamefont
  {Sch\"odel}}, \bibinfo {author} {\bibfnamefont {E.}~\bibnamefont {Davidson}},
  \ and\ \bibinfo {author} {\bibfnamefont {J.}~\bibnamefont {Cuadra}},\ }\href
  {\doibase 10.1093/mnras/stw433} {\bibfield  {journal} {\bibinfo  {journal}
  {Mon. Not. Roy. Astron. Soc.}\ }\textbf {\bibinfo {volume} {459}},\ \bibinfo
  {pages} {695} (\bibinfo {year} {2016})},\ \Eprint
  {http://arxiv.org/abs/1512.00456} {arXiv:1512.00456 [astro-ph.CO]}
  \BibitemShut {NoStop}%
\bibitem [{\citenamefont {Graham}\ \emph {et~al.}(2018)\citenamefont {Graham},
  \citenamefont {Janish}, \citenamefont {Narayan}, \citenamefont {Rajendran},\
  and\ \citenamefont {Riggins}}]{Graham:2018efk}%
  \BibitemOpen
  \bibfield  {author} {\bibinfo {author} {\bibfnamefont {P.~W.}\ \bibnamefont
  {Graham}}, \bibinfo {author} {\bibfnamefont {R.}~\bibnamefont {Janish}},
  \bibinfo {author} {\bibfnamefont {V.}~\bibnamefont {Narayan}}, \bibinfo
  {author} {\bibfnamefont {S.}~\bibnamefont {Rajendran}}, \ and\ \bibinfo
  {author} {\bibfnamefont {P.}~\bibnamefont {Riggins}},\ }\href {\doibase
  10.1103/PhysRevD.98.115027} {\bibfield  {journal} {\bibinfo  {journal} {Phys.
  Rev. D}\ }\textbf {\bibinfo {volume} {98}},\ \bibinfo {pages} {115027}
  (\bibinfo {year} {2018})},\ \Eprint {http://arxiv.org/abs/1805.07381}
  {arXiv:1805.07381 [hep-ph]} \BibitemShut {NoStop}%
\bibitem [{\citenamefont {Cerme\~no}\ and\ \citenamefont
  {P\'erez-Garc\'\i{}a}(2018)}]{Cermeno:2018qgu}%
  \BibitemOpen
  \bibfield  {author} {\bibinfo {author} {\bibfnamefont {M.}~\bibnamefont
  {Cerme\~no}}\ and\ \bibinfo {author} {\bibfnamefont {M.~A.}\ \bibnamefont
  {P\'erez-Garc\'\i{}a}},\ }\href {\doibase 10.1103/PhysRevD.98.063002}
  {\bibfield  {journal} {\bibinfo  {journal} {Phys. Rev. D}\ }\textbf {\bibinfo
  {volume} {98}},\ \bibinfo {pages} {063002} (\bibinfo {year} {2018})},\
  \Eprint {http://arxiv.org/abs/1807.03318} {arXiv:1807.03318 [hep-ph]}
  \BibitemShut {NoStop}%
\bibitem [{\citenamefont {Acevedo}\ and\ \citenamefont
  {Bramante}(2019)}]{Acevedo:2019gre}%
  \BibitemOpen
  \bibfield  {author} {\bibinfo {author} {\bibfnamefont {J.~F.}\ \bibnamefont
  {Acevedo}}\ and\ \bibinfo {author} {\bibfnamefont {J.}~\bibnamefont
  {Bramante}},\ }\href {\doibase 10.1103/PhysRevD.100.043020} {\bibfield
  {journal} {\bibinfo  {journal} {Phys. Rev.}\ }\textbf {\bibinfo {volume}
  {D100}},\ \bibinfo {pages} {043020} (\bibinfo {year} {2019})},\ \Eprint
  {http://arxiv.org/abs/1904.11993} {arXiv:1904.11993 [hep-ph]} \BibitemShut
  {NoStop}%
%%CITATION = ARXIV:1904.11993;%%
\bibitem [{\citenamefont {Acevedo}\ \emph
  {et~al.}(2021{\natexlab{b}})\citenamefont {Acevedo}, \citenamefont
  {Bramante},\ and\ \citenamefont {Goodman}}]{Acevedo:2020avd}%
  \BibitemOpen
  \bibfield  {author} {\bibinfo {author} {\bibfnamefont {J.~F.}\ \bibnamefont
  {Acevedo}}, \bibinfo {author} {\bibfnamefont {J.}~\bibnamefont {Bramante}}, \
  and\ \bibinfo {author} {\bibfnamefont {A.}~\bibnamefont {Goodman}},\ }\href
  {\doibase 10.1103/PhysRevD.103.123022} {\bibfield  {journal} {\bibinfo
  {journal} {Phys. Rev. D}\ }\textbf {\bibinfo {volume} {103}},\ \bibinfo
  {pages} {123022} (\bibinfo {year} {2021}{\natexlab{b}})},\ \Eprint
  {http://arxiv.org/abs/2012.10998} {arXiv:2012.10998 [hep-ph]} \BibitemShut
  {NoStop}%
\bibitem [{\citenamefont {Acevedo}\ \emph {et~al.}(2022)\citenamefont
  {Acevedo}, \citenamefont {Bramante},\ and\ \citenamefont
  {Goodman}}]{Acevedo:2021kly}%
  \BibitemOpen
  \bibfield  {author} {\bibinfo {author} {\bibfnamefont {J.~F.}\ \bibnamefont
  {Acevedo}}, \bibinfo {author} {\bibfnamefont {J.}~\bibnamefont {Bramante}}, \
  and\ \bibinfo {author} {\bibfnamefont {A.}~\bibnamefont {Goodman}},\ }\href
  {\doibase 10.1103/PhysRevD.105.023012} {\bibfield  {journal} {\bibinfo
  {journal} {Phys. Rev. D}\ }\textbf {\bibinfo {volume} {105}},\ \bibinfo
  {pages} {023012} (\bibinfo {year} {2022})},\ \Eprint
  {http://arxiv.org/abs/2108.10889} {arXiv:2108.10889 [hep-ph]} \BibitemShut
  {NoStop}%
\bibitem [{\citenamefont {Janish}\ \emph {et~al.}(2019)\citenamefont {Janish},
  \citenamefont {Narayan},\ and\ \citenamefont {Riggins}}]{Janish:2019nkk}%
  \BibitemOpen
  \bibfield  {author} {\bibinfo {author} {\bibfnamefont {R.}~\bibnamefont
  {Janish}}, \bibinfo {author} {\bibfnamefont {V.}~\bibnamefont {Narayan}}, \
  and\ \bibinfo {author} {\bibfnamefont {P.}~\bibnamefont {Riggins}},\ }\href
  {\doibase 10.1103/PhysRevD.100.035008} {\bibfield  {journal} {\bibinfo
  {journal} {Phys. Rev.}\ }\textbf {\bibinfo {volume} {D100}},\ \bibinfo
  {pages} {035008} (\bibinfo {year} {2019})},\ \Eprint
  {http://arxiv.org/abs/1905.00395} {arXiv:1905.00395 [hep-ph]} \BibitemShut
  {NoStop}%
%%CITATION = ARXIV:1905.00395;%%
\bibitem [{\citenamefont {Krall}\ and\ \citenamefont
  {Reece}(2018)}]{Krall:2017xij}%
  \BibitemOpen
  \bibfield  {author} {\bibinfo {author} {\bibfnamefont {R.}~\bibnamefont
  {Krall}}\ and\ \bibinfo {author} {\bibfnamefont {M.}~\bibnamefont {Reece}},\
  }\href {\doibase 10.1088/1674-1137/42/4/043105} {\bibfield  {journal}
  {\bibinfo  {journal} {Chin. Phys.}\ }\textbf {\bibinfo {volume} {C42}},\
  \bibinfo {pages} {043105} (\bibinfo {year} {2018})},\ \Eprint
  {http://arxiv.org/abs/1705.04843} {arXiv:1705.04843 [hep-ph]} \BibitemShut
  {NoStop}%
%%CITATION = ARXIV:1705.04843;%%
\bibitem [{\citenamefont {Panotopoulos}\ and\ \citenamefont
  {Lopes}(2020)}]{Panotopoulos:2020kuo}%
  \BibitemOpen
  \bibfield  {author} {\bibinfo {author} {\bibfnamefont {G.}~\bibnamefont
  {Panotopoulos}}\ and\ \bibinfo {author} {\bibfnamefont {I.}~\bibnamefont
  {Lopes}},\ }\href {\doibase 10.1142/S0218271820500583} {\bibfield  {journal}
  {\bibinfo  {journal} {Int. J. Mod. Phys. D}\ }\textbf {\bibinfo {volume}
  {29}},\ \bibinfo {pages} {2050058} (\bibinfo {year} {2020})},\ \Eprint
  {http://arxiv.org/abs/2005.11563} {arXiv:2005.11563 [hep-ph]} \BibitemShut
  {NoStop}%
\bibitem [{\citenamefont {Curtin}\ and\ \citenamefont
  {Setford}(2021)}]{Curtin:2020tkm}%
  \BibitemOpen
  \bibfield  {author} {\bibinfo {author} {\bibfnamefont {D.}~\bibnamefont
  {Curtin}}\ and\ \bibinfo {author} {\bibfnamefont {J.}~\bibnamefont
  {Setford}},\ }\href {\doibase 10.1007/JHEP03(2021)166} {\bibfield  {journal}
  {\bibinfo  {journal} {JHEP}\ }\textbf {\bibinfo {volume} {03}},\ \bibinfo
  {pages} {166} (\bibinfo {year} {2021})},\ \Eprint
  {http://arxiv.org/abs/2010.00601} {arXiv:2010.00601 [hep-ph]} \BibitemShut
  {NoStop}%
\bibitem [{\citenamefont {Bell}\ \emph
  {et~al.}(2021{\natexlab{b}})\citenamefont {Bell}, \citenamefont {Busoni},
  \citenamefont {Ramirez-Quezada}, \citenamefont {Robles},\ and\ \citenamefont
  {Virgato}}]{Bell:2021fye}%
  \BibitemOpen
  \bibfield  {author} {\bibinfo {author} {\bibfnamefont {N.~F.}\ \bibnamefont
  {Bell}}, \bibinfo {author} {\bibfnamefont {G.}~\bibnamefont {Busoni}},
  \bibinfo {author} {\bibfnamefont {M.~E.}\ \bibnamefont {Ramirez-Quezada}},
  \bibinfo {author} {\bibfnamefont {S.}~\bibnamefont {Robles}}, \ and\ \bibinfo
  {author} {\bibfnamefont {M.}~\bibnamefont {Virgato}},\ }\href {\doibase
  10.1088/1475-7516/2021/10/083} {\bibfield  {journal} {\bibinfo  {journal}
  {JCAP}\ }\textbf {\bibinfo {volume} {10}},\ \bibinfo {pages} {083} (\bibinfo
  {year} {2021}{\natexlab{b}})},\ \Eprint {http://arxiv.org/abs/2104.14367}
  {arXiv:2104.14367 [hep-ph]} \BibitemShut {NoStop}%
\bibitem [{\citenamefont {DeRocco}\ \emph {et~al.}(2022)\citenamefont
  {DeRocco}, \citenamefont {Galanis},\ and\ \citenamefont
  {Lasenby}}]{DeRocco:2022rze}%
  \BibitemOpen
  \bibfield  {author} {\bibinfo {author} {\bibfnamefont {W.}~\bibnamefont
  {DeRocco}}, \bibinfo {author} {\bibfnamefont {M.}~\bibnamefont {Galanis}}, \
  and\ \bibinfo {author} {\bibfnamefont {R.}~\bibnamefont {Lasenby}},\ }\href
  {\doibase 10.1088/1475-7516/2022/05/015} {\bibfield  {journal} {\bibinfo
  {journal} {JCAP}\ }\textbf {\bibinfo {volume} {05}},\ \bibinfo {pages} {015}
  (\bibinfo {year} {2022})},\ \Eprint {http://arxiv.org/abs/2201.05167}
  {arXiv:2201.05167 [hep-ph]} \BibitemShut {NoStop}%
\bibitem [{\citenamefont {Ramirez-Quezada}(2022)}]{Ramirez-Quezada:2022uou}%
  \BibitemOpen
  \bibfield  {author} {\bibinfo {author} {\bibfnamefont {M.~E.}\ \bibnamefont
  {Ramirez-Quezada}},\ }\href@noop {} {\  (\bibinfo {year} {2022})},\ \Eprint
  {http://arxiv.org/abs/2212.09785} {arXiv:2212.09785 [hep-ph]} \BibitemShut
  {NoStop}%
\bibitem [{\citenamefont {Smirnov}\ \emph {et~al.}(2022)\citenamefont
  {Smirnov}, \citenamefont {Goobar}, \citenamefont {Linden},\ and\
  \citenamefont {M\"ortsell}}]{Smirnov:2022zip}%
  \BibitemOpen
  \bibfield  {author} {\bibinfo {author} {\bibfnamefont {J.}~\bibnamefont
  {Smirnov}}, \bibinfo {author} {\bibfnamefont {A.}~\bibnamefont {Goobar}},
  \bibinfo {author} {\bibfnamefont {T.}~\bibnamefont {Linden}}, \ and\ \bibinfo
  {author} {\bibfnamefont {E.}~\bibnamefont {M\"ortsell}},\ }\href@noop {} {\
  (\bibinfo {year} {2022})},\ \Eprint {http://arxiv.org/abs/2211.00013}
  {arXiv:2211.00013 [astro-ph.CO]} \BibitemShut {NoStop}%
\bibitem [{\citenamefont {Garani}\ \emph {et~al.}(2023)\citenamefont {Garani},
  \citenamefont {Raj},\ and\ \citenamefont {Reynoso-Cordova}}]{Garani:2023esk}%
  \BibitemOpen
  \bibfield  {author} {\bibinfo {author} {\bibfnamefont {R.}~\bibnamefont
  {Garani}}, \bibinfo {author} {\bibfnamefont {N.}~\bibnamefont {Raj}}, \ and\
  \bibinfo {author} {\bibfnamefont {J.}~\bibnamefont {Reynoso-Cordova}},\
  }\href {\doibase 10.1088/1475-7516/2023/07/038} {\bibfield  {journal}
  {\bibinfo  {journal} {JCAP}\ }\textbf {\bibinfo {volume} {07}},\ \bibinfo
  {pages} {038} (\bibinfo {year} {2023})},\ \Eprint
  {http://arxiv.org/abs/2303.18009} {arXiv:2303.18009 [astro-ph.HE]}
  \BibitemShut {NoStop}%
\bibitem [{\citenamefont {Acevedo}\ \emph
  {et~al.}(2024{\natexlab{b}})\citenamefont {Acevedo}, \citenamefont {Leane},\
  and\ \citenamefont {Santos-Olmsted}}]{Acevedo:2023xnu}%
  \BibitemOpen
  \bibfield  {author} {\bibinfo {author} {\bibfnamefont {J.~F.}\ \bibnamefont
  {Acevedo}}, \bibinfo {author} {\bibfnamefont {R.~K.}\ \bibnamefont {Leane}},
  \ and\ \bibinfo {author} {\bibfnamefont {L.}~\bibnamefont {Santos-Olmsted}},\
  }\href {\doibase 10.1088/1475-7516/2024/03/042} {\bibfield  {journal}
  {\bibinfo  {journal} {JCAP}\ }\textbf {\bibinfo {volume} {03}},\ \bibinfo
  {pages} {042} (\bibinfo {year} {2024}{\natexlab{b}})},\ \Eprint
  {http://arxiv.org/abs/2309.10843} {arXiv:2309.10843 [hep-ph]} \BibitemShut
  {NoStop}%
\bibitem [{\citenamefont {Goldman}\ and\ \citenamefont
  {Nussinov}(1989)}]{Goldman:1989nd}%
  \BibitemOpen
  \bibfield  {author} {\bibinfo {author} {\bibfnamefont {I.}~\bibnamefont
  {Goldman}}\ and\ \bibinfo {author} {\bibfnamefont {S.}~\bibnamefont
  {Nussinov}},\ }\href {\doibase 10.1103/PhysRevD.40.3221} {\bibfield
  {journal} {\bibinfo  {journal} {Phys. Rev. D}\ }\textbf {\bibinfo {volume}
  {40}},\ \bibinfo {pages} {3221} (\bibinfo {year} {1989})}\BibitemShut
  {NoStop}%
\bibitem [{\citenamefont {Gould}\ \emph {et~al.}(1990)\citenamefont {Gould},
  \citenamefont {Draine}, \citenamefont {Romani},\ and\ \citenamefont
  {Nussinov}}]{Gould:1989gw}%
  \BibitemOpen
  \bibfield  {author} {\bibinfo {author} {\bibfnamefont {A.}~\bibnamefont
  {Gould}}, \bibinfo {author} {\bibfnamefont {B.~T.}\ \bibnamefont {Draine}},
  \bibinfo {author} {\bibfnamefont {R.~W.}\ \bibnamefont {Romani}}, \ and\
  \bibinfo {author} {\bibfnamefont {S.}~\bibnamefont {Nussinov}},\ }\href
  {\doibase 10.1016/0370-2693(90)91745-W} {\bibfield  {journal} {\bibinfo
  {journal} {Phys. Lett. B}\ }\textbf {\bibinfo {volume} {238}},\ \bibinfo
  {pages} {337} (\bibinfo {year} {1990})}\BibitemShut {NoStop}%
\bibitem [{\citenamefont {Kouvaris}(2008)}]{Kouvaris:2007ay}%
  \BibitemOpen
  \bibfield  {author} {\bibinfo {author} {\bibfnamefont {C.}~\bibnamefont
  {Kouvaris}},\ }\href {\doibase 10.1103/PhysRevD.77.023006} {\bibfield
  {journal} {\bibinfo  {journal} {Phys. Rev. D}\ }\textbf {\bibinfo {volume}
  {77}},\ \bibinfo {pages} {023006} (\bibinfo {year} {2008})},\ \Eprint
  {http://arxiv.org/abs/0708.2362} {arXiv:0708.2362 [astro-ph]} \BibitemShut
  {NoStop}%
\bibitem [{\citenamefont {de~Lavallaz}\ and\ \citenamefont
  {Fairbairn}(2010)}]{deLavallaz:2010wp}%
  \BibitemOpen
  \bibfield  {author} {\bibinfo {author} {\bibfnamefont {A.}~\bibnamefont
  {de~Lavallaz}}\ and\ \bibinfo {author} {\bibfnamefont {M.}~\bibnamefont
  {Fairbairn}},\ }\href {\doibase 10.1103/PhysRevD.81.123521} {\bibfield
  {journal} {\bibinfo  {journal} {Phys. Rev. D}\ }\textbf {\bibinfo {volume}
  {81}},\ \bibinfo {pages} {123521} (\bibinfo {year} {2010})},\ \Eprint
  {http://arxiv.org/abs/1004.0629} {arXiv:1004.0629 [astro-ph.GA]} \BibitemShut
  {NoStop}%
\bibitem [{\citenamefont {McDermott}\ \emph {et~al.}(2012)\citenamefont
  {McDermott}, \citenamefont {Yu},\ and\ \citenamefont
  {Zurek}}]{McDermott:2011jp}%
  \BibitemOpen
  \bibfield  {author} {\bibinfo {author} {\bibfnamefont {S.~D.}\ \bibnamefont
  {McDermott}}, \bibinfo {author} {\bibfnamefont {H.-B.}\ \bibnamefont {Yu}}, \
  and\ \bibinfo {author} {\bibfnamefont {K.~M.}\ \bibnamefont {Zurek}},\ }\href
  {\doibase 10.1103/PhysRevD.85.023519} {\bibfield  {journal} {\bibinfo
  {journal} {Phys. Rev. D}\ }\textbf {\bibinfo {volume} {85}},\ \bibinfo
  {pages} {023519} (\bibinfo {year} {2012})},\ \Eprint
  {http://arxiv.org/abs/1103.5472} {arXiv:1103.5472 [hep-ph]} \BibitemShut
  {NoStop}%
\bibitem [{\citenamefont {Kouvaris}\ and\ \citenamefont
  {Tinyakov}(2011{\natexlab{b}})}]{Kouvaris:2011fi}%
  \BibitemOpen
  \bibfield  {author} {\bibinfo {author} {\bibfnamefont {C.}~\bibnamefont
  {Kouvaris}}\ and\ \bibinfo {author} {\bibfnamefont {P.}~\bibnamefont
  {Tinyakov}},\ }\href {\doibase 10.1103/PhysRevLett.107.091301} {\bibfield
  {journal} {\bibinfo  {journal} {Phys. Rev. Lett.}\ }\textbf {\bibinfo
  {volume} {107}},\ \bibinfo {pages} {091301} (\bibinfo {year}
  {2011}{\natexlab{b}})},\ \Eprint {http://arxiv.org/abs/1104.0382}
  {arXiv:1104.0382 [astro-ph.CO]} \BibitemShut {NoStop}%
\bibitem [{\citenamefont {G\"uver}\ \emph {et~al.}(2014)\citenamefont
  {G\"uver}, \citenamefont {Erkoca}, \citenamefont {Hall~Reno},\ and\
  \citenamefont {Sarcevic}}]{Guver:2012ba}%
  \BibitemOpen
  \bibfield  {author} {\bibinfo {author} {\bibfnamefont {T.}~\bibnamefont
  {G\"uver}}, \bibinfo {author} {\bibfnamefont {A.~E.}\ \bibnamefont {Erkoca}},
  \bibinfo {author} {\bibfnamefont {M.}~\bibnamefont {Hall~Reno}}, \ and\
  \bibinfo {author} {\bibfnamefont {I.}~\bibnamefont {Sarcevic}},\ }\href
  {\doibase 10.1088/1475-7516/2014/05/013} {\bibfield  {journal} {\bibinfo
  {journal} {JCAP}\ }\textbf {\bibinfo {volume} {05}},\ \bibinfo {pages} {013}
  (\bibinfo {year} {2014})},\ \Eprint {http://arxiv.org/abs/1201.2400}
  {arXiv:1201.2400 [hep-ph]} \BibitemShut {NoStop}%
\bibitem [{\citenamefont {Bramante}\ \emph {et~al.}(2013)\citenamefont
  {Bramante}, \citenamefont {Fukushima},\ and\ \citenamefont
  {Kumar}}]{Bramante:2013hn}%
  \BibitemOpen
  \bibfield  {author} {\bibinfo {author} {\bibfnamefont {J.}~\bibnamefont
  {Bramante}}, \bibinfo {author} {\bibfnamefont {K.}~\bibnamefont {Fukushima}},
  \ and\ \bibinfo {author} {\bibfnamefont {J.}~\bibnamefont {Kumar}},\ }\href
  {\doibase 10.1103/PhysRevD.87.055012} {\bibfield  {journal} {\bibinfo
  {journal} {Phys. Rev. D}\ }\textbf {\bibinfo {volume} {87}},\ \bibinfo
  {pages} {055012} (\bibinfo {year} {2013})},\ \Eprint
  {http://arxiv.org/abs/1301.0036} {arXiv:1301.0036 [hep-ph]} \BibitemShut
  {NoStop}%
\bibitem [{\citenamefont {Bell}\ \emph {et~al.}(2013)\citenamefont {Bell},
  \citenamefont {Melatos},\ and\ \citenamefont {Petraki}}]{Bell:2013xk}%
  \BibitemOpen
  \bibfield  {author} {\bibinfo {author} {\bibfnamefont {N.~F.}\ \bibnamefont
  {Bell}}, \bibinfo {author} {\bibfnamefont {A.}~\bibnamefont {Melatos}}, \
  and\ \bibinfo {author} {\bibfnamefont {K.}~\bibnamefont {Petraki}},\ }\href
  {\doibase 10.1103/PhysRevD.87.123507} {\bibfield  {journal} {\bibinfo
  {journal} {Phys. Rev. D}\ }\textbf {\bibinfo {volume} {87}},\ \bibinfo
  {pages} {123507} (\bibinfo {year} {2013})},\ \Eprint
  {http://arxiv.org/abs/1301.6811} {arXiv:1301.6811 [hep-ph]} \BibitemShut
  {NoStop}%
\bibitem [{\citenamefont {Bramante}\ \emph {et~al.}(2014)\citenamefont
  {Bramante}, \citenamefont {Fukushima}, \citenamefont {Kumar},\ and\
  \citenamefont {Stopnitzky}}]{Bramante:2013nma}%
  \BibitemOpen
  \bibfield  {author} {\bibinfo {author} {\bibfnamefont {J.}~\bibnamefont
  {Bramante}}, \bibinfo {author} {\bibfnamefont {K.}~\bibnamefont {Fukushima}},
  \bibinfo {author} {\bibfnamefont {J.}~\bibnamefont {Kumar}}, \ and\ \bibinfo
  {author} {\bibfnamefont {E.}~\bibnamefont {Stopnitzky}},\ }\href {\doibase
  10.1103/PhysRevD.89.015010} {\bibfield  {journal} {\bibinfo  {journal} {Phys.
  Rev. D}\ }\textbf {\bibinfo {volume} {89}},\ \bibinfo {pages} {015010}
  (\bibinfo {year} {2014})},\ \Eprint {http://arxiv.org/abs/1310.3509}
  {arXiv:1310.3509 [hep-ph]} \BibitemShut {NoStop}%
\bibitem [{\citenamefont {Bertoni}\ \emph {et~al.}(2013)\citenamefont
  {Bertoni}, \citenamefont {Nelson},\ and\ \citenamefont
  {Reddy}}]{Bertoni:2013bsa}%
  \BibitemOpen
  \bibfield  {author} {\bibinfo {author} {\bibfnamefont {B.}~\bibnamefont
  {Bertoni}}, \bibinfo {author} {\bibfnamefont {A.~E.}\ \bibnamefont {Nelson}},
  \ and\ \bibinfo {author} {\bibfnamefont {S.}~\bibnamefont {Reddy}},\ }\href
  {\doibase 10.1103/PhysRevD.88.123505} {\bibfield  {journal} {\bibinfo
  {journal} {Phys. Rev. D}\ }\textbf {\bibinfo {volume} {88}},\ \bibinfo
  {pages} {123505} (\bibinfo {year} {2013})},\ \Eprint
  {http://arxiv.org/abs/1309.1721} {arXiv:1309.1721 [hep-ph]} \BibitemShut
  {NoStop}%
\bibitem [{\citenamefont {Angeles Perez-Garcia}\ and\ \citenamefont
  {Silk}(2015)}]{Perez-Garcia:2014dra}%
  \BibitemOpen
  \bibfield  {author} {\bibinfo {author} {\bibfnamefont {M.}~\bibnamefont
  {Angeles Perez-Garcia}}\ and\ \bibinfo {author} {\bibfnamefont
  {J.}~\bibnamefont {Silk}},\ }\href {\doibase 10.1016/j.physletb.2015.03.026}
  {\bibfield  {journal} {\bibinfo  {journal} {Phys. Lett.}\ }\textbf {\bibinfo
  {volume} {B744}},\ \bibinfo {pages} {13} (\bibinfo {year} {2015})},\ \Eprint
  {http://arxiv.org/abs/1403.6111} {arXiv:1403.6111 [astro-ph.SR]} \BibitemShut
  {NoStop}%
%%CITATION = ARXIV:1403.6111;%%
\bibitem [{\citenamefont {Cermeno}\ \emph {et~al.}(2016)\citenamefont
  {Cermeno}, \citenamefont {Perez-Garcia},\ and\ \citenamefont
  {Silk}}]{Cermeno:2016olb}%
  \BibitemOpen
  \bibfield  {author} {\bibinfo {author} {\bibfnamefont {M.}~\bibnamefont
  {Cermeno}}, \bibinfo {author} {\bibfnamefont {M.}~\bibnamefont
  {Perez-Garcia}}, \ and\ \bibinfo {author} {\bibfnamefont {J.}~\bibnamefont
  {Silk}},\ }\href {\doibase 10.1103/PhysRevD.94.063001} {\bibfield  {journal}
  {\bibinfo  {journal} {Phys. Rev.}\ }\textbf {\bibinfo {volume} {D94}},\
  \bibinfo {pages} {063001} (\bibinfo {year} {2016})},\ \Eprint
  {http://arxiv.org/abs/1607.06815} {arXiv:1607.06815 [astro-ph.HE]}
  \BibitemShut {NoStop}%
%%CITATION = ARXIV:1607.06815;%%
\bibitem [{\citenamefont {McKeen}\ \emph {et~al.}(2018)\citenamefont {McKeen},
  \citenamefont {Nelson}, \citenamefont {Reddy},\ and\ \citenamefont
  {Zhou}}]{McKeen:2018xwc}%
  \BibitemOpen
  \bibfield  {author} {\bibinfo {author} {\bibfnamefont {D.}~\bibnamefont
  {McKeen}}, \bibinfo {author} {\bibfnamefont {A.~E.}\ \bibnamefont {Nelson}},
  \bibinfo {author} {\bibfnamefont {S.}~\bibnamefont {Reddy}}, \ and\ \bibinfo
  {author} {\bibfnamefont {D.}~\bibnamefont {Zhou}},\ }\href {\doibase
  10.1103/PhysRevLett.121.061802} {\bibfield  {journal} {\bibinfo  {journal}
  {Phys. Rev. Lett.}\ }\textbf {\bibinfo {volume} {121}},\ \bibinfo {pages}
  {061802} (\bibinfo {year} {2018})},\ \Eprint
  {http://arxiv.org/abs/1802.08244} {arXiv:1802.08244 [hep-ph]} \BibitemShut
  {NoStop}%
%%CITATION = ARXIV:1802.08244;%%
\bibitem [{\citenamefont {Baryakhtar}\ \emph {et~al.}(2017)\citenamefont
  {Baryakhtar}, \citenamefont {Bramante}, \citenamefont {Li}, \citenamefont
  {Linden},\ and\ \citenamefont {Raj}}]{Baryakhtar:2017dbj}%
  \BibitemOpen
  \bibfield  {author} {\bibinfo {author} {\bibfnamefont {M.}~\bibnamefont
  {Baryakhtar}}, \bibinfo {author} {\bibfnamefont {J.}~\bibnamefont
  {Bramante}}, \bibinfo {author} {\bibfnamefont {S.~W.}\ \bibnamefont {Li}},
  \bibinfo {author} {\bibfnamefont {T.}~\bibnamefont {Linden}}, \ and\ \bibinfo
  {author} {\bibfnamefont {N.}~\bibnamefont {Raj}},\ }\href {\doibase
  10.1103/PhysRevLett.119.131801} {\bibfield  {journal} {\bibinfo  {journal}
  {Phys. Rev. Lett.}\ }\textbf {\bibinfo {volume} {119}},\ \bibinfo {pages}
  {131801} (\bibinfo {year} {2017})},\ \Eprint
  {http://arxiv.org/abs/1704.01577} {arXiv:1704.01577 [hep-ph]} \BibitemShut
  {NoStop}%
\bibitem [{\citenamefont {Raj}\ \emph {et~al.}(2018)\citenamefont {Raj},
  \citenamefont {Tanedo},\ and\ \citenamefont {Yu}}]{Raj:2017wrv}%
  \BibitemOpen
  \bibfield  {author} {\bibinfo {author} {\bibfnamefont {N.}~\bibnamefont
  {Raj}}, \bibinfo {author} {\bibfnamefont {P.}~\bibnamefont {Tanedo}}, \ and\
  \bibinfo {author} {\bibfnamefont {H.-B.}\ \bibnamefont {Yu}},\ }\href
  {\doibase 10.1103/PhysRevD.97.043006} {\bibfield  {journal} {\bibinfo
  {journal} {Phys. Rev.}\ }\textbf {\bibinfo {volume} {D97}},\ \bibinfo {pages}
  {043006} (\bibinfo {year} {2018})},\ \Eprint
  {http://arxiv.org/abs/1707.09442} {arXiv:1707.09442 [hep-ph]} \BibitemShut
  {NoStop}%
%%CITATION = ARXIV:1707.09442;%%
\bibitem [{\citenamefont {Bell}\ \emph {et~al.}(2018)\citenamefont {Bell},
  \citenamefont {Busoni},\ and\ \citenamefont {Robles}}]{Bell:2018pkk}%
  \BibitemOpen
  \bibfield  {author} {\bibinfo {author} {\bibfnamefont {N.~F.}\ \bibnamefont
  {Bell}}, \bibinfo {author} {\bibfnamefont {G.}~\bibnamefont {Busoni}}, \ and\
  \bibinfo {author} {\bibfnamefont {S.}~\bibnamefont {Robles}},\ }\href
  {\doibase 10.1088/1475-7516/2018/09/018} {\bibfield  {journal} {\bibinfo
  {journal} {JCAP}\ }\textbf {\bibinfo {volume} {09}},\ \bibinfo {pages} {018}
  (\bibinfo {year} {2018})},\ \Eprint {http://arxiv.org/abs/1807.02840}
  {arXiv:1807.02840 [hep-ph]} \BibitemShut {NoStop}%
\bibitem [{\citenamefont {Chen}\ and\ \citenamefont
  {Lin}(2018)}]{Chen:2018ohx}%
  \BibitemOpen
  \bibfield  {author} {\bibinfo {author} {\bibfnamefont {C.-S.}\ \bibnamefont
  {Chen}}\ and\ \bibinfo {author} {\bibfnamefont {Y.-H.}\ \bibnamefont {Lin}},\
  }\href {\doibase 10.1007/JHEP08(2018)069} {\bibfield  {journal} {\bibinfo
  {journal} {JHEP}\ }\textbf {\bibinfo {volume} {08}},\ \bibinfo {pages} {069}
  (\bibinfo {year} {2018})},\ \Eprint {http://arxiv.org/abs/1804.03409}
  {arXiv:1804.03409 [hep-ph]} \BibitemShut {NoStop}%
%%CITATION = ARXIV:1804.03409;%%
\bibitem [{\citenamefont {Hamaguchi}\ \emph {et~al.}(2019)\citenamefont
  {Hamaguchi}, \citenamefont {Nagata},\ and\ \citenamefont
  {Yanagi}}]{Hamaguchi:2019oev}%
  \BibitemOpen
  \bibfield  {author} {\bibinfo {author} {\bibfnamefont {K.}~\bibnamefont
  {Hamaguchi}}, \bibinfo {author} {\bibfnamefont {N.}~\bibnamefont {Nagata}}, \
  and\ \bibinfo {author} {\bibfnamefont {K.}~\bibnamefont {Yanagi}},\ }\href
  {\doibase 10.1016/j.physletb.2019.06.060} {\bibfield  {journal} {\bibinfo
  {journal} {Phys. Lett.}\ }\textbf {\bibinfo {volume} {B795}},\ \bibinfo
  {pages} {484} (\bibinfo {year} {2019})},\ \Eprint
  {http://arxiv.org/abs/1905.02991} {arXiv:1905.02991 [hep-ph]} \BibitemShut
  {NoStop}%
%%CITATION = ARXIV:1905.02991;%%
\bibitem [{\citenamefont {Camargo}\ \emph {et~al.}(2019)\citenamefont
  {Camargo}, \citenamefont {Queiroz},\ and\ \citenamefont
  {Sturani}}]{Camargo:2019wou}%
  \BibitemOpen
  \bibfield  {author} {\bibinfo {author} {\bibfnamefont {D.~A.}\ \bibnamefont
  {Camargo}}, \bibinfo {author} {\bibfnamefont {F.~S.}\ \bibnamefont
  {Queiroz}}, \ and\ \bibinfo {author} {\bibfnamefont {R.}~\bibnamefont
  {Sturani}},\ }\href {\doibase 10.1088/1475-7516/2019/09/051} {\bibfield
  {journal} {\bibinfo  {journal} {JCAP}\ }\textbf {\bibinfo {volume} {1909}},\
  \bibinfo {pages} {051} (\bibinfo {year} {2019})},\ \Eprint
  {http://arxiv.org/abs/1901.05474} {arXiv:1901.05474 [hep-ph]} \BibitemShut
  {NoStop}%
%%CITATION = ARXIV:1901.05474;%%
\bibitem [{\citenamefont {Garani}\ and\ \citenamefont
  {Heeck}(2019)}]{Garani:2019fpa}%
  \BibitemOpen
  \bibfield  {author} {\bibinfo {author} {\bibfnamefont {R.}~\bibnamefont
  {Garani}}\ and\ \bibinfo {author} {\bibfnamefont {J.}~\bibnamefont {Heeck}},\
  }\href {\doibase 10.1103/PhysRevD.100.035039} {\bibfield  {journal} {\bibinfo
   {journal} {Phys. Rev.}\ }\textbf {\bibinfo {volume} {D100}},\ \bibinfo
  {pages} {035039} (\bibinfo {year} {2019})},\ \Eprint
  {http://arxiv.org/abs/1906.10145} {arXiv:1906.10145 [hep-ph]} \BibitemShut
  {NoStop}%
%%CITATION = ARXIV:1906.10145;%%
\bibitem [{\citenamefont {Bell}\ \emph {et~al.}(2019)\citenamefont {Bell},
  \citenamefont {Busoni},\ and\ \citenamefont {Robles}}]{Bell:2019pyc}%
  \BibitemOpen
  \bibfield  {author} {\bibinfo {author} {\bibfnamefont {N.~F.}\ \bibnamefont
  {Bell}}, \bibinfo {author} {\bibfnamefont {G.}~\bibnamefont {Busoni}}, \ and\
  \bibinfo {author} {\bibfnamefont {S.}~\bibnamefont {Robles}},\ }\href
  {\doibase 10.1088/1475-7516/2019/06/054} {\bibfield  {journal} {\bibinfo
  {journal} {JCAP}\ }\textbf {\bibinfo {volume} {1906}},\ \bibinfo {pages}
  {054} (\bibinfo {year} {2019})},\ \Eprint {http://arxiv.org/abs/1904.09803}
  {arXiv:1904.09803 [hep-ph]} \BibitemShut {NoStop}%
%%CITATION = ARXIV:1904.09803;%%
\bibitem [{\citenamefont {Acevedo}\ \emph {et~al.}(2020)\citenamefont
  {Acevedo}, \citenamefont {Bramante}, \citenamefont {Leane},\ and\
  \citenamefont {Raj}}]{Acevedo:2019agu}%
  \BibitemOpen
  \bibfield  {author} {\bibinfo {author} {\bibfnamefont {J.~F.}\ \bibnamefont
  {Acevedo}}, \bibinfo {author} {\bibfnamefont {J.}~\bibnamefont {Bramante}},
  \bibinfo {author} {\bibfnamefont {R.~K.}\ \bibnamefont {Leane}}, \ and\
  \bibinfo {author} {\bibfnamefont {N.}~\bibnamefont {Raj}},\ }\href {\doibase
  10.1088/1475-7516/2020/03/038} {\bibfield  {journal} {\bibinfo  {journal}
  {JCAP}\ }\textbf {\bibinfo {volume} {03}},\ \bibinfo {pages} {038} (\bibinfo
  {year} {2020})},\ \Eprint {http://arxiv.org/abs/1911.06334} {arXiv:1911.06334
  [hep-ph]} \BibitemShut {NoStop}%
\bibitem [{\citenamefont {Joglekar}\ \emph {et~al.}(2019)\citenamefont
  {Joglekar}, \citenamefont {Raj}, \citenamefont {Tanedo},\ and\ \citenamefont
  {Yu}}]{Joglekar:2019vzy}%
  \BibitemOpen
  \bibfield  {author} {\bibinfo {author} {\bibfnamefont {A.}~\bibnamefont
  {Joglekar}}, \bibinfo {author} {\bibfnamefont {N.}~\bibnamefont {Raj}},
  \bibinfo {author} {\bibfnamefont {P.}~\bibnamefont {Tanedo}}, \ and\ \bibinfo
  {author} {\bibfnamefont {H.-B.}\ \bibnamefont {Yu}},\ }\href@noop {} {\
  (\bibinfo {year} {2019})},\ \Eprint {http://arxiv.org/abs/1911.13293}
  {arXiv:1911.13293 [hep-ph]} \BibitemShut {NoStop}%
%%CITATION = ARXIV:1911.13293;%%
\bibitem [{\citenamefont {Joglekar}\ \emph {et~al.}(2020)\citenamefont
  {Joglekar}, \citenamefont {Raj}, \citenamefont {Tanedo},\ and\ \citenamefont
  {Yu}}]{Joglekar:2020liw}%
  \BibitemOpen
  \bibfield  {author} {\bibinfo {author} {\bibfnamefont {A.}~\bibnamefont
  {Joglekar}}, \bibinfo {author} {\bibfnamefont {N.}~\bibnamefont {Raj}},
  \bibinfo {author} {\bibfnamefont {P.}~\bibnamefont {Tanedo}}, \ and\ \bibinfo
  {author} {\bibfnamefont {H.-B.}\ \bibnamefont {Yu}},\ }\href@noop {} {\
  (\bibinfo {year} {2020})},\ \Eprint {http://arxiv.org/abs/2004.09539}
  {arXiv:2004.09539 [hep-ph]} \BibitemShut {NoStop}%
\bibitem [{\citenamefont {Bell}\ \emph {et~al.}(2020)\citenamefont {Bell},
  \citenamefont {Busoni}, \citenamefont {Robles},\ and\ \citenamefont
  {Virgato}}]{Bell:2020jou}%
  \BibitemOpen
  \bibfield  {author} {\bibinfo {author} {\bibfnamefont {N.~F.}\ \bibnamefont
  {Bell}}, \bibinfo {author} {\bibfnamefont {G.}~\bibnamefont {Busoni}},
  \bibinfo {author} {\bibfnamefont {S.}~\bibnamefont {Robles}}, \ and\ \bibinfo
  {author} {\bibfnamefont {M.}~\bibnamefont {Virgato}},\ }\href@noop {} {\
  (\bibinfo {year} {2020})},\ \Eprint {http://arxiv.org/abs/2004.14888}
  {arXiv:2004.14888 [hep-ph]} \BibitemShut {NoStop}%
\bibitem [{\citenamefont {Garani}\ \emph {et~al.}(2020)\citenamefont {Garani},
  \citenamefont {Gupta},\ and\ \citenamefont {Raj}}]{Garani:2020wge}%
  \BibitemOpen
  \bibfield  {author} {\bibinfo {author} {\bibfnamefont {R.}~\bibnamefont
  {Garani}}, \bibinfo {author} {\bibfnamefont {A.}~\bibnamefont {Gupta}}, \
  and\ \bibinfo {author} {\bibfnamefont {N.}~\bibnamefont {Raj}},\ }\href@noop
  {} {\  (\bibinfo {year} {2020})},\ \Eprint {http://arxiv.org/abs/2009.10728}
  {arXiv:2009.10728 [hep-ph]} \BibitemShut {NoStop}%
\bibitem [{\citenamefont {Bose}\ \emph
  {et~al.}(2022{\natexlab{b}})\citenamefont {Bose}, \citenamefont {Maity},\
  and\ \citenamefont {Ray}}]{Bose:2021yhz}%
  \BibitemOpen
  \bibfield  {author} {\bibinfo {author} {\bibfnamefont {D.}~\bibnamefont
  {Bose}}, \bibinfo {author} {\bibfnamefont {T.~N.}\ \bibnamefont {Maity}}, \
  and\ \bibinfo {author} {\bibfnamefont {T.~S.}\ \bibnamefont {Ray}},\ }\href
  {\doibase 10.1088/1475-7516/2022/05/001} {\bibfield  {journal} {\bibinfo
  {journal} {JCAP}\ }\textbf {\bibinfo {volume} {05}},\ \bibinfo {pages} {001}
  (\bibinfo {year} {2022}{\natexlab{b}})},\ \Eprint
  {http://arxiv.org/abs/2108.12420} {arXiv:2108.12420 [hep-ph]} \BibitemShut
  {NoStop}%
\bibitem [{\citenamefont {Collier}\ \emph {et~al.}(2022)\citenamefont
  {Collier}, \citenamefont {Croon},\ and\ \citenamefont
  {Leane}}]{Collier:2022cpr}%
  \BibitemOpen
  \bibfield  {author} {\bibinfo {author} {\bibfnamefont {M.}~\bibnamefont
  {Collier}}, \bibinfo {author} {\bibfnamefont {D.}~\bibnamefont {Croon}}, \
  and\ \bibinfo {author} {\bibfnamefont {R.~K.}\ \bibnamefont {Leane}},\ }\href
  {\doibase 10.1103/PhysRevD.106.123027} {\bibfield  {journal} {\bibinfo
  {journal} {Phys. Rev. D}\ }\textbf {\bibinfo {volume} {106}},\ \bibinfo
  {pages} {123027} (\bibinfo {year} {2022})},\ \Eprint
  {http://arxiv.org/abs/2205.15337} {arXiv:2205.15337 [gr-qc]} \BibitemShut
  {NoStop}%
\bibitem [{\citenamefont {Nguyen}\ and\ \citenamefont
  {Tait}(2023)}]{Nguyen:2022zwb}%
  \BibitemOpen
  \bibfield  {author} {\bibinfo {author} {\bibfnamefont {T.~T.~Q.}\
  \bibnamefont {Nguyen}}\ and\ \bibinfo {author} {\bibfnamefont {T.~M.~P.}\
  \bibnamefont {Tait}},\ }\href {\doibase 10.1103/PhysRevD.107.115016}
  {\bibfield  {journal} {\bibinfo  {journal} {Phys. Rev. D}\ }\textbf {\bibinfo
  {volume} {107}},\ \bibinfo {pages} {115016} (\bibinfo {year} {2023})},\
  \Eprint {http://arxiv.org/abs/2212.12547} {arXiv:2212.12547 [hep-ph]}
  \BibitemShut {NoStop}%
\bibitem [{\citenamefont {Bose}\ \emph {et~al.}(2023)\citenamefont {Bose},
  \citenamefont {Chowdhury}, \citenamefont {Mondal},\ and\ \citenamefont
  {Ray}}]{Bose:2023yll}%
  \BibitemOpen
  \bibfield  {author} {\bibinfo {author} {\bibfnamefont {D.}~\bibnamefont
  {Bose}}, \bibinfo {author} {\bibfnamefont {D.}~\bibnamefont {Chowdhury}},
  \bibinfo {author} {\bibfnamefont {P.}~\bibnamefont {Mondal}}, \ and\ \bibinfo
  {author} {\bibfnamefont {T.~S.}\ \bibnamefont {Ray}},\ }\href@noop {} {\
  (\bibinfo {year} {2023})},\ \Eprint {http://arxiv.org/abs/2312.05131}
  {arXiv:2312.05131 [hep-ph]} \BibitemShut {NoStop}%
\bibitem [{\citenamefont {Alvarez}\ \emph {et~al.}(2023)\citenamefont
  {Alvarez}, \citenamefont {Joglekar}, \citenamefont {Phoroutan-Mehr},\ and\
  \citenamefont {Yu}}]{Alvarez:2023fjj}%
  \BibitemOpen
  \bibfield  {author} {\bibinfo {author} {\bibfnamefont {G.}~\bibnamefont
  {Alvarez}}, \bibinfo {author} {\bibfnamefont {A.}~\bibnamefont {Joglekar}},
  \bibinfo {author} {\bibfnamefont {M.}~\bibnamefont {Phoroutan-Mehr}}, \ and\
  \bibinfo {author} {\bibfnamefont {H.-B.}\ \bibnamefont {Yu}},\ }\href
  {\doibase 10.1103/PhysRevD.107.103024} {\bibfield  {journal} {\bibinfo
  {journal} {Phys. Rev. D}\ }\textbf {\bibinfo {volume} {107}},\ \bibinfo
  {pages} {103024} (\bibinfo {year} {2023})},\ \Eprint
  {http://arxiv.org/abs/2301.08767} {arXiv:2301.08767 [hep-ph]} \BibitemShut
  {NoStop}%
\bibitem [{\citenamefont {Acevedo}\ \emph
  {et~al.}(2024{\natexlab{c}})\citenamefont {Acevedo}, \citenamefont
  {Bramante}, \citenamefont {Liu},\ and\ \citenamefont
  {Tyagi}}]{Acevedo:2024ttq}%
  \BibitemOpen
  \bibfield  {author} {\bibinfo {author} {\bibfnamefont {J.~F.}\ \bibnamefont
  {Acevedo}}, \bibinfo {author} {\bibfnamefont {J.}~\bibnamefont {Bramante}},
  \bibinfo {author} {\bibfnamefont {Q.}~\bibnamefont {Liu}}, \ and\ \bibinfo
  {author} {\bibfnamefont {N.}~\bibnamefont {Tyagi}},\ }\href@noop {} {\
  (\bibinfo {year} {2024}{\natexlab{c}})},\ \Eprint
  {http://arxiv.org/abs/2404.10039} {arXiv:2404.10039 [hep-ph]} \BibitemShut
  {NoStop}%
\bibitem [{\citenamefont {Dasgupta}\ \emph {et~al.}(2020)\citenamefont
  {Dasgupta}, \citenamefont {Gupta},\ and\ \citenamefont
  {Ray}}]{Dasgupta:2020dik}%
  \BibitemOpen
  \bibfield  {author} {\bibinfo {author} {\bibfnamefont {B.}~\bibnamefont
  {Dasgupta}}, \bibinfo {author} {\bibfnamefont {A.}~\bibnamefont {Gupta}}, \
  and\ \bibinfo {author} {\bibfnamefont {A.}~\bibnamefont {Ray}},\ }\href
  {\doibase 10.1088/1475-7516/2020/10/023} {\bibfield  {journal} {\bibinfo
  {journal} {JCAP}\ }\textbf {\bibinfo {volume} {10}},\ \bibinfo {pages} {023}
  (\bibinfo {year} {2020})},\ \Eprint {http://arxiv.org/abs/2006.10773}
  {arXiv:2006.10773 [hep-ph]} \BibitemShut {NoStop}%
\bibitem [{\citenamefont {Dasgupta}\ \emph {et~al.}(2021)\citenamefont
  {Dasgupta}, \citenamefont {Laha},\ and\ \citenamefont
  {Ray}}]{Dasgupta:2020mqg}%
  \BibitemOpen
  \bibfield  {author} {\bibinfo {author} {\bibfnamefont {B.}~\bibnamefont
  {Dasgupta}}, \bibinfo {author} {\bibfnamefont {R.}~\bibnamefont {Laha}}, \
  and\ \bibinfo {author} {\bibfnamefont {A.}~\bibnamefont {Ray}},\ }\href
  {\doibase 10.1103/PhysRevLett.126.141105} {\bibfield  {journal} {\bibinfo
  {journal} {Phys. Rev. Lett.}\ }\textbf {\bibinfo {volume} {126}},\ \bibinfo
  {pages} {141105} (\bibinfo {year} {2021})},\ \Eprint
  {http://arxiv.org/abs/2009.01825} {arXiv:2009.01825 [astro-ph.HE]}
  \BibitemShut {NoStop}%
\bibitem [{\citenamefont {Bhattacharya}\ \emph {et~al.}(2023)\citenamefont
  {Bhattacharya}, \citenamefont {Dasgupta}, \citenamefont {Laha},\ and\
  \citenamefont {Ray}}]{Bhattacharya:2023stq}%
  \BibitemOpen
  \bibfield  {author} {\bibinfo {author} {\bibfnamefont {S.}~\bibnamefont
  {Bhattacharya}}, \bibinfo {author} {\bibfnamefont {B.}~\bibnamefont
  {Dasgupta}}, \bibinfo {author} {\bibfnamefont {R.}~\bibnamefont {Laha}}, \
  and\ \bibinfo {author} {\bibfnamefont {A.}~\bibnamefont {Ray}},\ }\href
  {\doibase 10.1103/PhysRevLett.131.091401} {\bibfield  {journal} {\bibinfo
  {journal} {Phys. Rev. Lett.}\ }\textbf {\bibinfo {volume} {131}},\ \bibinfo
  {pages} {091401} (\bibinfo {year} {2023})},\ \Eprint
  {http://arxiv.org/abs/2302.07898} {arXiv:2302.07898 [hep-ph]} \BibitemShut
  {NoStop}%
\bibitem [{\citenamefont {{Salati}}\ and\ \citenamefont
  {{Silk}}(1989)}]{1989ApJ...338...24S}%
  \BibitemOpen
  \bibfield  {author} {\bibinfo {author} {\bibfnamefont {P.}~\bibnamefont
  {{Salati}}}\ and\ \bibinfo {author} {\bibfnamefont {J.}~\bibnamefont
  {{Silk}}},\ }\href {\doibase 10.1086/167177} {\bibfield  {journal} {\bibinfo
  {journal} {Astrophys. J.}\ }\textbf {\bibinfo {volume} {338}},\ \bibinfo
  {pages} {24} (\bibinfo {year} {1989})}\BibitemShut {NoStop}%
\bibitem [{\citenamefont {Fairbairn}\ \emph {et~al.}(2008)\citenamefont
  {Fairbairn}, \citenamefont {Scott},\ and\ \citenamefont
  {Edsjo}}]{Fairbairn:2007bn}%
  \BibitemOpen
  \bibfield  {author} {\bibinfo {author} {\bibfnamefont {M.}~\bibnamefont
  {Fairbairn}}, \bibinfo {author} {\bibfnamefont {P.}~\bibnamefont {Scott}}, \
  and\ \bibinfo {author} {\bibfnamefont {J.}~\bibnamefont {Edsjo}},\ }\href
  {\doibase 10.1103/PhysRevD.77.047301} {\bibfield  {journal} {\bibinfo
  {journal} {Phys. Rev. D}\ }\textbf {\bibinfo {volume} {77}},\ \bibinfo
  {pages} {047301} (\bibinfo {year} {2008})},\ \Eprint
  {http://arxiv.org/abs/0710.3396} {arXiv:0710.3396 [astro-ph]} \BibitemShut
  {NoStop}%
\bibitem [{\citenamefont {Scott}\ \emph {et~al.}(2009)\citenamefont {Scott},
  \citenamefont {Fairbairn},\ and\ \citenamefont {Edsjo}}]{Scott:2008ns}%
  \BibitemOpen
  \bibfield  {author} {\bibinfo {author} {\bibfnamefont {P.}~\bibnamefont
  {Scott}}, \bibinfo {author} {\bibfnamefont {M.}~\bibnamefont {Fairbairn}}, \
  and\ \bibinfo {author} {\bibfnamefont {J.}~\bibnamefont {Edsjo}},\ }\href
  {\doibase 10.1111/j.1365-2966.2008.14282.x} {\bibfield  {journal} {\bibinfo
  {journal} {Mon. Not. Roy. Astron. Soc.}\ }\textbf {\bibinfo {volume} {394}},\
  \bibinfo {pages} {82} (\bibinfo {year} {2009})},\ \Eprint
  {http://arxiv.org/abs/0809.1871} {arXiv:0809.1871 [astro-ph]} \BibitemShut
  {NoStop}%
\bibitem [{\citenamefont {Iocco}(2008)}]{Iocco:2008xb}%
  \BibitemOpen
  \bibfield  {author} {\bibinfo {author} {\bibfnamefont {F.}~\bibnamefont
  {Iocco}},\ }\href {\doibase 10.1086/587959} {\bibfield  {journal} {\bibinfo
  {journal} {Astrophys. J. Lett.}\ }\textbf {\bibinfo {volume} {677}},\
  \bibinfo {pages} {L1} (\bibinfo {year} {2008})},\ \Eprint
  {http://arxiv.org/abs/0802.0941} {arXiv:0802.0941 [astro-ph]} \BibitemShut
  {NoStop}%
\bibitem [{\citenamefont {Freese}\ \emph {et~al.}(2009)\citenamefont {Freese},
  \citenamefont {Gondolo}, \citenamefont {Sellwood},\ and\ \citenamefont
  {Spolyar}}]{Freese:2008hb}%
  \BibitemOpen
  \bibfield  {author} {\bibinfo {author} {\bibfnamefont {K.}~\bibnamefont
  {Freese}}, \bibinfo {author} {\bibfnamefont {P.}~\bibnamefont {Gondolo}},
  \bibinfo {author} {\bibfnamefont {J.~A.}\ \bibnamefont {Sellwood}}, \ and\
  \bibinfo {author} {\bibfnamefont {D.}~\bibnamefont {Spolyar}},\ }\href
  {\doibase 10.1088/0004-637X/693/2/1563} {\bibfield  {journal} {\bibinfo
  {journal} {Astrophys. J.}\ }\textbf {\bibinfo {volume} {693}},\ \bibinfo
  {pages} {1563} (\bibinfo {year} {2009})},\ \Eprint
  {http://arxiv.org/abs/0805.3540} {arXiv:0805.3540 [astro-ph]} \BibitemShut
  {NoStop}%
\bibitem [{\citenamefont {Taoso}\ \emph {et~al.}(2008)\citenamefont {Taoso},
  \citenamefont {Bertone}, \citenamefont {Meynet},\ and\ \citenamefont
  {Ekstrom}}]{Taoso:2008kw}%
  \BibitemOpen
  \bibfield  {author} {\bibinfo {author} {\bibfnamefont {M.}~\bibnamefont
  {Taoso}}, \bibinfo {author} {\bibfnamefont {G.}~\bibnamefont {Bertone}},
  \bibinfo {author} {\bibfnamefont {G.}~\bibnamefont {Meynet}}, \ and\ \bibinfo
  {author} {\bibfnamefont {S.}~\bibnamefont {Ekstrom}},\ }\href {\doibase
  10.1103/PhysRevD.78.123510} {\bibfield  {journal} {\bibinfo  {journal} {Phys.
  Rev. D}\ }\textbf {\bibinfo {volume} {78}},\ \bibinfo {pages} {123510}
  (\bibinfo {year} {2008})},\ \Eprint {http://arxiv.org/abs/0806.2681}
  {arXiv:0806.2681 [astro-ph]} \BibitemShut {NoStop}%
\bibitem [{\citenamefont {Sivertsson}\ and\ \citenamefont
  {Gondolo}(2011)}]{Sivertsson:2010zm}%
  \BibitemOpen
  \bibfield  {author} {\bibinfo {author} {\bibfnamefont {S.}~\bibnamefont
  {Sivertsson}}\ and\ \bibinfo {author} {\bibfnamefont {P.}~\bibnamefont
  {Gondolo}},\ }\href {\doibase 10.1088/0004-637X/729/1/51} {\bibfield
  {journal} {\bibinfo  {journal} {Astrophys. J.}\ }\textbf {\bibinfo {volume}
  {729}},\ \bibinfo {pages} {51} (\bibinfo {year} {2011})},\ \Eprint
  {http://arxiv.org/abs/1006.0025} {arXiv:1006.0025 [astro-ph.CO]} \BibitemShut
  {NoStop}%
\bibitem [{\citenamefont {Freese}\ \emph {et~al.}(2016)\citenamefont {Freese},
  \citenamefont {Rindler-Daller}, \citenamefont {Spolyar},\ and\ \citenamefont
  {Valluri}}]{Freese:2015mta}%
  \BibitemOpen
  \bibfield  {author} {\bibinfo {author} {\bibfnamefont {K.}~\bibnamefont
  {Freese}}, \bibinfo {author} {\bibfnamefont {T.}~\bibnamefont
  {Rindler-Daller}}, \bibinfo {author} {\bibfnamefont {D.}~\bibnamefont
  {Spolyar}}, \ and\ \bibinfo {author} {\bibfnamefont {M.}~\bibnamefont
  {Valluri}},\ }\href {\doibase 10.1088/0034-4885/79/6/066902} {\bibfield
  {journal} {\bibinfo  {journal} {Rept. Prog. Phys.}\ }\textbf {\bibinfo
  {volume} {79}},\ \bibinfo {pages} {066902} (\bibinfo {year} {2016})},\
  \Eprint {http://arxiv.org/abs/1501.02394} {arXiv:1501.02394 [astro-ph.CO]}
  \BibitemShut {NoStop}%
\bibitem [{\citenamefont {Ilie}\ \emph
  {et~al.}(2020{\natexlab{a}})\citenamefont {Ilie}, \citenamefont {Levy},
  \citenamefont {Pilawa},\ and\ \citenamefont {Zhang}}]{Ilie:2020iup}%
  \BibitemOpen
  \bibfield  {author} {\bibinfo {author} {\bibfnamefont {C.}~\bibnamefont
  {Ilie}}, \bibinfo {author} {\bibfnamefont {C.}~\bibnamefont {Levy}}, \bibinfo
  {author} {\bibfnamefont {J.}~\bibnamefont {Pilawa}}, \ and\ \bibinfo {author}
  {\bibfnamefont {S.}~\bibnamefont {Zhang}},\ }\href@noop {} {\  (\bibinfo
  {year} {2020}{\natexlab{a}})},\ \Eprint {http://arxiv.org/abs/2009.11478}
  {arXiv:2009.11478 [astro-ph.CO]} \BibitemShut {NoStop}%
\bibitem [{\citenamefont {Ilie}\ \emph
  {et~al.}(2020{\natexlab{b}})\citenamefont {Ilie}, \citenamefont {Levy},
  \citenamefont {Pilawa},\ and\ \citenamefont {Zhang}}]{Ilie:2020nzp}%
  \BibitemOpen
  \bibfield  {author} {\bibinfo {author} {\bibfnamefont {C.}~\bibnamefont
  {Ilie}}, \bibinfo {author} {\bibfnamefont {C.}~\bibnamefont {Levy}}, \bibinfo
  {author} {\bibfnamefont {J.}~\bibnamefont {Pilawa}}, \ and\ \bibinfo {author}
  {\bibfnamefont {S.}~\bibnamefont {Zhang}},\ }\href@noop {} {\  (\bibinfo
  {year} {2020}{\natexlab{b}})},\ \Eprint {http://arxiv.org/abs/2009.11474}
  {arXiv:2009.11474 [astro-ph.CO]} \BibitemShut {NoStop}%
\bibitem [{\citenamefont {Lopes}\ and\ \citenamefont
  {Lopes}(2021)}]{Lopes:2021jcy}%
  \BibitemOpen
  \bibfield  {author} {\bibinfo {author} {\bibfnamefont {J.}~\bibnamefont
  {Lopes}}\ and\ \bibinfo {author} {\bibfnamefont {I.}~\bibnamefont {Lopes}},\
  }\href {\doibase 10.1051/0004-6361/202140750} {\bibfield  {journal} {\bibinfo
   {journal} {Astron. Astrophys.}\ }\textbf {\bibinfo {volume} {651}},\
  \bibinfo {pages} {A101} (\bibinfo {year} {2021})},\ \Eprint
  {http://arxiv.org/abs/2107.13885} {arXiv:2107.13885 [astro-ph.SR]}
  \BibitemShut {NoStop}%
\bibitem [{\citenamefont {Ellis}(2021)}]{Ellis:2021ztw}%
  \BibitemOpen
  \bibfield  {author} {\bibinfo {author} {\bibfnamefont {S.~A.~R.}\
  \bibnamefont {Ellis}},\ }\href@noop {} {\  (\bibinfo {year} {2021})},\
  \Eprint {http://arxiv.org/abs/2111.02414} {arXiv:2111.02414 [astro-ph.CO]}
  \BibitemShut {NoStop}%
\bibitem [{\citenamefont {John}\ \emph {et~al.}(2023)\citenamefont {John},
  \citenamefont {Leane},\ and\ \citenamefont {Linden}}]{John:2023knt}%
  \BibitemOpen
  \bibfield  {author} {\bibinfo {author} {\bibfnamefont {I.}~\bibnamefont
  {John}}, \bibinfo {author} {\bibfnamefont {R.~K.}\ \bibnamefont {Leane}}, \
  and\ \bibinfo {author} {\bibfnamefont {T.}~\bibnamefont {Linden}},\
  }\href@noop {} {\  (\bibinfo {year} {2023})},\ \Eprint
  {http://arxiv.org/abs/2311.16228} {arXiv:2311.16228 [astro-ph.HE]}
  \BibitemShut {NoStop}%
\bibitem [{\citenamefont {Croon}\ and\ \citenamefont
  {Sakstein}(2023)}]{Croon:2023trk}%
  \BibitemOpen
  \bibfield  {author} {\bibinfo {author} {\bibfnamefont {D.}~\bibnamefont
  {Croon}}\ and\ \bibinfo {author} {\bibfnamefont {J.}~\bibnamefont
  {Sakstein}},\ }\href@noop {} {\  (\bibinfo {year} {2023})},\ \Eprint
  {http://arxiv.org/abs/2310.20044} {arXiv:2310.20044 [astro-ph.HE]}
  \BibitemShut {NoStop}%
\bibitem [{\citenamefont {Ilie}\ \emph {et~al.}(2023)\citenamefont {Ilie},
  \citenamefont {Paulin},\ and\ \citenamefont {Freese}}]{Ilie:2023zfv}%
  \BibitemOpen
  \bibfield  {author} {\bibinfo {author} {\bibfnamefont {C.}~\bibnamefont
  {Ilie}}, \bibinfo {author} {\bibfnamefont {J.}~\bibnamefont {Paulin}}, \ and\
  \bibinfo {author} {\bibfnamefont {K.}~\bibnamefont {Freese}},\ }\href
  {\doibase 10.1073/pnas.2305762120} {\bibfield  {journal} {\bibinfo  {journal}
  {Proc. Nat. Acad. Sci.}\ }\textbf {\bibinfo {volume} {120}},\ \bibinfo
  {pages} {e2305762120} (\bibinfo {year} {2023})},\ \Eprint
  {http://arxiv.org/abs/2304.01173} {arXiv:2304.01173 [astro-ph.CO]}
  \BibitemShut {NoStop}%
\bibitem [{\citenamefont {Freese}\ \emph {et~al.}(2010)\citenamefont {Freese},
  \citenamefont {Ilie}, \citenamefont {Spolyar}, \citenamefont {Valluri},\ and\
  \citenamefont {Bodenheimer}}]{Freese:2010re}%
  \BibitemOpen
  \bibfield  {author} {\bibinfo {author} {\bibfnamefont {K.}~\bibnamefont
  {Freese}}, \bibinfo {author} {\bibfnamefont {C.}~\bibnamefont {Ilie}},
  \bibinfo {author} {\bibfnamefont {D.}~\bibnamefont {Spolyar}}, \bibinfo
  {author} {\bibfnamefont {M.}~\bibnamefont {Valluri}}, \ and\ \bibinfo
  {author} {\bibfnamefont {P.}~\bibnamefont {Bodenheimer}},\ }\href {\doibase
  10.1088/0004-637X/716/2/1397} {\bibfield  {journal} {\bibinfo  {journal}
  {Astrophys. J.}\ }\textbf {\bibinfo {volume} {716}},\ \bibinfo {pages} {1397}
  (\bibinfo {year} {2010})},\ \Eprint {http://arxiv.org/abs/1002.2233}
  {arXiv:1002.2233 [astro-ph.CO]} \BibitemShut {NoStop}%
\bibitem [{\citenamefont {Bhattacharya}\ \emph {et~al.}(2024)\citenamefont
  {Bhattacharya}, \citenamefont {Miller},\ and\ \citenamefont
  {Ray}}]{Bhattacharya:2024pmp}%
  \BibitemOpen
  \bibfield  {author} {\bibinfo {author} {\bibfnamefont {S.}~\bibnamefont
  {Bhattacharya}}, \bibinfo {author} {\bibfnamefont {A.~L.}\ \bibnamefont
  {Miller}}, \ and\ \bibinfo {author} {\bibfnamefont {A.}~\bibnamefont {Ray}},\
  }\href@noop {} {\  (\bibinfo {year} {2024})},\ \Eprint
  {http://arxiv.org/abs/2403.13886} {arXiv:2403.13886 [hep-ph]} \BibitemShut
  {NoStop}%
\bibitem [{\citenamefont {John}\ \emph {et~al.}(2024)\citenamefont {John},
  \citenamefont {Leane},\ and\ \citenamefont {Linden}}]{John:2024thz}%
  \BibitemOpen
  \bibfield  {author} {\bibinfo {author} {\bibfnamefont {I.}~\bibnamefont
  {John}}, \bibinfo {author} {\bibfnamefont {R.~K.}\ \bibnamefont {Leane}}, \
  and\ \bibinfo {author} {\bibfnamefont {T.}~\bibnamefont {Linden}},\
  }\href@noop {} {\  (\bibinfo {year} {2024})},\ \Eprint
  {http://arxiv.org/abs/2405.12267} {arXiv:2405.12267 [astro-ph.HE]}
  \BibitemShut {NoStop}%
\bibitem [{\citenamefont {Read}(2014)}]{Read:2014qva}%
  \BibitemOpen
  \bibfield  {author} {\bibinfo {author} {\bibfnamefont {J.~I.}\ \bibnamefont
  {Read}},\ }\href {\doibase 10.1088/0954-3899/41/6/063101} {\bibfield
  {journal} {\bibinfo  {journal} {J. Phys. G}\ }\textbf {\bibinfo {volume}
  {41}},\ \bibinfo {pages} {063101} (\bibinfo {year} {2014})},\ \Eprint
  {http://arxiv.org/abs/1404.1938} {arXiv:1404.1938 [astro-ph.GA]} \BibitemShut
  {NoStop}%
\bibitem [{\citenamefont {Pieri}\ \emph {et~al.}(2011)\citenamefont {Pieri},
  \citenamefont {Lavalle}, \citenamefont {Bertone},\ and\ \citenamefont
  {Branchini}}]{Pieri:2009je}%
  \BibitemOpen
  \bibfield  {author} {\bibinfo {author} {\bibfnamefont {L.}~\bibnamefont
  {Pieri}}, \bibinfo {author} {\bibfnamefont {J.}~\bibnamefont {Lavalle}},
  \bibinfo {author} {\bibfnamefont {G.}~\bibnamefont {Bertone}}, \ and\
  \bibinfo {author} {\bibfnamefont {E.}~\bibnamefont {Branchini}},\ }\href
  {\doibase 10.1103/PhysRevD.83.023518} {\bibfield  {journal} {\bibinfo
  {journal} {Phys. Rev. D}\ }\textbf {\bibinfo {volume} {83}},\ \bibinfo
  {pages} {023518} (\bibinfo {year} {2011})},\ \Eprint
  {http://arxiv.org/abs/0908.0195} {arXiv:0908.0195 [astro-ph.HE]} \BibitemShut
  {NoStop}%
\bibitem [{\citenamefont {{Gnedin}}\ \emph {et~al.}(2011)\citenamefont
  {{Gnedin}}, \citenamefont {{Ceverino}}, \citenamefont {{Gnedin}},
  \citenamefont {{Klypin}}, \citenamefont {{Kravtsov}}, \citenamefont
  {{Levine}}, \citenamefont {{Nagai}},\ and\ \citenamefont
  {{Yepes}}}]{2011arXiv1108.5736G}%
  \BibitemOpen
  \bibfield  {author} {\bibinfo {author} {\bibfnamefont {O.~Y.}\ \bibnamefont
  {{Gnedin}}}, \bibinfo {author} {\bibfnamefont {D.}~\bibnamefont
  {{Ceverino}}}, \bibinfo {author} {\bibfnamefont {N.~Y.}\ \bibnamefont
  {{Gnedin}}}, \bibinfo {author} {\bibfnamefont {A.~A.}\ \bibnamefont
  {{Klypin}}}, \bibinfo {author} {\bibfnamefont {A.~V.}\ \bibnamefont
  {{Kravtsov}}}, \bibinfo {author} {\bibfnamefont {R.}~\bibnamefont
  {{Levine}}}, \bibinfo {author} {\bibfnamefont {D.}~\bibnamefont {{Nagai}}}, \
  and\ \bibinfo {author} {\bibfnamefont {G.}~\bibnamefont {{Yepes}}},\ }\href
  {\doibase 10.48550/arXiv.1108.5736} {\bibfield  {journal} {\bibinfo
  {journal} {arXiv e-prints}\ ,\ \bibinfo {eid} {arXiv:1108.5736}} (\bibinfo
  {year} {2011})},\ \Eprint {http://arxiv.org/abs/1108.5736} {arXiv:1108.5736
  [astro-ph.CO]} \BibitemShut {NoStop}%
\bibitem [{\citenamefont {Di~Cintio}\ \emph {et~al.}(2014)\citenamefont
  {Di~Cintio}, \citenamefont {Brook}, \citenamefont {Dutton}, \citenamefont
  {Macci\`o}, \citenamefont {Stinson},\ and\ \citenamefont
  {Knebe}}]{DiCintio:2014xia}%
  \BibitemOpen
  \bibfield  {author} {\bibinfo {author} {\bibfnamefont {A.}~\bibnamefont
  {Di~Cintio}}, \bibinfo {author} {\bibfnamefont {C.~B.}\ \bibnamefont
  {Brook}}, \bibinfo {author} {\bibfnamefont {A.~A.}\ \bibnamefont {Dutton}},
  \bibinfo {author} {\bibfnamefont {A.~V.}\ \bibnamefont {Macci\`o}}, \bibinfo
  {author} {\bibfnamefont {G.~S.}\ \bibnamefont {Stinson}}, \ and\ \bibinfo
  {author} {\bibfnamefont {A.}~\bibnamefont {Knebe}},\ }\href {\doibase
  10.1093/mnras/stu729} {\bibfield  {journal} {\bibinfo  {journal} {Mon. Not.
  Roy. Astron. Soc.}\ }\textbf {\bibinfo {volume} {441}},\ \bibinfo {pages}
  {2986} (\bibinfo {year} {2014})},\ \Eprint {http://arxiv.org/abs/1404.5959}
  {arXiv:1404.5959 [astro-ph.CO]} \BibitemShut {NoStop}%
\bibitem [{\citenamefont {Navarro}\ \emph {et~al.}(1997)\citenamefont
  {Navarro}, \citenamefont {Frenk},\ and\ \citenamefont
  {White}}]{Navarro:1996gj}%
  \BibitemOpen
  \bibfield  {author} {\bibinfo {author} {\bibfnamefont {J.~F.}\ \bibnamefont
  {Navarro}}, \bibinfo {author} {\bibfnamefont {C.~S.}\ \bibnamefont {Frenk}},
  \ and\ \bibinfo {author} {\bibfnamefont {S.~D.~M.}\ \bibnamefont {White}},\
  }\href {\doibase 10.1086/304888} {\bibfield  {journal} {\bibinfo  {journal}
  {Astrophys. J.}\ }\textbf {\bibinfo {volume} {490}},\ \bibinfo {pages} {493}
  (\bibinfo {year} {1997})},\ \Eprint {http://arxiv.org/abs/astro-ph/9611107}
  {arXiv:astro-ph/9611107} \BibitemShut {NoStop}%
\bibitem [{\citenamefont {Salucci}\ \emph {et~al.}(2010)\citenamefont
  {Salucci}, \citenamefont {Nesti}, \citenamefont {Gentile},\ and\
  \citenamefont {Martins}}]{Salucci:2010qr}%
  \BibitemOpen
  \bibfield  {author} {\bibinfo {author} {\bibfnamefont {P.}~\bibnamefont
  {Salucci}}, \bibinfo {author} {\bibfnamefont {F.}~\bibnamefont {Nesti}},
  \bibinfo {author} {\bibfnamefont {G.}~\bibnamefont {Gentile}}, \ and\
  \bibinfo {author} {\bibfnamefont {C.~F.}\ \bibnamefont {Martins}},\ }\href
  {\doibase 10.1051/0004-6361/201014385} {\bibfield  {journal} {\bibinfo
  {journal} {Astron. Astrophys.}\ }\textbf {\bibinfo {volume} {523}},\ \bibinfo
  {pages} {A83} (\bibinfo {year} {2010})},\ \Eprint
  {http://arxiv.org/abs/1003.3101} {arXiv:1003.3101 [astro-ph.GA]} \BibitemShut
  {NoStop}%
\bibitem [{\citenamefont {Sofue}(2013)}]{Sofue:2013kja}%
  \BibitemOpen
  \bibfield  {author} {\bibinfo {author} {\bibfnamefont {Y.}~\bibnamefont
  {Sofue}},\ }\href {\doibase 10.1093/pasj/65.6.118} {\bibfield  {journal}
  {\bibinfo  {journal} {Publ. Astron. Soc. Jap.}\ }\textbf {\bibinfo {volume}
  {65}},\ \bibinfo {pages} {118} (\bibinfo {year} {2013})},\ \Eprint
  {http://arxiv.org/abs/1307.8241} {arXiv:1307.8241 [astro-ph.GA]} \BibitemShut
  {NoStop}%
\bibitem [{\citenamefont {Bramante}\ \emph {et~al.}(2017)\citenamefont
  {Bramante}, \citenamefont {Delgado},\ and\ \citenamefont
  {Martin}}]{Bramante:2017xlb}%
  \BibitemOpen
  \bibfield  {author} {\bibinfo {author} {\bibfnamefont {J.}~\bibnamefont
  {Bramante}}, \bibinfo {author} {\bibfnamefont {A.}~\bibnamefont {Delgado}}, \
  and\ \bibinfo {author} {\bibfnamefont {A.}~\bibnamefont {Martin}},\ }\href
  {\doibase 10.1103/PhysRevD.96.063002} {\bibfield  {journal} {\bibinfo
  {journal} {Phys. Rev. D}\ }\textbf {\bibinfo {volume} {96}},\ \bibinfo
  {pages} {063002} (\bibinfo {year} {2017})},\ \Eprint
  {http://arxiv.org/abs/1703.04043} {arXiv:1703.04043 [hep-ph]} \BibitemShut
  {NoStop}%
\bibitem [{\citenamefont {Zaharijas}\ and\ \citenamefont
  {Farrar}(2005)}]{Zaharijas:2004jv}%
  \BibitemOpen
  \bibfield  {author} {\bibinfo {author} {\bibfnamefont {G.}~\bibnamefont
  {Zaharijas}}\ and\ \bibinfo {author} {\bibfnamefont {G.~R.}\ \bibnamefont
  {Farrar}},\ }\href {\doibase 10.1103/PhysRevD.72.083502} {\bibfield
  {journal} {\bibinfo  {journal} {Phys. Rev. D}\ }\textbf {\bibinfo {volume}
  {72}},\ \bibinfo {pages} {083502} (\bibinfo {year} {2005})},\ \Eprint
  {http://arxiv.org/abs/astro-ph/0406531} {arXiv:astro-ph/0406531} \BibitemShut
  {NoStop}%
\bibitem [{\citenamefont {Leane}\ and\ \citenamefont
  {Smirnov}(2023{\natexlab{a}})}]{Leane:2023woh}%
  \BibitemOpen
  \bibfield  {author} {\bibinfo {author} {\bibfnamefont {R.~K.}\ \bibnamefont
  {Leane}}\ and\ \bibinfo {author} {\bibfnamefont {J.}~\bibnamefont
  {Smirnov}},\ }\href {\doibase 10.1088/1475-7516/2023/12/040} {\bibfield
  {journal} {\bibinfo  {journal} {JCAP}\ }\textbf {\bibinfo {volume} {12}},\
  \bibinfo {pages} {040} (\bibinfo {year} {2023}{\natexlab{a}})},\ \Eprint
  {http://arxiv.org/abs/2309.00669} {arXiv:2309.00669 [hep-ph]} \BibitemShut
  {NoStop}%
\bibitem [{\citenamefont {Bell}\ \emph
  {et~al.}(2021{\natexlab{c}})\citenamefont {Bell}, \citenamefont {Busoni},
  \citenamefont {Motta}, \citenamefont {Robles}, \citenamefont {Thomas},\ and\
  \citenamefont {Virgato}}]{Bell:2020obw}%
  \BibitemOpen
  \bibfield  {author} {\bibinfo {author} {\bibfnamefont {N.~F.}\ \bibnamefont
  {Bell}}, \bibinfo {author} {\bibfnamefont {G.}~\bibnamefont {Busoni}},
  \bibinfo {author} {\bibfnamefont {T.~F.}\ \bibnamefont {Motta}}, \bibinfo
  {author} {\bibfnamefont {S.}~\bibnamefont {Robles}}, \bibinfo {author}
  {\bibfnamefont {A.~W.}\ \bibnamefont {Thomas}}, \ and\ \bibinfo {author}
  {\bibfnamefont {M.}~\bibnamefont {Virgato}},\ }\href {\doibase
  10.1103/PhysRevLett.127.111803} {\bibfield  {journal} {\bibinfo  {journal}
  {Phys. Rev. Lett.}\ }\textbf {\bibinfo {volume} {127}},\ \bibinfo {pages}
  {111803} (\bibinfo {year} {2021}{\natexlab{c}})},\ \Eprint
  {http://arxiv.org/abs/2012.08918} {arXiv:2012.08918 [hep-ph]} \BibitemShut
  {NoStop}%
\bibitem [{\citenamefont {Ilie}\ \emph
  {et~al.}(2020{\natexlab{c}})\citenamefont {Ilie}, \citenamefont {Pilawa},\
  and\ \citenamefont {Zhang}}]{Ilie:2020vec}%
  \BibitemOpen
  \bibfield  {author} {\bibinfo {author} {\bibfnamefont {C.}~\bibnamefont
  {Ilie}}, \bibinfo {author} {\bibfnamefont {J.}~\bibnamefont {Pilawa}}, \ and\
  \bibinfo {author} {\bibfnamefont {S.}~\bibnamefont {Zhang}},\ }\href
  {\doibase 10.1103/PhysRevD.102.048301} {\bibfield  {journal} {\bibinfo
  {journal} {Phys. Rev. D}\ }\textbf {\bibinfo {volume} {102}},\ \bibinfo
  {pages} {048301} (\bibinfo {year} {2020}{\natexlab{c}})},\ \Eprint
  {http://arxiv.org/abs/2005.05946} {arXiv:2005.05946 [astro-ph.CO]}
  \BibitemShut {NoStop}%
\bibitem [{\citenamefont {Ilie}\ and\ \citenamefont
  {Levy}(2021)}]{Ilie:2021iyh}%
  \BibitemOpen
  \bibfield  {author} {\bibinfo {author} {\bibfnamefont {C.}~\bibnamefont
  {Ilie}}\ and\ \bibinfo {author} {\bibfnamefont {C.}~\bibnamefont {Levy}},\
  }\href {\doibase 10.1103/PhysRevD.104.083033} {\bibfield  {journal} {\bibinfo
   {journal} {Phys. Rev. D}\ }\textbf {\bibinfo {volume} {104}},\ \bibinfo
  {pages} {083033} (\bibinfo {year} {2021})},\ \Eprint
  {http://arxiv.org/abs/2105.09765} {arXiv:2105.09765 [astro-ph.CO]}
  \BibitemShut {NoStop}%
\bibitem [{\citenamefont {Digman}\ \emph {et~al.}(2019)\citenamefont {Digman},
  \citenamefont {Cappiello}, \citenamefont {Beacom}, \citenamefont {Hirata},\
  and\ \citenamefont {Peter}}]{Digman:2019wdm}%
  \BibitemOpen
  \bibfield  {author} {\bibinfo {author} {\bibfnamefont {M.~C.}\ \bibnamefont
  {Digman}}, \bibinfo {author} {\bibfnamefont {C.~V.}\ \bibnamefont
  {Cappiello}}, \bibinfo {author} {\bibfnamefont {J.~F.}\ \bibnamefont
  {Beacom}}, \bibinfo {author} {\bibfnamefont {C.~M.}\ \bibnamefont {Hirata}},
  \ and\ \bibinfo {author} {\bibfnamefont {A.~H.~G.}\ \bibnamefont {Peter}},\
  }\href {\doibase 10.1103/PhysRevD.100.063013} {\bibfield  {journal} {\bibinfo
   {journal} {Phys. Rev. D}\ }\textbf {\bibinfo {volume} {100}},\ \bibinfo
  {pages} {063013} (\bibinfo {year} {2019})},\ \bibinfo {note} {[Erratum:
  Phys.Rev.D 106, 089902 (2022)]},\ \Eprint {http://arxiv.org/abs/1907.10618}
  {arXiv:1907.10618 [hep-ph]} \BibitemShut {NoStop}%
\bibitem [{\citenamefont {Xu}\ and\ \citenamefont {Farrar}(2023)}]{Xu:2020qjk}%
  \BibitemOpen
  \bibfield  {author} {\bibinfo {author} {\bibfnamefont {X.}~\bibnamefont
  {Xu}}\ and\ \bibinfo {author} {\bibfnamefont {G.~R.}\ \bibnamefont
  {Farrar}},\ }\href {\doibase 10.1103/PhysRevD.107.095028} {\bibfield
  {journal} {\bibinfo  {journal} {Phys. Rev. D}\ }\textbf {\bibinfo {volume}
  {107}},\ \bibinfo {pages} {095028} (\bibinfo {year} {2023})},\ \Eprint
  {http://arxiv.org/abs/2101.00142} {arXiv:2101.00142 [hep-ph]} \BibitemShut
  {NoStop}%
\bibitem [{\citenamefont {Asplund}\ \emph {et~al.}(2009)\citenamefont
  {Asplund}, \citenamefont {Grevesse}, \citenamefont {Sauval},\ and\
  \citenamefont {Scott}}]{Asplund:2009fu}%
  \BibitemOpen
  \bibfield  {author} {\bibinfo {author} {\bibfnamefont {M.}~\bibnamefont
  {Asplund}}, \bibinfo {author} {\bibfnamefont {N.}~\bibnamefont {Grevesse}},
  \bibinfo {author} {\bibfnamefont {A.~J.}\ \bibnamefont {Sauval}}, \ and\
  \bibinfo {author} {\bibfnamefont {P.}~\bibnamefont {Scott}},\ }\href
  {\doibase 10.1146/annurev.astro.46.060407.145222} {\bibfield  {journal}
  {\bibinfo  {journal} {Ann. Rev. Astron. Astrophys.}\ }\textbf {\bibinfo
  {volume} {47}},\ \bibinfo {pages} {481} (\bibinfo {year} {2009})},\ \Eprint
  {http://arxiv.org/abs/0909.0948} {arXiv:0909.0948 [astro-ph.SR]} \BibitemShut
  {NoStop}%
\bibitem [{\citenamefont {Bellinger}\ \emph {et~al.}(2019)\citenamefont
  {Bellinger}, \citenamefont {Hekker}, \citenamefont {Angelou}, \citenamefont
  {Stokholm},\ and\ \citenamefont {Basu}}]{Bellinger_2019}%
  \BibitemOpen
  \bibfield  {author} {\bibinfo {author} {\bibfnamefont {E.~P.}\ \bibnamefont
  {Bellinger}}, \bibinfo {author} {\bibfnamefont {S.}~\bibnamefont {Hekker}},
  \bibinfo {author} {\bibfnamefont {G.~C.}\ \bibnamefont {Angelou}}, \bibinfo
  {author} {\bibfnamefont {A.}~\bibnamefont {Stokholm}}, \ and\ \bibinfo
  {author} {\bibfnamefont {S.}~\bibnamefont {Basu}},\ }\href {\doibase
  10.1051/0004-6361/201834461} {\bibfield  {journal} {\bibinfo  {journal}
  {Astronomy \& Astrophysics}\ }\textbf {\bibinfo {volume} {622}},\ \bibinfo
  {pages} {A130} (\bibinfo {year} {2019})}\BibitemShut {NoStop}%
\bibitem [{\citenamefont {Sch\"odel}\ \emph {et~al.}(2018)\citenamefont
  {Sch\"odel}, \citenamefont {Gallego-Cano}, \citenamefont {Dong},
  \citenamefont {Nogueras-Lara}, \citenamefont {Gallego-Calvente},
  \citenamefont {Amaro-Seoane},\ and\ \citenamefont
  {Baumgardt}}]{Schodel:2017vjf}%
  \BibitemOpen
  \bibfield  {author} {\bibinfo {author} {\bibfnamefont {R.}~\bibnamefont
  {Sch\"odel}}, \bibinfo {author} {\bibfnamefont {E.}~\bibnamefont
  {Gallego-Cano}}, \bibinfo {author} {\bibfnamefont {H.}~\bibnamefont {Dong}},
  \bibinfo {author} {\bibfnamefont {F.}~\bibnamefont {Nogueras-Lara}}, \bibinfo
  {author} {\bibfnamefont {A.~T.}\ \bibnamefont {Gallego-Calvente}}, \bibinfo
  {author} {\bibfnamefont {P.}~\bibnamefont {Amaro-Seoane}}, \ and\ \bibinfo
  {author} {\bibfnamefont {H.}~\bibnamefont {Baumgardt}},\ }\href {\doibase
  10.1051/0004-6361/201730452} {\bibfield  {journal} {\bibinfo  {journal}
  {Astron. Astrophys.}\ }\textbf {\bibinfo {volume} {609}},\ \bibinfo {pages}
  {A27} (\bibinfo {year} {2018})},\ \Eprint {http://arxiv.org/abs/1701.03817}
  {arXiv:1701.03817 [astro-ph.GA]} \BibitemShut {NoStop}%
\bibitem [{\citenamefont {Generozov}\ \emph {et~al.}(2018)\citenamefont
  {Generozov}, \citenamefont {Stone}, \citenamefont {Metzger},\ and\
  \citenamefont {Ostriker}}]{Generozov:2018niv}%
  \BibitemOpen
  \bibfield  {author} {\bibinfo {author} {\bibfnamefont {A.}~\bibnamefont
  {Generozov}}, \bibinfo {author} {\bibfnamefont {N.~C.}\ \bibnamefont
  {Stone}}, \bibinfo {author} {\bibfnamefont {B.~D.}\ \bibnamefont {Metzger}},
  \ and\ \bibinfo {author} {\bibfnamefont {J.~P.}\ \bibnamefont {Ostriker}},\
  }\href {\doibase 10.1093/mnras/sty1262} {\bibfield  {journal} {\bibinfo
  {journal} {Mon. Not. Roy. Astron. Soc.}\ }\textbf {\bibinfo {volume} {478}},\
  \bibinfo {pages} {4030} (\bibinfo {year} {2018})},\ \Eprint
  {http://arxiv.org/abs/1804.01543} {arXiv:1804.01543 [astro-ph.HE]}
  \BibitemShut {NoStop}%
\bibitem [{\citenamefont {Moore}(1996)}]{Moore:1995pb}%
  \BibitemOpen
  \bibfield  {author} {\bibinfo {author} {\bibfnamefont {B.}~\bibnamefont
  {Moore}},\ }\href {\doibase 10.1086/309998} {\bibfield  {journal} {\bibinfo
  {journal} {Astrophys. J. Lett.}\ }\textbf {\bibinfo {volume} {461}},\
  \bibinfo {pages} {L13} (\bibinfo {year} {1996})},\ \Eprint
  {http://arxiv.org/abs/astro-ph/9511147} {arXiv:astro-ph/9511147} \BibitemShut
  {NoStop}%
\bibitem [{\citenamefont {Saitoh}\ \emph {et~al.}(2006)\citenamefont {Saitoh},
  \citenamefont {Koda}, \citenamefont {Okamoto}, \citenamefont {Wada},\ and\
  \citenamefont {Habe}}]{Saitoh:2005tt}%
  \BibitemOpen
  \bibfield  {author} {\bibinfo {author} {\bibfnamefont {T.~R.}\ \bibnamefont
  {Saitoh}}, \bibinfo {author} {\bibfnamefont {J.}~\bibnamefont {Koda}},
  \bibinfo {author} {\bibfnamefont {T.}~\bibnamefont {Okamoto}}, \bibinfo
  {author} {\bibfnamefont {K.}~\bibnamefont {Wada}}, \ and\ \bibinfo {author}
  {\bibfnamefont {A.}~\bibnamefont {Habe}},\ }\href {\doibase 10.1086/500104}
  {\bibfield  {journal} {\bibinfo  {journal} {Astrophys. J.}\ }\textbf
  {\bibinfo {volume} {640}},\ \bibinfo {pages} {22} (\bibinfo {year} {2006})},\
  \Eprint {http://arxiv.org/abs/astro-ph/0511692} {arXiv:astro-ph/0511692}
  \BibitemShut {NoStop}%
\bibitem [{\citenamefont {Kouvaris}\ and\ \citenamefont
  {Tinyakov}(2014)}]{Kouvaris:2013kra}%
  \BibitemOpen
  \bibfield  {author} {\bibinfo {author} {\bibfnamefont {C.}~\bibnamefont
  {Kouvaris}}\ and\ \bibinfo {author} {\bibfnamefont {P.}~\bibnamefont
  {Tinyakov}},\ }\href {\doibase 10.1103/PhysRevD.90.043512} {\bibfield
  {journal} {\bibinfo  {journal} {Phys. Rev. D}\ }\textbf {\bibinfo {volume}
  {90}},\ \bibinfo {pages} {043512} (\bibinfo {year} {2014})},\ \Eprint
  {http://arxiv.org/abs/1312.3764} {arXiv:1312.3764 [astro-ph.SR]} \BibitemShut
  {NoStop}%
\bibitem [{\citenamefont {Bramante}\ and\ \citenamefont
  {Linden}(2014)}]{Bramante:2014zca}%
  \BibitemOpen
  \bibfield  {author} {\bibinfo {author} {\bibfnamefont {J.}~\bibnamefont
  {Bramante}}\ and\ \bibinfo {author} {\bibfnamefont {T.}~\bibnamefont
  {Linden}},\ }\href {\doibase 10.1103/PhysRevLett.113.191301} {\bibfield
  {journal} {\bibinfo  {journal} {Phys. Rev. Lett.}\ }\textbf {\bibinfo
  {volume} {113}},\ \bibinfo {pages} {191301} (\bibinfo {year} {2014})},\
  \Eprint {http://arxiv.org/abs/1405.1031} {arXiv:1405.1031 [astro-ph.HE]}
  \BibitemShut {NoStop}%
\bibitem [{\citenamefont {Garani}\ \emph {et~al.}(2019)\citenamefont {Garani},
  \citenamefont {Genolini},\ and\ \citenamefont {Hambye}}]{Garani:2018kkd}%
  \BibitemOpen
  \bibfield  {author} {\bibinfo {author} {\bibfnamefont {R.}~\bibnamefont
  {Garani}}, \bibinfo {author} {\bibfnamefont {Y.}~\bibnamefont {Genolini}}, \
  and\ \bibinfo {author} {\bibfnamefont {T.}~\bibnamefont {Hambye}},\ }\href
  {\doibase 10.1088/1475-7516/2019/05/035} {\bibfield  {journal} {\bibinfo
  {journal} {JCAP}\ }\textbf {\bibinfo {volume} {05}},\ \bibinfo {pages} {035}
  (\bibinfo {year} {2019})},\ \Eprint {http://arxiv.org/abs/1812.08773}
  {arXiv:1812.08773 [hep-ph]} \BibitemShut {NoStop}%
\bibitem [{\citenamefont {Kouvaris}\ \emph {et~al.}(2018)\citenamefont
  {Kouvaris}, \citenamefont {Tinyakov},\ and\ \citenamefont
  {Tytgat}}]{Kouvaris:2018wnh}%
  \BibitemOpen
  \bibfield  {author} {\bibinfo {author} {\bibfnamefont {C.}~\bibnamefont
  {Kouvaris}}, \bibinfo {author} {\bibfnamefont {P.}~\bibnamefont {Tinyakov}},
  \ and\ \bibinfo {author} {\bibfnamefont {M.~H.~G.}\ \bibnamefont {Tytgat}},\
  }\href {\doibase 10.1103/PhysRevLett.121.221102} {\bibfield  {journal}
  {\bibinfo  {journal} {Phys. Rev. Lett.}\ }\textbf {\bibinfo {volume} {121}},\
  \bibinfo {pages} {221102} (\bibinfo {year} {2018})},\ \Eprint
  {http://arxiv.org/abs/1804.06740} {arXiv:1804.06740 [astro-ph.HE]}
  \BibitemShut {NoStop}%
\bibitem [{\citenamefont {Lin}\ and\ \citenamefont {Lin}(2020)}]{Lin:2020zmm}%
  \BibitemOpen
  \bibfield  {author} {\bibinfo {author} {\bibfnamefont {G.-L.}\ \bibnamefont
  {Lin}}\ and\ \bibinfo {author} {\bibfnamefont {Y.-H.}\ \bibnamefont {Lin}},\
  }\href {\doibase 10.1088/1475-7516/2020/08/022} {\bibfield  {journal}
  {\bibinfo  {journal} {JCAP}\ }\textbf {\bibinfo {volume} {08}},\ \bibinfo
  {pages} {022} (\bibinfo {year} {2020})},\ \Eprint
  {http://arxiv.org/abs/2004.05312} {arXiv:2004.05312 [hep-ph]} \BibitemShut
  {NoStop}%
\bibitem [{\citenamefont {Takhistov}\ \emph {et~al.}(2021)\citenamefont
  {Takhistov}, \citenamefont {Fuller},\ and\ \citenamefont
  {Kusenko}}]{Takhistov:2020vxs}%
  \BibitemOpen
  \bibfield  {author} {\bibinfo {author} {\bibfnamefont {V.}~\bibnamefont
  {Takhistov}}, \bibinfo {author} {\bibfnamefont {G.~M.}\ \bibnamefont
  {Fuller}}, \ and\ \bibinfo {author} {\bibfnamefont {A.}~\bibnamefont
  {Kusenko}},\ }\href {\doibase 10.1103/PhysRevLett.126.071101} {\bibfield
  {journal} {\bibinfo  {journal} {Phys. Rev. Lett.}\ }\textbf {\bibinfo
  {volume} {126}},\ \bibinfo {pages} {071101} (\bibinfo {year} {2021})},\
  \Eprint {http://arxiv.org/abs/2008.12780} {arXiv:2008.12780 [astro-ph.HE]}
  \BibitemShut {NoStop}%
\bibitem [{\citenamefont {Garani}\ \emph {et~al.}(2022)\citenamefont {Garani},
  \citenamefont {Levkov},\ and\ \citenamefont {Tinyakov}}]{Garani:2021gvc}%
  \BibitemOpen
  \bibfield  {author} {\bibinfo {author} {\bibfnamefont {R.}~\bibnamefont
  {Garani}}, \bibinfo {author} {\bibfnamefont {D.}~\bibnamefont {Levkov}}, \
  and\ \bibinfo {author} {\bibfnamefont {P.}~\bibnamefont {Tinyakov}},\ }\href
  {\doibase 10.1103/PhysRevD.105.063019} {\bibfield  {journal} {\bibinfo
  {journal} {Phys. Rev. D}\ }\textbf {\bibinfo {volume} {105}},\ \bibinfo
  {pages} {063019} (\bibinfo {year} {2022})},\ \Eprint
  {http://arxiv.org/abs/2112.09716} {arXiv:2112.09716 [hep-ph]} \BibitemShut
  {NoStop}%
\bibitem [{\citenamefont {Steigerwald}\ \emph {et~al.}(2022)\citenamefont
  {Steigerwald}, \citenamefont {Marra},\ and\ \citenamefont
  {Profumo}}]{Steigerwald:2022pjo}%
  \BibitemOpen
  \bibfield  {author} {\bibinfo {author} {\bibfnamefont {H.}~\bibnamefont
  {Steigerwald}}, \bibinfo {author} {\bibfnamefont {V.}~\bibnamefont {Marra}},
  \ and\ \bibinfo {author} {\bibfnamefont {S.}~\bibnamefont {Profumo}},\ }\href
  {\doibase 10.1103/PhysRevD.105.083507} {\bibfield  {journal} {\bibinfo
  {journal} {Phys. Rev. D}\ }\textbf {\bibinfo {volume} {105}},\ \bibinfo
  {pages} {083507} (\bibinfo {year} {2022})},\ \Eprint
  {http://arxiv.org/abs/2203.09054} {arXiv:2203.09054 [astro-ph.CO]}
  \BibitemShut {NoStop}%
\bibitem [{\citenamefont {Singh}\ \emph {et~al.}(2023)\citenamefont {Singh},
  \citenamefont {Gupta}, \citenamefont {Berti}, \citenamefont {Reddy},\ and\
  \citenamefont {Sathyaprakash}}]{Singh:2022wvw}%
  \BibitemOpen
  \bibfield  {author} {\bibinfo {author} {\bibfnamefont {D.}~\bibnamefont
  {Singh}}, \bibinfo {author} {\bibfnamefont {A.}~\bibnamefont {Gupta}},
  \bibinfo {author} {\bibfnamefont {E.}~\bibnamefont {Berti}}, \bibinfo
  {author} {\bibfnamefont {S.}~\bibnamefont {Reddy}}, \ and\ \bibinfo {author}
  {\bibfnamefont {B.~S.}\ \bibnamefont {Sathyaprakash}},\ }\href {\doibase
  10.1103/PhysRevD.107.083037} {\bibfield  {journal} {\bibinfo  {journal}
  {Phys. Rev. D}\ }\textbf {\bibinfo {volume} {107}},\ \bibinfo {pages}
  {083037} (\bibinfo {year} {2023})},\ \Eprint
  {http://arxiv.org/abs/2210.15739} {arXiv:2210.15739 [gr-qc]} \BibitemShut
  {NoStop}%
\bibitem [{\citenamefont {Starkman}\ \emph {et~al.}(1990)\citenamefont
  {Starkman}, \citenamefont {Gould}, \citenamefont {Esmailzadeh},\ and\
  \citenamefont {Dimopoulos}}]{Starkman:1990nj}%
  \BibitemOpen
  \bibfield  {author} {\bibinfo {author} {\bibfnamefont {G.~D.}\ \bibnamefont
  {Starkman}}, \bibinfo {author} {\bibfnamefont {A.}~\bibnamefont {Gould}},
  \bibinfo {author} {\bibfnamefont {R.}~\bibnamefont {Esmailzadeh}}, \ and\
  \bibinfo {author} {\bibfnamefont {S.}~\bibnamefont {Dimopoulos}},\ }\href
  {\doibase 10.1103/PhysRevD.41.3594} {\bibfield  {journal} {\bibinfo
  {journal} {Phys. Rev. D}\ }\textbf {\bibinfo {volume} {41}},\ \bibinfo
  {pages} {3594} (\bibinfo {year} {1990})}\BibitemShut {NoStop}%
\bibitem [{\citenamefont {Kurita}\ and\ \citenamefont
  {Nakano}(2016)}]{Kurita:2015vga}%
  \BibitemOpen
  \bibfield  {author} {\bibinfo {author} {\bibfnamefont {Y.}~\bibnamefont
  {Kurita}}\ and\ \bibinfo {author} {\bibfnamefont {H.}~\bibnamefont
  {Nakano}},\ }\href {\doibase 10.1103/PhysRevD.93.023508} {\bibfield
  {journal} {\bibinfo  {journal} {Phys. Rev. D}\ }\textbf {\bibinfo {volume}
  {93}},\ \bibinfo {pages} {023508} (\bibinfo {year} {2016})},\ \Eprint
  {http://arxiv.org/abs/1510.00893} {arXiv:1510.00893 [gr-qc]} \BibitemShut
  {NoStop}%
\bibitem [{\citenamefont {Banks}\ \emph {et~al.}(2022)\citenamefont {Banks},
  \citenamefont {Ansari}, \citenamefont {Vincent},\ and\ \citenamefont
  {Scott}}]{Banks:2021sba}%
  \BibitemOpen
  \bibfield  {author} {\bibinfo {author} {\bibfnamefont {H.}~\bibnamefont
  {Banks}}, \bibinfo {author} {\bibfnamefont {S.}~\bibnamefont {Ansari}},
  \bibinfo {author} {\bibfnamefont {A.~C.}\ \bibnamefont {Vincent}}, \ and\
  \bibinfo {author} {\bibfnamefont {P.}~\bibnamefont {Scott}},\ }\href
  {\doibase 10.1088/1475-7516/2022/04/002} {\bibfield  {journal} {\bibinfo
  {journal} {JCAP}\ }\textbf {\bibinfo {volume} {04}},\ \bibinfo {pages} {002}
  (\bibinfo {year} {2022})},\ \Eprint {http://arxiv.org/abs/2111.06895}
  {arXiv:2111.06895 [hep-ph]} \BibitemShut {NoStop}%
\bibitem [{\citenamefont {Steigman}\ \emph {et~al.}(1978)\citenamefont
  {Steigman}, \citenamefont {Sarazin}, \citenamefont {Quintana},\ and\
  \citenamefont {Faulkner}}]{Steigman:1978wqb}%
  \BibitemOpen
  \bibfield  {author} {\bibinfo {author} {\bibfnamefont {G.}~\bibnamefont
  {Steigman}}, \bibinfo {author} {\bibfnamefont {C.~L.}\ \bibnamefont
  {Sarazin}}, \bibinfo {author} {\bibfnamefont {H.}~\bibnamefont {Quintana}}, \
  and\ \bibinfo {author} {\bibfnamefont {J.}~\bibnamefont {Faulkner}},\ }\href
  {\doibase 10.1086/112290} {\bibfield  {journal} {\bibinfo  {journal} {Astron.
  J.}\ }\textbf {\bibinfo {volume} {83}},\ \bibinfo {pages} {1050} (\bibinfo
  {year} {1978})}\BibitemShut {NoStop}%
\bibitem [{\citenamefont {Spergel}\ and\ \citenamefont
  {Press}(1985)}]{Spergel:1984re}%
  \BibitemOpen
  \bibfield  {author} {\bibinfo {author} {\bibfnamefont {D.~N.}\ \bibnamefont
  {Spergel}}\ and\ \bibinfo {author} {\bibfnamefont {W.~H.}\ \bibnamefont
  {Press}},\ }\href {\doibase 10.1086/163336} {\bibfield  {journal} {\bibinfo
  {journal} {Astrophys. J.}\ }\textbf {\bibinfo {volume} {294}},\ \bibinfo
  {pages} {663} (\bibinfo {year} {1985})}\BibitemShut {NoStop}%
\bibitem [{\citenamefont {Faulkner}\ and\ \citenamefont
  {Gilliland}(1985)}]{Faulkner:1985rm}%
  \BibitemOpen
  \bibfield  {author} {\bibinfo {author} {\bibfnamefont {J.}~\bibnamefont
  {Faulkner}}\ and\ \bibinfo {author} {\bibfnamefont {R.~L.}\ \bibnamefont
  {Gilliland}},\ }\href {\doibase 10.1086/163766} {\bibfield  {journal}
  {\bibinfo  {journal} {Astrophys. J.}\ }\textbf {\bibinfo {volume} {299}},\
  \bibinfo {pages} {994} (\bibinfo {year} {1985})}\BibitemShut {NoStop}%
\bibitem [{\citenamefont {Gould}\ and\ \citenamefont
  {Raffelt}(1990)}]{Gould:1989hm}%
  \BibitemOpen
  \bibfield  {author} {\bibinfo {author} {\bibfnamefont {A.}~\bibnamefont
  {Gould}}\ and\ \bibinfo {author} {\bibfnamefont {G.}~\bibnamefont
  {Raffelt}},\ }\href {\doibase 10.1086/168568} {\bibfield  {journal} {\bibinfo
   {journal} {Astrophys. J.}\ }\textbf {\bibinfo {volume} {352}},\ \bibinfo
  {pages} {654} (\bibinfo {year} {1990})}\BibitemShut {NoStop}%
\bibitem [{\citenamefont {Lopes}\ \emph
  {et~al.}(2002{\natexlab{a}})\citenamefont {Lopes}, \citenamefont {Silk},\
  and\ \citenamefont {Hansen}}]{Lopes:2001ra}%
  \BibitemOpen
  \bibfield  {author} {\bibinfo {author} {\bibfnamefont {I.~P.}\ \bibnamefont
  {Lopes}}, \bibinfo {author} {\bibfnamefont {J.}~\bibnamefont {Silk}}, \ and\
  \bibinfo {author} {\bibfnamefont {S.~H.}\ \bibnamefont {Hansen}},\ }\href
  {\doibase 10.1046/j.1365-8711.2002.05238.x} {\bibfield  {journal} {\bibinfo
  {journal} {Mon. Not. Roy. Astron. Soc.}\ }\textbf {\bibinfo {volume} {331}},\
  \bibinfo {pages} {361} (\bibinfo {year} {2002}{\natexlab{a}})},\ \Eprint
  {http://arxiv.org/abs/astro-ph/0111530} {arXiv:astro-ph/0111530} \BibitemShut
  {NoStop}%
\bibitem [{\citenamefont {Lopes}\ \emph
  {et~al.}(2002{\natexlab{b}})\citenamefont {Lopes}, \citenamefont {Bertone},\
  and\ \citenamefont {Silk}}]{Lopes:2002gp}%
  \BibitemOpen
  \bibfield  {author} {\bibinfo {author} {\bibfnamefont {I.~P.}\ \bibnamefont
  {Lopes}}, \bibinfo {author} {\bibfnamefont {G.}~\bibnamefont {Bertone}}, \
  and\ \bibinfo {author} {\bibfnamefont {J.}~\bibnamefont {Silk}},\ }\href
  {\doibase 10.1046/j.1365-8711.2002.05835.x} {\bibfield  {journal} {\bibinfo
  {journal} {Mon. Not. Roy. Astron. Soc.}\ }\textbf {\bibinfo {volume} {337}},\
  \bibinfo {pages} {1179} (\bibinfo {year} {2002}{\natexlab{b}})},\ \Eprint
  {http://arxiv.org/abs/astro-ph/0205066} {arXiv:astro-ph/0205066} \BibitemShut
  {NoStop}%
\bibitem [{\citenamefont {Bottino}\ \emph {et~al.}(2002)\citenamefont
  {Bottino}, \citenamefont {Fiorentini}, \citenamefont {Fornengo},
  \citenamefont {Ricci}, \citenamefont {Scopel},\ and\ \citenamefont
  {Villante}}]{Bottino:2002pd}%
  \BibitemOpen
  \bibfield  {author} {\bibinfo {author} {\bibfnamefont {A.}~\bibnamefont
  {Bottino}}, \bibinfo {author} {\bibfnamefont {G.}~\bibnamefont {Fiorentini}},
  \bibinfo {author} {\bibfnamefont {N.}~\bibnamefont {Fornengo}}, \bibinfo
  {author} {\bibfnamefont {B.}~\bibnamefont {Ricci}}, \bibinfo {author}
  {\bibfnamefont {S.}~\bibnamefont {Scopel}}, \ and\ \bibinfo {author}
  {\bibfnamefont {F.~L.}\ \bibnamefont {Villante}},\ }\href {\doibase
  10.1103/PhysRevD.66.053005} {\bibfield  {journal} {\bibinfo  {journal} {Phys.
  Rev. D}\ }\textbf {\bibinfo {volume} {66}},\ \bibinfo {pages} {053005}
  (\bibinfo {year} {2002})},\ \Eprint {http://arxiv.org/abs/hep-ph/0206211}
  {arXiv:hep-ph/0206211} \BibitemShut {NoStop}%
\bibitem [{\citenamefont {Lopes}\ and\ \citenamefont
  {Silk}(2012)}]{Lopes:2012af}%
  \BibitemOpen
  \bibfield  {author} {\bibinfo {author} {\bibfnamefont {I.}~\bibnamefont
  {Lopes}}\ and\ \bibinfo {author} {\bibfnamefont {J.}~\bibnamefont {Silk}},\
  }\href {\doibase 10.1088/0004-637X/757/2/130} {\bibfield  {journal} {\bibinfo
   {journal} {Astrophys. J.}\ }\textbf {\bibinfo {volume} {757}},\ \bibinfo
  {pages} {130} (\bibinfo {year} {2012})},\ \Eprint
  {http://arxiv.org/abs/1209.3631} {arXiv:1209.3631 [astro-ph.SR]} \BibitemShut
  {NoStop}%
\bibitem [{\citenamefont {Lopes}\ \emph
  {et~al.}(2014{\natexlab{a}})\citenamefont {Lopes}, \citenamefont {Kadota},\
  and\ \citenamefont {Silk}}]{Lopes:2013xua}%
  \BibitemOpen
  \bibfield  {author} {\bibinfo {author} {\bibfnamefont {I.}~\bibnamefont
  {Lopes}}, \bibinfo {author} {\bibfnamefont {K.}~\bibnamefont {Kadota}}, \
  and\ \bibinfo {author} {\bibfnamefont {J.}~\bibnamefont {Silk}},\ }\href
  {\doibase 10.1088/2041-8205/780/2/L15} {\bibfield  {journal} {\bibinfo
  {journal} {Astrophys. J. Lett.}\ }\textbf {\bibinfo {volume} {780}},\
  \bibinfo {pages} {L15} (\bibinfo {year} {2014}{\natexlab{a}})},\ \Eprint
  {http://arxiv.org/abs/1310.0673} {arXiv:1310.0673 [astro-ph.SR]} \BibitemShut
  {NoStop}%
\bibitem [{\citenamefont {Lopes}\ \emph
  {et~al.}(2014{\natexlab{b}})\citenamefont {Lopes}, \citenamefont {Panci},\
  and\ \citenamefont {Silk}}]{Lopes:2014aoa}%
  \BibitemOpen
  \bibfield  {author} {\bibinfo {author} {\bibfnamefont {I.}~\bibnamefont
  {Lopes}}, \bibinfo {author} {\bibfnamefont {P.}~\bibnamefont {Panci}}, \ and\
  \bibinfo {author} {\bibfnamefont {J.}~\bibnamefont {Silk}},\ }\href {\doibase
  10.1088/0004-637X/795/2/162} {\bibfield  {journal} {\bibinfo  {journal}
  {Astrophys. J.}\ }\textbf {\bibinfo {volume} {795}},\ \bibinfo {pages} {162}
  (\bibinfo {year} {2014}{\natexlab{b}})},\ \Eprint
  {http://arxiv.org/abs/1402.0682} {arXiv:1402.0682 [astro-ph.SR]} \BibitemShut
  {NoStop}%
\bibitem [{\citenamefont {Geytenbeek}\ \emph {et~al.}(2017)\citenamefont
  {Geytenbeek}, \citenamefont {Rao}, \citenamefont {Scott}, \citenamefont
  {Serenelli}, \citenamefont {Vincent}, \citenamefont {White},\ and\
  \citenamefont {Williams}}]{Geytenbeek:2016nfg}%
  \BibitemOpen
  \bibfield  {author} {\bibinfo {author} {\bibfnamefont {B.}~\bibnamefont
  {Geytenbeek}}, \bibinfo {author} {\bibfnamefont {S.}~\bibnamefont {Rao}},
  \bibinfo {author} {\bibfnamefont {P.}~\bibnamefont {Scott}}, \bibinfo
  {author} {\bibfnamefont {A.}~\bibnamefont {Serenelli}}, \bibinfo {author}
  {\bibfnamefont {A.~C.}\ \bibnamefont {Vincent}}, \bibinfo {author}
  {\bibfnamefont {M.}~\bibnamefont {White}}, \ and\ \bibinfo {author}
  {\bibfnamefont {A.~G.}\ \bibnamefont {Williams}},\ }\href {\doibase
  10.1088/1475-7516/2017/03/029} {\bibfield  {journal} {\bibinfo  {journal}
  {JCAP}\ }\textbf {\bibinfo {volume} {03}},\ \bibinfo {pages} {029} (\bibinfo
  {year} {2017})},\ \Eprint {http://arxiv.org/abs/1610.06737} {arXiv:1610.06737
  [hep-ph]} \BibitemShut {NoStop}%
\bibitem [{\citenamefont {Frandsen}\ and\ \citenamefont
  {Sarkar}(2010)}]{Frandsen:2010yj}%
  \BibitemOpen
  \bibfield  {author} {\bibinfo {author} {\bibfnamefont {M.~T.}\ \bibnamefont
  {Frandsen}}\ and\ \bibinfo {author} {\bibfnamefont {S.}~\bibnamefont
  {Sarkar}},\ }\href {\doibase 10.1103/PhysRevLett.105.011301} {\bibfield
  {journal} {\bibinfo  {journal} {Phys. Rev. Lett.}\ }\textbf {\bibinfo
  {volume} {105}},\ \bibinfo {pages} {011301} (\bibinfo {year} {2010})},\
  \Eprint {http://arxiv.org/abs/1003.4505} {arXiv:1003.4505 [hep-ph]}
  \BibitemShut {NoStop}%
\bibitem [{\citenamefont {Cumberbatch}\ \emph {et~al.}(2010)\citenamefont
  {Cumberbatch}, \citenamefont {Guzik}, \citenamefont {Silk}, \citenamefont
  {Watson},\ and\ \citenamefont {West}}]{Cumberbatch:2010hh}%
  \BibitemOpen
  \bibfield  {author} {\bibinfo {author} {\bibfnamefont {D.~T.}\ \bibnamefont
  {Cumberbatch}}, \bibinfo {author} {\bibfnamefont {J.~A.}\ \bibnamefont
  {Guzik}}, \bibinfo {author} {\bibfnamefont {J.}~\bibnamefont {Silk}},
  \bibinfo {author} {\bibfnamefont {L.~S.}\ \bibnamefont {Watson}}, \ and\
  \bibinfo {author} {\bibfnamefont {S.~M.}\ \bibnamefont {West}},\ }\href
  {\doibase 10.1103/PhysRevD.82.103503} {\bibfield  {journal} {\bibinfo
  {journal} {Phys. Rev. D}\ }\textbf {\bibinfo {volume} {82}},\ \bibinfo
  {pages} {103503} (\bibinfo {year} {2010})},\ \Eprint
  {http://arxiv.org/abs/1005.5102} {arXiv:1005.5102 [astro-ph.SR]} \BibitemShut
  {NoStop}%
\bibitem [{\citenamefont {Taoso}\ \emph {et~al.}(2010)\citenamefont {Taoso},
  \citenamefont {Iocco}, \citenamefont {Meynet}, \citenamefont {Bertone},\ and\
  \citenamefont {Eggenberger}}]{Taoso:2010tg}%
  \BibitemOpen
  \bibfield  {author} {\bibinfo {author} {\bibfnamefont {M.}~\bibnamefont
  {Taoso}}, \bibinfo {author} {\bibfnamefont {F.}~\bibnamefont {Iocco}},
  \bibinfo {author} {\bibfnamefont {G.}~\bibnamefont {Meynet}}, \bibinfo
  {author} {\bibfnamefont {G.}~\bibnamefont {Bertone}}, \ and\ \bibinfo
  {author} {\bibfnamefont {P.}~\bibnamefont {Eggenberger}},\ }\href {\doibase
  10.1103/PhysRevD.82.083509} {\bibfield  {journal} {\bibinfo  {journal} {Phys.
  Rev. D}\ }\textbf {\bibinfo {volume} {82}},\ \bibinfo {pages} {083509}
  (\bibinfo {year} {2010})},\ \Eprint {http://arxiv.org/abs/1005.5711}
  {arXiv:1005.5711 [astro-ph.CO]} \BibitemShut {NoStop}%
\bibitem [{\citenamefont {Vincent}\ \emph
  {et~al.}(2015{\natexlab{a}})\citenamefont {Vincent}, \citenamefont {Scott},\
  and\ \citenamefont {Serenelli}}]{Vincent:2014jia}%
  \BibitemOpen
  \bibfield  {author} {\bibinfo {author} {\bibfnamefont {A.~C.}\ \bibnamefont
  {Vincent}}, \bibinfo {author} {\bibfnamefont {P.}~\bibnamefont {Scott}}, \
  and\ \bibinfo {author} {\bibfnamefont {A.}~\bibnamefont {Serenelli}},\ }\href
  {\doibase 10.1103/PhysRevLett.114.081302} {\bibfield  {journal} {\bibinfo
  {journal} {Phys. Rev. Lett.}\ }\textbf {\bibinfo {volume} {114}},\ \bibinfo
  {pages} {081302} (\bibinfo {year} {2015}{\natexlab{a}})},\ \Eprint
  {http://arxiv.org/abs/1411.6626} {arXiv:1411.6626 [hep-ph]} \BibitemShut
  {NoStop}%
\bibitem [{\citenamefont {Vincent}\ \emph
  {et~al.}(2015{\natexlab{b}})\citenamefont {Vincent}, \citenamefont
  {Serenelli},\ and\ \citenamefont {Scott}}]{Vincent:2015gqa}%
  \BibitemOpen
  \bibfield  {author} {\bibinfo {author} {\bibfnamefont {A.~C.}\ \bibnamefont
  {Vincent}}, \bibinfo {author} {\bibfnamefont {A.}~\bibnamefont {Serenelli}},
  \ and\ \bibinfo {author} {\bibfnamefont {P.}~\bibnamefont {Scott}},\ }\href
  {\doibase 10.1088/1475-7516/2015/08/040} {\bibfield  {journal} {\bibinfo
  {journal} {JCAP}\ }\textbf {\bibinfo {volume} {08}},\ \bibinfo {pages} {040}
  (\bibinfo {year} {2015}{\natexlab{b}})},\ \Eprint
  {http://arxiv.org/abs/1504.04378} {arXiv:1504.04378 [hep-ph]} \BibitemShut
  {NoStop}%
\bibitem [{\citenamefont {Vincent}\ \emph {et~al.}(2016)\citenamefont
  {Vincent}, \citenamefont {Scott},\ and\ \citenamefont
  {Serenelli}}]{Vincent:2016dcp}%
  \BibitemOpen
  \bibfield  {author} {\bibinfo {author} {\bibfnamefont {A.~C.}\ \bibnamefont
  {Vincent}}, \bibinfo {author} {\bibfnamefont {P.}~\bibnamefont {Scott}}, \
  and\ \bibinfo {author} {\bibfnamefont {A.}~\bibnamefont {Serenelli}},\ }\href
  {\doibase 10.1088/1475-7516/2016/11/007} {\bibfield  {journal} {\bibinfo
  {journal} {JCAP}\ }\textbf {\bibinfo {volume} {11}},\ \bibinfo {pages} {007}
  (\bibinfo {year} {2016})},\ \Eprint {http://arxiv.org/abs/1605.06502}
  {arXiv:1605.06502 [hep-ph]} \BibitemShut {NoStop}%
\bibitem [{\citenamefont {{Paxton}}\ \emph {et~al.}(2011)\citenamefont
  {{Paxton}}, \citenamefont {{Bildsten}}, \citenamefont {{Dotter}},
  \citenamefont {{Herwig}}, \citenamefont {{Lesaffre}},\ and\ \citenamefont
  {{Timmes}}}]{2011ApJS..192....3P}%
  \BibitemOpen
  \bibfield  {author} {\bibinfo {author} {\bibfnamefont {B.}~\bibnamefont
  {{Paxton}}}, \bibinfo {author} {\bibfnamefont {L.}~\bibnamefont
  {{Bildsten}}}, \bibinfo {author} {\bibfnamefont {A.}~\bibnamefont
  {{Dotter}}}, \bibinfo {author} {\bibfnamefont {F.}~\bibnamefont {{Herwig}}},
  \bibinfo {author} {\bibfnamefont {P.}~\bibnamefont {{Lesaffre}}}, \ and\
  \bibinfo {author} {\bibfnamefont {F.}~\bibnamefont {{Timmes}}},\ }\href
  {\doibase 10.1088/0067-0049/192/1/3} {\bibfield  {journal} {\bibinfo
  {journal} {ApJS}\ }\textbf {\bibinfo {volume} {192}},\ \bibinfo {eid} {3}
  (\bibinfo {year} {2011})},\ \Eprint {http://arxiv.org/abs/1009.1622}
  {arXiv:1009.1622 [astro-ph.SR]} \BibitemShut {NoStop}%
\bibitem [{\citenamefont {{Paxton}}\ \emph {et~al.}(2013)\citenamefont
  {{Paxton}}, \citenamefont {{Cantiello}}, \citenamefont {{Arras}},
  \citenamefont {{Bildsten}}, \citenamefont {{Brown}}, \citenamefont
  {{Dotter}}, \citenamefont {{Mankovich}}, \citenamefont {{Montgomery}},
  \citenamefont {{Stello}}, \citenamefont {{Timmes}},\ and\ \citenamefont
  {{Townsend}}}]{2013ApJS..208....4P}%
  \BibitemOpen
  \bibfield  {author} {\bibinfo {author} {\bibfnamefont {B.}~\bibnamefont
  {{Paxton}}}, \bibinfo {author} {\bibfnamefont {M.}~\bibnamefont
  {{Cantiello}}}, \bibinfo {author} {\bibfnamefont {P.}~\bibnamefont
  {{Arras}}}, \bibinfo {author} {\bibfnamefont {L.}~\bibnamefont {{Bildsten}}},
  \bibinfo {author} {\bibfnamefont {E.~F.}\ \bibnamefont {{Brown}}}, \bibinfo
  {author} {\bibfnamefont {A.}~\bibnamefont {{Dotter}}}, \bibinfo {author}
  {\bibfnamefont {C.}~\bibnamefont {{Mankovich}}}, \bibinfo {author}
  {\bibfnamefont {M.~H.}\ \bibnamefont {{Montgomery}}}, \bibinfo {author}
  {\bibfnamefont {D.}~\bibnamefont {{Stello}}}, \bibinfo {author}
  {\bibfnamefont {F.~X.}\ \bibnamefont {{Timmes}}}, \ and\ \bibinfo {author}
  {\bibfnamefont {R.}~\bibnamefont {{Townsend}}},\ }\href {\doibase
  10.1088/0067-0049/208/1/4} {\bibfield  {journal} {\bibinfo  {journal} {ApJS}\
  }\textbf {\bibinfo {volume} {208}},\ \bibinfo {eid} {4} (\bibinfo {year}
  {2013})},\ \Eprint {http://arxiv.org/abs/1301.0319} {arXiv:1301.0319
  [astro-ph.SR]} \BibitemShut {NoStop}%
\bibitem [{\citenamefont {{Paxton}}\ \emph {et~al.}(2015)\citenamefont
  {{Paxton}}, \citenamefont {{Marchant}}, \citenamefont {{Schwab}},
  \citenamefont {{Bauer}}, \citenamefont {{Bildsten}}, \citenamefont
  {{Cantiello}}, \citenamefont {{Dessart}}, \citenamefont {{Farmer}},
  \citenamefont {{Hu}}, \citenamefont {{Langer}}, \citenamefont {{Townsend}},
  \citenamefont {{Townsley}},\ and\ \citenamefont
  {{Timmes}}}]{2015ApJS..220...15P}%
  \BibitemOpen
  \bibfield  {author} {\bibinfo {author} {\bibfnamefont {B.}~\bibnamefont
  {{Paxton}}}, \bibinfo {author} {\bibfnamefont {P.}~\bibnamefont
  {{Marchant}}}, \bibinfo {author} {\bibfnamefont {J.}~\bibnamefont
  {{Schwab}}}, \bibinfo {author} {\bibfnamefont {E.~B.}\ \bibnamefont
  {{Bauer}}}, \bibinfo {author} {\bibfnamefont {L.}~\bibnamefont {{Bildsten}}},
  \bibinfo {author} {\bibfnamefont {M.}~\bibnamefont {{Cantiello}}}, \bibinfo
  {author} {\bibfnamefont {L.}~\bibnamefont {{Dessart}}}, \bibinfo {author}
  {\bibfnamefont {R.}~\bibnamefont {{Farmer}}}, \bibinfo {author}
  {\bibfnamefont {H.}~\bibnamefont {{Hu}}}, \bibinfo {author} {\bibfnamefont
  {N.}~\bibnamefont {{Langer}}}, \bibinfo {author} {\bibfnamefont {R.~H.~D.}\
  \bibnamefont {{Townsend}}}, \bibinfo {author} {\bibfnamefont {D.~M.}\
  \bibnamefont {{Townsley}}}, \ and\ \bibinfo {author} {\bibfnamefont {F.~X.}\
  \bibnamefont {{Timmes}}},\ }\href {\doibase 10.1088/0067-0049/220/1/15}
  {\bibfield  {journal} {\bibinfo  {journal} {ApJS}\ }\textbf {\bibinfo
  {volume} {220}},\ \bibinfo {eid} {15} (\bibinfo {year} {2015})},\ \Eprint
  {http://arxiv.org/abs/1506.03146} {arXiv:1506.03146 [astro-ph.SR]}
  \BibitemShut {NoStop}%
\bibitem [{\citenamefont {{Paxton}}\ \emph {et~al.}(2018)\citenamefont
  {{Paxton}}, \citenamefont {{Schwab}}, \citenamefont {{Bauer}}, \citenamefont
  {{Bildsten}}, \citenamefont {{Blinnikov}}, \citenamefont {{Duffell}},
  \citenamefont {{Farmer}}, \citenamefont {{Goldberg}}, \citenamefont
  {{Marchant}}, \citenamefont {{Sorokina}}, \citenamefont {{Thoul}},
  \citenamefont {{Townsend}},\ and\ \citenamefont
  {{Timmes}}}]{2018ApJS..234...34P}%
  \BibitemOpen
  \bibfield  {author} {\bibinfo {author} {\bibfnamefont {B.}~\bibnamefont
  {{Paxton}}}, \bibinfo {author} {\bibfnamefont {J.}~\bibnamefont {{Schwab}}},
  \bibinfo {author} {\bibfnamefont {E.~B.}\ \bibnamefont {{Bauer}}}, \bibinfo
  {author} {\bibfnamefont {L.}~\bibnamefont {{Bildsten}}}, \bibinfo {author}
  {\bibfnamefont {S.}~\bibnamefont {{Blinnikov}}}, \bibinfo {author}
  {\bibfnamefont {P.}~\bibnamefont {{Duffell}}}, \bibinfo {author}
  {\bibfnamefont {R.}~\bibnamefont {{Farmer}}}, \bibinfo {author}
  {\bibfnamefont {J.~A.}\ \bibnamefont {{Goldberg}}}, \bibinfo {author}
  {\bibfnamefont {P.}~\bibnamefont {{Marchant}}}, \bibinfo {author}
  {\bibfnamefont {E.}~\bibnamefont {{Sorokina}}}, \bibinfo {author}
  {\bibfnamefont {A.}~\bibnamefont {{Thoul}}}, \bibinfo {author} {\bibfnamefont
  {R.~H.~D.}\ \bibnamefont {{Townsend}}}, \ and\ \bibinfo {author}
  {\bibfnamefont {F.~X.}\ \bibnamefont {{Timmes}}},\ }\href {\doibase
  10.3847/1538-4365/aaa5a8} {\bibfield  {journal} {\bibinfo  {journal} {ApJS}\
  }\textbf {\bibinfo {volume} {234}},\ \bibinfo {eid} {34} (\bibinfo {year}
  {2018})},\ \Eprint {http://arxiv.org/abs/1710.08424} {arXiv:1710.08424
  [astro-ph.SR]} \BibitemShut {NoStop}%
\bibitem [{\citenamefont {{Paxton}}\ \emph {et~al.}(2019)\citenamefont
  {{Paxton}}, \citenamefont {{Smolec}}, \citenamefont {{Schwab}}, \citenamefont
  {{Gautschy}}, \citenamefont {{Bildsten}}, \citenamefont {{Cantiello}},
  \citenamefont {{Dotter}}, \citenamefont {{Farmer}}, \citenamefont
  {{Goldberg}}, \citenamefont {{Jermyn}}, \citenamefont {{Kanbur}},
  \citenamefont {{Marchant}}, \citenamefont {{Thoul}}, \citenamefont
  {{Townsend}}, \citenamefont {{Wolf}}, \citenamefont {{Zhang}},\ and\
  \citenamefont {{Timmes}}}]{2019ApJS..243...10P}%
  \BibitemOpen
  \bibfield  {author} {\bibinfo {author} {\bibfnamefont {B.}~\bibnamefont
  {{Paxton}}}, \bibinfo {author} {\bibfnamefont {R.}~\bibnamefont {{Smolec}}},
  \bibinfo {author} {\bibfnamefont {J.}~\bibnamefont {{Schwab}}}, \bibinfo
  {author} {\bibfnamefont {A.}~\bibnamefont {{Gautschy}}}, \bibinfo {author}
  {\bibfnamefont {L.}~\bibnamefont {{Bildsten}}}, \bibinfo {author}
  {\bibfnamefont {M.}~\bibnamefont {{Cantiello}}}, \bibinfo {author}
  {\bibfnamefont {A.}~\bibnamefont {{Dotter}}}, \bibinfo {author}
  {\bibfnamefont {R.}~\bibnamefont {{Farmer}}}, \bibinfo {author}
  {\bibfnamefont {J.~A.}\ \bibnamefont {{Goldberg}}}, \bibinfo {author}
  {\bibfnamefont {A.~S.}\ \bibnamefont {{Jermyn}}}, \bibinfo {author}
  {\bibfnamefont {S.~M.}\ \bibnamefont {{Kanbur}}}, \bibinfo {author}
  {\bibfnamefont {P.}~\bibnamefont {{Marchant}}}, \bibinfo {author}
  {\bibfnamefont {A.}~\bibnamefont {{Thoul}}}, \bibinfo {author} {\bibfnamefont
  {R.~H.~D.}\ \bibnamefont {{Townsend}}}, \bibinfo {author} {\bibfnamefont
  {W.~M.}\ \bibnamefont {{Wolf}}}, \bibinfo {author} {\bibfnamefont
  {M.}~\bibnamefont {{Zhang}}}, \ and\ \bibinfo {author} {\bibfnamefont
  {F.~X.}\ \bibnamefont {{Timmes}}},\ }\href {\doibase
  10.3847/1538-4365/ab2241} {\bibfield  {journal} {\bibinfo  {journal} {ApJS}\
  }\textbf {\bibinfo {volume} {243}},\ \bibinfo {eid} {10} (\bibinfo {year}
  {2019})},\ \Eprint {http://arxiv.org/abs/1903.01426} {arXiv:1903.01426
  [astro-ph.SR]} \BibitemShut {NoStop}%
\bibitem [{\citenamefont {{Jermyn}}\ \emph {et~al.}(2023)\citenamefont
  {{Jermyn}}, \citenamefont {{Bauer}}, \citenamefont {{Schwab}}, \citenamefont
  {{Farmer}}, \citenamefont {{Ball}}, \citenamefont {{Bellinger}},
  \citenamefont {{Dotter}}, \citenamefont {{Joyce}}, \citenamefont
  {{Marchant}}, \citenamefont {{Mombarg}}, \citenamefont {{Wolf}},
  \citenamefont {{Sunny Wong}}, \citenamefont {{Cinquegrana}}, \citenamefont
  {{Farrell}}, \citenamefont {{Smolec}}, \citenamefont {{Thoul}}, \citenamefont
  {{Cantiello}}, \citenamefont {{Herwig}}, \citenamefont {{Toloza}},
  \citenamefont {{Bildsten}}, \citenamefont {{Townsend}},\ and\ \citenamefont
  {{Timmes}}}]{2023ApJS..265...15J}%
  \BibitemOpen
  \bibfield  {author} {\bibinfo {author} {\bibfnamefont {A.~S.}\ \bibnamefont
  {{Jermyn}}}, \bibinfo {author} {\bibfnamefont {E.~B.}\ \bibnamefont
  {{Bauer}}}, \bibinfo {author} {\bibfnamefont {J.}~\bibnamefont {{Schwab}}},
  \bibinfo {author} {\bibfnamefont {R.}~\bibnamefont {{Farmer}}}, \bibinfo
  {author} {\bibfnamefont {W.~H.}\ \bibnamefont {{Ball}}}, \bibinfo {author}
  {\bibfnamefont {E.~P.}\ \bibnamefont {{Bellinger}}}, \bibinfo {author}
  {\bibfnamefont {A.}~\bibnamefont {{Dotter}}}, \bibinfo {author}
  {\bibfnamefont {M.}~\bibnamefont {{Joyce}}}, \bibinfo {author} {\bibfnamefont
  {P.}~\bibnamefont {{Marchant}}}, \bibinfo {author} {\bibfnamefont {J.~S.~G.}\
  \bibnamefont {{Mombarg}}}, \bibinfo {author} {\bibfnamefont {W.~M.}\
  \bibnamefont {{Wolf}}}, \bibinfo {author} {\bibfnamefont {T.~L.}\
  \bibnamefont {{Sunny Wong}}}, \bibinfo {author} {\bibfnamefont {G.~C.}\
  \bibnamefont {{Cinquegrana}}}, \bibinfo {author} {\bibfnamefont
  {E.}~\bibnamefont {{Farrell}}}, \bibinfo {author} {\bibfnamefont
  {R.}~\bibnamefont {{Smolec}}}, \bibinfo {author} {\bibfnamefont
  {A.}~\bibnamefont {{Thoul}}}, \bibinfo {author} {\bibfnamefont
  {M.}~\bibnamefont {{Cantiello}}}, \bibinfo {author} {\bibfnamefont
  {F.}~\bibnamefont {{Herwig}}}, \bibinfo {author} {\bibfnamefont
  {O.}~\bibnamefont {{Toloza}}}, \bibinfo {author} {\bibfnamefont
  {L.}~\bibnamefont {{Bildsten}}}, \bibinfo {author} {\bibfnamefont {R.~H.~D.}\
  \bibnamefont {{Townsend}}}, \ and\ \bibinfo {author} {\bibfnamefont {F.~X.}\
  \bibnamefont {{Timmes}}},\ }\href {\doibase 10.3847/1538-4365/acae8d}
  {\bibfield  {journal} {\bibinfo  {journal} {ApJS}\ }\textbf {\bibinfo
  {volume} {265}},\ \bibinfo {eid} {15} (\bibinfo {year} {2023})},\ \Eprint
  {http://arxiv.org/abs/2208.03651} {arXiv:2208.03651 [astro-ph.SR]}
  \BibitemShut {NoStop}%
\bibitem [{\citenamefont {Beardsmore}\ and\ \citenamefont
  {Cull}(2001)}]{beardsmore2001crustal}%
  \BibitemOpen
  \bibfield  {author} {\bibinfo {author} {\bibfnamefont {G.~R.}\ \bibnamefont
  {Beardsmore}}\ and\ \bibinfo {author} {\bibfnamefont {J.~P.}\ \bibnamefont
  {Cull}},\ }\href@noop {} {\emph {\bibinfo {title} {Crustal heat flow: a guide
  to measurement and modelling}}}\ (\bibinfo  {publisher} {Cambridge university
  press},\ \bibinfo {year} {2001})\BibitemShut {NoStop}%
\bibitem [{\citenamefont {Pollack}\ \emph {et~al.}(1993)\citenamefont
  {Pollack}, \citenamefont {Hurter},\ and\ \citenamefont
  {Johnson}}]{pollack1993heat}%
  \BibitemOpen
  \bibfield  {author} {\bibinfo {author} {\bibfnamefont {H.~N.}\ \bibnamefont
  {Pollack}}, \bibinfo {author} {\bibfnamefont {S.~J.}\ \bibnamefont {Hurter}},
  \ and\ \bibinfo {author} {\bibfnamefont {J.~R.}\ \bibnamefont {Johnson}},\
  }\href {\doibase https://doi.org/10.1029/93RG01249} {\bibfield  {journal}
  {\bibinfo  {journal} {Reviews of Geophysics}\ }\textbf {\bibinfo {volume}
  {31}},\ \bibinfo {pages} {267} (\bibinfo {year} {1993})},\ \Eprint
  {http://arxiv.org/abs/https://agupubs.onlinelibrary.wiley.com/doi/pdf/10.1029/93RG01249}
  {https://agupubs.onlinelibrary.wiley.com/doi/pdf/10.1029/93RG01249}
  \BibitemShut {NoStop}%
\bibitem [{\citenamefont {Davies}\ and\ \citenamefont
  {Davies}(2010)}]{se-1-5-2010}%
  \BibitemOpen
  \bibfield  {author} {\bibinfo {author} {\bibfnamefont {J.~H.}\ \bibnamefont
  {Davies}}\ and\ \bibinfo {author} {\bibfnamefont {D.~R.}\ \bibnamefont
  {Davies}},\ }\href {\doibase 10.5194/se-1-5-2010} {\bibfield  {journal}
  {\bibinfo  {journal} {Solid Earth}\ }\textbf {\bibinfo {volume} {1}},\
  \bibinfo {pages} {5} (\bibinfo {year} {2010})}\BibitemShut {NoStop}%
\bibitem [{\citenamefont {{Global Heat Flow Data Assessment Group}}\ \emph
  {et~al.}(2024)\citenamefont {{Global Heat Flow Data Assessment Group}},
  \citenamefont {Fuchs}, \citenamefont {Neumann}, \citenamefont {Norden},
  \citenamefont {Balkan-Pazvantoglu}, \citenamefont {Elbarbary}, \citenamefont
  {Petrunin}, \citenamefont {Beardsmore}, \citenamefont {Harris}, \citenamefont
  {Negrete-Aranda}, \citenamefont {Poort}, \citenamefont {Verdoya},
  \citenamefont {Liu}, \citenamefont {Chambers}, \citenamefont
  {Fuentes-Bustillos}, \citenamefont {Sidagam}, \citenamefont {Matiz-Leon},
  \citenamefont {Bencharef}, \citenamefont {Mino}, \citenamefont {Khaled},
  \citenamefont {Verch}, \citenamefont {Berger}, \citenamefont {Chishti},
  \citenamefont {Dergunova}, \citenamefont {Liebing}, \citenamefont {Schulz},
  \citenamefont {Schuppe}, \citenamefont {Trepalova}, \citenamefont {Chiozzi},
  \citenamefont {Duque}, \citenamefont {Forster}, \citenamefont {Leveni},\ and\
  \citenamefont {Staal}}]{global_heat_flow_data_assessment_group_global_2024}%
  \BibitemOpen
  \bibfield  {author} {\bibinfo {author} {\bibnamefont {{Global Heat Flow Data
  Assessment Group}}}, \bibinfo {author} {\bibfnamefont {S.}~\bibnamefont
  {Fuchs}}, \bibinfo {author} {\bibfnamefont {F.}~\bibnamefont {Neumann}},
  \bibinfo {author} {\bibfnamefont {B.}~\bibnamefont {Norden}}, \bibinfo
  {author} {\bibfnamefont {E.}~\bibnamefont {Balkan-Pazvantoglu}}, \bibinfo
  {author} {\bibfnamefont {S.}~\bibnamefont {Elbarbary}}, \bibinfo {author}
  {\bibfnamefont {A.}~\bibnamefont {Petrunin}}, \bibinfo {author}
  {\bibfnamefont {G.}~\bibnamefont {Beardsmore}}, \bibinfo {author}
  {\bibfnamefont {R.}~\bibnamefont {Harris}}, \bibinfo {author} {\bibfnamefont
  {R.}~\bibnamefont {Negrete-Aranda}}, \bibinfo {author} {\bibfnamefont
  {J.}~\bibnamefont {Poort}}, \bibinfo {author} {\bibfnamefont
  {M.}~\bibnamefont {Verdoya}}, \bibinfo {author} {\bibfnamefont
  {S.}~\bibnamefont {Liu}}, \bibinfo {author} {\bibfnamefont {E.}~\bibnamefont
  {Chambers}}, \bibinfo {author} {\bibfnamefont {K.}~\bibnamefont
  {Fuentes-Bustillos}}, \bibinfo {author} {\bibfnamefont {E.~R.}\ \bibnamefont
  {Sidagam}}, \bibinfo {author} {\bibfnamefont {J.~C.}\ \bibnamefont
  {Matiz-Leon}}, \bibinfo {author} {\bibfnamefont {M.~H.}\ \bibnamefont
  {Bencharef}}, \bibinfo {author} {\bibfnamefont {B.~G.}\ \bibnamefont {Mino}},
  \bibinfo {author} {\bibfnamefont {M.~S.}\ \bibnamefont {Khaled}}, \bibinfo
  {author} {\bibfnamefont {D.}~\bibnamefont {Verch}}, \bibinfo {author}
  {\bibfnamefont {L.}~\bibnamefont {Berger}}, \bibinfo {author} {\bibfnamefont
  {S.~F.}\ \bibnamefont {Chishti}}, \bibinfo {author} {\bibfnamefont
  {V.}~\bibnamefont {Dergunova}}, \bibinfo {author} {\bibfnamefont
  {H.}~\bibnamefont {Liebing}}, \bibinfo {author} {\bibfnamefont
  {M.}~\bibnamefont {Schulz}}, \bibinfo {author} {\bibfnamefont
  {P.}~\bibnamefont {Schuppe}}, \bibinfo {author} {\bibfnamefont
  {Z.}~\bibnamefont {Trepalova}}, \bibinfo {author} {\bibfnamefont
  {P.}~\bibnamefont {Chiozzi}}, \bibinfo {author} {\bibfnamefont {M.~R.~A.}\
  \bibnamefont {Duque}}, \bibinfo {author} {\bibfnamefont {F.}~\bibnamefont
  {Forster}}, \bibinfo {author} {\bibfnamefont {M.}~\bibnamefont {Leveni}}, \
  and\ \bibinfo {author} {\bibfnamefont {T.}~\bibnamefont {Staal}},\ }\href
  {\doibase 10.5880/FIDGEO.2024.014} {\enquote {\bibinfo {title} {The {Global}
  {Heat} {Flow} {Database}: {Release} 2024},}\ } (\bibinfo {year}
  {2024})\BibitemShut {NoStop}%
\bibitem [{\citenamefont {Windhorst}\ \emph {et~al.}(2006)\citenamefont
  {Windhorst}, \citenamefont {Cohen}, \citenamefont {Jansen}, \citenamefont
  {Conselice},\ and\ \citenamefont {Yan}}]{Windhorst:2005as}%
  \BibitemOpen
  \bibfield  {author} {\bibinfo {author} {\bibfnamefont {R.~A.}\ \bibnamefont
  {Windhorst}}, \bibinfo {author} {\bibfnamefont {S.~H.}\ \bibnamefont
  {Cohen}}, \bibinfo {author} {\bibfnamefont {R.~A.}\ \bibnamefont {Jansen}},
  \bibinfo {author} {\bibfnamefont {C.}~\bibnamefont {Conselice}}, \ and\
  \bibinfo {author} {\bibfnamefont {H.-J.}\ \bibnamefont {Yan}},\ }\href
  {\doibase 10.1016/j.newar.2005.11.018} {\bibfield  {journal} {\bibinfo
  {journal} {New Astron. Rev.}\ }\textbf {\bibinfo {volume} {50}},\ \bibinfo
  {pages} {113} (\bibinfo {year} {2006})},\ \Eprint
  {http://arxiv.org/abs/astro-ph/0506253} {arXiv:astro-ph/0506253} \BibitemShut
  {NoStop}%
\bibitem [{\citenamefont {Abell}\ \emph {et~al.}(2009)\citenamefont {Abell}
  \emph {et~al.}}]{LSSTScience:2009jmu}%
  \BibitemOpen
  \bibfield  {author} {\bibinfo {author} {\bibfnamefont {P.~A.}\ \bibnamefont
  {Abell}} \emph {et~al.} (\bibinfo {collaboration} {LSST Science, LSST
  Project}),\ }\href@noop {} {\  (\bibinfo {year} {2009})},\ \Eprint
  {http://arxiv.org/abs/0912.0201} {arXiv:0912.0201 [astro-ph.IM]} \BibitemShut
  {NoStop}%
\bibitem [{\citenamefont {Wang}\ \emph {et~al.}(2023)\citenamefont {Wang} \emph
  {et~al.}}]{Wang:2022qov}%
  \BibitemOpen
  \bibfield  {author} {\bibinfo {author} {\bibfnamefont {K.~X.}\ \bibnamefont
  {Wang}} \emph {et~al.},\ }\href {\doibase 10.1093/mnras/stad1652} {\bibfield
  {journal} {\bibinfo  {journal} {Mon. Not. Roy. Astron. Soc.}\ }\textbf
  {\bibinfo {volume} {523}},\ \bibinfo {pages} {3874} (\bibinfo {year}
  {2023})},\ \Eprint {http://arxiv.org/abs/2204.13553} {arXiv:2204.13553
  [astro-ph.CO]} \BibitemShut {NoStop}%
\bibitem [{\citenamefont {Stauffer}\ \emph {et~al.}(2018)\citenamefont
  {Stauffer} \emph {et~al.}}]{stauffer2018science}%
  \BibitemOpen
  \bibfield  {author} {\bibinfo {author} {\bibfnamefont {J.}~\bibnamefont
  {Stauffer}} \emph {et~al.},\ }\href@noop {} {\enquote {\bibinfo {title} {The
  science advantage of a redder filter for wfirst},}\ } (\bibinfo {year}
  {2018}),\ \Eprint {http://arxiv.org/abs/1806.00554} {arXiv:1806.00554
  [astro-ph.GA]} \BibitemShut {NoStop}%
\bibitem [{\citenamefont {Bedin}\ \emph {et~al.}(2009)\citenamefont {Bedin},
  \citenamefont {Salaris}, \citenamefont {Piotto}, \citenamefont {Anderson},
  \citenamefont {King},\ and\ \citenamefont {Cassisi}}]{Bedin:2009it}%
  \BibitemOpen
  \bibfield  {author} {\bibinfo {author} {\bibfnamefont {L.~R.}\ \bibnamefont
  {Bedin}}, \bibinfo {author} {\bibfnamefont {M.}~\bibnamefont {Salaris}},
  \bibinfo {author} {\bibfnamefont {G.}~\bibnamefont {Piotto}}, \bibinfo
  {author} {\bibfnamefont {J.}~\bibnamefont {Anderson}}, \bibinfo {author}
  {\bibfnamefont {I.~R.}\ \bibnamefont {King}}, \ and\ \bibinfo {author}
  {\bibfnamefont {S.}~\bibnamefont {Cassisi}},\ }\href {\doibase
  10.1088/0004-637X/697/2/965} {\bibfield  {journal} {\bibinfo  {journal}
  {Astrophys. J.}\ }\textbf {\bibinfo {volume} {697}},\ \bibinfo {pages} {965}
  (\bibinfo {year} {2009})},\ \Eprint {http://arxiv.org/abs/0903.2839}
  {arXiv:0903.2839 [astro-ph.GA]} \BibitemShut {NoStop}%
\bibitem [{\citenamefont {Ghez}\ \emph {et~al.}(2005)\citenamefont {Ghez},
  \citenamefont {Salim}, \citenamefont {Hornstein}, \citenamefont {Tanner},
  \citenamefont {Morris}, \citenamefont {Becklin},\ and\ \citenamefont
  {Duchene}}]{Ghez:2003qj}%
  \BibitemOpen
  \bibfield  {author} {\bibinfo {author} {\bibfnamefont {A.~M.}\ \bibnamefont
  {Ghez}}, \bibinfo {author} {\bibfnamefont {S.}~\bibnamefont {Salim}},
  \bibinfo {author} {\bibfnamefont {S.~D.}\ \bibnamefont {Hornstein}}, \bibinfo
  {author} {\bibfnamefont {A.}~\bibnamefont {Tanner}}, \bibinfo {author}
  {\bibfnamefont {M.}~\bibnamefont {Morris}}, \bibinfo {author} {\bibfnamefont
  {E.~E.}\ \bibnamefont {Becklin}}, \ and\ \bibinfo {author} {\bibfnamefont
  {G.}~\bibnamefont {Duchene}},\ }\href {\doibase 10.1086/427175} {\bibfield
  {journal} {\bibinfo  {journal} {Astrophys. J.}\ }\textbf {\bibinfo {volume}
  {620}},\ \bibinfo {pages} {744} (\bibinfo {year} {2005})},\ \Eprint
  {http://arxiv.org/abs/astro-ph/0306130} {arXiv:astro-ph/0306130} \BibitemShut
  {NoStop}%
\bibitem [{\citenamefont {Martins}\ \emph {et~al.}(2008)\citenamefont
  {Martins}, \citenamefont {Gillessen}, \citenamefont {Eisenhauer},
  \citenamefont {Genzel}, \citenamefont {Ott},\ and\ \citenamefont
  {Trippe}}]{Martins:2007rv}%
  \BibitemOpen
  \bibfield  {author} {\bibinfo {author} {\bibfnamefont {F.}~\bibnamefont
  {Martins}}, \bibinfo {author} {\bibfnamefont {S.}~\bibnamefont {Gillessen}},
  \bibinfo {author} {\bibfnamefont {F.}~\bibnamefont {Eisenhauer}}, \bibinfo
  {author} {\bibfnamefont {R.}~\bibnamefont {Genzel}}, \bibinfo {author}
  {\bibfnamefont {T.}~\bibnamefont {Ott}}, \ and\ \bibinfo {author}
  {\bibfnamefont {S.}~\bibnamefont {Trippe}},\ }\href {\doibase 10.1086/526768}
  {\bibfield  {journal} {\bibinfo  {journal} {Astrophys. J. Lett.}\ }\textbf
  {\bibinfo {volume} {672}},\ \bibinfo {pages} {L119} (\bibinfo {year}
  {2008})},\ \Eprint {http://arxiv.org/abs/0711.3344} {arXiv:0711.3344
  [astro-ph]} \BibitemShut {NoStop}%
\bibitem [{\citenamefont {Ghez}\ \emph {et~al.}(2008)\citenamefont {Ghez} \emph
  {et~al.}}]{Ghez:2008ms}%
  \BibitemOpen
  \bibfield  {author} {\bibinfo {author} {\bibfnamefont {A.~M.}\ \bibnamefont
  {Ghez}} \emph {et~al.},\ }\href {\doibase 10.1086/592738} {\bibfield
  {journal} {\bibinfo  {journal} {Astrophys. J.}\ }\textbf {\bibinfo {volume}
  {689}},\ \bibinfo {pages} {1044} (\bibinfo {year} {2008})},\ \Eprint
  {http://arxiv.org/abs/0808.2870} {arXiv:0808.2870 [astro-ph]} \BibitemShut
  {NoStop}%
\bibitem [{\citenamefont {Gillessen}\ \emph {et~al.}(2009)\citenamefont
  {Gillessen}, \citenamefont {Eisenhauer}, \citenamefont {Trippe},
  \citenamefont {Alexander}, \citenamefont {Genzel}, \citenamefont {Martins},\
  and\ \citenamefont {Ott}}]{Gillessen:2008qv}%
  \BibitemOpen
  \bibfield  {author} {\bibinfo {author} {\bibfnamefont {S.}~\bibnamefont
  {Gillessen}}, \bibinfo {author} {\bibfnamefont {F.}~\bibnamefont
  {Eisenhauer}}, \bibinfo {author} {\bibfnamefont {S.}~\bibnamefont {Trippe}},
  \bibinfo {author} {\bibfnamefont {T.}~\bibnamefont {Alexander}}, \bibinfo
  {author} {\bibfnamefont {R.}~\bibnamefont {Genzel}}, \bibinfo {author}
  {\bibfnamefont {F.}~\bibnamefont {Martins}}, \ and\ \bibinfo {author}
  {\bibfnamefont {T.}~\bibnamefont {Ott}},\ }\href {\doibase
  10.1088/0004-637X/692/2/1075} {\bibfield  {journal} {\bibinfo  {journal}
  {Astrophys. J.}\ }\textbf {\bibinfo {volume} {692}},\ \bibinfo {pages} {1075}
  (\bibinfo {year} {2009})},\ \Eprint {http://arxiv.org/abs/0810.4674}
  {arXiv:0810.4674 [astro-ph]} \BibitemShut {NoStop}%
\bibitem [{\citenamefont {{Genzel}}\ \emph {et~al.}(2010)\citenamefont
  {{Genzel}}, \citenamefont {{Eisenhauer}},\ and\ \citenamefont
  {{Gillessen}}}]{2010RvMP...82.3121G}%
  \BibitemOpen
  \bibfield  {author} {\bibinfo {author} {\bibfnamefont {R.}~\bibnamefont
  {{Genzel}}}, \bibinfo {author} {\bibfnamefont {F.}~\bibnamefont
  {{Eisenhauer}}}, \ and\ \bibinfo {author} {\bibfnamefont {S.}~\bibnamefont
  {{Gillessen}}},\ }\href {\doibase 10.1103/RevModPhys.82.3121} {\bibfield
  {journal} {\bibinfo  {journal} {Reviews of Modern Physics}\ }\textbf
  {\bibinfo {volume} {82}},\ \bibinfo {pages} {3121} (\bibinfo {year}
  {2010})},\ \Eprint {http://arxiv.org/abs/1006.0064} {arXiv:1006.0064
  [astro-ph.GA]} \BibitemShut {NoStop}%
\bibitem [{\citenamefont {{Habibi}}\ \emph {et~al.}(2017)\citenamefont
  {{Habibi}}, \citenamefont {{Gillessen}}, \citenamefont {{Martins}},
  \citenamefont {{Eisenhauer}}, \citenamefont {{Plewa}}, \citenamefont
  {{Pfuhl}}, \citenamefont {{George}}, \citenamefont {{Dexter}}, \citenamefont
  {{Waisberg}}, \citenamefont {{Ott}}, \citenamefont {{von Fellenberg}},
  \citenamefont {{Baub{\"o}ck}}, \citenamefont {{Jimenez-Rosales}},\ and\
  \citenamefont {{Genzel}}}]{2017ApJ...847..120H}%
  \BibitemOpen
  \bibfield  {author} {\bibinfo {author} {\bibfnamefont {M.}~\bibnamefont
  {{Habibi}}}, \bibinfo {author} {\bibfnamefont {S.}~\bibnamefont
  {{Gillessen}}}, \bibinfo {author} {\bibfnamefont {F.}~\bibnamefont
  {{Martins}}}, \bibinfo {author} {\bibfnamefont {F.}~\bibnamefont
  {{Eisenhauer}}}, \bibinfo {author} {\bibfnamefont {P.~M.}\ \bibnamefont
  {{Plewa}}}, \bibinfo {author} {\bibfnamefont {O.}~\bibnamefont {{Pfuhl}}},
  \bibinfo {author} {\bibfnamefont {E.}~\bibnamefont {{George}}}, \bibinfo
  {author} {\bibfnamefont {J.}~\bibnamefont {{Dexter}}}, \bibinfo {author}
  {\bibfnamefont {I.}~\bibnamefont {{Waisberg}}}, \bibinfo {author}
  {\bibfnamefont {T.}~\bibnamefont {{Ott}}}, \bibinfo {author} {\bibfnamefont
  {S.}~\bibnamefont {{von Fellenberg}}}, \bibinfo {author} {\bibfnamefont
  {M.}~\bibnamefont {{Baub{\"o}ck}}}, \bibinfo {author} {\bibfnamefont
  {A.}~\bibnamefont {{Jimenez-Rosales}}}, \ and\ \bibinfo {author}
  {\bibfnamefont {R.}~\bibnamefont {{Genzel}}},\ }\href {\doibase
  10.3847/1538-4357/aa876f} {\bibfield  {journal} {\bibinfo  {journal} {\apj}\
  }\textbf {\bibinfo {volume} {847}},\ \bibinfo {eid} {120} (\bibinfo {year}
  {2017})},\ \Eprint {http://arxiv.org/abs/1708.06353} {arXiv:1708.06353
  [astro-ph.SR]} \BibitemShut {NoStop}%
\bibitem [{\citenamefont {Pei{\ss}ker}\ \emph {et~al.}(2020)\citenamefont
  {Pei{\ss}ker}, \citenamefont {Eckart}, \citenamefont {Zaja{\v{c} }ek},
  \citenamefont {Ali},\ and\ \citenamefont {Parsa}}]{Pei_ker_2020}%
  \BibitemOpen
  \bibfield  {author} {\bibinfo {author} {\bibfnamefont {F.}~\bibnamefont
  {Pei{\ss}ker}}, \bibinfo {author} {\bibfnamefont {A.}~\bibnamefont {Eckart}},
  \bibinfo {author} {\bibfnamefont {M.}~\bibnamefont {Zaja{\v{c} }ek}},
  \bibinfo {author} {\bibfnamefont {B.}~\bibnamefont {Ali}}, \ and\ \bibinfo
  {author} {\bibfnamefont {M.}~\bibnamefont {Parsa}},\ }\href {\doibase
  10.3847/1538-4357/ab9c1c} {\bibfield  {journal} {\bibinfo  {journal} {The
  Astrophysical Journal}\ }\textbf {\bibinfo {volume} {899}},\ \bibinfo {pages}
  {50} (\bibinfo {year} {2020})}\BibitemShut {NoStop}%
\bibitem [{\citenamefont {Collar}(2018)}]{Collar:2018ydf}%
  \BibitemOpen
  \bibfield  {author} {\bibinfo {author} {\bibfnamefont {J.~I.}\ \bibnamefont
  {Collar}},\ }\href {\doibase 10.1103/PhysRevD.98.023005} {\bibfield
  {journal} {\bibinfo  {journal} {Phys. Rev. D}\ }\textbf {\bibinfo {volume}
  {98}},\ \bibinfo {pages} {023005} (\bibinfo {year} {2018})},\ \Eprint
  {http://arxiv.org/abs/1805.02646} {arXiv:1805.02646 [astro-ph.CO]}
  \BibitemShut {NoStop}%
\bibitem [{\citenamefont {Angloher}\ \emph {et~al.}(2022)\citenamefont
  {Angloher} \emph {et~al.}}]{CRESST:2022dtl}%
  \BibitemOpen
  \bibfield  {author} {\bibinfo {author} {\bibfnamefont {G.}~\bibnamefont
  {Angloher}} \emph {et~al.} (\bibinfo {collaboration} {CRESST}),\ }\href
  {\doibase 10.1103/PhysRevD.106.092008} {\bibfield  {journal} {\bibinfo
  {journal} {Phys. Rev. D}\ }\textbf {\bibinfo {volume} {106}},\ \bibinfo
  {pages} {092008} (\bibinfo {year} {2022})},\ \Eprint
  {http://arxiv.org/abs/2207.07640} {arXiv:2207.07640 [astro-ph.CO]}
  \BibitemShut {NoStop}%
\bibitem [{\citenamefont {Amole}\ \emph {et~al.}(2019)\citenamefont {Amole}
  \emph {et~al.}}]{PICO:2019vsc}%
  \BibitemOpen
  \bibfield  {author} {\bibinfo {author} {\bibfnamefont {C.}~\bibnamefont
  {Amole}} \emph {et~al.} (\bibinfo {collaboration} {PICO}),\ }\href {\doibase
  10.1103/PhysRevD.100.022001} {\bibfield  {journal} {\bibinfo  {journal}
  {Phys. Rev. D}\ }\textbf {\bibinfo {volume} {100}},\ \bibinfo {pages}
  {022001} (\bibinfo {year} {2019})},\ \Eprint
  {http://arxiv.org/abs/1902.04031} {arXiv:1902.04031 [astro-ph.CO]}
  \BibitemShut {NoStop}%
\bibitem [{\citenamefont {Aalbers}\ \emph {et~al.}(2023)\citenamefont {Aalbers}
  \emph {et~al.}}]{LZ:2022lsv}%
  \BibitemOpen
  \bibfield  {author} {\bibinfo {author} {\bibfnamefont {J.}~\bibnamefont
  {Aalbers}} \emph {et~al.} (\bibinfo {collaboration} {LZ}),\ }\href {\doibase
  10.1103/PhysRevLett.131.041002} {\bibfield  {journal} {\bibinfo  {journal}
  {Phys. Rev. Lett.}\ }\textbf {\bibinfo {volume} {131}},\ \bibinfo {pages}
  {041002} (\bibinfo {year} {2023})},\ \Eprint
  {http://arxiv.org/abs/2207.03764} {arXiv:2207.03764 [hep-ex]} \BibitemShut
  {NoStop}%
\bibitem [{\citenamefont {Franco}(2023)}]{Franco:2023sjx}%
  \BibitemOpen
  \bibfield  {author} {\bibinfo {author} {\bibfnamefont {D.}~\bibnamefont
  {Franco}} (\bibinfo {collaboration} {DarkSide-50}),\ }in\ \href@noop {}
  {\emph {\bibinfo {booktitle} {{57th Rencontres de Moriond on Electroweak
  Interactions and Unified Theories}}}}\ (\bibinfo {year} {2023})\ \Eprint
  {http://arxiv.org/abs/2306.12151} {arXiv:2306.12151 [hep-ex]} \BibitemShut
  {NoStop}%
\bibitem [{\citenamefont {Abe}\ \emph {et~al.}(2023)\citenamefont {Abe} \emph
  {et~al.}}]{Super-Kamiokande:2022ncz}%
  \BibitemOpen
  \bibfield  {author} {\bibinfo {author} {\bibfnamefont {K.}~\bibnamefont
  {Abe}} \emph {et~al.} (\bibinfo {collaboration} {Super-Kamiokande}),\ }\href
  {\doibase 10.1103/PhysRevLett.130.031802} {\bibfield  {journal} {\bibinfo
  {journal} {Phys. Rev. Lett.}\ }\textbf {\bibinfo {volume} {130}},\ \bibinfo
  {pages} {031802} (\bibinfo {year} {2023})},\ \bibinfo {note} {[Erratum:
  Phys.Rev.Lett. 131, 159903 (2023)]},\ \Eprint
  {http://arxiv.org/abs/2209.14968} {arXiv:2209.14968 [hep-ex]} \BibitemShut
  {NoStop}%
\bibitem [{\citenamefont {Bringmann}\ and\ \citenamefont
  {Pospelov}(2019)}]{Bringmann:2018cvk}%
  \BibitemOpen
  \bibfield  {author} {\bibinfo {author} {\bibfnamefont {T.}~\bibnamefont
  {Bringmann}}\ and\ \bibinfo {author} {\bibfnamefont {M.}~\bibnamefont
  {Pospelov}},\ }\href {\doibase 10.1103/PhysRevLett.122.171801} {\bibfield
  {journal} {\bibinfo  {journal} {Phys. Rev. Lett.}\ }\textbf {\bibinfo
  {volume} {122}},\ \bibinfo {pages} {171801} (\bibinfo {year} {2019})},\
  \Eprint {http://arxiv.org/abs/1810.10543} {arXiv:1810.10543 [hep-ph]}
  \BibitemShut {NoStop}%
\bibitem [{\citenamefont {Maity}\ and\ \citenamefont
  {Laha}(2024)}]{Maity:2022exk}%
  \BibitemOpen
  \bibfield  {author} {\bibinfo {author} {\bibfnamefont {T.~N.}\ \bibnamefont
  {Maity}}\ and\ \bibinfo {author} {\bibfnamefont {R.}~\bibnamefont {Laha}},\
  }\href {\doibase 10.1140/epjc/s10052-024-12464-8} {\bibfield  {journal}
  {\bibinfo  {journal} {Eur. Phys. J. C}\ }\textbf {\bibinfo {volume} {84}},\
  \bibinfo {pages} {117} (\bibinfo {year} {2024})},\ \Eprint
  {http://arxiv.org/abs/2210.01815} {arXiv:2210.01815 [hep-ph]} \BibitemShut
  {NoStop}%
\bibitem [{\citenamefont {Leane}\ and\ \citenamefont
  {Smirnov}(2023{\natexlab{b}})}]{Leane:2022hkk}%
  \BibitemOpen
  \bibfield  {author} {\bibinfo {author} {\bibfnamefont {R.~K.}\ \bibnamefont
  {Leane}}\ and\ \bibinfo {author} {\bibfnamefont {J.}~\bibnamefont
  {Smirnov}},\ }\href {\doibase 10.1088/1475-7516/2023/10/057} {\bibfield
  {journal} {\bibinfo  {journal} {JCAP}\ }\textbf {\bibinfo {volume} {10}},\
  \bibinfo {pages} {057} (\bibinfo {year} {2023}{\natexlab{b}})},\ \Eprint
  {http://arxiv.org/abs/2209.09834} {arXiv:2209.09834 [hep-ph]} \BibitemShut
  {NoStop}%
\bibitem [{\citenamefont {Garani}\ and\ \citenamefont
  {Palomares-Ruiz}(2022)}]{Garani:2021feo}%
  \BibitemOpen
  \bibfield  {author} {\bibinfo {author} {\bibfnamefont {R.}~\bibnamefont
  {Garani}}\ and\ \bibinfo {author} {\bibfnamefont {S.}~\bibnamefont
  {Palomares-Ruiz}},\ }\href {\doibase 10.1088/1475-7516/2022/05/042}
  {\bibfield  {journal} {\bibinfo  {journal} {JCAP}\ }\textbf {\bibinfo
  {volume} {05}},\ \bibinfo {pages} {042} (\bibinfo {year} {2022})},\ \Eprint
  {http://arxiv.org/abs/2104.12757} {arXiv:2104.12757 [hep-ph]} \BibitemShut
  {NoStop}%
\bibitem [{\citenamefont {Acevedo}\ \emph
  {et~al.}(2024{\natexlab{d}})\citenamefont {Acevedo}, \citenamefont {Leane},\
  and\ \citenamefont {Smirnov}}]{Acevedo:2023owd}%
  \BibitemOpen
  \bibfield  {author} {\bibinfo {author} {\bibfnamefont {J.~F.}\ \bibnamefont
  {Acevedo}}, \bibinfo {author} {\bibfnamefont {R.~K.}\ \bibnamefont {Leane}},
  \ and\ \bibinfo {author} {\bibfnamefont {J.}~\bibnamefont {Smirnov}},\ }\href
  {\doibase 10.1088/1475-7516/2024/04/038} {\bibfield  {journal} {\bibinfo
  {journal} {JCAP}\ }\textbf {\bibinfo {volume} {04}},\ \bibinfo {pages} {038}
  (\bibinfo {year} {2024}{\natexlab{d}})},\ \Eprint
  {http://arxiv.org/abs/2303.01516} {arXiv:2303.01516 [hep-ph]} \BibitemShut
  {NoStop}%
\bibitem [{\citenamefont {Perryman}\ \emph {et~al.}(2014)\citenamefont
  {Perryman}, \citenamefont {Hartman}, \citenamefont {Bakos},\ and\
  \citenamefont {Lindegren}}]{Perryman_2014}%
  \BibitemOpen
  \bibfield  {author} {\bibinfo {author} {\bibfnamefont {M.}~\bibnamefont
  {Perryman}}, \bibinfo {author} {\bibfnamefont {J.}~\bibnamefont {Hartman}},
  \bibinfo {author} {\bibfnamefont {G.~A.}\ \bibnamefont {Bakos}}, \ and\
  \bibinfo {author} {\bibfnamefont {L.}~\bibnamefont {Lindegren}},\ }\href
  {\doibase 10.1088/0004-637x/797/1/14} {\bibfield  {journal} {\bibinfo
  {journal} {The Astrophysical Journal}\ }\textbf {\bibinfo {volume} {797}},\
  \bibinfo {pages} {14} (\bibinfo {year} {2014})}\BibitemShut {NoStop}%
\bibitem [{\citenamefont {Sj\"ostrand}\ \emph {et~al.}(2015)\citenamefont
  {Sj\"ostrand}, \citenamefont {Ask}, \citenamefont {Christiansen},
  \citenamefont {Corke}, \citenamefont {Desai}, \citenamefont {Ilten},
  \citenamefont {Mrenna}, \citenamefont {Prestel}, \citenamefont {Rasmussen},\
  and\ \citenamefont {Skands}}]{Sjostrand:2014zea}%
  \BibitemOpen
  \bibfield  {author} {\bibinfo {author} {\bibfnamefont {T.}~\bibnamefont
  {Sj\"ostrand}}, \bibinfo {author} {\bibfnamefont {S.}~\bibnamefont {Ask}},
  \bibinfo {author} {\bibfnamefont {J.~R.}\ \bibnamefont {Christiansen}},
  \bibinfo {author} {\bibfnamefont {R.}~\bibnamefont {Corke}}, \bibinfo
  {author} {\bibfnamefont {N.}~\bibnamefont {Desai}}, \bibinfo {author}
  {\bibfnamefont {P.}~\bibnamefont {Ilten}}, \bibinfo {author} {\bibfnamefont
  {S.}~\bibnamefont {Mrenna}}, \bibinfo {author} {\bibfnamefont
  {S.}~\bibnamefont {Prestel}}, \bibinfo {author} {\bibfnamefont {C.~O.}\
  \bibnamefont {Rasmussen}}, \ and\ \bibinfo {author} {\bibfnamefont {P.~Z.}\
  \bibnamefont {Skands}},\ }\href {\doibase 10.1016/j.cpc.2015.01.024}
  {\bibfield  {journal} {\bibinfo  {journal} {Comput. Phys. Commun.}\ }\textbf
  {\bibinfo {volume} {191}},\ \bibinfo {pages} {159} (\bibinfo {year}
  {2015})},\ \Eprint {http://arxiv.org/abs/1410.3012} {arXiv:1410.3012
  [hep-ph]} \BibitemShut {NoStop}%
\bibitem [{\citenamefont {Tang}\ \emph {et~al.}(2018)\citenamefont {Tang},
  \citenamefont {Ng}, \citenamefont {Linden}, \citenamefont {Zhou},
  \citenamefont {Beacom},\ and\ \citenamefont {Peter}}]{Tang:2018wqp}%
  \BibitemOpen
  \bibfield  {author} {\bibinfo {author} {\bibfnamefont {Q.-W.}\ \bibnamefont
  {Tang}}, \bibinfo {author} {\bibfnamefont {K.~C.~Y.}\ \bibnamefont {Ng}},
  \bibinfo {author} {\bibfnamefont {T.}~\bibnamefont {Linden}}, \bibinfo
  {author} {\bibfnamefont {B.}~\bibnamefont {Zhou}}, \bibinfo {author}
  {\bibfnamefont {J.~F.}\ \bibnamefont {Beacom}}, \ and\ \bibinfo {author}
  {\bibfnamefont {A.~H.~G.}\ \bibnamefont {Peter}},\ }\href {\doibase
  10.1103/PhysRevD.98.063019} {\bibfield  {journal} {\bibinfo  {journal} {Phys.
  Rev. D}\ }\textbf {\bibinfo {volume} {98}},\ \bibinfo {pages} {063019}
  (\bibinfo {year} {2018})},\ \Eprint {http://arxiv.org/abs/1804.06846}
  {arXiv:1804.06846 [astro-ph.HE]} \BibitemShut {NoStop}%
\bibitem [{\citenamefont {Malyshev}\ \emph {et~al.}(2015)\citenamefont
  {Malyshev}, \citenamefont {Chernyakova}, \citenamefont {Neronov},\ and\
  \citenamefont {Walter}}]{Malyshev:2015hqa}%
  \BibitemOpen
  \bibfield  {author} {\bibinfo {author} {\bibfnamefont {D.}~\bibnamefont
  {Malyshev}}, \bibinfo {author} {\bibfnamefont {M.}~\bibnamefont
  {Chernyakova}}, \bibinfo {author} {\bibfnamefont {A.}~\bibnamefont
  {Neronov}}, \ and\ \bibinfo {author} {\bibfnamefont {R.}~\bibnamefont
  {Walter}},\ }\href {\doibase 10.1051/0004-6361/201526120} {\bibfield
  {journal} {\bibinfo  {journal} {Astron. Astrophys.}\ }\textbf {\bibinfo
  {volume} {582}},\ \bibinfo {pages} {A11} (\bibinfo {year} {2015})},\ \Eprint
  {http://arxiv.org/abs/1503.05120} {arXiv:1503.05120 [astro-ph.HE]}
  \BibitemShut {NoStop}%
\bibitem [{\citenamefont {Abdo}\ \emph {et~al.}(2011)\citenamefont {Abdo} \emph
  {et~al.}}]{Fermi-LAT:2011nwz}%
  \BibitemOpen
  \bibfield  {author} {\bibinfo {author} {\bibfnamefont {A.~A.}\ \bibnamefont
  {Abdo}} \emph {et~al.} (\bibinfo {collaboration} {Fermi-LAT}),\ }\href
  {\doibase 10.1088/0004-637X/734/2/116} {\bibfield  {journal} {\bibinfo
  {journal} {Astrophys. J.}\ }\textbf {\bibinfo {volume} {734}},\ \bibinfo
  {pages} {116} (\bibinfo {year} {2011})},\ \Eprint
  {http://arxiv.org/abs/1104.2093} {arXiv:1104.2093 [astro-ph.HE]} \BibitemShut
  {NoStop}%
\bibitem [{\citenamefont {Aharonian}(2009)}]{Aharonian:2009zk}%
  \BibitemOpen
  \bibfield  {author} {\bibinfo {author} {\bibfnamefont {F.}~\bibnamefont
  {Aharonian}} (\bibinfo {collaboration} {H.E.S.S.}),\ }\href {\doibase
  10.1051/0004-6361/200811569} {\bibfield  {journal} {\bibinfo  {journal}
  {Astron. Astrophys.}\ }\textbf {\bibinfo {volume} {503}},\ \bibinfo {pages}
  {817} (\bibinfo {year} {2009})},\ \Eprint {http://arxiv.org/abs/0906.1247}
  {arXiv:0906.1247 [astro-ph.GA]} \BibitemShut {NoStop}%
\bibitem [{\citenamefont {Maggio}(2021)}]{Maggio:2021pca}%
  \BibitemOpen
  \bibfield  {author} {\bibinfo {author} {\bibfnamefont {C.}~\bibnamefont
  {Maggio}},\ }\emph {\bibinfo {title} {{Indirect Search for WIMP Dark Matter
  with the MAGIC Telescopes}}},\ \href@noop {} {Ph.D. thesis},\ \bibinfo
  {school} {Barcelona, Autonoma U.} (\bibinfo {year} {2021})\BibitemShut
  {NoStop}%
\bibitem [{\citenamefont {Read}(2002)}]{Read:2002hq}%
  \BibitemOpen
  \bibfield  {author} {\bibinfo {author} {\bibfnamefont {A.~L.}\ \bibnamefont
  {Read}},\ }\href {\doibase 10.1088/0954-3899/28/10/313} {\bibfield  {journal}
  {\bibinfo  {journal} {J. Phys. G}\ }\textbf {\bibinfo {volume} {28}},\
  \bibinfo {pages} {2693} (\bibinfo {year} {2002})}\BibitemShut {NoStop}%
\bibitem [{\citenamefont {Cowan}\ \emph {et~al.}(2011)\citenamefont {Cowan},
  \citenamefont {Cranmer}, \citenamefont {Gross},\ and\ \citenamefont
  {Vitells}}]{Cowan:2010js}%
  \BibitemOpen
  \bibfield  {author} {\bibinfo {author} {\bibfnamefont {G.}~\bibnamefont
  {Cowan}}, \bibinfo {author} {\bibfnamefont {K.}~\bibnamefont {Cranmer}},
  \bibinfo {author} {\bibfnamefont {E.}~\bibnamefont {Gross}}, \ and\ \bibinfo
  {author} {\bibfnamefont {O.}~\bibnamefont {Vitells}},\ }\href {\doibase
  10.1140/epjc/s10052-011-1554-0} {\bibfield  {journal} {\bibinfo  {journal}
  {Eur. Phys. J. C}\ }\textbf {\bibinfo {volume} {71}},\ \bibinfo {pages}
  {1554} (\bibinfo {year} {2011})},\ \bibinfo {note} {[Erratum: Eur.Phys.J.C
  73, 2501 (2013)]},\ \Eprint {http://arxiv.org/abs/1007.1727} {arXiv:1007.1727
  [physics.data-an]} \BibitemShut {NoStop}%
\bibitem [{\citenamefont {Choi}\ \emph {et~al.}(2015)\citenamefont {Choi} \emph
  {et~al.}}]{Super-Kamiokande:2015xms}%
  \BibitemOpen
  \bibfield  {author} {\bibinfo {author} {\bibfnamefont {K.}~\bibnamefont
  {Choi}} \emph {et~al.} (\bibinfo {collaboration} {Super-Kamiokande}),\ }\href
  {\doibase 10.1103/PhysRevLett.114.141301} {\bibfield  {journal} {\bibinfo
  {journal} {Phys. Rev. Lett.}\ }\textbf {\bibinfo {volume} {114}},\ \bibinfo
  {pages} {141301} (\bibinfo {year} {2015})},\ \Eprint
  {http://arxiv.org/abs/1503.04858} {arXiv:1503.04858 [hep-ex]} \BibitemShut
  {NoStop}%
\bibitem [{\citenamefont {Abe}\ \emph {et~al.}(2011)\citenamefont {Abe} \emph
  {et~al.}}]{Abe:2011ts}%
  \BibitemOpen
  \bibfield  {author} {\bibinfo {author} {\bibfnamefont {K.}~\bibnamefont
  {Abe}} \emph {et~al.},\ }\href@noop {} {\  (\bibinfo {year} {2011})},\
  \Eprint {http://arxiv.org/abs/1109.3262} {arXiv:1109.3262 [hep-ex]}
  \BibitemShut {NoStop}%
\bibitem [{\citenamefont {Adrian-Martinez}\ \emph {et~al.}(2016)\citenamefont
  {Adrian-Martinez} \emph {et~al.}}]{ANTARES:2016xuh}%
  \BibitemOpen
  \bibfield  {author} {\bibinfo {author} {\bibfnamefont {S.}~\bibnamefont
  {Adrian-Martinez}} \emph {et~al.} (\bibinfo {collaboration} {ANTARES}),\
  }\href {\doibase 10.1016/j.physletb.2016.05.019} {\bibfield  {journal}
  {\bibinfo  {journal} {Phys. Lett. B}\ }\textbf {\bibinfo {volume} {759}},\
  \bibinfo {pages} {69} (\bibinfo {year} {2016})},\ \Eprint
  {http://arxiv.org/abs/1603.02228} {arXiv:1603.02228 [astro-ph.HE]}
  \BibitemShut {NoStop}%
\bibitem [{\citenamefont {Albert}\ \emph {et~al.}(2017)\citenamefont {Albert}
  \emph {et~al.}}]{ANTARES:2016bxz}%
  \BibitemOpen
  \bibfield  {author} {\bibinfo {author} {\bibfnamefont {A.}~\bibnamefont
  {Albert}} \emph {et~al.} (\bibinfo {collaboration} {ANTARES}),\ }\href
  {\doibase 10.1016/j.dark.2017.04.005} {\bibfield  {journal} {\bibinfo
  {journal} {Phys. Dark Univ.}\ }\textbf {\bibinfo {volume} {16}},\ \bibinfo
  {pages} {41} (\bibinfo {year} {2017})},\ \Eprint
  {http://arxiv.org/abs/1612.06792} {arXiv:1612.06792 [hep-ex]} \BibitemShut
  {NoStop}%
\bibitem [{\citenamefont {Aartsen}\ \emph {et~al.}(2016)\citenamefont {Aartsen}
  \emph {et~al.}}]{IceCube:2016yoy}%
  \BibitemOpen
  \bibfield  {author} {\bibinfo {author} {\bibfnamefont {M.~G.}\ \bibnamefont
  {Aartsen}} \emph {et~al.} (\bibinfo {collaboration} {IceCube}),\ }\href
  {\doibase 10.1088/1475-7516/2016/04/022} {\bibfield  {journal} {\bibinfo
  {journal} {JCAP}\ }\textbf {\bibinfo {volume} {04}},\ \bibinfo {pages} {022}
  (\bibinfo {year} {2016})},\ \Eprint {http://arxiv.org/abs/1601.00653}
  {arXiv:1601.00653 [hep-ph]} \BibitemShut {NoStop}%
\bibitem [{\citenamefont {Abbasi}\ \emph {et~al.}(2020)\citenamefont {Abbasi}
  \emph {et~al.}}]{IceCube:2020wxa}%
  \BibitemOpen
  \bibfield  {author} {\bibinfo {author} {\bibfnamefont {R.}~\bibnamefont
  {Abbasi}} \emph {et~al.} (\bibinfo {collaboration} {IceCube}),\ }\href
  {\doibase 10.22323/1.358.0541} {\bibfield  {journal} {\bibinfo  {journal}
  {PoS}\ }\textbf {\bibinfo {volume} {ICRC2019}},\ \bibinfo {pages} {541}
  (\bibinfo {year} {2020})},\ \Eprint {http://arxiv.org/abs/1908.07255}
  {arXiv:1908.07255 [astro-ph.HE]} \BibitemShut {NoStop}%
\bibitem [{\citenamefont {Aguilar}\ \emph {et~al.}(2021)\citenamefont {Aguilar}
  \emph {et~al.}}]{AMS:2021nhj}%
  \BibitemOpen
  \bibfield  {author} {\bibinfo {author} {\bibfnamefont {M.}~\bibnamefont
  {Aguilar}} \emph {et~al.} (\bibinfo {collaboration} {AMS}),\ }\href {\doibase
  10.1016/j.physrep.2020.09.003} {\bibfield  {journal} {\bibinfo  {journal}
  {Phys. Rept.}\ }\textbf {\bibinfo {volume} {894}},\ \bibinfo {pages} {1}
  (\bibinfo {year} {2021})}\BibitemShut {NoStop}%
\bibitem [{\citenamefont {Ambrosi}\ \emph {et~al.}(2017)\citenamefont {Ambrosi}
  \emph {et~al.}}]{DAMPE:2017fbg}%
  \BibitemOpen
  \bibfield  {author} {\bibinfo {author} {\bibfnamefont {G.}~\bibnamefont
  {Ambrosi}} \emph {et~al.} (\bibinfo {collaboration} {DAMPE}),\ }\href
  {\doibase 10.1038/nature24475} {\bibfield  {journal} {\bibinfo  {journal}
  {Nature}\ }\textbf {\bibinfo {volume} {552}},\ \bibinfo {pages} {63}
  (\bibinfo {year} {2017})},\ \Eprint {http://arxiv.org/abs/1711.10981}
  {arXiv:1711.10981 [astro-ph.HE]} \BibitemShut {NoStop}%
\bibitem [{\citenamefont {Ajello}\ \emph {et~al.}(2011)\citenamefont {Ajello}
  \emph {et~al.}}]{FermiLAT:2011ozd}%
  \BibitemOpen
  \bibfield  {author} {\bibinfo {author} {\bibfnamefont {M.}~\bibnamefont
  {Ajello}} \emph {et~al.} (\bibinfo {collaboration} {Fermi LAT}),\ }\href
  {\doibase 10.1103/PhysRevD.84.032007} {\bibfield  {journal} {\bibinfo
  {journal} {Phys. Rev. D}\ }\textbf {\bibinfo {volume} {84}},\ \bibinfo
  {pages} {032007} (\bibinfo {year} {2011})},\ \Eprint
  {http://arxiv.org/abs/1107.4272} {arXiv:1107.4272 [astro-ph.HE]} \BibitemShut
  {NoStop}%
\bibitem [{\citenamefont {Kawasaki}\ \emph {et~al.}(2005)\citenamefont
  {Kawasaki}, \citenamefont {Kohri},\ and\ \citenamefont
  {Moroi}}]{Kawasaki:2004yh}%
  \BibitemOpen
  \bibfield  {author} {\bibinfo {author} {\bibfnamefont {M.}~\bibnamefont
  {Kawasaki}}, \bibinfo {author} {\bibfnamefont {K.}~\bibnamefont {Kohri}}, \
  and\ \bibinfo {author} {\bibfnamefont {T.}~\bibnamefont {Moroi}},\ }\href
  {\doibase 10.1016/j.physletb.2005.08.045} {\bibfield  {journal} {\bibinfo
  {journal} {Phys. Lett. B}\ }\textbf {\bibinfo {volume} {625}},\ \bibinfo
  {pages} {7} (\bibinfo {year} {2005})},\ \Eprint
  {http://arxiv.org/abs/astro-ph/0402490} {arXiv:astro-ph/0402490} \BibitemShut
  {NoStop}%
\bibitem [{\citenamefont {Jedamzik}(2006)}]{Jedamzik:2006xz}%
  \BibitemOpen
  \bibfield  {author} {\bibinfo {author} {\bibfnamefont {K.}~\bibnamefont
  {Jedamzik}},\ }\href {\doibase 10.1103/PhysRevD.74.103509} {\bibfield
  {journal} {\bibinfo  {journal} {Phys. Rev. D}\ }\textbf {\bibinfo {volume}
  {74}},\ \bibinfo {pages} {103509} (\bibinfo {year} {2006})},\ \Eprint
  {http://arxiv.org/abs/hep-ph/0604251} {arXiv:hep-ph/0604251} \BibitemShut
  {NoStop}%
\bibitem [{\citenamefont {Pospelov}\ and\ \citenamefont
  {Pradler}(2010)}]{Pospelov:2010hj}%
  \BibitemOpen
  \bibfield  {author} {\bibinfo {author} {\bibfnamefont {M.}~\bibnamefont
  {Pospelov}}\ and\ \bibinfo {author} {\bibfnamefont {J.}~\bibnamefont
  {Pradler}},\ }\href {\doibase 10.1146/annurev.nucl.012809.104521} {\bibfield
  {journal} {\bibinfo  {journal} {Ann. Rev. Nucl. Part. Sci.}\ }\textbf
  {\bibinfo {volume} {60}},\ \bibinfo {pages} {539} (\bibinfo {year} {2010})},\
  \Eprint {http://arxiv.org/abs/1011.1054} {arXiv:1011.1054 [hep-ph]}
  \BibitemShut {NoStop}%
\end{thebibliography}%
\end{document}